\documentclass{pasj00}

\usepackage[OT2,T1]{fontenc}

\def\commenta{$^*$}
\def\commentb{$^\dagger$}
\def\commentc{$^\ddagger$}
\def\commentd{$^\S$}

\newcounter{author}
\def\authorcount#1#2{\refstepcounter{author}\label{#1}
                     \altaffiltext{\ref{#1}}{#2}}

\begin{document}
\SetRunningHead{T. Kato et al.}{Period Variations in SU UMa-Type Dwarf Novae VI}

\Received{201X/XX/XX}
\Accepted{201X/XX/XX}

\title{Survey of Period Variations of Superhumps in SU UMa-Type Dwarf Novae.
    VI: The Sixth Year (2013--2014)}

\author{Taichi~\textsc{Kato},\altaffilmark{\ref{affil:Kyoto}*}
        Pavol~A.~\textsc{Dubovsky},\altaffilmark{\ref{affil:Dubovsky}}
        Igor~\textsc{Kudzej},\altaffilmark{\ref{affil:Dubovsky}}
        Franz-Josef~\textsc{Hambsch},\altaffilmark{\ref{affil:GEOS}}$^,$\altaffilmark{\ref{affil:BAV}}$^,$\altaffilmark{\ref{affil:Hambsch}}
        Ian~\textsc{Miller},\altaffilmark{\ref{affil:Miller}}
        Tomohito~\textsc{Ohshima},\altaffilmark{\ref{affil:Kyoto}}
        Chikako~\textsc{Nakata},\altaffilmark{\ref{affil:Kyoto}}
        Miho~\textsc{Kawabata},\altaffilmark{\ref{affil:OKU}}
        Hirochika~\textsc{Nishino},\altaffilmark{\ref{affil:OKU}}
        Kazunari~\textsc{Masumoto},\altaffilmark{\ref{affil:OKU}}
        Sahori~\textsc{Mizoguchi},\altaffilmark{\ref{affil:Sendai}}$^,$\altaffilmark{\ref{affil:OKU}}
        Masayuki~\textsc{Yamanaka},\altaffilmark{\ref{affil:HidaKwasan}}$^,$\altaffilmark{\ref{affil:OKU}}
        Katsura~\textsc{Matsumoto},\altaffilmark{\ref{affil:OKU}}
        Daisuke~\textsc{Sakai},\altaffilmark{\ref{affil:OKU}}
        Daiki~\textsc{Fukushima},\altaffilmark{\ref{affil:OKU}}
        Minami~\textsc{Matsuura},\altaffilmark{\ref{affil:OKU}}
        Genki~\textsc{Bouno},\altaffilmark{\ref{affil:OKU}}
        Megumi~\textsc{Takenaka},\altaffilmark{\ref{affil:OKU}}
        Shinichi~\textsc{Nakagawa},\altaffilmark{\ref{affil:OKU}}
        Ryo~\textsc{Noguchi},\altaffilmark{\ref{affil:OKU}}
        Eriko~\textsc{Iino},\altaffilmark{\ref{affil:OKU}}
        Roger~D.~\textsc{Pickard},\altaffilmark{\ref{affil:BAAVSS}}$^,$\altaffilmark{\ref{affil:Pickard}}
        Yutaka~\textsc{Maeda},\altaffilmark{\ref{affil:Mdy}}
        Arne~\textsc{Henden},\altaffilmark{\ref{affil:AAVSO}}
        Kiyoshi~\textsc{Kasai},\altaffilmark{\ref{affil:Kai}}
        Seiichiro~\textsc{Kiyota},\altaffilmark{\ref{affil:Kis}}
        Hidehiko~\textsc{Akazawa},\altaffilmark{\ref{affil:OUS}}
        Kazuyoshi~\textsc{Imamura},\altaffilmark{\ref{affil:OUS}}
        Enrique~de~\textsc{Miguel},\altaffilmark{\ref{affil:Miguel}}$^,$\altaffilmark{\ref{affil:Miguel2}}
        Hiroyuki~\textsc{Maehara},\altaffilmark{\ref{affil:Kiso}}
        Berto~\textsc{Monard},\altaffilmark{\ref{affil:Monard}}$^,$\altaffilmark{\ref{affil:Monard2}}
        Elena~P.~\textsc{Pavlenko},\altaffilmark{\ref{affil:CrAO}}
        Kirill~\textsc{Antonyuk},\altaffilmark{\ref{affil:CrAO}}
        Nikolaj~\textsc{Pit},\altaffilmark{\ref{affil:CrAO}}
        Oksana~I.~\textsc{Antonyuk},\altaffilmark{\ref{affil:CrAO}}
        Aleksei~V.~\textsc{Baklanov},\altaffilmark{\ref{affil:CrAO}}
        Javier~\textsc{Ruiz},\altaffilmark{\ref{affil:Ruiz1}}$^,$\altaffilmark{\ref{affil:Ruiz2}}$^,$\altaffilmark{\ref{affil:Ruiz3}}
        Michael~\textsc{Richmond},\altaffilmark{\ref{affil:RIT}}
        Arto~\textsc{Oksanen},\altaffilmark{\ref{affil:Nyrola}}
        Caisey~\textsc{Harlingten},\altaffilmark{\ref{affil:Harlingten}}
        Sergey~Yu.~\textsc{Shugarov},\altaffilmark{\ref{affil:Sternberg}}$^,$\altaffilmark{\ref{affil:Slovak}}
        Drahomir~\textsc{Chochol},\altaffilmark{\ref{affil:Slovak}}
        Gianluca~\textsc{Masi},\altaffilmark{\ref{affil:Masi}}
        Francesca~\textsc{Nocentini},\altaffilmark{\ref{affil:Masi}}
        Patrick~\textsc{Schmeer},\altaffilmark{\ref{affil:Schmeer}}
        Greg~\textsc{Bolt},\altaffilmark{\ref{affil:Bolt}}
        Peter~\textsc{Nelson},\altaffilmark{\ref{affil:Nelson}}
        Joseph~\textsc{Ulowetz},\altaffilmark{\ref{affil:Ulowetz}}
        Richard~\textsc{Sabo},\altaffilmark{\ref{affil:Sabo}}
        William~N.~\textsc{Goff},\altaffilmark{\ref{affil:Goff}}
        William~\textsc{Stein},\altaffilmark{\ref{affil:Stein}}
        Ra\'ul~\textsc{Michel},\altaffilmark{\ref{affil:UNAM}}
        Shawn~\textsc{Dvorak},\altaffilmark{\ref{affil:Dvorak}}
        Irina~B.~\textsc{Voloshina},\altaffilmark{\ref{affil:Sternberg}}
        Vladimir~{Metlov},\altaffilmark{\ref{affil:Sternberg}}
        Natalia~\textsc{Katysheva},\altaffilmark{\ref{affil:Sternberg}}
        Vitaly~V.~\textsc{Neustroev},\altaffilmark{\ref{affil:Neustroev}}
        George~\textsc{Sjoberg},\altaffilmark{\ref{affil:Sjoberg}}$^,$\altaffilmark{\ref{affil:AAVSO}}
        Colin~\textsc{Littlefield},\altaffilmark{\ref{affil:LCO}}
        Bart{\l}omiej~\textsc{D\k{e}bski},\altaffilmark{\ref{affil:Debski}}
        Paulina~\textsc{Sowicka},\altaffilmark{\ref{affil:Debski}}
        Marcin~\textsc{Klimaszewski},\altaffilmark{\ref{affil:Debski}}
        Ma{\l}gorzata~\textsc{Cury{\l}o},\altaffilmark{\ref{affil:Debski}}
        Etienne~\textsc{Morelle},\altaffilmark{\ref{affil:Morelle}}
        Ivan~A.~\textsc{Curtis},\altaffilmark{\ref{affil:Curtis}}
        Hidetoshi~\textsc{Iwamatsu},\altaffilmark{\ref{affil:Iwamatsu}}$^,$\altaffilmark{\ref{affil:Kyoto}}
        Neil~D.~\textsc{Butterworth},\altaffilmark{\ref{affil:Butterworth}}
        Maksim~V.~\textsc{Andreev},\altaffilmark{\ref{affil:Terskol}}$^,$\altaffilmark{\ref{affil:ICUkraine}}
        Nikolai~\textsc{Parakhin},\altaffilmark{\ref{affil:Terskol}}
        Aleksandr~\textsc{Sklyanov},\altaffilmark{\ref{affil:Kazan}} 
        Kazuhiko~\textsc{Shiokawa},\altaffilmark{\ref{affil:Siz}} 
        Rudolf~\textsc{Nov\'ak},\altaffilmark{\ref{affil:Novak}}
        Tat'yana~R.~\textsc{Irsmambetova},\altaffilmark{\ref{affil:Sternberg}}
        Hiroshi~\textsc{Itoh},\altaffilmark{\ref{affil:Ioh}}
        Yoshiharu~\textsc{Ito},\altaffilmark{\ref{affil:Iha}}
        Kenji~\textsc{Hirosawa},\altaffilmark{\ref{affil:Hsk}}
        Denis~\textsc{Denisenko},\altaffilmark{\ref{affil:Denisenko}}
        Christopher~S.~\textsc{Kochanek},\altaffilmark{\ref{affil:Ohio}}
        Benjamin~\textsc{Shappee},\altaffilmark{\ref{affil:Ohio}}
        Krzysztof~Z.~\textsc{Stanek},\altaffilmark{\ref{affil:Ohio}}
        Jos\'e~L.~\textsc{Prieto},\altaffilmark{\ref{affil:Princeton}}
        Koh-ichi~\textsc{Itagaki},\altaffilmark{\ref{affil:Itagaki}}
        Rod~\textsc{Stubbings},\altaffilmark{\ref{affil:Stubbings}}
        Jose~\textsc{Ripero},\altaffilmark{\ref{affil:Ripero}}
        Eddy~\textsc{Muyllaert},\altaffilmark{\ref{affil:VVSBelgium}}
        Gary~\textsc{Poyner},\altaffilmark{\ref{affil:Poyner}}
}

\authorcount{affil:Kyoto}{
     Department of Astronomy, Kyoto University, Kyoto 606-8502}
\email{$^*$tkato@kusastro.kyoto-u.ac.jp}

\authorcount{affil:Dubovsky}{
     Vihorlat Observatory, Mierova 4, Humenne, Slovakia}

\authorcount{affil:GEOS}{
     Groupe Europ\'een d'Observations Stellaires (GEOS),
     23 Parc de Levesville, 28300 Bailleau l'Ev\^eque, France}

\authorcount{affil:BAV}{
     Bundesdeutsche Arbeitsgemeinschaft f\"ur Ver\"anderliche Sterne
     (BAV), Munsterdamm 90, 12169 Berlin, Germany}

\authorcount{affil:Hambsch}{
     Vereniging Voor Sterrenkunde (VVS), Oude Bleken 12, 2400 Mol, Belgium}

\authorcount{affil:Miller}{
     Furzehill House, Ilston, Swansea, SA2 7LE, UK}

\authorcount{affil:OKU}{
     Osaka Kyoiku University, 4-698-1 Asahigaoka, Osaka 582-8582}

\authorcount{affil:Sendai}{
     Sendai Astronomical Observatory, Nishikigaoka, Aoba-ku, Sendai 989-3123}

\authorcount{affil:HidaKwasan}{
     Kwasan and Hida Observatories, Kyoto University, Yamashina,
     Kyoto 607-8471}

\authorcount{affil:BAAVSS}{
     The British Astronomical Association, Variable Star Section (BAA VSS),
     Burlington House, Piccadilly, London, W1J 0DU, UK}

\authorcount{affil:Pickard}{
     3 The Birches, Shobdon, Leominster, Herefordshire, HR6 9NG, UK}

\authorcount{affil:Mdy}{
     Kaminishiyamamachi 12-14, Nagasaki, Nagasaki 850-0006}

\authorcount{affil:AAVSO}{
     American Association of Variable Star Observers, 49 Bay State Rd.,
     Cambridge, MA 02138, USA}

\authorcount{affil:Kai}{
     Baselstrasse 133D, CH-4132 Muttenz, Switzerland}

\authorcount{affil:Kis}{
     Variable Star Observers League in Japan (VSOLJ),
     7-1 Kitahatsutomi, Kamagaya, Chiba 273-0126}

\authorcount{affil:OUS}{
     Department of Biosphere-Geosphere System Science, Faculty of Informatics,
     Okayama University of Science, 1-1 Ridai-cho, Okayama, Okayama 700-0005}

\authorcount{affil:Miguel}{
     Departamento de F\'isica Aplicada, Facultad de Ciencias
     Experimentales, Universidad de Huelva,
     21071 Huelva, Spain}

\authorcount{affil:Miguel2}{
     Center for Backyard Astrophysics, Observatorio del CIECEM,
     Parque Dunar, Matalasca\~nas, 21760 Almonte, Huelva, Spain}

\authorcount{affil:Kiso}{
     Kiso Observatory, Institute of Astronomy, School of Science, 
     The University of Tokyo 10762-30, Mitake, Kiso-machi, Kiso-gun,
     Nagano 397-0101}

\authorcount{affil:Monard}{
     Bronberg Observatory, Center for Backyard Astronomy Pretoria,
     PO Box 11426, Tiegerpoort 0056, South Africa}

\authorcount{affil:Monard2}{
     Kleinkaroo Observatory, Center for Backyard Astronomy Kleinkaroo,
     Sint Helena 1B, PO Box 281, Calitzdorp 6660, South Africa}

\authorcount{affil:CrAO}{
     Crimean Astrophysical Observatory, Kyiv Shevchenko 
     National University, 98409, Nauchny, Crimea, Ukraine}

\authorcount{affil:Ruiz1}{
     Observatorio de C\'antabria, Ctra. de Rocamundo s/n, Valderredible, 
     Cantabria, Spain}

\authorcount{affil:Ruiz2}{
     Instituto de F\'{\i}sica de Cantabria (CSIC-UC), Avenida Los Castros s/n, 
     E-39005 Santander, Cantabria, Spain}

\authorcount{affil:Ruiz3}{
     Agrupaci\'on Astron\'omica C\'antabra, Apartado 573,
     39080, Santander, Spain}

\authorcount{affil:RIT}{
     Physics Department, Rochester Institute of Technology, Rochester,
     New York 14623, USA}

\authorcount{affil:Nyrola}{
     Hankasalmi observatory, Jyvaskylan Sirius ry, Vertaalantie
     419, FI-40270 Palokka, Finland}

\authorcount{affil:Harlingten}{
     Searchlight Observatory Network, The Grange, Erpingham,
     Norfolk, NR11 7QX, UK}

\authorcount{affil:Sternberg}{
     Sternberg Astronomical Institute, Lomonosov Moscow University, 
     Universitetsky Ave., 13, Moscow 119992, Russia}

\authorcount{affil:Slovak}{
     Astronomical Institute of the Slovak Academy of Sciences, 05960,
     Tatranska Lomnica, the Slovak Republic}

\authorcount{affil:Masi}{
     The Virtual Telescope Project, Via Madonna del Loco 47, 03023
     Ceccano (FR), Italy}

\authorcount{affil:Schmeer}{
     Bischmisheim, Am Probstbaum 10, 66132 Saarbr\"{u}cken, Germany}

\authorcount{affil:Bolt}{
     Camberwarra Drive, Craigie, Western Australia 6025, Australia}

\authorcount{affil:Nelson}{
     1105 Hazeldean Rd, Ellinbank 3820, Australia}

\authorcount{affil:Ulowetz}{
     Center for Backyard Astrophysics Illinois,
     Northbrook Meadow Observatory, 855 Fair Ln, Northbrook,
     Illinois 60062, USA}

\authorcount{affil:Sabo}{
     2336 Trailcrest Dr., Bozeman, Montana 59718, USA}

\authorcount{affil:Goff}{
     13508 Monitor Ln., Sutter Creek, California 95685, USA}

\authorcount{affil:Stein}{
     6025 Calle Paraiso, Las Cruces, New Mexico 88012, USA}

\authorcount{affil:UNAM}{
     Instituto de Astronom\'{\i}a UNAM, Apartado Postal 877, 22800 Ensenada
     B.C., M\'{e}xico}

\authorcount{affil:Dvorak}{
     Rolling Hills Observatory, 1643 Nightfall Drive,
     Clermont, Florida 34711, USA}

\authorcount{affil:Neustroev}{
     Astronomy Division, Department of Physics, PO Box 3000,
     FIN-90014 University of Oulu, Finland}

\authorcount{affil:Sjoberg}{
     The George-Elma Observatory, 9 Contentment Crest, \#182,
     Mayhill, New Mexico 88339, USA}

\authorcount{affil:LCO}{
     Department of Physics, University of Notre Dame, Notre Dame,
     Indiana 46556, USA}

\authorcount{affil:Debski}{
     Astronomical Observatory, Jagiellonian University,
     ul. Orla 171 30-244 Krak\'ow, Poland}

\authorcount{affil:Morelle}{
     9 rue Vasco de GAMA, 59553 Lauwin Planque, France}

\authorcount{affil:Curtis}{
     2 Yandra Street, Vale Park, Adelaide, South Australia 5081, Australia}

\authorcount{affil:Iwamatsu}{
     St. Dominic Junior and Senior High School
     2-2-14, Tsunogorou, Aoba Ward, Sendai, Miyagi 980-0874}

\authorcount{affil:Butterworth}{
     24 Payne Street, Mount Louisa, Queensland 4814, Australia}

\authorcount{affil:Terskol}{
     Institute of Astronomy, Russian Academy of Sciences, 361605 Peak Terskol,
     Kabardino-Balkaria, Russia}

\authorcount{affil:ICUkraine}{
     International Center for Astronomical, Medical and Ecological Research
     of NASU, Ukraine 27 Akademika Zabolotnoho Str. 03680 Kyiv,
     Ukraine}

\authorcount{affil:Kazan}{
     Kazan Federal University, Kremlevskaya str., 18, Kazan, 420008, Russia}

\authorcount{affil:Siz}{
     Moriyama 810, Komoro, Nagano 384-0085}

\authorcount{affil:Novak}{
     Research Centre for Toxic Compounds in the Environment, Faculty of 
     Science, Masaryk University, Kamenice 3, 625 00 Brno, Czech Republic}

\authorcount{affil:Ioh}{
     VSOLJ, 1001-105 Nishiterakata, Hachioji, Tokyo 192-0153}

\authorcount{affil:Iha}{
     VSOLJ, Kamisugi 6-2-69, Aoba-ku, Sendai, Miyagi 980-0011}

\authorcount{affil:Hsk}{
     216-4 Maeda, Inazawa-cho, Inazawa-shi, Aichi 492-8217}

\authorcount{affil:Denisenko}{
     Space Research Institute (IKI), Russian Academy of Sciences, Moscow,
     Russia}

\authorcount{affil:Ohio}{
     Department of Astronomy, the Ohio State University, Columbia,
     OH 43210, USA}

\authorcount{affil:Princeton}{
     Department of Astrophysical Sciences, Princeton University,
     NJ 08544, USA}

\authorcount{affil:Itagaki}{
     Itagaki Astronomical Observatory, Teppo-cho, Yamagata 990-2492}

\authorcount{affil:Stubbings}{
     Tetoora Observatory, Tetoora Road, Victoria, Australia}

\authorcount{affil:Ripero}{
     President of CAA (Centro Astronomico de Avila) and Variable and SNe
     Group M1, Buenavista 7, Ciudad Sto. Domingo, 28110 Algete/Madrid, Spain}

\authorcount{affil:VVSBelgium}{
     Vereniging Voor Sterrenkunde (VVS), Moffelstraat 13 3370
     Boutersem, Belgium}

\authorcount{affil:Poyner}{
     BAA Variable Star Section, 67 Ellerton Road, Kingstanding,
     Birmingham B44 0QE, UK}


\KeyWords{accretion, accretion disks
          --- stars: novae, cataclysmic variables
          --- stars: dwarf novae
         }

\maketitle

\begin{abstract}
Continuing the project described by \citet{Pdot}, we collected
times of superhump maxima for 56 SU UMa-type dwarf novae 
mainly observed during the 2013--2014 season and characterized
these objects.  We detected negative superhumps in VW Hyi
and indicated that the low number of normal outbursts
in some supercycle can be interpreted as a result of
the disk tilt.  This finding, combined with the Kepler
observation of V1504 Cyg and V344 Lyr, suggests that 
the disk tilt is responsible for modulating the outburst
pattern in SU UMa-type dwarf novae.  We also studied
the deeply eclipsing WZ Sge-type dwarf nova
MASTER OT J005740.99$+$443101.5 and found evidence of
a sharp eclipse during the phase of early superhumps.
The profile can be reproduced by a combination of the eclipse
of the axisymmetric disk and the uneclipsed light source
of early superhumps.  This finding confirms the lack
of evince of a greatly enhanced hot spot during the early
stage of WZ Sge-type outburst.  We detected growing
(stage A) superhumps in MN Dra and give a suggestion
that some of SU UMa-type dwarf novae situated near the
critical condition of tidal instability may show long-lasting
stage A superhumps.  The large negative period derivatives
reported in such systems can be understood a result of
the combination of stage A and B superhumps.
The WZ Sge-type dwarf novae AL Com and ASASSN-13ck 
showed a long-lasting (plateau-type) rebrightening.
In the early phase of the rebrightening, both objects showed
a precursor-like outburst, suggesting that the long-lasting
rebrightening is triggered by a precursor outburst.
\end{abstract}

\section{Introduction}

   Cataclysmic variables (CVs) are close binary systems
transferring matter from a low-mass dwarf secondary to
a white dwarf.  The transferred matter forms an accretion
disk.  In dwarf novae (DNe), a class of CVs, thermal-viscous
instability in the accretion disk causes outbursts.
SU UMa-type dwarf novae, a subclass of DNe,
show long outbursts (superoutburst) in addition to
ordinary short outbursts, and semi-periodic modulations
called superhumps are detected during the superoutbursts.
The superhumps have periods a few percent longer
than the orbital period.  It is now widely believed
that the 3:1 resonance in the accretion disk causes
the eccentric deformation of the disk, resulting
superhumps [tidal instability: \citet{whi88tidal};
\citet{hir90SHexcess}; \citet{lub91SHa}].
The superoutburst can be understood as a result of
the increased tidal dissipation
and removal of the angular momentum
when the tidal instability works [thermal tidal instability
(TTI) model: \citet{osa89suuma}; \citet{osa96review}].
This process produces a relaxation
oscillation in the total angular momentum of the disk,
and the superoutbursts recur.  The interval between
the successive superoutbursts is called supercycle.
[For general information of CVs, DNe, SU UMa-type 
dwarf novae and superhumps, see e.g. \citet{war95book}].

   In a series of papers \citet{Pdot}, \citet{Pdot2}, \citet{Pdot3}
\citet{Pdot4} and \citet{Pdot5}, we systematically surveyed
SU UMa-type dwarf novae particularly on period variations
of superhumps.  The change in the superhump period reflects 
the precession angular velocity of the eccentric (or flexing) 
disk, and is expected to be an excellent probe for studying 
the structure of the accretion disk during dwarf nova outbursts.
In recent series of papers, we dealt with a various of
topics related to SU UMa-type dwarf novae and superhumps:
in \citet{Pdot3}, we also studied Kepler data and made
a pilot study of variations of superhumps amplitudes
motivated by \citet{sma10SHamp}.  In \citet{Pdot4},
we systematically studied ER UMa-type dwarf novae
(a class of SU UMa-type dwarf novae with very short
supercycles, see e.g. \cite{kat95eruma}; \cite{rob95eruma};
\cite{kat99erumareview}) and helium dwarf novae
(AM CVn-type objects).  In \citet{Pdot5}, we made
a pilot study of the decline rate of the superoutburst
motivated by \citet{can10v344lyr} and studied negative
superhumps (superhumps having periods shorter than
the orbital period and are considered as a manifestation
of a tilted disk, see e.g. \cite{har95v503cyg};
\cite{pat97v603aql}; \cite{woo07negSH}) particularly in
BK Lyn, which displays both states of a novalike variable
(a thermally stable CV) and an ER UMa-type dwarf nova.

   We continue this extended, comprehensive research of
SU UMa-type dwarf novae and superhumps in general
in this paper.  Since most of the objects treated in this
paper is little documented in the past, and since a compilation
of historical descriptions of dwarf novae has not been
issued for long since \citet{GlasbyDNbook}, we intend
these series of papers to be also a source of compiled
information of individual dwarf novae.

   The present advances in understanding of the superhump
periods and their variations started with the new interpretation
of the Kepler observation \citep{osa13v1504cygKepler},
who used the negative superhumps as a probe for
the variation of the disk radius over the supercycle,
and confirmed the radius variation predicted by the TTI
model.  Combined with \citet{osa13v344lyrv1504cyg},
\citet{osa14v1504cygv344lyrpaper3}, the TTI model
is currently the only viable model to explain
the SU UMa-type phenomenon.

   In \citet{Pdot}, we demonstrated that most of
the $O-C$ diagrams of superhumps in SU UMa-type dwarf novae
can be expressed by three distinct stages: initial growing 
stage (stage A) wuth a long period and fully developed 
stage (stage B) with a systematically varying period
and later stage C with a shorter, almost constant period.
[see \citet{Pdot} for the notation of stages A-B-C
of superhumps].  The origin of these three stages of
superhumps was a mystery when it was documented
in \citet{Pdot}.  After examination of the Kepler data,
\citet{osa13v344lyrv1504cyg} proposed that the appearance
of the pressure effect is responsible for the transition
from stage A to stage B.  In this interpretation, stage A
reflects the state when the 3:1 resonance is confined
to the resonance region.  This interpretation allowed
a new method to determine the mass ratio ($q = M_2/M_1$)
only from superhump observations and the orbital period
\citep{kat13qfromstageA}.  This method is particularly
suitable for measuring mass ratios in WZ Sge-type
dwarf novae (SU UMa-type dwarf novae with very long
supercycles, and are considered to be the terminal
stage of the CV evolution) and for discriminating hitherto
very poorly known period bouncers (CVs which have
passed the minimum orbital period in  evolution:
\cite{kat13j1222}; \cite{nak13j2112j2037}).
Superhumps are now not only a powerful tool for diagnosing 
the accretion disk, but also a powerful tool for illuminating
the CV evolution.  

   The material and methods of analysis are given in
section \ref{sec:obs}, observations and analysis of
individual objects including short discussions on
individual objects are given in section \ref{sec:individual},
the general discussion is given in section
\ref{sec:discuss} and the summary is given in section
\ref{sec:summary}.

\section{Observation and Analysis}\label{sec:obs}

   The data were obtained under campaigns led by 
the VSNET Collaboration \citep{VSNET}.
For some objects, we used the public data from 
the AAVSO International Database\footnote{
   $<$http://www.aavso.org/data-download$>$.
}.

   The majority of the data were acquired
by time-resolved CCD photometry by using 30 cm-class telescopes
located world-wide, whose observational details will be
presented in future papers dealing with analysis and discussion
on individual objects of interest.
The list of outbursts and observers is summarized in 
table \ref{tab:outobs}.
The data analysis was performed just in the same way described
in \citet{Pdot} and \citet{Pdot5} and we mainly used
R software\footnote{
   The R Foundation for Statistical Computing:\\
   $<$http://cran.r-project.org/$>$.
} for data analysis.
In de-trending the data, we used both lower (1--5th order)
polynomial fitting and locally-weighted polynomial regression 
(LOWESS: \cite{LOWESS}).
The times of superhumps maxima were determined by
the template fitting method as described in \citet{Pdot}.
The times of all observations are expressed in 
Barycentric Julian Dates (BJD).

   The abbreviations used in this paper are the same
as in \citet{Pdot5}: $P_{\rm orb}$ means
the orbital period and $\varepsilon \equiv P_{\rm SH}/P_{\rm orb}-1$ for 
the fractional superhump excess.   Since \citet{osa13v1504cygKepler},
the alternative fractional superhump excess in the frequency unit
$\varepsilon^* \equiv 1-P_{\rm orb}/P_{\rm SH}-1 = \varepsilon/(1+\varepsilon)$
has been introduced because this fractional superhump excess
can be directly compared to the precession rate.  We therefore
used $\varepsilon^*$ in referring the precession rate.

   We used phase dispersion minimization (PDM; \cite{PDM})
for period analysis and 1$\sigma$ errors for the PDM analysis
was estimated by the methods of \citet{fer89error} and \citet{Pdot2}.
We also used least absolute shrinkage and selection operator (Lasso)
method (\cite{lasso}; \cite{kat12perlasso}), which has been 
proven to yield very sharp signals.  In this paper, we used
the two-dimensional Lasso power spectra as introduced
in the analysis of the Kepler data such as in \citet{kat13j1924}; 
\citet{osa13v344lyrv1504cyg};
\citet{kat13j1939v585lyrv516lyr}.  These two-dimensional
Lasso power spectra have been proven to be helpful in detecting
negative superhumps (cf. \cite{osa13v344lyrv1504cyg}) as well as
superhumps with varying frequencies (cf. \cite{kat13j1924}).
Although the application of two-dimensional Lasso power spectra
to the Kepler data dealt with almost uniformly sampled data,
we have demonstrated in \citet{Pdot5} and \citet{ohs14eruma}
that two-dimensional Lasso power spectra are also effective
in detecting multiple signals and their variations in
non-uniformly sampled ground-based data.

   The resultant $P_{\rm SH}$, $P_{\rm dot}$ and other parameters
are listed in table \ref{tab:perlist} in same format as in
\citet{Pdot}.  The definitions of parameters $P_1, P_2, E_1, E_2$
and $P_{\rm dot}$ are the same as in \citet{Pdot}.
We also presented comparisons of $O-C$ diagrams between different
superoutbursts since this has been one of the motivations of
these surveys (cf. \cite{uem05tvcrv}) and it has been
demonstrated that a combination of $O-C$ diagrams between
different superoutbursts can better describe the overall
pattern of the period variation \citep{Pdot}.
In drawing combined $O-C$ diagrams, we usually used
$E=0$ for the start of the superoutburst, which usually
refers to the first positive detection of the outburst.
This epoch usually has an accuracy of $\sim$1 d for
well-observed objects, and if the outburst was not sufficiently
observed, we mentioned how to estimated $E$ in such an outburst.
We also present representative $O-C$ diagrams and 
light curves especially for WZ Sge-type dwarf novae, 
which are not expected to undergo outbursts in the near future.
In all figures, the binned magnitudes and $O-C$ values
are accompnied by 1$\sigma$ error bars, which are omitted
when the error is smaller than the plot mark.

   We used the same terminology of superhumps summarized in
\citet{Pdot3}.  We especially call attention to
the term ``late superhumps''.  We only used ``traditional''
late superhumps when an $\sim$0.5 phase shift is confirmed
[\citet{vog83lateSH}; see also table 1 in \citet{Pdot3} 
for various types of superhumps], 
since we suspect that many of the past
claims of detections of ``late superhumps'' were likely
stage C superhumps [cf. \citet{Pdot}; note that the Kepler 
observation of V585 Lyr also demonstrated this persistent stage C
superhumps without a phase shift \citep{kat13j1939v585lyrv516lyr}].

   Early superhumps are double-wave humps seen during the early stages
of WZ Sge-type dwarf novae, and have period close to the orbital
periods (\cite{kat96alcom}; \cite{kat02wzsgeESH}; 
\cite{osa02wzsgehump}).
We used the period of early superhumps as approximate
orbital period.  The validity of this assumption is also
reviewed in this paper.

   As in \citet{Pdot}, we have used coordinate-based 
optical transient (OT) designations for some objects, such as 
Catalina Real-time Transient Survey (CRTS; \cite{CRTS})\footnote{
   $<$http://nesssi.cacr.caltech.edu/catalina/$>$.
   For the information of the individual Catalina CVs, see
   $<$http://nesssi.cacr.caltech.edu/catalina/AllCV.html$>$.
} transients
and listed the original identifiers in table \ref{tab:outobs}.
When available, we preferably used the International Astronomical
Union (IAU)-format names provided by the CRTS team 
in the public data release\footnote{
  $<$http://nesssi.cacr.caltech.edu/DataRelease/$>$.
}

\begin{table*}
\caption{List of Superoutbursts.}\label{tab:outobs}
\begin{center}
\begin{tabular}{ccccl}
\hline
Subsection & Object & Year & Observers or references\commenta & ID\commentb \\
\hline
\ref{obj:foand}    & FO And       & 2013 & Aka & \\
\ref{obj:dhaql}    & DH Aql       & 2000 & Oud, Btw & \\
\ref{obj:bbari}    & BB Ari       & 2013 & GFB, OkC, HaC, Mhh & \\
\ref{obj:uzboo}    & UZ Boo       & 2013 & KU, GFB, DPV, RIT, Rui, Mhh, & \\
                   &              &      & AAVSO, RPc, Kai, CRI, IMi, & \\
                   &              &      & Mdy, Iak, Ioh \\ 
\ref{obj:v342cam}  & V342 Cam     & 2013 & DPV & \\
\ref{obj:v452cas}  & V452 Cas     & 2013 & IMi, RPc, DPV & \\
\ref{obj:v359cen}  & V359 Cen     & 2014 & HaC & \\
\ref{obj:yzcnc}    & YZ Cnc       & 2014 & Aka, Mdy, HaC & \\
\ref{obj:gzcnc}    & GZ Cnc       & 2014 & Kis, Mdy, IMi, Kai, Aka & \\
\ref{obj:alcom}    & AL Com       & 2013 & OKU, NKa, KU, Irs, CRI, & \\
                   &              &      & DKS, Kis, IMi, AAVSO & \\
\ref{obj:v503cyg}  & V503 Cyg     & 2013 & RPc & \\
\ref{obj:ixdra}    & IX Dra       & 2012 & MEV, AAVSO, UJH & \\
\ref{obj:mndra}    & MN Dra       & 2012 & Ast, CRI & \\
                   &              & 2013 & CRI, Kai, Ter, KU & \\
\ref{obj:cperi}    & CP Eri       & 2013 & OkC, SWI & \\
\ref{obj:v1239her} & V1239 Her    & 2013 & RPc, IMi & \\
\ref{obj:cthya}    & CT Hya       & 2014 & Mdy & \\
\ref{obj:vwhyi}    & VW Hyi       & 2012 & HaC, CTA, Hen & \\
\ref{obj:wxhyi}    & WX Hyi       & 1977 & \citet{bai79wxhyiv436cen} & \\
                   &              & 2014 & HaC & \\
\hline
  \multicolumn{5}{l}{\parbox{500pt}{\commenta Key to observers:
Aka (H. Akazawa, OUS),
Ast (Astrotel telescope, by A. Sklyanov),
Btw (N. Butterworth),
CRI (Crimean Astrophys. Obs.),
CTA (I. Curtis),
deM (E. de Miguel),
DKS\commentc (S. Dvorak),
DPV (P. Dubovsky),
GBo (G. Bolt),
GFB\commentc (W. Goff),
HaC (F.-J. Hambsch, remote obs. in Chile), 
Han (Hankasalmi Obs., by A. Oksanen)
Hsk (K. Hirosawa),
Iha (Y. Ito),
IMi\commentc (I. Miller),
Ioh (H. Itoh),
Irs (T. Irsmambetova),
Kai (K. Kasai),
Kis (S. Kiyota),
Krw (Krakow team),
KU (Kyoto U., campus obs.),
LCO (C. Littlefield),
Mas (G. Masi team),
Mdy (Y. Maeda),
MEV\commentc (E. Morelle),
Mhh (H. Maehara),
Mic (R. Michel-Murillo team),
Nel\commentc (P. Nelson),
Neu (V. Neustroev),
NKa (N. Katysheva),
Nov (R. Nov\'ak),
OkC \commentc (A. Oksanen, remote obs. in Chile),
OKU (Osaya Kyoiku U.),
Oud (Ouda station, Kyoto U.),
RIT (M. Richmond),
RPc\commentc (R. Pickard),
Rui (J. Ruiz),
Shu (S. Shugarov),
Siz (K. Shiokawa),
SRI\commentc (R. Sabo),
SWI\commentc (W. Stein),
Ter (Terskol Obs.),
UJH\commentc (J. Ulowetz),
Vol (I. Voloshina),
AAVSO (AAVSO database)
}} \\
  \multicolumn{5}{l}{\commentb Original identifications, discoverers or data source.} \\
  \multicolumn{5}{l}{\commentc Inclusive of observations from the AAVSO database.} \\
\end{tabular}
\end{center}
\end{table*}

\addtocounter{table}{-1}
\begin{table*}
\caption{List of Superoutbursts (continued).}
\begin{center}
\begin{tabular}{ccccl}
\hline
Subsection & Object & Year & Observers or references\commenta & ID\commentb \\
\hline
\ref{obj:aylyr}    & AY Lyr       & 2013 & Aka & \\
\ref{obj:aooct}    & AO Oct       & 2013 & HaC & \\
\ref{obj:dtoct}    & DT Oct       & 2014 & OkC & \\
\ref{obj:v521peg}  & V521 Peg     & 2013 & KU, Mdy, Aka, DPV, RPc, & \\
                   &              &      & IMi, Hsk, Mhh & \\
\ref{obj:typsc}    & TY Psc       & 2013 & Aka & \\
\ref{obj:v893sco}  & V893 Sco     & 2007 & MLF, OKU & \\
                   &              & 2008 & GBo & \\
                   &              & 2010 & GBo, OKU & \\
                   &              & 2013 & HaC, MLF & \\
\ref{obj:rzsge}    & RZ Sge       & 2013 & AAVSO, IMi, Rui & \\
\ref{obj:awsge}    & AW Sge       & 2013 & Vol, SRI & \\
\ref{obj:v1265tau} & V1265 Tau    & 2013 & GFB, HaC, KU & \\
\ref{obj:suuma}    & SU UMa       & 2013 & OKU, DPV, Kis, Iha & \\
\ref{obj:ssumi}    & SS UMi       & 2013 & DPV, Kai, Krw & \\
\ref{obj:cuvel}    & CU Vel       & 2013 & HaC, Nel & \\
\ref{obj:j231935}  & 1RXS J231935 & 2013 & DPV & \\
\ref{obj:asas2243} & ASAS J224349 & 2013 & Krw, HaC, Mdy, DPV, Shu, IMi, & ASAS J224349$+$0809.5 \\
                   &              &      & Rui, Mhh & \\
\ref{obj:asassn13cf} & ASASSN-13cf & 2013 & LCO, IMi, RPc & \\
\ref{obj:asassn13cg} & ASASSN-13cg & 2013 & LCO, CRI & \\
\ref{obj:asassn13ck} & ASASSN-13ck & 2013 & deM, OkC, HaC, Mas, Han, & \\
                   &              &      & DPV, UJH, IMi, RPc, AAVSO, &  \\
                   &              &      & Nel, Mdy, Mic, RIT & \\
\ref{obj:asassn13cv} & ASASSN-13cv & 2013 & Kai & \\
\ref{obj:asassn13cz} & ASASSN-13cz & 2013 & IMi & \\
\ref{obj:asassn13da} & ASASSN-13da & 2013 & HaC & \\
\ref{obj:asassn14ac} & ASASSN-14ac & 2014 & OKU, Kis, Mdy, DPV, RPc, & \\
                   &              &      & KU, RIT, Mic & \\
\ref{obj:j024354}  & CSS J024354  & 2013 & MLF & CSS131026:024354$-$160314 \\
\ref{obj:dde31}    & DDE 31       & 2014 & Mdy & \\
\ref{obj:j004527}  & MASTER J004527 & 2013 & OKU, UJH, deM, AAVSO, IMi, & MASTER OT J004527.52$+$503213.8 \\
                   &              &      & Kai, SRI, RPc, DKS, DPV, & \\
                   &              &      & Neu, RIT & \\
\hline
\end{tabular}
\end{center}
\end{table*}

\addtocounter{table}{-1}
\begin{table*}
\caption{List of Superoutbursts (continued).}
\begin{center}
\begin{tabular}{ccccl}
\hline
Subsection & Object & Year & Observers or references\commenta & ID\commentb \\
\hline
\ref{obj:j005740}  & MASTER J005740 & 2013 & deM, NKa, Shu, Mas, SWI, & MASTER OT J005740.99$+$443101.5 \\
                   &              &      & UJH, AAVSO, KU, SRI, OKU, & \\
                   &              &      & DPV, RPc & \\
\ref{obj:j024847}  & MASTER J024847 & 2013 & IMi & MASTER OT J024847.86+501239.7 \\
\ref{obj:j061335}  & MASTER J061335 & 2013 & OKU, Ter, DPV, KU, Mas, Nov & MASTER OT J061335.30$+$395714.7 \\
\ref{obj:j073208}  & MASTER J073208 & 2013 & Shu & MASTER OT J073208.11$+$064149.5 \\
\ref{obj:j095018}  & MASTER J095018 & 2013 & HaC, OKU & MASTER OT J095018.04$-$063921.9 \\
\ref{obj:j141143}  & MASTER J141143 & 2014 & DPV & MASTER OT J141143.46$+$262051.5 \\
\ref{obj:j162323}  & MASTER J162323 & 2013 & Shu, IMi, Neu, Rui, DPV & MASTER OT J162323.48$+$782603.3 \\
\ref{obj:j234843}  & MASTER J234843 & 2013 & OKU, Mas, DPV & MASTER OT J234843.23$+$250250.4 \\
\ref{obj:j013741}  & OT J013741   & 2014 & deM, Kai & CSS140104:013741$+$220312 \\
\ref{obj:j210016}  & OT J210016   & 2013 & MLF, deM, OKU, DKS & CSS130905:210016$-$024258 \\
\ref{obj:j191501}  & PNV J191501  & 2013 & KU, deM, AAVSO, GBo, Vol, & PNV J19150199$+$0719471 \\
                   &              &      & OkC, SWI, OKU, Mic, HaC, & \\
                   &              &      & DPV, SRI, Mdy, MEV, RIT, & \\
                   &              &      & Aka, Siz, GFB, RPc, Kis, & \\
                   &              &      & IMi, Mhh, Nel, Ioh  & \\
\ref{obj:j094327}  & SSS J094327  & 2013 & Kis & SSS J094327.3$-$272038 \\
                   &              & 2014 & GBo, Kis, HaC & \\
\ref{obj:j233822}  & TCP J233822  & 2013 & MLF, HaC, Nel & TCP J23382254$-$2049518 \\
\hline
\end{tabular}
\end{center}
\end{table*}

\begin{table*}
\caption{Superhump Periods and Period Derivatives}\label{tab:perlist}
\begin{center}
\begin{tabular}{c@{\hspace{7pt}}c@{\hspace{7pt}}c@{\hspace{7pt}}c@{\hspace{7pt}}c@{\hspace{7pt}}c@{\hspace{7pt}}c@{\hspace{7pt}}c@{\hspace{7pt}}c@{\hspace{7pt}}c@{\hspace{7pt}}c@{\hspace{7pt}}c@{\hspace{7pt}}c@{\hspace{7pt}}c}
\hline
Object & Year & $P_1$ (d) & err & \multicolumn{2}{c}{$E_1$\commenta} & $P_{\rm dot}$\commentb & err\commentb & $P_2$ (d) & err & \multicolumn{2}{c}{$E_2$\commenta} & $P_{\rm orb}$ (d)\commentc & Q\commentd \\
\hline
FO And & 2013 & 0.074412 & 0.000070 & 1 & 70 & -- & -- & -- & -- & -- & -- & 0.07161 & CGM \\
DH Aql & 2000 & 0.080005 & 0.000067 & 0 & 39 & -- & -- & -- & -- & -- & -- & -- & C \\
BB Ari & 2013 & 0.072544 & 0.000097 & 0 & 42 & -- & -- & 0.072135 & 0.000046 & 40 & 111 & -- & B \\
UZ Boo & 2013 & 0.062066 & 0.000029 & 15 & 85 & 5.1 & 5.1 & -- & -- & -- & -- & -- & B \\
V342 Cam & 2013 & -- & -- & -- & -- & -- & -- & 0.078067 & 0.000080 & 0 & 65 & 0.07531 & C \\
V452 Cas & 2013 & 0.088596 & 0.000068 & 0 & 92 & $-$14.9 & 2.9 & -- & -- & -- & -- & -- & CG \\
V359 Cen & 2014 & 0.081064 & 0.000026 & 0 & 49 & $-$6.3 & 4.2 & 0.080744 & 0.000052 & 60 & 86 & -- & B \\
FZ Cet & 2014 & 0.058547 & 0.000062 & 0 & 43 & -- & -- & -- & -- & -- & -- & -- & C \\
YZ Cnc & 2014 & 0.090422 & 0.000069 & 0 & 41 & $-$3.9 & 11.2 & -- & -- & -- & -- & 0.0868 & CG \\
GZ Cnc & 2014 & 0.092699 & 0.000056 & 12 & 92 & $-$2.5 & 4.8 & -- & -- & -- & -- & 0.08825 & C \\
AL Com & 2013 & 0.057323 & 0.000022 & 87 & 210 & 4.9 & 1.9 & -- & -- & -- & -- & 0.056669 & B \\
IX Dra & 2012 & 0.066955 & 0.000021 & 0 & 146 & 0.4 & 1.5 & -- & -- & -- & -- & -- & B \\
MN Dra & 2012 & 0.105299 & 0.000061 & 47 & 115 & -- & -- & -- & -- & -- & -- & 0.0998 & C \\
MN Dra & 2013 & 0.105040 & 0.000066 & 26 & 66 & $-$14.8 & 9.5 & -- & -- & -- & -- & 0.0998 & C \\
CP Eri & 2013 & 0.019897 & 0.000003 & 0 & 111 & 3.1 & 0.9 & -- & -- & -- & -- & 0.019692 & B \\
CT Hya & 2014 & -- & -- & -- & -- & -- & -- & 0.066178 & 0.000064 & 0 & 46 & -- & C \\
VW Hyi & 2012 & 0.076916 & 0.000014 & 11 & 90 & 2.9 & 1.3 & 0.076579 & 0.000019 & 87 & 159 & 0.074271 & A \\
WX Hyi & 1977 & 0.077612 & 0.000113 & 0 & 14 & -- & -- & 0.077106 & 0.000091 & 26 & 40 & 0.074813 & C \\
WX Hyi & 2014 & 0.077616 & 0.000037 & 0 & 52 & -- & -- & 0.077367 & 0.000072 & 52 & 103 & 0.074813 & C \\
AO Oct & 2013 & 0.067326 & 0.000046 & 0 & 59 & 19.6 & 6.4 & 0.066776 & 0.000064 & 59 & 91 & 0.06535 & B \\
DT Oct & 2014 & -- & -- & -- & -- & -- & -- & 0.074022 & 0.000113 & 54 & 81 & -- & C \\
V521 Peg & 2013 & 0.061503 & 0.000032 & 0 & 67 & 13.8 & 5.8 & 0.061006 & 0.000029 & 67 & 199 & 0.0599 & B \\
TY Psc & 2013 & -- & -- & -- & -- & -- & -- & 0.070381 & 0.000030 & 27 & 57 & 0.068348 & C \\
\hline
  \multicolumn{13}{l}{\commenta Interval used for calculating the period (corresponding to $E$ in section \ref{sec:individual}).} \\
  \multicolumn{13}{l}{\commentb Unit $10^{-5}$.} \\
  \multicolumn{13}{l}{\parbox{500pt}{\commentc References: 
FO And \citep{tho96Porb},
V342 Cam \citep{she11j0423},
YZ Cnc \citep{sha88yzcnc},
GZ Cnc \citep{tap03gzcnc},
AL Com (this work),
MN Dra \citep{pav10mndra},
CP Eri \citep{arm12cperi},
VW Hyi (this work),
WX Hyi \citep{sch81vwhyiwxhyi},
AO Oct \citep{wou04CV4},
V521 Peg \citep{rod05hs2219},
TY Psc \citep{tho96Porb},
V893 Sco (this work),
RZ Sge \citep{pat03suumas},
SU UMa \citep{tho86suuma},
SS UMi \citep{tho96Porb},
CU Vel (this work),
ASASSN-13ck -- TCP J233822 (this work)
}}\\
  \multicolumn{13}{l}{\parbox{500pt}{\commentd Data quality and comments. A: excellent, B: partial coverage or slightly low quality, C: insufficient coverage or observations with large scatter, G: $P_{\rm dot}$ denotes global $P_{\rm dot}$, M: observational gap in middle stage, 2: late-stage coverage, the listed period may refer to $P_2$, E: $P_{\rm orb}$ refers to the period of early superhumps, P: $P_{\rm orb}$ refers to a shorter stable periodicity recorded in outburst.}} \\
\end{tabular}
\end{center}
\end{table*}

\addtocounter{table}{-1}
\begin{table*}
\caption{Superhump Periods and Period Derivatives (continued)}
\begin{center}
\begin{tabular}{c@{\hspace{7pt}}c@{\hspace{7pt}}c@{\hspace{7pt}}c@{\hspace{7pt}}c@{\hspace{7pt}}c@{\hspace{7pt}}c@{\hspace{7pt}}c@{\hspace{7pt}}c@{\hspace{7pt}}c@{\hspace{7pt}}c@{\hspace{7pt}}c@{\hspace{7pt}}c@{\hspace{7pt}}c}
\hline
Object & Year & $P_1$ & err & \multicolumn{2}{c}{$E_1$} & $P_{\rm dot}$ & err & $P_2$ & err & \multicolumn{2}{c}{$E_2$} & $P_{\rm orb}$ & Q \\
\hline
V893 Sco & 2013 & 0.078675 & 0.000025 & 0 & 52 & -- & -- & 0.078288 & 0.000054 & 51 & 153 & 0.075961 & B \\
RZ Sge & 2013 & 0.070642 & 0.000026 & 0 & 58 & 11.5 & 4.2 & -- & -- & -- & -- & 0.06828 & C \\
AW Sge & 2013 & -- & -- & -- & -- & -- & -- & 0.074293 & 0.000025 & 62 & 90 & -- & C \\
V1265 Tau & 2013 & 0.053428 & 0.000024 & 0 & 187 & 1.9 & 1.9 & 0.053086 & 0.000043 & 186 & 319 & -- & B \\
SU UMa & 2013 & -- & -- & -- & -- & -- & -- & 0.078775 & 0.000131 & 0 & 51 & 0.07635 & C \\
SS UMi & 2013 & -- & -- & -- & -- & -- & -- & 0.069936 & 0.000095 & 0 & 59 & 0.06778 & C \\
CU Vel & 2013 & -- & -- & -- & -- & -- & -- & 0.080573 & 0.000043 & 0 & 63 & 0.078054 & B \\
ASAS J224349 & 2013 & 0.069719 & 0.000048 & 0 & 55 & 25.2 & 7.5 & 0.069513 & 0.000015 & 87 & 149 & -- & B \\
ASASSN-13cf & 2013 & 0.058407 & 0.000028 & 0 & 115 & 7.1 & 1.9 & -- & -- & -- & -- & -- & B \\
ASASSN-13cg & 2013 & 0.060228 & 0.000037 & 0 & 63 & 24.4 & 6.9 & -- & -- & -- & -- & -- & C \\
ASASSN-13ck & 2013 & 0.056186 & 0.000010 & 35 & 185 & 5.6 & 0.4 & -- & -- & -- & -- & 0.055348 & AE \\
ASASSN-13cz & 2013 & 0.079773 & 0.000058 & 0 & 13 & -- & -- & -- & -- & -- & -- & -- & C \\
ASASSN-13da & 2013 & 0.071781 & 0.000037 & 56 & 140 & 6.5 & 4.1 & 0.071259 & 0.000309 & 154 & 182 & -- & C \\
ASASSN-14ac & 2014 & 0.058550 & 0.000009 & 57 & 188 & $-$1.7 & 1.2 & -- & -- & -- & -- & -- & B \\
CSS J024354 & 2013 & 0.062076 & 0.000042 & 0 & 129 & -- & -- & -- & -- & -- & -- & -- & CGM \\
MASTER J004527 & 2013 & 0.080365 & 0.000020 & 12 & 50 & $-$3.8 & 4.8 & 0.080004 & 0.000012 & 50 & 144 & -- & A \\
MASTER J005740 & 2013 & 0.057067 & 0.000011 & 14 & 144 & 4.0 & 1.0 & -- & -- & -- & -- & 0.056190 & B \\
MASTER J024847 & 2013 & 0.0644 & 0.0003 & 0 & 2 & -- & -- & -- & -- & -- & -- & -- & C \\
MASTER J061335 & 2013 & 0.056091 & 0.000021 & 61 & 269 & 5.1 & 0.6 & 0.055950 & 0.000072 & 268 & 321 & -- & B \\
MASTER J073208 & 2013 & 0.058836 & 0.000081 & 0 & 38 & -- & -- & -- & -- & -- & -- & -- & C \\
MASTER J162323 & 2013 & 0.088661 & 0.000020 & 35 & 192 & 3.9 & 0.9 & -- & -- & -- & -- & -- & B \\
MASTER J234843 & 2013 & 0.032007 & 0.000005 & 0 & 255 & 1.3 & 0.5 & 0.031977 & 0.000010 & 249 & 438 & -- & C \\
OT J210016 & 2013 & 0.058502 & 0.000020 & 17 & 160 & 2.3 & 1.5 & -- & -- & -- & -- & 0.05787 & CE \\
PNV J191501 & 2013 & 0.058382 & 0.000010 & 58 & 297 & 5.2 & 0.2 & 0.058176 & 0.000033 & 292 & 366 & 0.05706 & AE \\
SSS J094327 & 2014 & 0.070500 & 0.000010 & 14 & 60 & 5.6 & 2.3 & 0.070241 & 0.000048 & 57 & 88 & -- & C \\
TCP J233822 & 2013 & 0.057868 & 0.000014 & 39 & 206 & 2.7 & 1.1 & -- & -- & -- & -- & 0.057255 & AE \\
\hline
\end{tabular}
\end{center}
\end{table*}

\section{Individual Objects}\label{sec:individual}

\subsection{FO Andromedae}\label{obj:foand}

   FO And was discovered as a dwarf nova by \citet{hof67an289205}.
\citet{mei84betaand} showed that the outbursts occur with
intervals of 10--30~d, and there was already likely
a superoutburst.  \citet{mei84foand} reported the detection
of three superoutbursts and the intervals between normal outbursts
were 15--23~d.  This object has been monitored by amateur
observers since 1982, and both AAVSO and VSOLJ observers
detected superoutbursts.  \citet{bru89CVspec1} obtained
a spectrum in quiescence and detected Balmer and He\textsc{ii}
emission lines.  \citet{szk89faintCV2} reported time-resolved
photometry in quiescence without detecting a significant period.
According to \citet{szk89faintCV2}, Grauer and Bond (1986)
detected superhumps with a period of $\sim$105~d, but this
result was not published.

   The first solid publication of the superhumps in this object
was \citet{kat95foand}, whose result was refined in \citet{Pdot}.
\citet{tho96Porb} determined the orbital period by a radial-velocity
study.  \citet{Pdot3} reported superhumps in the 2010 and 2011
superoutbursts.

\begin{figure}
  \begin{center}
    \FigureFile(88mm,70mm){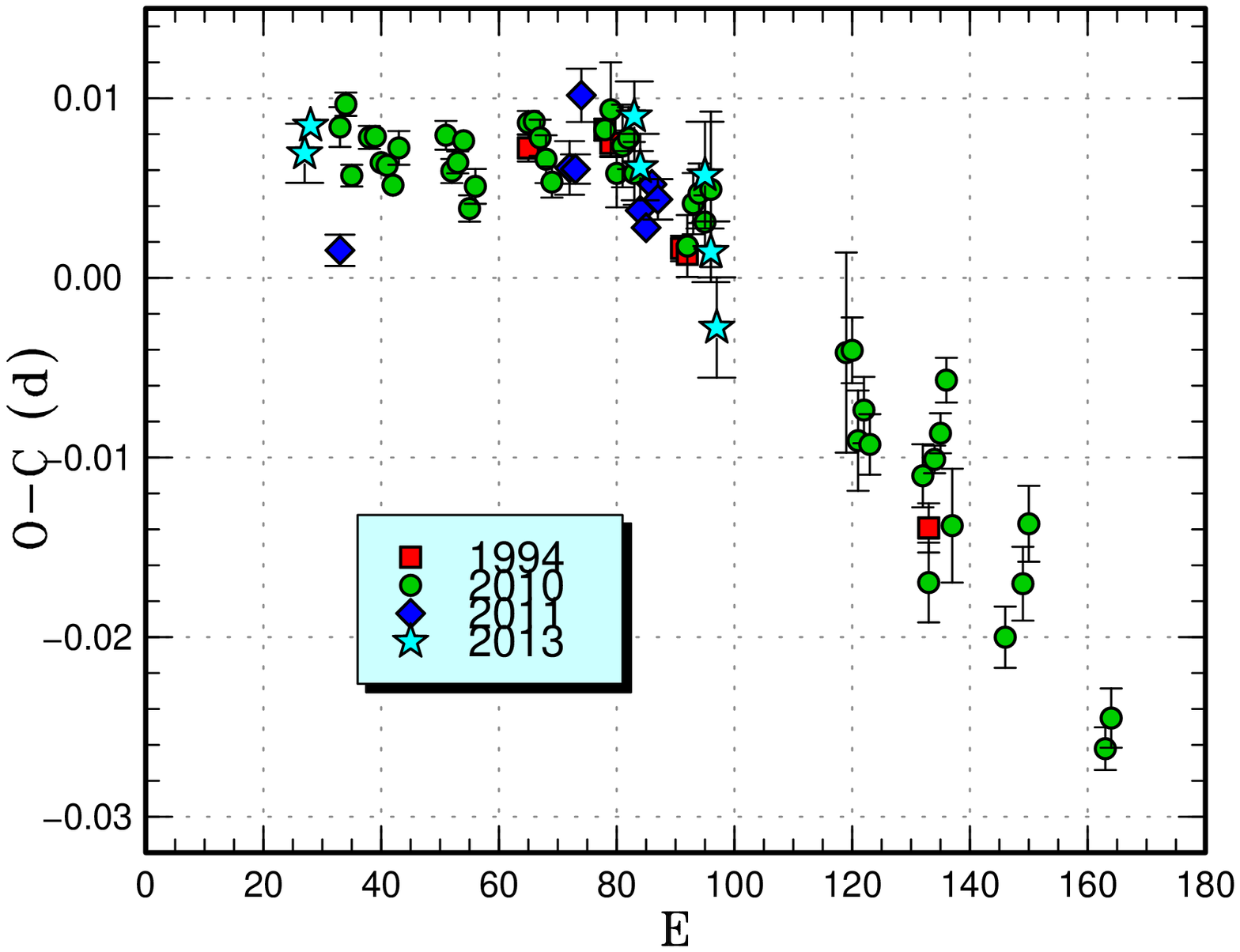}
  \end{center}
  \caption{Comparison of $O-C$ diagrams of FO And between different
  superoutbursts.  A period of 0.07451~d was used to draw this figure.
  Approximate cycle counts ($E$) after the start of the superoutburst
  were used.}
  \label{fig:foandcomp2}
\end{figure}

   The 2013 November--December superoutburst was detected
on November 24 by J. Ripero (vsnet-alert 16647).  Time-series
observations started two nights later, and a total of
three-night observation was obtained (vsnet-alert 16696).
The times of superhump maxima are listed in table
\ref{tab:foandoc2013}.  A comparison of the $O-C$ diagrams
(figure \ref{fig:foandcomp2}) suggests that the obtained
global period is a mixture of different stages.

\begin{table}
\caption{Superhump maxima of FO And (2013)}\label{tab:foandoc2013}
\begin{center}
\begin{tabular}{rp{55pt}p{40pt}r@{.}lr}
\hline
\multicolumn{1}{c}{$E$} & \multicolumn{1}{c}{max\commenta} & \multicolumn{1}{c}{error} & \multicolumn{2}{c}{$O-C$\commentb} & \multicolumn{1}{c}{$N$\commentc} \\
\hline
0 & 56622.9587 & 0.0016 & $-$0&0015 & 51 \\
1 & 56623.0348 & 0.0006 & 0&0001 & 48 \\
56 & 56627.1334 & 0.0019 & 0&0048 & 56 \\
57 & 56627.2050 & 0.0018 & 0&0020 & 50 \\
68 & 56628.0242 & 0.0030 & 0&0024 & 55 \\
69 & 56628.0944 & 0.0017 & $-$0&0018 & 56 \\
70 & 56628.1647 & 0.0028 & $-$0&0060 & 55 \\
\hline
  \multicolumn{6}{l}{\commenta BJD$-$2400000.} \\
  \multicolumn{6}{l}{\commentb Against max $= 2456622.9602 + 0.074435 E$.} \\
  \multicolumn{6}{l}{\commentc Number of points used to determine the maximum.} \\
\end{tabular}
\end{center}
\end{table}

\subsection{DH Aquilae}\label{obj:dhaql}

   DH Aql was discovered as a Mira-type variable
(=HV 3899) with a range of 12.5 to fainter than 16
in photographic range \citep{can25dhaql}.  This element
was also used in \citet{GCVS}.  \citet{tse69dhaqlgmasql}
recorded three outbursts and identified this object
as a dwarf nova.  The durations of outbursts were 3--4~d.
\citet{zhu72dhaql} also reported another outburst and
confirmed the periodicity suggested by \citet{tse69dhaqlgmasql}.
This object has been regularly monitored since 1980
[e.g \citet{bat82DNe3}; \citet{bat82DNe4}; also summarized
in \citet{mas03faintCV}].

   \citet{nog95dhaql} detected superhumps during the 1994
September outburst and established the SU UMa-type
classification.  \citet{bat98dhaql} reported seven superoutbursts
and the supercycle lengths of 259--450~d (361~d in average).

\begin{figure}
  \begin{center}
    \FigureFile(88mm,70mm){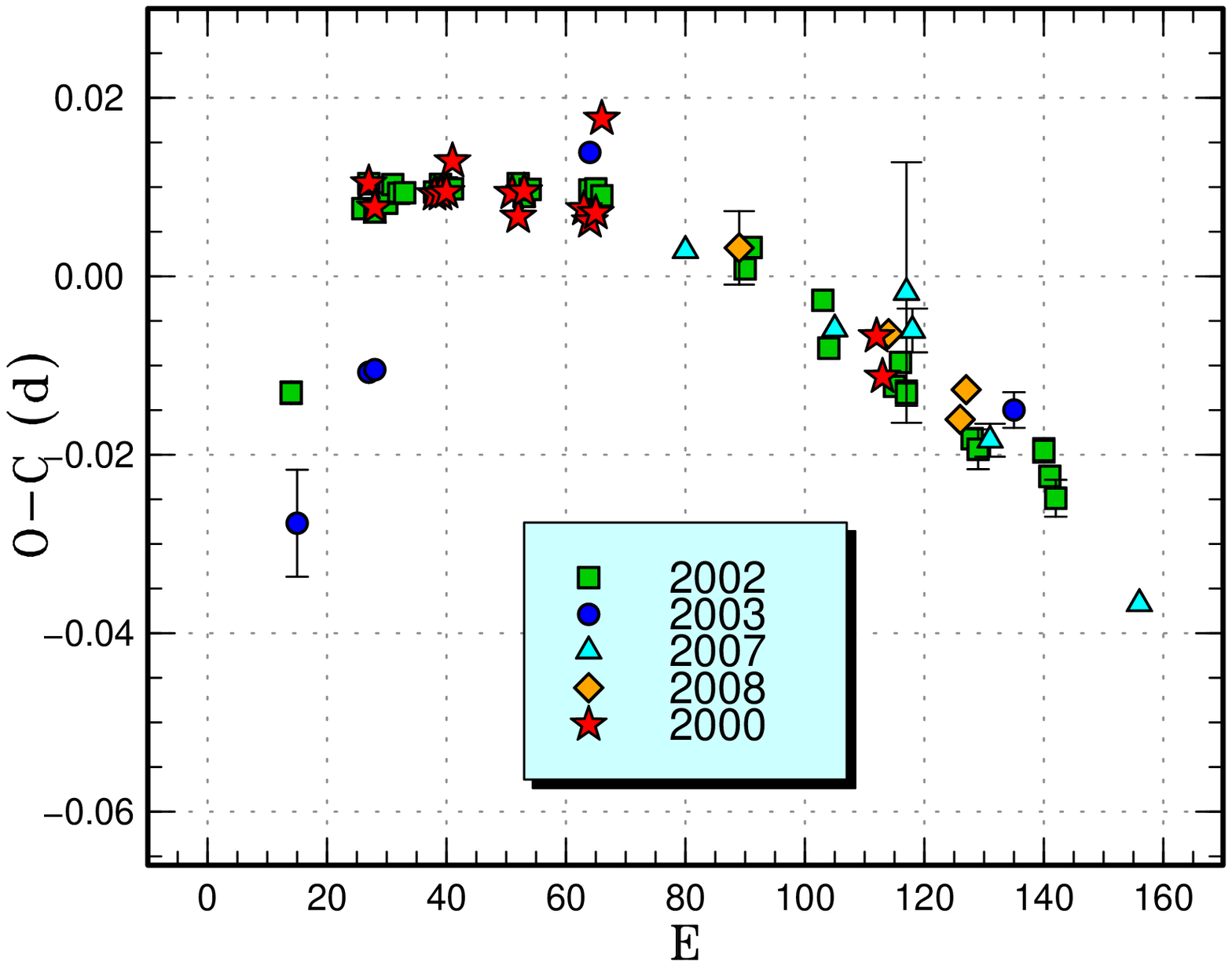}
  \end{center}
  \caption{Comparison of $O-C$ diagrams of DH Aql between different
  superoutbursts.  A period of 0.08000~d was used to draw this figure.
  Approximate cycle counts ($E$) after the start of the outburst
  wes used.
  }
  \label{fig:dhaqlcomp2}
\end{figure}

   \citet{Pdot} reported observations of superoutbursts
in 2002, 2003 and 2008.  We found the data of the unreported
2000 superoutburst (vsnet-alert 5163) and summarize the
result here.  The times of superhump maxima are listed in table
\ref{tab:dhaqloc2000}.  This superoutburst was observed
in relatively early phase and stages B and C can be
recognized.  The resultant $O-C$ diagram agrees with
the others well (figure \ref{fig:dhaqlcomp2}).

\begin{table}
\caption{Superhump maxima of DH Aql (2000)}\label{tab:dhaqloc2000}
\begin{center}
\begin{tabular}{rp{55pt}p{40pt}r@{.}lr}
\hline
\multicolumn{1}{c}{$E$} & \multicolumn{1}{c}{max\commenta} & \multicolumn{1}{c}{error} & \multicolumn{2}{c}{$O-C$\commentb} & \multicolumn{1}{c}{$N$\commentc} \\
\hline
0 & 51759.1254 & 0.0005 & $-$0&0030 & 101 \\
1 & 51759.2027 & 0.0005 & $-$0&0055 & 161 \\
11 & 51760.0042 & 0.0002 & $-$0&0019 & 171 \\
12 & 51760.0842 & 0.0002 & $-$0&0016 & 172 \\
13 & 51760.1645 & 0.0003 & $-$0&0012 & 122 \\
14 & 51760.2479 & 0.0009 & 0&0025 & 49 \\
24 & 51761.0444 & 0.0008 & 0&0011 & 174 \\
25 & 51761.1217 & 0.0003 & $-$0&0013 & 173 \\
26 & 51761.2045 & 0.0011 & 0&0017 & 150 \\
36 & 51762.0026 & 0.0002 & 0&0019 & 138 \\
37 & 51762.0812 & 0.0003 & 0&0007 & 170 \\
38 & 51762.1621 & 0.0004 & 0&0019 & 171 \\
39 & 51762.2527 & 0.0011 & 0&0127 & 44 \\
85 & 51765.9083 & 0.0011 & $-$0&0017 & 54 \\
86 & 51765.9838 & 0.0011 & $-$0&0061 & 66 \\
\hline
  \multicolumn{6}{l}{\commenta BJD$-$2400000.} \\
  \multicolumn{6}{l}{\commentb Against max $= 2451759.1285 + 0.079783 E$.} \\
  \multicolumn{6}{l}{\commentc Number of points used to determine the maximum.} \\
\end{tabular}
\end{center}
\end{table}

\subsection{BB Arietis}\label{obj:bbari}

   This object (=NSV 907) was suspected as a dwarf nova because
this suspected variable is located close to an ROSAT X-ray
source (Kato, vsnet-chat 3317).  In 2004, two outbursts
were detected P. Schmeer, confirming the dwarf nova-type nature.
Superhumps were detected during the second outburst \citep{Pdot}.

\begin{figure}
  \begin{center}
    \FigureFile(88mm,70mm){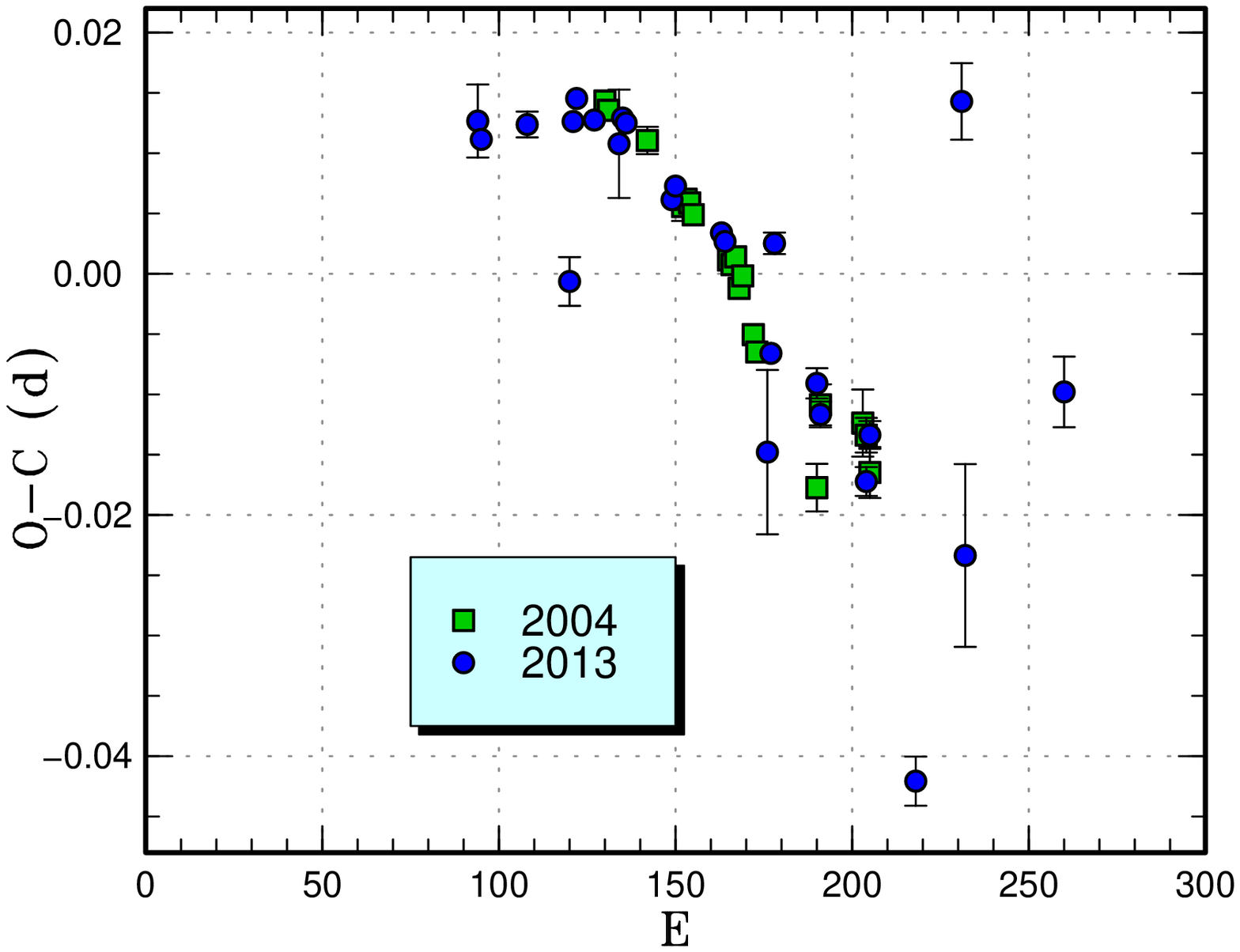}
  \end{center}
  \caption{Comparison of $O-C$ diagrams of BB Ari between different
  superoutbursts.  A period of 0.07254~d was used to draw this figure.
  Approximate cycle counts ($E$) after the precursor outburst
  were used (2013).  Since the start of the 2004 superoutburst
  was not well constrained, we shifted the $O-C$ diagram
  to best fit the 2013 one.
  }
  \label{fig:bbaricomp}
\end{figure}

   On 2013 August 3, ASAS-SN \citep{ASASSN} team detected this object
in outburst (vsnet-alert 16111).  Although the object
once faded rapidly, it brightened five days later showing
superhumps and the initial ASAS-SN detection turned out to be
a precursor outburst (vsnet-alert 16169, 16170).
The times of ssuperhump maxima during the main superoutburst
are listed in table \ref{tab:bbarioc2013}.
There was a stage B-C transition in the $O-C$ data.
Although stage A superhumps were likely recorded between
the precursor and the main superoutburst, we could not
determine the period due to the insufficiency of observations.
The present determination of superhump period confirmed
the suggestion that we observed only stage C superhumps
in 2004 \citep{Pdot} (figure \ref{fig:bbaricomp}).

\begin{table}
\caption{Superhump maxima of BB Ari (2013)}\label{tab:bbarioc2013}
\begin{center}
\begin{tabular}{rp{55pt}p{40pt}r@{.}lr}
\hline
\multicolumn{1}{c}{$E$} & \multicolumn{1}{c}{max\commenta} & \multicolumn{1}{c}{error} & \multicolumn{2}{c}{$O-C$\commentb} & \multicolumn{1}{c}{$N$\commentc} \\
\hline
0 & 56514.8763 & 0.0030 & $-$0&0029 & 17 \\
1 & 56514.9473 & 0.0009 & $-$0&0042 & 24 \\
14 & 56515.8916 & 0.0011 & $-$0&0001 & 44 \\
26 & 56516.7491 & 0.0020 & $-$0&0104 & 25 \\
27 & 56516.8349 & 0.0004 & 0&0031 & 71 \\
28 & 56516.9093 & 0.0004 & 0&0053 & 99 \\
33 & 56517.2702 & 0.0007 & 0&0046 & 69 \\
40 & 56517.7760 & 0.0045 & 0&0042 & 24 \\
41 & 56517.8507 & 0.0007 & 0&0066 & 58 \\
42 & 56517.9228 & 0.0004 & 0&0064 & 74 \\
55 & 56518.8595 & 0.0003 & 0&0029 & 110 \\
56 & 56518.9332 & 0.0005 & 0&0043 & 35 \\
69 & 56519.8723 & 0.0006 & 0&0033 & 52 \\
70 & 56519.9441 & 0.0008 & 0&0029 & 25 \\
82 & 56520.7971 & 0.0068 & $-$0&0119 & 37 \\
83 & 56520.8779 & 0.0004 & $-$0&0035 & 121 \\
84 & 56520.9595 & 0.0009 & 0&0059 & 8 \\
96 & 56521.8184 & 0.0013 & $-$0&0031 & 23 \\
97 & 56521.8884 & 0.0011 & $-$0&0054 & 27 \\
110 & 56522.8258 & 0.0012 & $-$0&0080 & 24 \\
111 & 56522.9022 & 0.0011 & $-$0&0040 & 47 \\
124 & 56523.8165 & 0.0020 & $-$0&0297 & 25 \\
137 & 56524.8159 & 0.0032 & 0&0296 & 56 \\
138 & 56524.8508 & 0.0076 & $-$0&0079 & 67 \\
166 & 56526.8955 & 0.0029 & 0&0120 & 48 \\
\hline
  \multicolumn{6}{l}{\commenta BJD$-$2400000.} \\
  \multicolumn{6}{l}{\commentb Against max $= 2456514.8792 + 0.072315 E$.} \\
  \multicolumn{6}{l}{\commentc Number of points used to determine the maximum.} \\
\end{tabular}
\end{center}
\end{table}

\subsection{UZ Bootis}\label{obj:uzboo}

   UZ Boo is a renowned object selected for a small group of
WZ Sge-type dwarf novae when this subclass was proposed
\citep{bai79wzsge}.  Only a small number of outbursts were
recorded [1929 April, 1937 June, 1938 May, 1978 September
\citep{ric86CVamplitudecyclelength}, 1994 August
\citep{iid94uzbooiauc}, 2003 December \citep{Pdot}].
Although superhumps were first recorded During the 1994
superoutburst, the period was only marginally estimated
to be 0.0619~d \citep{kat01hvvir} due to the very unfavorable
condition.  During the 2003 superoutburst, the superhump
period was established to be 0.06192(3)~d (stage B,
\cite{Pdot}) despite the unfavorable seasonal observing
condition.  The presence of multiple post-superoutburst rebrightenings
was suspected during the 1994 superoutburst \citep{kuu96TOAD},
although the detections of these rebrightenings were based on
visual observations.  It took additional two years before
the phenomenon of multiple post-superoutburst rebrightenings
was established in dwarf novae (EG Cnc: cf. \cite{osa97egcnc},
\cite{pat98egcnc}, \cite{kat04egcnc}).  During the 2003
superoutburst, UZ Boo showed four post-superoutburst
rebrightenings \citep{Pdot} and the close resemblance with
EG Cnc was highlighted.

   The object was again detected in outburst on 2013 July 26
by C. Chiselbrook visually and confirmed by W. MacDonald II
using a CCD.  Since the object was not detected by the
same observer on July 25, the outburst appeared to be
a young one.  Although this outburst was supposed to provide
an opportunity to detect early superhumps, no meaningful
coherent period was detected (vsnet-alert 16064, 16065,
16075) partly due to the high air mass and the short
visibility in the evening.  Only three days after the outburst
detection, likely growing ordinary superhumps were detected
(vsnet-alert 16080, 16087).  The period of early superhumps,
if there was any, was very short compared to other WZ Sge-type
dwarf novae.

   Since the comparison star was much redder than the variable
and some observations were done at high airmasses, 
we corrected observations by using a second-order atmospheric 
extinction whose coefficients were experimentally determined.
The times of superhump maxima during the plateau phase
are listed in table \ref{tab:uzboooc2013}.  The times of
superhumps were not determined by the fitting method
on July 29 (BJD 2456503).  Although the amplitudes of
superhumps grew during the initial three nights and
they were likely stage A superhumps, the stages were not
distinct on the $O-C$ diagram.  This was probably due to
the shortness of stage A itself and limited observation.
We used $E \ge 15$ for determining the period in table
\ref{tab:perlist} to avoid the inclusion of stage A
superhumps.  Using the data for BJD 2456503--2456506,
we obtained a period of 0.06210(5)~d with the PDM method.
We regard it the likely period of stage A superhumps.
The superhump period was almost constant during stage B
and no clear transition to stage C was recorded.

   After the rapid fading from the superoutburst, individual
times of superhumps could not be measured due to the
faintness.  We could, however, detect signals with the
PDM method.  During the interval BJD 2456511--2456515
(the ``dip'' after the fading), we obtained a period of
0.0610(1)~d.  During the interval BJD 2456515--2456522
(the initial two rebrightenings), we detected a period
of 0.06182(4)~d.  During the interval BJD 2456522--2456532
(the last two rebrightenings), we detected a period
of 0.06197(4)~d.  These periods suggest that superhumps
persisted during the entire rebrightening phase.

\begin{figure}
  \begin{center}
    \FigureFile(88mm,70mm){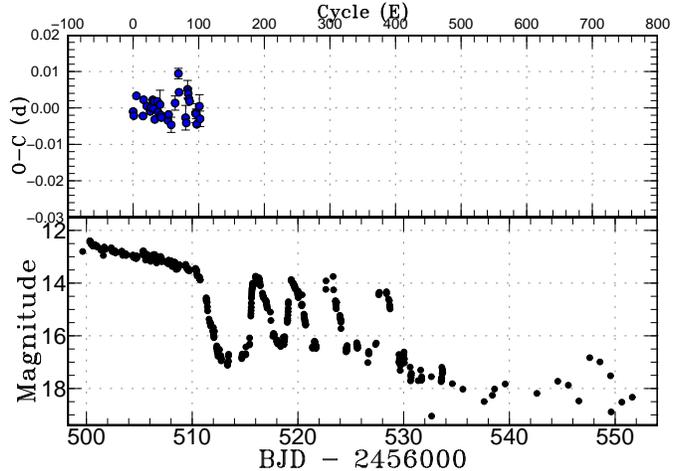}
  \end{center}
  \caption{$O-C$ diagram of superhumps in UZ Boo (2013).
     (Upper): $O-C$ diagram.  A period of 0.062032~d
     was used to draw this figure.
     (Lower): Light curve.  The observations were binned to 0.012~d.}
  \label{fig:uzboo2013humpall}
\end{figure}

   The object underwent four post-superoutburst
rebrightenings as in the 2003 superoutburst
(figure \ref{fig:uzboo2013humpall}).  The object slightly
brightened when superhump appeared.  This phenomenon
is common to what was observed in objects with
multiple rebrightenings [EG Cnc \citep{pat98egcnc};
EZ Lyn \citep{Pdot3}; 
MASTER OT J211258.65$+$242145.4 and
MASTER OT J203749.39$+$552210.3 \citep{nak13j2112j2037};
for a complete list, see \citet{nak13j2112j2037}].
It looks like this phenomenon is more apparent
in systems with multiple rebrightenings.

   The mean length of the supercycle has been updated
3170(110)~d assuming that one superoutburst escaped
detection around 1986.

\begin{table}
\caption{Superhump maxima of UZ Boo (2013)}\label{tab:uzboooc2013}
\begin{center}
\begin{tabular}{rp{55pt}p{40pt}r@{.}lr}
\hline
\multicolumn{1}{c}{$E$} & \multicolumn{1}{c}{max\commenta} & \multicolumn{1}{c}{error} & \multicolumn{2}{c}{$O-C$\commentb} & \multicolumn{1}{c}{$N$\commentc} \\
\hline
0 & 56504.4170 & 0.0005 & $-$0&0007 & 136 \\
1 & 56504.4762 & 0.0006 & $-$0&0034 & 110 \\
5 & 56504.7297 & 0.0003 & 0&0020 & 90 \\
15 & 56505.3481 & 0.0002 & 0&0004 & 106 \\
16 & 56505.4120 & 0.0004 & 0&0024 & 52 \\
21 & 56505.7209 & 0.0003 & 0&0013 & 91 \\
26 & 56506.0283 & 0.0005 & $-$0&0013 & 266 \\
27 & 56506.0914 & 0.0009 & $-$0&0002 & 171 \\
30 & 56506.2797 & 0.0004 & 0&0022 & 127 \\
31 & 56506.3391 & 0.0006 & $-$0&0004 & 125 \\
32 & 56506.4039 & 0.0004 & 0&0024 & 222 \\
33 & 56506.4631 & 0.0008 & $-$0&0004 & 52 \\
37 & 56506.7129 & 0.0004 & 0&0014 & 106 \\
38 & 56506.7725 & 0.0006 & $-$0&0010 & 58 \\
41 & 56506.9606 & 0.0039 & 0&0011 & 117 \\
42 & 56507.0198 & 0.0002 & $-$0&0017 & 424 \\
43 & 56507.0811 & 0.0003 & $-$0&0024 & 445 \\
53 & 56507.7009 & 0.0005 & $-$0&0026 & 91 \\
54 & 56507.7644 & 0.0006 & $-$0&0011 & 62 \\
58 & 56508.0069 & 0.0024 & $-$0&0065 & 93 \\
64 & 56508.3879 & 0.0019 & 0&0025 & 25 \\
69 & 56508.7038 & 0.0015 & 0&0084 & 84 \\
70 & 56508.7620 & 0.0011 & 0&0046 & 88 \\
80 & 56509.3604 & 0.0015 & $-$0&0169 & 28 \\
81 & 56509.4368 & 0.0008 & $-$0&0025 & 129 \\
83 & 56509.5668 & 0.0021 & 0&0035 & 63 \\
84 & 56509.6294 & 0.0015 & 0&0041 & 76 \\
85 & 56509.6910 & 0.0009 & 0&0038 & 69 \\
86 & 56509.7519 & 0.0009 & 0&0026 & 89 \\
95 & 56510.3080 & 0.0035 & 0&0007 & 64 \\
96 & 56510.3681 & 0.0008 & $-$0&0011 & 132 \\
97 & 56510.4276 & 0.0013 & $-$0&0037 & 61 \\
101 & 56510.6821 & 0.0032 & 0&0029 & 55 \\
102 & 56510.7407 & 0.0021 & $-$0&0005 & 67 \\
\hline
  \multicolumn{6}{l}{\commenta BJD$-$2400000.} \\
  \multicolumn{6}{l}{\commentb Against max $= 2456504.4177 + 0.0619954 E$.} \\
  \multicolumn{6}{l}{\commentc Number of points used to determine the maximum.} \\
\end{tabular}
\end{center}
\end{table}

\subsection{V342 Camelopardalis}\label{obj:v342cam}

   V342 Cam (=1RXS J042332$+$745300 =HS 0417$+$7445) was
selected as a CV as an ROSAT X-ray source \citep{wu01j0209j0423}
and spectroscopic survey \citep{aun06HSCV}.
\citet{Pdot} and \citet{Pdot2} reported observations of superhumps
during the 2008 and 2010 superoutburst, respectively.
\citet{she11j0423} also presented analysis of the 2008
superoutburst and examined outburst behavior during 
the 2005--2010 interval.  \citet{she11j0423} photometrically
obtained an orbital period of 0.07531(8)~d.

\begin{figure}
  \begin{center}
    \FigureFile(88mm,70mm){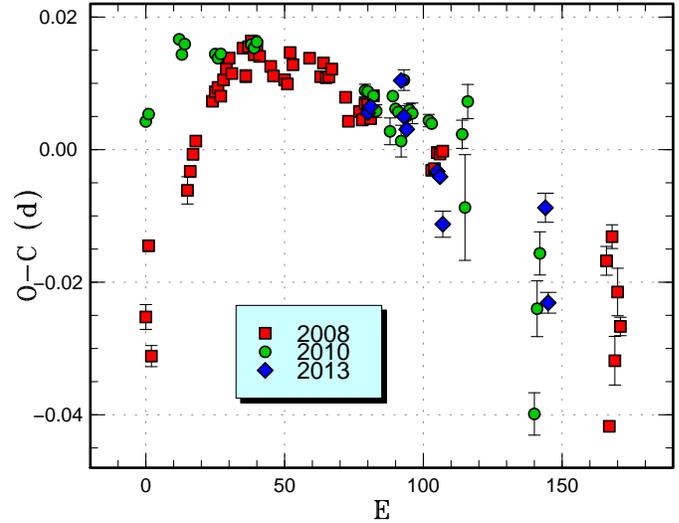}
  \end{center}
  \caption{Comparison of $O-C$ diagrams of V342 Cam between different
  superoutbursts.  A period of 0.07845~d was used to draw this figure.
  Approximate cycle counts ($E$) after the start of the superoutburst
  were used.  Since the start of the 2013 superoutburst
  was not well constrained, we shifted the $O-C$ diagram
  to best fit the others.
  }
  \label{fig:v342camcomp2}
\end{figure}

   On 2013 August 1, ASAS-SN detected an outburst
(vsnet-alert 16096).  The outburst was not young enough
and only stage C superhumps were recorded
(table \ref{tab:v342camoc2013}).  One the final night,
the profile was double-humped.  We showed the maxima
on the smooth extension of stage C superhumps in the table.
A comparison of $O-C$ diagrams of between different
superoutbursts is shown in figure \ref{fig:v342camcomp2}.

\begin{table}
\caption{Superhump maxima of V342 Cam (2013)}\label{tab:v342camoc2013}
\begin{center}
\begin{tabular}{rp{55pt}p{40pt}r@{.}lr}
\hline
\multicolumn{1}{c}{$E$} & \multicolumn{1}{c}{max\commenta} & \multicolumn{1}{c}{error} & \multicolumn{2}{c}{$O-C$\commentb} & \multicolumn{1}{c}{$N$\commentc} \\
\hline
0 & 56511.4451 & 0.0005 & $-$0&0019 & 78 \\
1 & 56511.5245 & 0.0005 & $-$0&0006 & 72 \\
12 & 56512.3914 & 0.0011 & 0&0076 & 70 \\
13 & 56512.4645 & 0.0008 & 0&0025 & 78 \\
14 & 56512.5410 & 0.0005 & 0&0010 & 70 \\
25 & 56513.3975 & 0.0006 & $-$0&0012 & 79 \\
26 & 56513.4752 & 0.0006 & $-$0&0016 & 73 \\
27 & 56513.5465 & 0.0020 & $-$0&0084 & 55 \\
64 & 56516.4516 & 0.0022 & 0&0083 & 83 \\
65 & 56516.5157 & 0.0016 & $-$0&0057 & 83 \\
\hline
  \multicolumn{6}{l}{\commenta BJD$-$2400000.} \\
  \multicolumn{6}{l}{\commentb Against max $= 2456511.4471 + 0.078067 E$.} \\
  \multicolumn{6}{l}{\commentc Number of points used to determine the maximum.} \\
\end{tabular}
\end{center}
\end{table}

\subsection{V452 Cassiopeiae}\label{obj:v452cas}

   V452 Cas was discovered as a dwarf nova (=S 10453)
with a range of 14--17.5 (photographic magnitudes)
by \citet{ric69v452cas}.  \citet{BruchCVatlas} also detected
an outburst.  \citet{liu00CVspec3} obtained a spectrum
in quiescence and detected H$\alpha$ line in emission.

   Although the object has been monitored visually by
amateur observers since 1992, no secure outburst had been
detected since 1999 October 8, when P. Schmeer detected
an outburst of 16.2 mag (unfiltered CCD; vsnet-alert 3561).
Schmeer suspected that past possible visual detections
were likely the close companion star rather than true
outbursts.  The object further brightened to 15.52 mag
on the next night.  On 1999 November 9, P. Schmeer reported
(private communication) a bright (around 15.0 mag) outburst
and suspected it a superoutburst.  This bright outburst
was also confirmed visually at magnitude 14.7--14.9
(vsnet-alert 3684).  Schmeer suspected that this object
is an SU UMa-type dwarf nova with (rather) frequent
small-amplitude outbursts (vsnet-alert 3687).
T. Vanmunster detected superhumps with a period of 0.0891~d
(vsnet-alert 3698, 3707).  Superhumps were also detected
during the 2000 September outburst (vsnet-alert 5276).

   \citet{she09v452cas} systematically studied this object
between 2005 and 2008, and obtained supercycle lengths
of 146$\pm$16~d.  \citet{she09v452cas} also reported
superhumps during the 2007 September superoutburst.
The long-period superhumps detected in the early stage
were likely stage A superhumps.
An analysis of the 1999 and 2008 superoutburst was reported
in \citet{Pdot}.

\begin{figure}
  \begin{center}
    \FigureFile(88mm,70mm){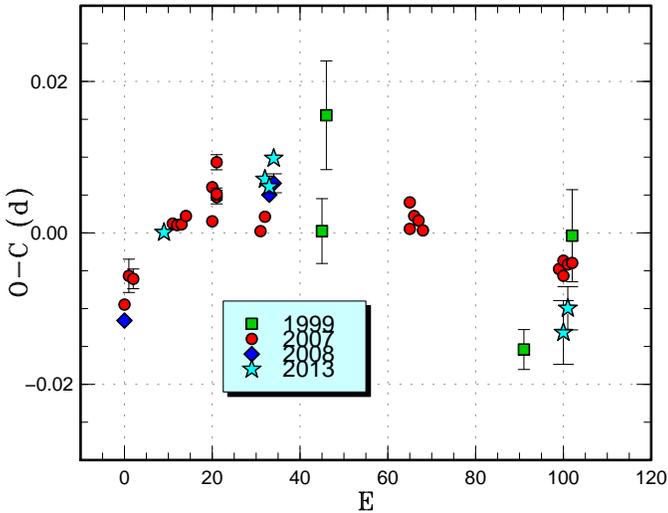}
  \end{center}
  \caption{Comparison of $O-C$ diagrams of V452 Cas between different
  superoutbursts.  A period of 0.08880~d was used to draw this figure.
  Approximate cycle counts ($E$) after the start of the superoutburst
  were used.}
  \label{fig:v452cascomp2}
\end{figure}

   I. Miller detected a superoutburst on 2013 Novmeber 19
and detected a superhump (vsnet-alert 16632).
The times of superhump maxima are listed in table
\ref{tab:v452casoc2013}.  On BJD 2456624, secondary maxima
of superhumps became strong.  In the table, we listed
maxima which are in a smooth extension of earlier times
of maxima.  A comparison of $O-C$ diagram (figure
\ref{fig:v452cascomp2}) indicates that the evolution of
superhumps during this superoutburst followed the trend
previously recorded.  The initial epoch likely detected
the time near stage A-B transition.  In table \ref{tab:perlist},
a global $P_{\rm dot}$ including all stages is given.

   The preceding superoutbursts occurred in 2013 January
and June (BJD 2456305 and 2456445).  The supercycles
between these three superoutburst were 140~d and 171~d,
suggesting that the supercycle significantly varies.

\begin{table}
\caption{Superhump maxima of V452 Cas (2013)}\label{tab:v452casoc2013}
\begin{center}
\begin{tabular}{rp{55pt}p{40pt}r@{.}lr}
\hline
\multicolumn{1}{c}{$E$} & \multicolumn{1}{c}{max\commenta} & \multicolumn{1}{c}{error} & \multicolumn{2}{c}{$O-C$\commentb} & \multicolumn{1}{c}{$N$\commentc} \\
\hline
0 & 56616.2975 & 0.0004 & $-$0&0086 & 117 \\
23 & 56618.3469 & 0.0007 & 0&0031 & 92 \\
24 & 56618.4347 & 0.0005 & 0&0023 & 131 \\
25 & 56618.5273 & 0.0010 & 0&0063 & 69 \\
91 & 56624.3651 & 0.0042 & $-$0&0033 & 41 \\
92 & 56624.4570 & 0.0029 & 0&0001 & 38 \\
\hline
  \multicolumn{6}{l}{\commenta BJD$-$2400000.} \\
  \multicolumn{6}{l}{\commentb Against max $= 2456616.3061 + 0.088596 E$.} \\
  \multicolumn{6}{l}{\commentc Number of points used to determine the maximum.} \\
\end{tabular}
\end{center}
\end{table}

\subsection{V359 Centauri}\label{obj:v359cen}

   V359 Cen was originally discovered as a possible nova by
A. Opolski [originally in Lw\'ow Contr. 4
\citep{pra41VScatalog}.  The object was visible between
1930 April 20 and 27 and faded from 13.8 to 15.0 mag.
Assuming a typical absolute maximum for a nova,
a large distance of 160 kpc was inferred \citep{mcl45novadist}.
\citet{mcl45novadist} already discussed the possibility
of a dwarf nova.  \citet{due87novaatlas} proposed a 21.0 mag
quiescent counterpart.  \citet{mun98CVspec5} tried to study
the proposed quiescent counterpart spectroscopically, but
the attempt failed due to the faintness.  \citet{gil98v359cen}
could not find a nova shell in a deep image.

   The second historical outburst was recorded by R. Stubbings
on 1999 July 13.  \citet{wou01v359cenxzeriyytel} obtained
time-resolved CCD photometry following the 1999 July outburst
and detected a period of 0.0779 d.  The object underwent
further outbursts in 2000 May, 2001 April and 2002 June,
and the object was recognized as an SU UMa-type dwarf nova.
The detection of superhumps during the 1999 and 2002
superoutbursts was reported in \citet{kat02v359cen}.

\begin{figure}
  \begin{center}
    \FigureFile(88mm,70mm){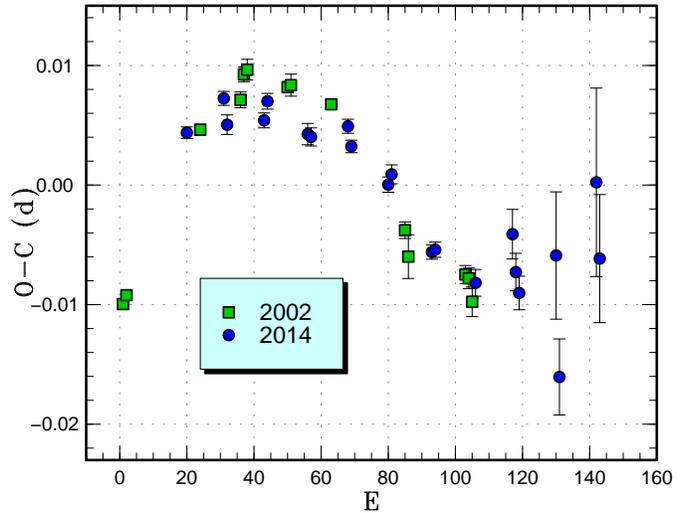}
  \end{center}
  \caption{Comparison of $O-C$ diagrams of V359 Cen between different
  superoutbursts.  A period of 0.08110~d was used to draw this figure.
  Approximate cycle counts ($E$) after the start of the superoutburst
  were used.  Since the start of the 2014 superoutburst
  was not well constrained, we shifted the $O-C$ diagram
  to best fit the 2002 one.}
  \label{fig:v359cencomp}
\end{figure}

   The 2014 superoutburst was detected by R. Stubbings
(cf. vsnet-alert 16941).  Subsequent observations detected
superhumps (vsnet-alert 16946, 16948, 16952).
The times of superhump maxima are listed in table
\ref{tab:v359cenoc2014}.  The observation recorded
the middle to later part of the superoutburst.
The epochs for $E \ge 97$ correspond to the rapidly
fading part of the superoutburst.  The times of maxima
and the identification of the superhumps was not secure
due to the faintness of the object.
The combined $O-C$ diagrams of the 2002 and 2014
superoutburst agree with each other (figure \ref{fig:v359cencomp}).

\begin{table}
\caption{Superhump maxima of V359 Cen (2014)}\label{tab:v359cenoc2014}
\begin{center}
\begin{tabular}{rp{55pt}p{40pt}r@{.}lr}
\hline
\multicolumn{1}{c}{$E$} & \multicolumn{1}{c}{max\commenta} & \multicolumn{1}{c}{error} & \multicolumn{2}{c}{$O-C$\commentb} & \multicolumn{1}{c}{$N$\commentc} \\
\hline
0 & 56712.8358 & 0.0005 & $-$0&0034 & 17 \\
11 & 56713.7308 & 0.0006 & 0&0009 & 19 \\
12 & 56713.8097 & 0.0008 & $-$0&0011 & 20 \\
23 & 56714.7021 & 0.0006 & 0&0007 & 16 \\
24 & 56714.7848 & 0.0007 & 0&0025 & 19 \\
36 & 56715.7553 & 0.0009 & 0&0014 & 20 \\
37 & 56715.8362 & 0.0008 & 0&0013 & 19 \\
48 & 56716.7291 & 0.0006 & 0&0037 & 20 \\
49 & 56716.8086 & 0.0005 & 0&0021 & 20 \\
60 & 56717.6975 & 0.0006 & 0&0004 & 17 \\
61 & 56717.7794 & 0.0008 & 0&0014 & 20 \\
73 & 56718.7461 & 0.0006 & $-$0&0034 & 20 \\
74 & 56718.8274 & 0.0006 & $-$0&0031 & 20 \\
86 & 56719.7978 & 0.0011 & $-$0&0042 & 20 \\
97 & 56720.6940 & 0.0021 & 0&0013 & 19 \\
98 & 56720.7719 & 0.0016 & $-$0&0017 & 19 \\
99 & 56720.8513 & 0.0014 & $-$0&0033 & 19 \\
110 & 56721.7465 & 0.0053 & 0&0013 & 20 \\
111 & 56721.8175 & 0.0032 & $-$0&0087 & 20 \\
122 & 56722.7259 & 0.0079 & 0&0091 & 15 \\
123 & 56722.8006 & 0.0054 & 0&0029 & 16 \\
\hline
  \multicolumn{6}{l}{\commenta BJD$-$2400000.} \\
  \multicolumn{6}{l}{\commentb Against max $= 2456712.8393 + 0.080963 E$.} \\
  \multicolumn{6}{l}{\commentc Number of points used to determine the maximum.} \\
\end{tabular}
\end{center}
\end{table}

\subsection{FZ Ceti}\label{obj:fzcet}

   This object was discovered as a variable star of unknown
type (=BV 1187, NSV 601) with a range of 12.2 to fainter than 14.4
in photographic magnitude \citep{ave68fzcet}.  The object
was also selected as a faint blue star (=PHL 3637)\footnote{
   The name in the Downes CV catalog \citep{DownesCVatlas3}
   is incorrect.
} with a photographic magnitude of 18.7 and $U-V=-0.2$
The identification in \citet{dem96VSpos} was incorrect.
\citep{har62PHL}.  S. Otero found in 2005 that this object
is a dwarf nova based on ASAS-3 \citep{ASAS3} 
observations (vsnet-alert 8620).
The ASAS-3 light curve immediately suggested that
this object is an SU UMa-type dwarf nova showing
superoutbursts (vsnet-alert 8621).  The object was given
the GCVS name FZ Cet in \citet{NameList79}.
Despite its brightness in outburst, the two superoutbursts
occurred in unfavorable seasonal condition
(2010 February and 2012 February) and superhumps had not
been detected until 2014.

\begin{figure}
  \begin{center}
    \FigureFile(88mm,110mm){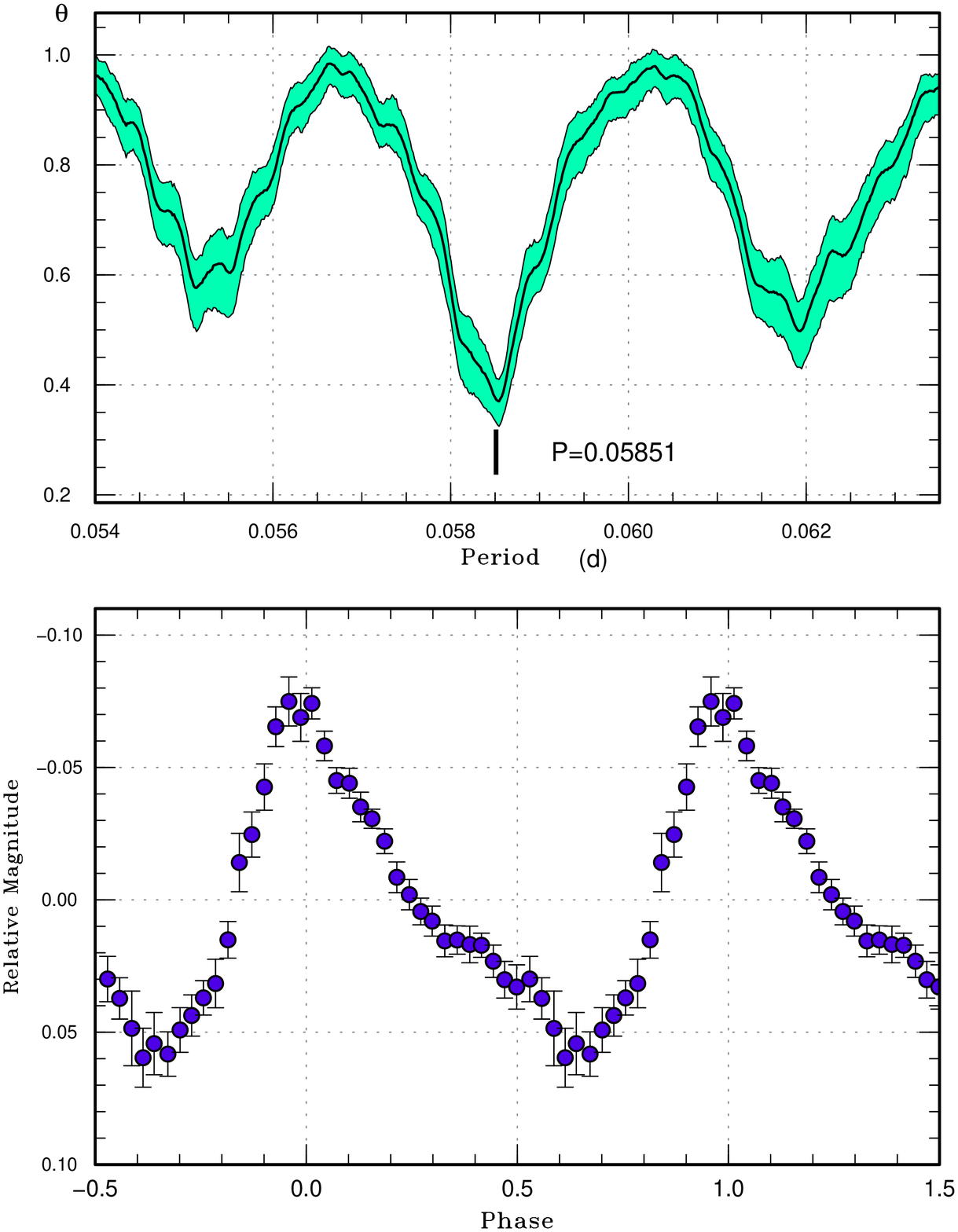}
  \end{center}
  \caption{Superhumps in FZ Cet (2013).  (Upper): PDM analysis.
     (Lower): Phase-averaged profile.}
  \label{fig:fzcetshpdm}
\end{figure}

   The 2014 outburst was detected by R. Stubbings on
January 19 (vsnet-alert 16797).  This outburst was in better
seasonal condition than the preceding two superoutbursts,
and subsequent observations detected superhumps
(vsnet-alert 16803, 16808; figure \ref{fig:fzcetshpdm}).
The times of superhump maxima are listed in table
\ref{tab:fzcetoc2014}.
The superoutburst lasted more than 11~d (vsnet-alert 16845).

\begin{table}
\caption{Superhump maxima of FZ Cet (2014)}\label{tab:fzcetoc2014}
\begin{center}
\begin{tabular}{rp{55pt}p{40pt}r@{.}lr}
\hline
\multicolumn{1}{c}{$E$} & \multicolumn{1}{c}{max\commenta} & \multicolumn{1}{c}{error} & \multicolumn{2}{c}{$O-C$\commentb} & \multicolumn{1}{c}{$N$\commentc} \\
\hline
0 & 56678.0315 & 0.0002 & 0&0012 & 106 \\
1 & 56678.0889 & 0.0003 & 0&0000 & 114 \\
9 & 56678.5564 & 0.0006 & $-$0&0008 & 17 \\
17 & 56679.0233 & 0.0020 & $-$0&0023 & 56 \\
18 & 56679.0821 & 0.0003 & $-$0&0020 & 107 \\
19 & 56679.1463 & 0.0015 & 0&0036 & 33 \\
43 & 56680.5482 & 0.0023 & 0&0004 & 12 \\
\hline
  \multicolumn{6}{l}{\commenta BJD$-$2400000.} \\
  \multicolumn{6}{l}{\commentb Against max $= 2456678.0303 + 0.058547 E$.} \\
  \multicolumn{6}{l}{\commentc Number of points used to determine the maximum.} \\
\end{tabular}
\end{center}
\end{table}

\subsection{YZ Cancri}\label{obj:yzcnc}

\begin{figure}
  \begin{center}
    \FigureFile(88mm,70mm){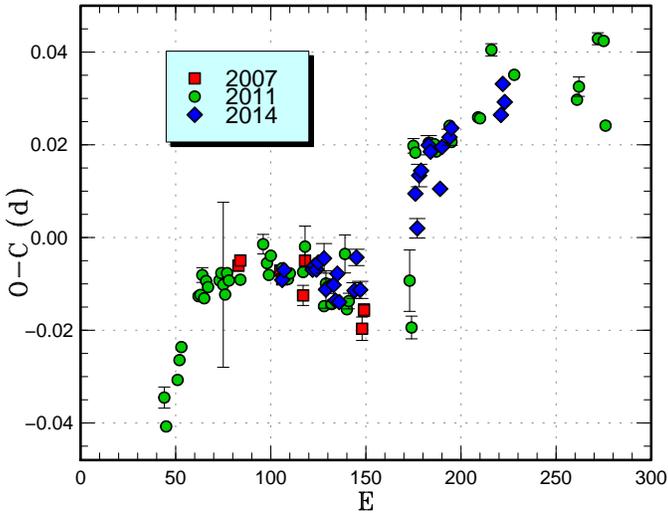}
  \end{center}
  \caption{Comparison of $O-C$ diagrams of YZ Cnc between different
  superoutbursts.  A period of 0.09050~d was used to draw this figure.
  Approximate cycle counts ($E$) after the start of the superoutburst
  were used (in the case of YZ Cnc,
  this refers to the precursor outburst).
  Since the start of the 2014 superoutburst
  was not well constrained, we shifted the $O-C$ diagram
  to best fit the others.}
  \label{fig:yzcnccomp}
\end{figure}

   YZ Cnc is a well-known active SU UMa-type dwarf nova
(e.g. \cite{szk84AAVSO}).  Although \citet{pat79SH} detected
superhumps, the identification of the period was incorrect
\citep{Pdot}.  The 2007 and 2011 superoutbursts were reported
in \citet{Pdot}.  The 2014 superoutburst was reported
in \citet{Pdot4}.  We here reported another superoutburst
in 2014 January.
The times of superhump maxima during the main superoutburst
are listed in table \ref{tab:yzcncoc2014}.
Since we could not determine whether there was a jump
in the phase (as in traditional late superhumps)
after the fading from the main superoutburst, we listed
the times of superhumps after the superoutburst
separately (table \ref{tab:yzcncoc2014b}).
A comparison of thr $O-C$ diagrams (figure \ref{fig:yzcnccomp})
suggest that post-superoutburst superhumps in the 2014
superoutburst were indeed traditional late superhumps.

\begin{table}
\caption{Superhump maxima of YZ Cnc (2014)}\label{tab:yzcncoc2014}
\begin{center}
\begin{tabular}{rp{55pt}p{40pt}r@{.}lr}
\hline
\multicolumn{1}{c}{$E$} & \multicolumn{1}{c}{max\commenta} & \multicolumn{1}{c}{error} & \multicolumn{2}{c}{$O-C$\commentb} & \multicolumn{1}{c}{$N$\commentc} \\
\hline
0 & 56678.6122 & 0.0016 & $-$0&0023 & 23 \\
1 & 56678.7048 & 0.0011 & $-$0&0001 & 21 \\
16 & 56680.0624 & 0.0006 & 0&0011 & 100 \\
17 & 56680.1538 & 0.0007 & 0&0021 & 201 \\
18 & 56680.2434 & 0.0005 & 0&0013 & 201 \\
19 & 56680.3355 & 0.0013 & 0&0029 & 48 \\
22 & 56680.6079 & 0.0032 & 0&0041 & 21 \\
23 & 56680.6916 & 0.0040 & $-$0&0026 & 20 \\
27 & 56681.0546 & 0.0011 & $-$0&0013 & 100 \\
28 & 56681.1417 & 0.0011 & $-$0&0046 & 101 \\
29 & 56681.2380 & 0.0011 & 0&0013 & 99 \\
30 & 56681.3224 & 0.0009 & $-$0&0047 & 82 \\
38 & 56682.0489 & 0.0017 & $-$0&0016 & 55 \\
39 & 56682.1466 & 0.0018 & 0&0056 & 48 \\
41 & 56682.3205 & 0.0018 & $-$0&0013 & 38 \\
\hline
  \multicolumn{6}{l}{\commenta BJD$-$2400000.} \\
  \multicolumn{6}{l}{\commentb Against max $= 2456678.6145 + 0.090422 E$.} \\
  \multicolumn{6}{l}{\commentc Number of points used to determine the maximum.} \\
\end{tabular}
\end{center}
\end{table}

\begin{table}
\caption{Superhump maxima of YZ Cnc (2014) (post-superoutburst)}\label{tab:yzcncoc2014b}
\begin{center}
\begin{tabular}{rp{55pt}p{40pt}r@{.}lr}
\hline
\multicolumn{1}{c}{$E$} & \multicolumn{1}{c}{max\commenta} & \multicolumn{1}{c}{error} & \multicolumn{2}{c}{$O-C$\commentb} & \multicolumn{1}{c}{$N$\commentc} \\
\hline
0 & 56684.9658 & 0.0015 & $-$0&0020 & 68 \\
1 & 56685.0488 & 0.0021 & $-$0&0098 & 65 \\
2 & 56685.1507 & 0.0024 & 0&0011 & 66 \\
3 & 56685.2422 & 0.0016 & 0&0017 & 68 \\
7 & 56685.6098 & 0.0021 & 0&0056 & 24 \\
8 & 56685.6989 & 0.0011 & 0&0038 & 22 \\
13 & 56686.1433 & 0.0015 & $-$0&0064 & 68 \\
14 & 56686.2429 & 0.0007 & 0&0023 & 68 \\
18 & 56686.6070 & 0.0017 & 0&0027 & 19 \\
19 & 56686.6994 & 0.0011 & 0&0042 & 22 \\
45 & 56689.0553 & 0.0010 & $-$0&0038 & 66 \\
46 & 56689.1525 & 0.0015 & 0&0025 & 68 \\
47 & 56689.2391 & 0.0012 & $-$0&0019 & 67 \\
\hline
  \multicolumn{6}{l}{\commenta BJD$-$2400000.} \\
  \multicolumn{6}{l}{\commentb Against max $= 2456684.9678 + 0.090918 E$.} \\
  \multicolumn{6}{l}{\commentc Number of points used to determine the maximum.} \\
\end{tabular}
\end{center}
\end{table}

\subsection{GZ Cancri}\label{obj:gzcnc}

   GZ Cnc is a variable star (=TmzV34) discovered by
K. Takamizawa.  The object turned out to be an active
dwarf nova (\cite{kat01gzcnc}; \cite{kat01gzcnc};
\cite{kat02gzcncnsv10934}).  \citet{tap03gzcnc}
obtained the orbital period of 0.08825(28)~d by
radial-velocity observations.  This period indicates
that GZ Cnc is located in the period gap, and it became
an interesting question whether GZ Cnc is an SU UMa-type
dwarf nova.  In 2010 March, a long outburst turned out
to be a superoutburst \citep{Pdot2}.  Another superoutburst
was recorded in 2013 February \citep{Pdot5}.

\begin{figure}
  \begin{center}
    \FigureFile(88mm,70mm){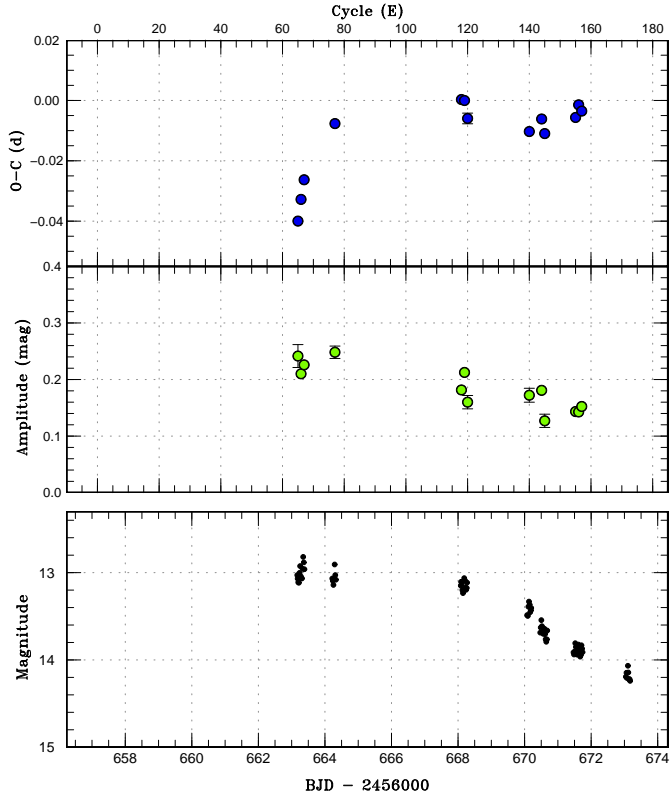}
  \end{center}
  \caption{$O-C$ diagram of superhumps in GZ Cnc (2014).
     (Upper): $O-C$ diagram.  A period of 0.09270~d
     was used to draw this figure.  Approximate cycle counts ($E$)
     after the start of the superoutburst were used.  
     (Middle): Amplitudes of the superhumps.
     (Lower): Light curve.  The observations were binned to 0.019~d.}
  \label{fig:gzcnc2014humpamp}
\end{figure}

\begin{figure}
  \begin{center}
    \FigureFile(88mm,70mm){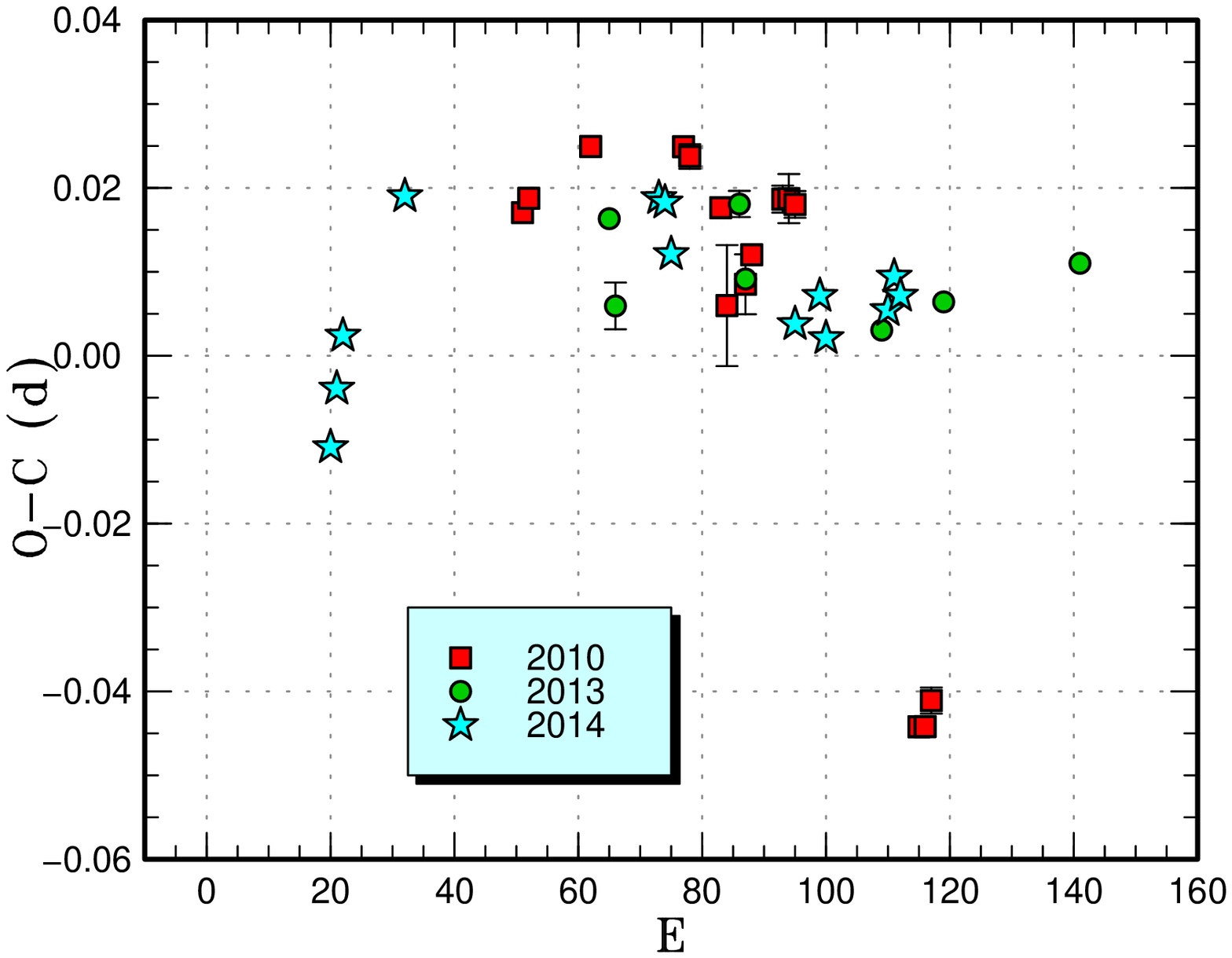}
  \end{center}
  \caption{Comparison of $O-C$ diagrams of GZ Cnc between different
  superoutbursts.  A period of 0.09290~d was used to draw this figure.
  Approximate cycle counts ($E$) after the start of the superoutburst
  were used.
  }
  \label{fig:gzcnccomp2}
\end{figure}

   The 2014 January superoutburst was detected by
R. Stubbings (vsnet-alert 16758).  The bright magnitude
immediately suggested a superoutburst.  The initial observation
recorded a long superhump period (vsnet-alert 16782).
The times of superhump maxima are listed in table
\ref{tab:gzcncoc2014}.  The maxima $E \le 2$ correspond to
stage A superhumps (see also figures \ref{fig:gzcnc2014humpamp};
\ref{fig:gzcnccomp2}). 
A PDM analysis of this part of the
data yielded a period of 0.0969(3)~d (the period in
vsnet-alert 16782 referred to $E \le 12$).  Due to the shortness
of the run, the accuracy od this period of stage A superhumps
is limited.  This $\varepsilon$ corresponds to $q$=0.30(2)
(\cite{kat13qfromstageA}; see subsection \ref{sec:stagea}).
Although this estimate is based on very limited observations,
this observation seems to support that this object has
a mass ratio near the borderline of the condition for
the 3:1 resonance ($q$ range 0.25--0.33 depending on
simulations).  The object is indeed likely a ``borderline''
SU UMa-type dwarf nova.

\begin{table}
\caption{Superhump maxima of GZ Cnc (2014)}\label{tab:gzcncoc2014}
\begin{center}
\begin{tabular}{rp{55pt}p{40pt}r@{.}lr}
\hline
\multicolumn{1}{c}{$E$} & \multicolumn{1}{c}{max\commenta} & \multicolumn{1}{c}{error} & \multicolumn{2}{c}{$O-C$\commentb} & \multicolumn{1}{c}{$N$\commentc} \\
\hline
0 & 56663.1487 & 0.0015 & $-$0&0142 & 60 \\
1 & 56663.2486 & 0.0003 & $-$0&0073 & 181 \\
2 & 56663.3478 & 0.0003 & $-$0&0010 & 181 \\
12 & 56664.2934 & 0.0005 & 0&0149 & 69 \\
53 & 56668.1021 & 0.0004 & 0&0118 & 127 \\
54 & 56668.1945 & 0.0004 & 0&0112 & 162 \\
55 & 56668.2812 & 0.0018 & 0&0050 & 64 \\
75 & 56670.1309 & 0.0008 & $-$0&0048 & 160 \\
79 & 56670.5059 & 0.0005 & $-$0&0017 & 102 \\
80 & 56670.5937 & 0.0008 & $-$0&0068 & 76 \\
90 & 56671.5261 & 0.0005 & $-$0&0042 & 96 \\
91 & 56671.6229 & 0.0006 & $-$0&0002 & 91 \\
92 & 56671.7136 & 0.0006 & $-$0&0026 & 89 \\
\hline
  \multicolumn{6}{l}{\commenta BJD$-$2400000.} \\
  \multicolumn{6}{l}{\commentb Against max $= 2456663.1629 + 0.092970 E$.} \\
  \multicolumn{6}{l}{\commentc Number of points used to determine the maximum.} \\
\end{tabular}
\end{center}
\end{table}

\subsection{AL Comae Berenices}\label{obj:alcom}

   AL Com is one of the renowned high-amplitude dwarf novae
since the discovery in 1962 by Rosino \citep{ber64alcom}.
\citet{szk87shortPCV} showed a large amplitude variation with 
a period near 40 min and suggested that AL Com may belong to
either DQ Her-type magnetic systems or AM CVn-type double
degenerate systems.  Further spectroscopic observation by 
\citet{muk90faintCV} precluded the latter possibility.
More extensive photometry by \citet{abb92alcomcperi}
showed two distinct periodicities of 41 min and 87--90 min;
the latter was suggested to be the orbital period, and the former
the rotation period of the white dwarf.  From these periods,
\citet{abb92alcomcperi} concluded that AL Com bears the properties 
of both an enigmatic dwarf nova WZ Sge and a unique intermediate 
polar EX Hya.  The 1995 superoutburst clarified that
this object is a WZ Sge-type dwarf nova, and it shows
double-wave modulations (now called early superhumps)
during the early stage of the superoutburst \citep{kat96alcom}.
Based on the outburst light curve and the orbital parameters,
AL Com was considered as a ``twin'' of WZ Sge.
This superoutburst was particularly well documented
(\cite{pat96alcom}; \cite{how96alcom}; \cite{nog97alcom}).

   Another superoutburst was recorded in 2001
\citep{ish02wzsgeletter}.  A less observed superoutburst in
2007 indicated that the post-superoutburst rebrightenings
are different between different superoutbursts
\citep{uem08alcom}.  There was an outburst in 2003 detected
in SDSS, which was missed by visual observers.

   The 2013 superoutburst was detected on December 6
by C. Gualdoni (cvnet-outburst 5738).  Subsequent observations
detected early superhumps (vsnet-alert 16695, 16712).
The evolution of superhumps in the early stage was
well observed (vsnet-alert 16718, 16724).

\begin{figure}
  \begin{center}
    \FigureFile(88mm,110mm){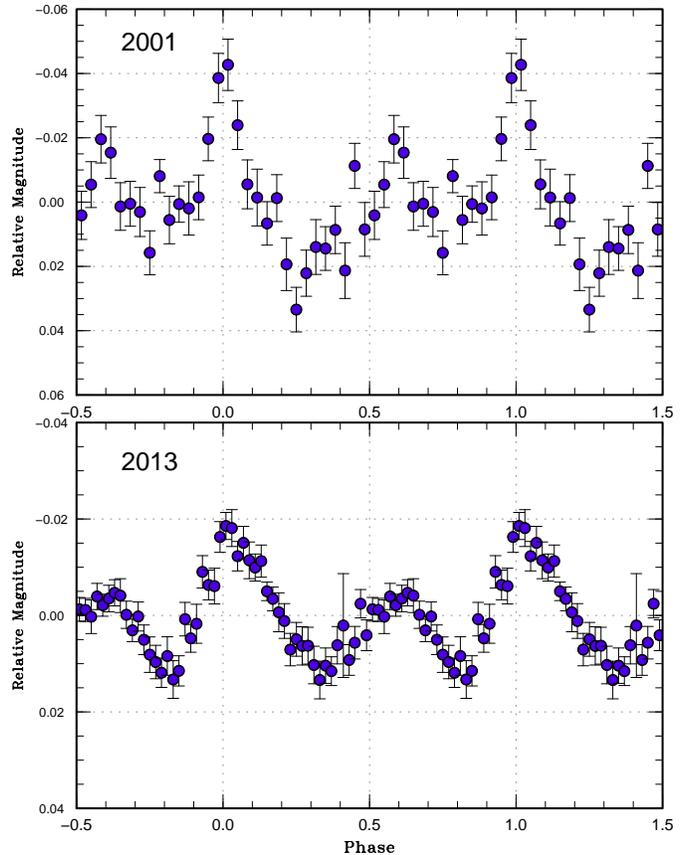}
  \end{center}
  \caption{Comparison of early superhumps in AL Com between 2001 and 2013.}
  \label{fig:alcomesh2013}
\end{figure}

   The period of early superhumps was determined
to be 0.056660(8)~d with the PDM method
from observations before BJD 2456639.5.
This value is in very good agreement with the 2001
measurement of early superhumps
(figure \ref{fig:alcomesh2013}; see subsection \ref{sec:earlysh}).
The commonly accepted mechanism of early superhumps
(cf. \cite{osa02wzsgehump}) suggests that the phases of
early superhumps is constant between superoutburst
(which is defined by the orientation of the observer
against the binary orbit).  Assuming that the phases
of early superhumps are the same between the 1995, 2001
and 2013 superoutbursts, we can determine the orbital period.
The refined orbital period of 0.056668589(9)~d well expresses
all the observations.  An alias by one cycle is 0.056667180(9)~d
is not favored if we believe the identification of
the period by \citet{pat96alcom}.

   The times of superhump maxima during the plateau phase
of the superoutburst are listed in table \ref{tab:alcomoc2013}.
Stage A superhumps were better detected during the 
2013 superoutburst than during the previous ones.
The 2013 superoutburst showed a positive $P_{\rm dot}$ of
$+4.9(19) \times 10^{-5}$.  This value is larger than
the 1995 and 2001 measurements.  A comparison of the $O-C$
diagrams between different superoutbursts is presented
in figure \ref{fig:alcomcomp2}.

\begin{figure}
  \begin{center}
    \FigureFile(88mm,70mm){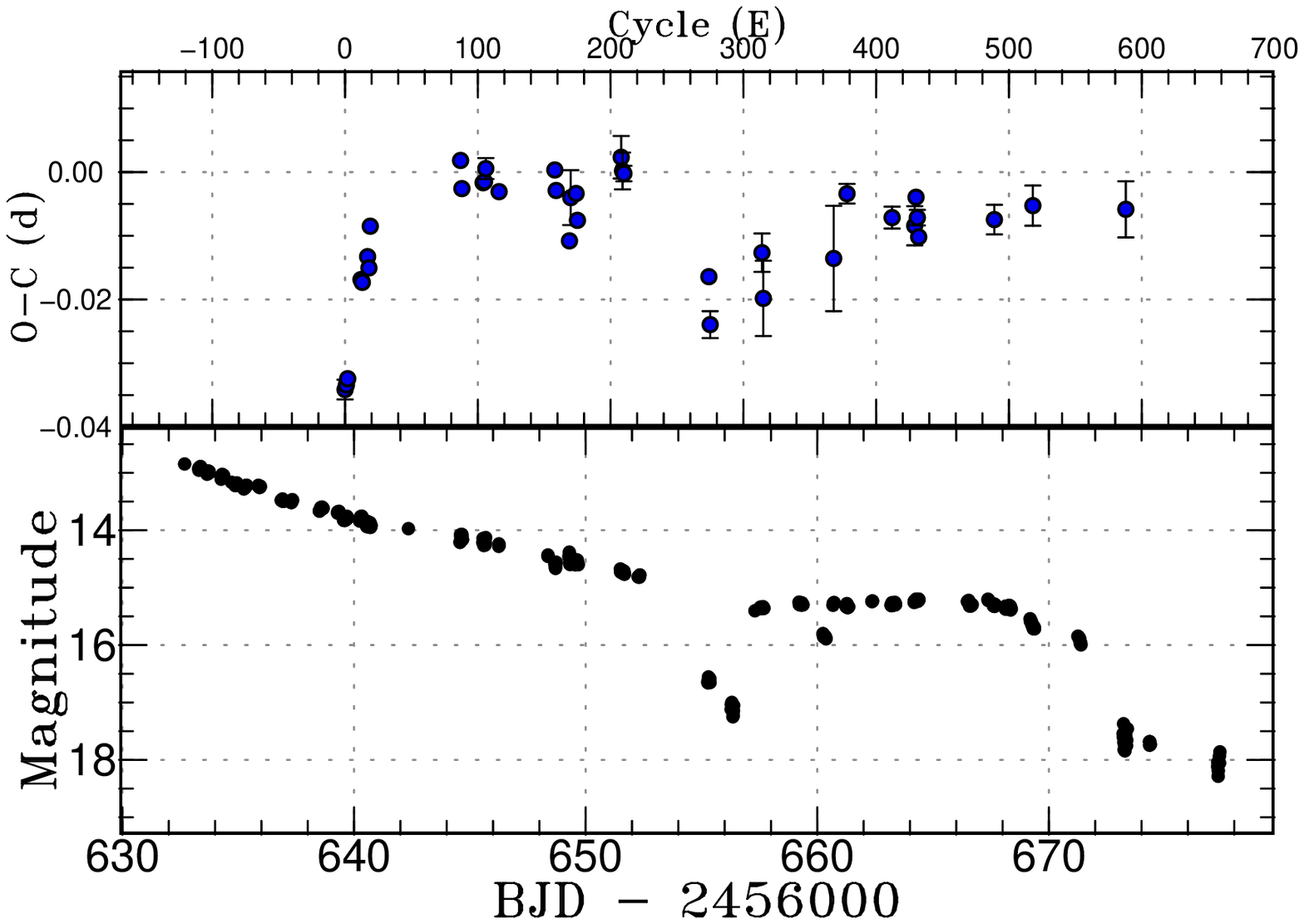}
  \end{center}
  \caption{$O-C$ diagram of superhumps in AL Com (2013).
     (Upper): $O-C$ diagram.  A period of 0.05733~d
     was used to draw this figure.
     (Lower): Light curve.  The observations were binned to 0.011~d.}
  \label{fig:alcom2013humpall}
\end{figure}

\begin{figure}
  \begin{center}
    \FigureFile(88mm,70mm){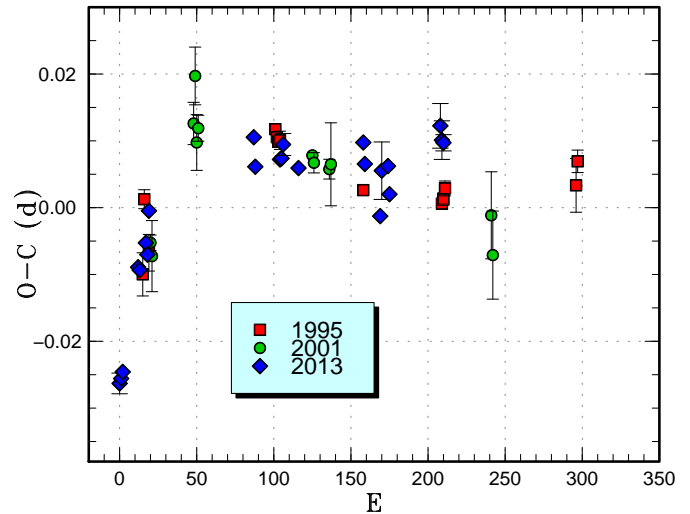}
  \end{center}
  \caption{Comparison of $O-C$ diagrams of AL Com between different
  superoutbursts.  A period of 0.05732~d was used to draw this figure.
  Approximate cycle counts ($E$) after the emergence of superhump
  were used.  Assuming that the stage A was best observed in
  2013, the 1995 and 2001 $O-C$ diagrams were shifted within
  20 cycles to best match the stage A-B transition in 2013.}
  \label{fig:alcomcomp2}
\end{figure}

   The times of superhump maxima after the rapid fading
are listed in table \ref{tab:alcomoc2013b}.  A PDM analysis
of the combined data in this interval yielded a period
of 0.057393(10)~d.  Although the overall $O-C$ diagram
(figure \ref{fig:alcom2013humpall}) suggests the presence
of a phase jump around the dip, the later times of
superhump maxima appear to be on a smooth extension
of stage B superhumps.

\begin{figure}
  \begin{center}
    \FigureFile(88mm,70mm){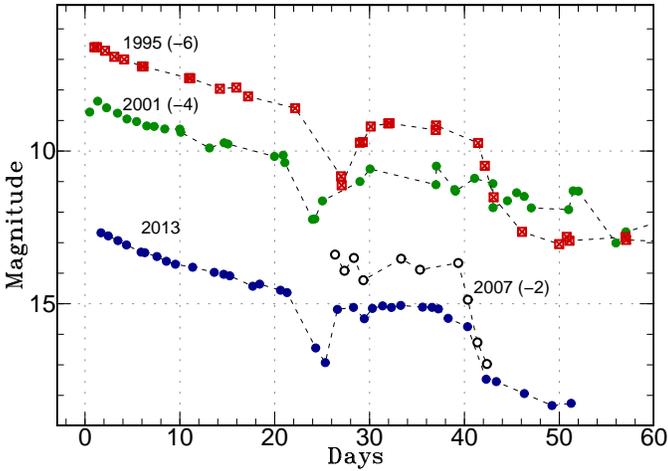}
  \end{center}
  \caption{Comparison of superoutbursts of AL Com.
  The data were binned to 1~d and shifted in magnitude.
  The dashed lines are added to aid recognizing the variation.
  The data for the 2007 superoutburst were from \citet{uem08alcom}.}
  \label{fig:alcomlccomp}
\end{figure}

   A comparison of light curves of different superoutbursts
is shown in figure \ref{fig:alcomlccomp}.  All superoutbursts
(except the poorly observed 2007 one) showed a dip and
a plateau-type rebrightening.  The second (rebrightening)
plateau after the dip was almost flat for the 1995 superoutburst 
and was associated by an initial small dip in the 2013 one.
The 2007 superoutburst was more structured, as reported 
in \citet{uem08alcom}, but it is still unlike discrete 
rebrightenings such as EG Cnc (\cite{pat98egcnc}; \cite{kat04egcnc}).
We consider that the 2007 superoutburst resembled the 2001
superoutburst of WZ Sge or with small brightenings with
amplitudes less than 1 mag.  Although the duration of
the rebrightening plateau in the 2001 superoutburst appears
longer than in other superoutbursts, this part of
the observation was of low quality and it needs to be
interpreted carefully.
A variation in the rebrightening was also observed in
EZ Lyn \citep{Pdot3} and suspected in WZ Sge \citep{pat81wzsge}.
Although there is subtle difference between the pattern
of rebrightenings in the same object, the present comparison
of the light curves in AL Com suggests that the pattern
of rebrightening is in general reproducible.

   The  known outbursts of AL Com are listed in table
\ref{tab:alcomout}.  The supercycle appears to be 6--7~yr.
Since the 2003 outburst escaped detection by visual
observers, the true frequency of normal outburst may be
higher than this table suggests.

\begin{table}
\caption{Superhump maxima of AL Com (2013)}\label{tab:alcomoc2013}
\begin{center}
\begin{tabular}{rp{55pt}p{40pt}r@{.}lr}
\hline
\multicolumn{1}{c}{$E$} & \multicolumn{1}{c}{max\commenta} & \multicolumn{1}{c}{error} & \multicolumn{2}{c}{$O-C$\commentb} & \multicolumn{1}{c}{$N$\commentc} \\
\hline
0 & 56639.5761 & 0.0015 & $-$0&0141 & 92 \\
1 & 56639.6341 & 0.0010 & $-$0&0135 & 93 \\
2 & 56639.6924 & 0.0006 & $-$0&0126 & 95 \\
12 & 56640.2813 & 0.0004 & 0&0018 & 70 \\
13 & 56640.3382 & 0.0003 & 0&0013 & 50 \\
17 & 56640.5715 & 0.0005 & 0&0049 & 73 \\
18 & 56640.6271 & 0.0006 & 0&0030 & 89 \\
19 & 56640.6910 & 0.0004 & 0&0094 & 80 \\
87 & 56644.5997 & 0.0004 & 0&0122 & 28 \\
88 & 56644.6527 & 0.0008 & 0&0077 & 19 \\
104 & 56645.5709 & 0.0003 & 0&0068 & 24 \\
105 & 56645.6283 & 0.0005 & 0&0069 & 26 \\
106 & 56645.6877 & 0.0016 & 0&0088 & 12 \\
116 & 56646.2574 & 0.0006 & 0&0041 & 50 \\
158 & 56648.6687 & 0.0006 & 0&0028 & 72 \\
159 & 56648.7228 & 0.0006 & $-$0&0005 & 67 \\
169 & 56649.2882 & 0.0010 & $-$0&0095 & 90 \\
170 & 56649.3523 & 0.0043 & $-$0&0028 & 77 \\
174 & 56649.5823 & 0.0005 & $-$0&0026 & 29 \\
175 & 56649.6354 & 0.0007 & $-$0&0070 & 28 \\
208 & 56651.5372 & 0.0033 & $-$0&0007 & 26 \\
209 & 56651.5924 & 0.0029 & $-$0&0030 & 52 \\
210 & 56651.6493 & 0.0012 & $-$0&0035 & 73 \\
\hline
  \multicolumn{6}{l}{\commenta BJD$-$2400000.} \\
  \multicolumn{6}{l}{\commentb Against max $= 2456639.5902 + 0.057441 E$.} \\
  \multicolumn{6}{l}{\commentc Number of points used to determine the maximum.} \\
\end{tabular}
\end{center}
\end{table}

\begin{table}
\caption{Superhump maxima of AL Com (2013) (post-superoutburst)}\label{tab:alcomoc2013b}
\begin{center}
\begin{tabular}{rp{55pt}p{40pt}r@{.}lr}
\hline
\multicolumn{1}{c}{$E$} & \multicolumn{1}{c}{max\commenta} & \multicolumn{1}{c}{error} & \multicolumn{2}{c}{$O-C$\commentb} & \multicolumn{1}{c}{$N$\commentc} \\
\hline
0 & 56655.3022 & 0.0011 & 0&0006 & 42 \\
1 & 56655.3520 & 0.0021 & $-$0&0070 & 39 \\
40 & 56657.5992 & 0.0030 & 0&0023 & 65 \\
41 & 56657.6493 & 0.0059 & $-$0&0049 & 71 \\
94 & 56660.6941 & 0.0083 & $-$0&0014 & 28 \\
104 & 56661.2776 & 0.0015 & 0&0083 & 59 \\
138 & 56663.2230 & 0.0017 & 0&0028 & 16 \\
155 & 56664.1963 & 0.0031 & 0&0007 & 55 \\
156 & 56664.2582 & 0.0010 & 0&0051 & 60 \\
157 & 56664.3123 & 0.0012 & 0&0018 & 60 \\
158 & 56664.3666 & 0.0009 & $-$0&0012 & 59 \\
215 & 56667.6371 & 0.0023 & $-$0&0014 & 58 \\
244 & 56669.3019 & 0.0032 & $-$0&0007 & 41 \\
314 & 56673.3144 & 0.0044 & $-$0&0048 & 30 \\
\hline
  \multicolumn{6}{l}{\commenta BJD$-$2400000.} \\
  \multicolumn{6}{l}{\commentb Against max $= 2456655.3017 + 0.057381 E$.} \\
  \multicolumn{6}{l}{\commentc Number of points used to determine the maximum.} \\
\end{tabular}
\end{center}
\end{table}

\begin{table*}
\caption{Outbursts of AL Com}\label{tab:alcomout}
\begin{center}
\begin{tabular}{lccc}
\hline
Date\commenta & Maximum\commentb & Type & Reference \\
\hline
1892 April 26 & 14.3p & -- & \citet{ber64alcom} \\
1941 June 25  & 13.8p & -- & \citet{luc72alcom} \\
1961 November 17--December 20 & 13.8p & super & \citet{ros61alcomiauc}; \citet{ber64alcom} \\
1965 March 26--27 & 13.4p & super & \citet{zwi65alcom}; \citet{ber65alcom} \\
1974 April 19--20 & 14.1v & normal? & AAVSO \\
1975 March 16--June 29 & 12.8v & super (two outbursts?) & AAVSO \\ 
1995 April 5--May 19 & 12.4V & super & \citet{pat96alcom}; \citet{nog97alcom} \\
2001 May 18--June 9? & 12.6v & super & \citet{ish02wzsgeletter} \\
2003 January 28 & 15.5g & normal & SDSS \\
2007 November 6--24\commentc & 15.4V & super & \citet{uem08alcom} \\
2013 December 6--2014 January 15 & 12.7V & super & this work \\
\hline
  \multicolumn{4}{l}{\commenta For modern data, the end date refers to the end of the (second) plateau phase.} \\
  \multicolumn{4}{l}{\commentb p: photographic, v: visual} \\
  \multicolumn{4}{l}{\commentc Rebrightening part only.} \\
\\
\end{tabular}
\end{center}
\end{table*}

\subsection{V503 Cygni}\label{obj:v503cyg}

   This SU UMa-type dwarf nova is notable for its unusually
short (89~d) supercycle and the occasional presence of
negative superhumps \citep{har95v503cyg}.
\citet{kat02v503cyg} reported a dramatic variation in
the number of normal outbursts, and this finding led to
the discovery of the state with negative superhumps
suppressing the number of normal outbursts
(\cite{ohs12eruma}; \cite{zem13eruma}; 
\cite{osa13v1504cygKepler}; \cite{osa13v344lyrv1504cyg}).

   The superoutburst in 2013 August was observed only
for two nights during in the final part.  We obtained
maxima of BJD 2456527.4717(11) ($N=88$),
2456527.5580(27) ($N=47$), 2456532.3896(20) ($N=67$),
2456532.4696(13) ($N=63$).  Since the phases of the latter
two maxima are $\sim$0.5 phase different from the earlier
ones, they may be traditional late superhumps.

\subsection{IX Draconis}\label{obj:ixdra}

   IX Dra was selected as an ultraviolet-excess object
(=KUV 18126$+$6704) by \citet{nog80KUV}.
\citet{nog82KUVpropermotion} detected its variability.
\citet{weg88KUVspec4} spectroscopically classified
the object as a B subdwarf.  \citet{liu99CVspec1}
classified the object as a dwarf nova by spectroscopy.
\citet{klo95exdraixdra} detected variability of this
object on photographic plates and obtained a period
of $\sim$45.7~d, although it was not strictly periodic.
T. Vanmunster detected superhumps in 2000
(vsnet-alert 5368, 5369).  \citet{ish01ixdra} studied
this object and clarified the it is a new ER UMa-type
dwarf nova with a supercycle of 53~d.
\citet{ole04ixdra} studied the 2003 superoutburst
and suggested a period of 0.06646(6)~d, which they
proposed to be the orbital period.  This period
led to a very small fractional superhump excess,
from which \citet{ole04ixdra} suggested IX Dra
to be a period bouncer.  \citet{otu13ixdra} further
studied this object and obtained a longer supercycle.
\citet{otu13ixdra} discussed the secular increase of
the supercycles in ER UMa-type dwarf novae.
The identification of the orbital period, however,
was less convincing as in \citet{ole04ixdra} and
there is a possibility of another period which places
IX Dra in a region of ordinary dwarf nova before
the period minimum.  \citet{Pdot4} did not regard the
orbital period by \citet{otu13ixdra} as the true one
based on the high similarity of IX Dra to ER UMa.

\begin{figure}
  \begin{center}
    \FigureFile(88mm,70mm){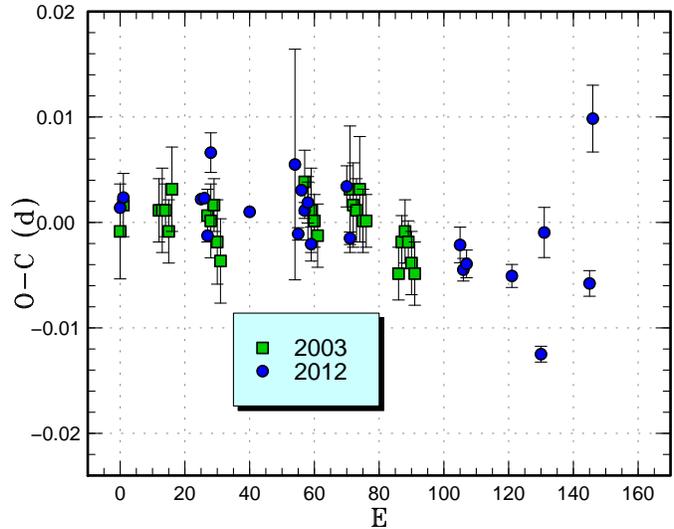}
  \end{center}
  \caption{Comparison of $O-C$ diagrams of IX Dra between different
  superoutbursts.  A period of 0.06700~d was used to draw this figure.
  Approximate cycle counts ($E$) after the start of the observation
  were used.  The starts of the superoutbursts were not
  well constrained.
  }
  \label{fig:ixdracomp}
\end{figure}

   We analyzed the superoutburst in 2012 July--August.
The times of superhump maxima are listed in table
\ref{tab:ixdraoc2012}.  The resultant $O-C$ data indicate
that the superhumps can be expressed by a single period
without a strong period variation.  A phase reversal
as seen in ER UMa \citep{kat96erumaSH} was not apparent.
\citet{ole04ixdra} also reported only small period
variations.  Although \citet{Pdot} recognized stages B and C
in the data of \citet{ole04ixdra}, the present data did
not show a strong sign of a stage transition
(see also figure \ref{fig:ixdracomp}).

\begin{table}
\caption{Superhump maxima of IX Dra (2012)}\label{tab:ixdraoc2012}
\begin{center}
\begin{tabular}{rp{55pt}p{40pt}r@{.}lr}
\hline
\multicolumn{1}{c}{$E$} & \multicolumn{1}{c}{max\commenta} & \multicolumn{1}{c}{error} & \multicolumn{2}{c}{$O-C$\commentb} & \multicolumn{1}{c}{$N$\commentc} \\
\hline
0 & 56130.7403 & 0.0003 & $-$0&0017 & 58 \\
1 & 56130.8082 & 0.0003 & $-$0&0007 & 56 \\
25 & 56132.4161 & 0.0003 & 0&0002 & 126 \\
26 & 56132.4832 & 0.0003 & 0&0003 & 144 \\
27 & 56132.5466 & 0.0003 & $-$0&0032 & 144 \\
28 & 56132.6215 & 0.0019 & 0&0047 & 45 \\
40 & 56133.4199 & 0.0005 & $-$0&0003 & 75 \\
54 & 56134.3624 & 0.0109 & 0&0048 & 24 \\
55 & 56134.4228 & 0.0006 & $-$0&0018 & 70 \\
56 & 56134.4939 & 0.0005 & 0&0024 & 75 \\
57 & 56134.5590 & 0.0008 & 0&0005 & 75 \\
58 & 56134.6268 & 0.0019 & 0&0013 & 58 \\
59 & 56134.6898 & 0.0016 & $-$0&0026 & 44 \\
70 & 56135.4323 & 0.0019 & 0&0034 & 75 \\
71 & 56135.4944 & 0.0006 & $-$0&0015 & 66 \\
105 & 56137.7717 & 0.0017 & $-$0&0006 & 66 \\
106 & 56137.8364 & 0.0011 & $-$0&0029 & 68 \\
107 & 56137.9039 & 0.0013 & $-$0&0023 & 68 \\
121 & 56138.8408 & 0.0011 & $-$0&0028 & 88 \\
130 & 56139.4364 & 0.0008 & $-$0&0098 & 70 \\
131 & 56139.5149 & 0.0024 & 0&0017 & 75 \\
145 & 56140.4481 & 0.0012 & $-$0&0025 & 61 \\
146 & 56140.5307 & 0.0032 & 0&0132 & 59 \\
\hline
  \multicolumn{6}{l}{\commenta BJD$-$2400000.} \\
  \multicolumn{6}{l}{\commentb Against max $= 2456130.7420 + 0.066955 E$.} \\
  \multicolumn{6}{l}{\commentc Number of points used to determine the maximum.} \\
\end{tabular}
\end{center}
\end{table}

\subsection{MN Draconis}\label{obj:mndra}

   This object is an SU UMa-type dwarf nova in the period
gap (\cite{ant02var73dra}; \cite{nog03var73dra}).
It is notable that this object showed negative superhump
in quiescence (\cite{pav10mndra}; \cite{sam10mndra}).

\begin{figure}
  \begin{center}
    \FigureFile(88mm,70mm){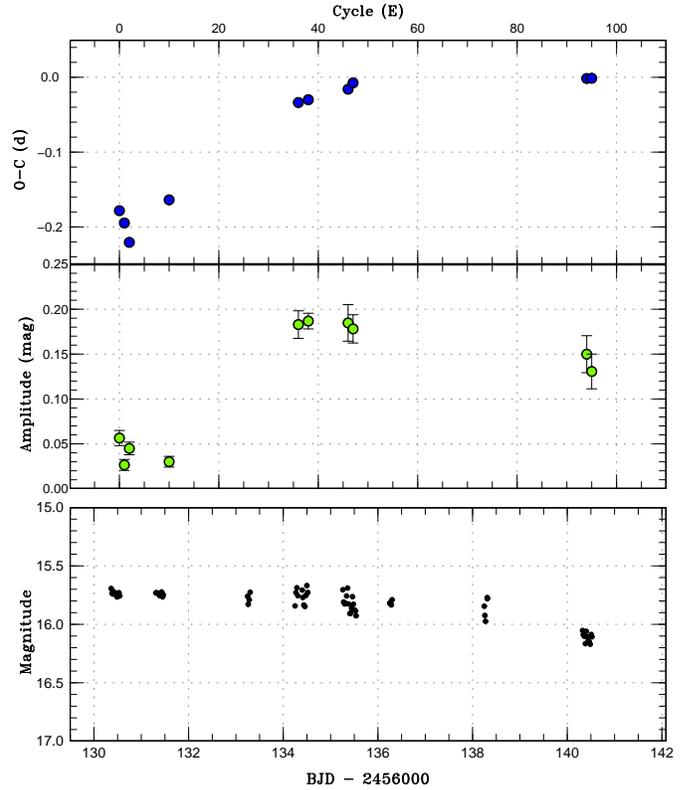}
  \end{center}
  \caption{$O-C$ diagram of superhumps in MN Dra (2012).
     (Upper): $O-C$ diagram.  A period of 0.10504~d
     was used to draw this figure.  Cycle counts ($E$) after
     the start of the observation were used.
     (Middle): Amplitudes of the superhumps.
     (Lower): Light curve.  The observations were binned to 0.021~d.}
  \label{fig:mndra2012humpamp}
\end{figure}

   The 2012 July-August superoutburst was observed
starting from the growing stage of superhumps.
The times of superhump maxima are listed in table
\ref{tab:mndraoc2012}.  Although the individual times
of maxima were not well determined before $E \le 10$,
the $O-C$ diagram suggests that the interval $E \le 39$
was stage A (figure \ref{fig:mndra2012humpamp}).
A PDM analysis of this segment yielded
a period of 0.10993(9)~d.

\begin{figure}
  \begin{center}
    \FigureFile(88mm,70mm){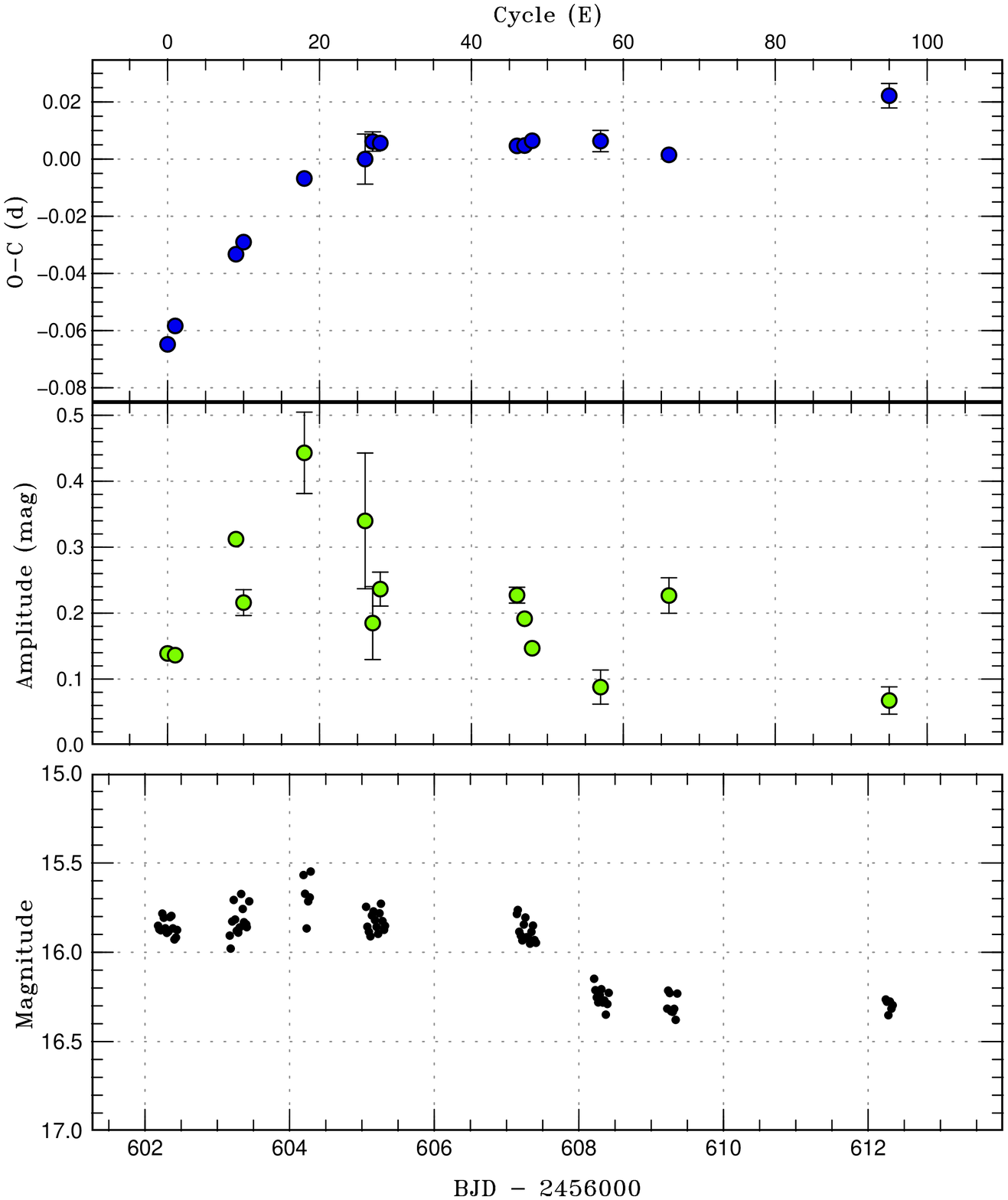}
  \end{center}
  \caption{$O-C$ diagram of superhumps in MN Dra (2013).
     (Upper): $O-C$ diagram.  A period of 0.10504~d
     was used to draw this figure.  Cycle counts ($E$) after
     the start of the observation were used.
     (Middle): Amplitudes of the superhumps.
     (Lower): Light curve.  The observations were binned to 0.021~d.}
  \label{fig:mndra2013humpamp}
\end{figure}

   The 2013 November superoutburst (vsnet-alert 16611)
was observed for eight nights.
The times of superhump maxima are listed in table
\ref{tab:mndraoc2013}.  A large period variation was
detected.  Up to $E=18$, the amplitudes of the superhumps
grew, and stage A superhumps were likely recorded
(figure \ref{fig:mndra2013humpamp}).
Although we identified the following phase as stage B,
the identification for $E=95$ is uncertain due to the
lower quality of the data.  In table \ref{tab:perlist},
we list a period derived for the interval $26 \le E \le 66$.

\begin{figure}
  \begin{center}
    \FigureFile(88mm,70mm){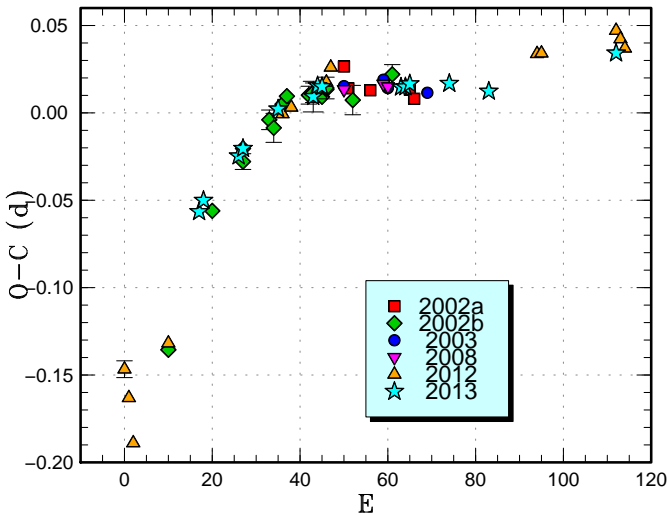}
  \end{center}
  \caption{Comparison of $O-C$ diagrams of MN Dra between different
  superoutbursts.  A period of 0.1050~d was used to draw this figure.
  Approximate cycle counts ($E$) after the start of the outburst
  were used (2012).  Since the start of the other superoutbursts
  was not well constrained, we shifted the $O-C$ diagram
  to best fit the 2012 one.
  }
  \label{fig:mndracomp2}
\end{figure}

   Assuming the orbital period of 0.0998(2)~d \citep{pav10mndra},
the values of $\varepsilon$ for stage A superhumps are 
9.2(1)\% for 2012 and 7.8(1)\% for 2013,
which correspond to $q$=0.327(5) and $q$=0.258(5).
Since both observations were not ideal (the lack of well
measured maxima for the 2012 observation and the lack
of data for the early stage of the 2013 observation),
we simply make an average of these values to obtain
$q$=0.029(5).  A comparison of the
$O-C$ diagrams between different superoutburst
(figure \ref{fig:mndracomp2}) suggests that the large
negative $P_{\rm dot}$ in the 2002b superoutburst
reported in \citet{nog03var73dra}; \citet{Pdot}
reflected stage A-B transition.

\begin{table}
\caption{Superhump maxima of MN Dra (2012)}\label{tab:mndraoc2012}
\begin{center}
\begin{tabular}{rp{55pt}p{40pt}r@{.}lr}
\hline
\multicolumn{1}{c}{$E$} & \multicolumn{1}{c}{max\commenta} & \multicolumn{1}{c}{error} & \multicolumn{2}{c}{$O-C$\commentb} & \multicolumn{1}{c}{$N$\commentc} \\
\hline
0 & 56130.3597 & 0.0048 & $-$0&0239 & 42 \\
1 & 56130.4483 & 0.0030 & $-$0&0420 & 80 \\
2 & 56130.5274 & 0.0017 & $-$0&0696 & 56 \\
10 & 56131.4246 & 0.0026 & $-$0&0261 & 80 \\
36 & 56134.2857 & 0.0011 & 0&0608 & 30 \\
38 & 56134.4995 & 0.0006 & 0&0612 & 28 \\
46 & 56135.3539 & 0.0015 & 0&0619 & 28 \\
47 & 56135.4674 & 0.0012 & 0&0688 & 27 \\
94 & 56140.4101 & 0.0017 & $-$0&0035 & 27 \\
95 & 56140.5155 & 0.0027 & $-$0&0049 & 21 \\
112 & 56142.3134 & 0.0042 & $-$0&0209 & 32 \\
113 & 56142.4136 & 0.0036 & $-$0&0274 & 30 \\
114 & 56142.5133 & 0.0045 & $-$0&0344 & 28 \\
\hline
  \multicolumn{6}{l}{\commenta BJD$-$2400000.} \\
  \multicolumn{6}{l}{\commentb Against max $= 2456130.3836 + 0.106703 E$.} \\
  \multicolumn{6}{l}{\commentc Number of points used to determine the maximum.} \\
\end{tabular}
\end{center}
\end{table}

\begin{table}
\caption{Superhump maxima of MN Dra (2013)}\label{tab:mndraoc2013}
\begin{center}
\begin{tabular}{rp{55pt}p{40pt}r@{.}lr}
\hline
\multicolumn{1}{c}{$E$} & \multicolumn{1}{c}{max\commenta} & \multicolumn{1}{c}{error} & \multicolumn{2}{c}{$O-C$\commentb} & \multicolumn{1}{c}{$N$\commentc} \\
\hline
0 & 56602.2479 & 0.0011 & $-$0&0287 & 114 \\
1 & 56602.3594 & 0.0010 & $-$0&0231 & 116 \\
9 & 56603.2248 & 0.0004 & $-$0&0042 & 198 \\
10 & 56603.3341 & 0.0014 & $-$0&0007 & 90 \\
18 & 56604.1967 & 0.0019 & 0&0154 & 13 \\
26 & 56605.0438 & 0.0088 & 0&0159 & 99 \\
27 & 56605.1550 & 0.0034 & 0&0213 & 166 \\
28 & 56605.2595 & 0.0015 & 0&0200 & 20 \\
46 & 56607.1492 & 0.0011 & 0&0051 & 64 \\
47 & 56607.2544 & 0.0006 & 0&0044 & 106 \\
48 & 56607.3611 & 0.0007 & 0&0053 & 131 \\
57 & 56608.3064 & 0.0037 & $-$0&0018 & 59 \\
66 & 56609.2469 & 0.0012 & $-$0&0136 & 95 \\
95 & 56612.3137 & 0.0043 & $-$0&0153 & 37 \\
\hline
  \multicolumn{6}{l}{\commenta BJD$-$2400000.} \\
  \multicolumn{6}{l}{\commentb Against max $= 2456602.2767 + 0.105815 E$.} \\
  \multicolumn{6}{l}{\commentc Number of points used to determine the maximum.} \\
\end{tabular}
\end{center}
\end{table}

\subsection{CP Eridani}\label{obj:cperi}

   CP Eri was discovered as a faint blue variable showing
an outburst \citep{luy59egaqrehaqr}.  \citet{szk89faintCV2}
obtained time-series photometry in quiescence and detected
sporadic variations of 0.2--0.4 mag without clear periodicity.
\citet{how91faintCV4} observed this object again in quiescence
and detected modulations with periods of 28.6--29.5 min.
Due to its shortness, \citet{how91faintCV4} considered
this period to be the spin period of the magnetic white
dwarf.  \citet{abb92alcomcperi} obtained higher quality
time-series photometry and identified a period od 1724(4)~s
(28.6 min).  Furthermore, \citet{abb92alcomcperi} obtained
spectra both in high and low states, and clarified that
this object lacks hydrogen lines.  The helium lines were
in emission in low state and in absorption in high state,
and this behavior was very similar to that of CR Boo
\citep{woo87crboo}.  \citet{abb92alcomcperi} concluded that
CP Eri is an interacting binary white dwarf (IBWD, or
AM CVn-type star).  \citet{pat93v603aql}, however, suggested
that this photometric period is the superhump period
since most of AM CVn-type stars show superhumps.
\citet{zwi95CVspec2} obtained a spectrum with
a featureless blue continuum.

   Although this object was discovered as an outbursting
object, its outburst behavior was not clarified for
a long time.  Although J. Patterson (cba-info message on
1998 January 1) reported an outburst of 16.5 mag and
0.2 superhumps, the result was published only in
\citet{arm12cperi}.  Starting from 2003, B. Monard regularly 
monitored this object and detected several outbursts between
16.0 and 17.5 mag.  More recently, CRTS data suggested
a cycle length of $\sim$100~d \citep{Pdot3}.
\citet{ram12amcvnLC} presented the result of long-term 
monitoring of AM CVn-type stars and detected three outbursts
in CP Eri.  The duration of the outbursts was 15~d and the
outburst duty cycle was 27\%.  As judged from the duration,
these outbursts were likely superoutbursts.

   \citet{arm12cperi}, following the interpretation
in \citet{pat93v603aql}, identified the orbital and superhump
period from the 1998 data.  According to this interpretation,
the orbital modulation [1701.4(2)~s] has doubly humped.

   Om 2013 October 3, ASAS-SN team detected an outburst
of CP Eri (vsnet-alert 16501).  Subsequent observations
recorded superhumps (vsnet-alert 16510, 16515).
This outbursting state near the peak was observed for 
three nights (the outburst lasted at least 5~d including 
the ASAS-SN detection) and followed by a dip (vsnet-alert 16526).
After the dip, the superhump signal once became weaker,
but became detectable again (vsnet-alert 16530, 16537).
The later part of the outburst consisted of oscillations
as reported in \citet{arm12cperi}.

\begin{figure}
  \begin{center}
    \FigureFile(88mm,70mm){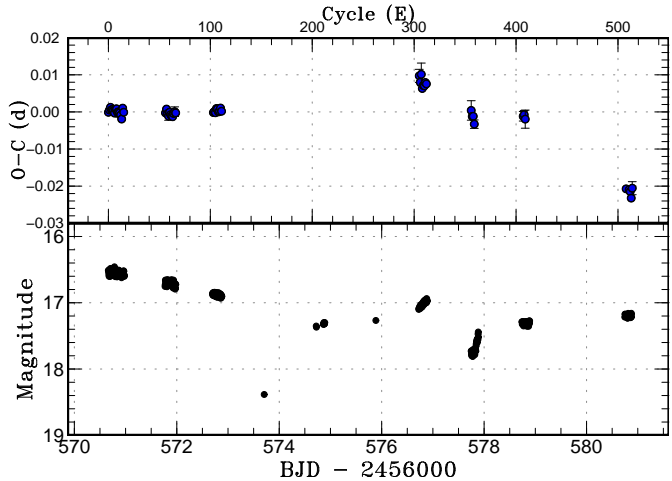}
  \end{center}
  \caption{$O-C$ diagram of superhumps in CP Eri (2013).
     (Upper): $O-C$ diagram.  A period of 0.019897~d
     was used to draw this figure.
     (Lower): Light curve.  The observations were binned to 0.004~d.}
  \label{fig:cperi2013humpall}
\end{figure}

   The times of superhump maxima during the initial peak
are listed in table \ref{tab:cperioc2013}.  A positive
$P_{\rm dot}$ of $+3.1(9) \times 10^{-5}$ was first time
recorded in CP Eri.  Superhumps after the dips had a shorter
period [0.019752(4)~d with the PDM method].  The times
of maxima of these superhump are listed in table
\ref{tab:cperioc2013b}.  The interpretation of these
superhumps (whether they are superhumps as in the initial
peak or ``traditional'' late superhumps) is not clear.
It was, however, likely these superhumps were excited again
after the dip phenomenon.  The combined $O-C$ diagram
(figure \ref{fig:cperi2013humpall}) may suggest that
the phase of the superhumps was continuous if the
superhump period continued to increase as in stage B
of hydrogen-rich dwarf novae.

\begin{table}
\caption{Superhump maxima of CP Eri (2013)}\label{tab:cperioc2013}
\begin{center}
\begin{tabular}{rp{55pt}p{40pt}r@{.}lr}
\hline
\multicolumn{1}{c}{$E$} & \multicolumn{1}{c}{max\commenta} & \multicolumn{1}{c}{error} & \multicolumn{2}{c}{$O-C$\commentb} & \multicolumn{1}{c}{$N$\commentc} \\
\hline
0 & 56570.6649 & 0.0008 & $-$0&0001 & 14 \\
1 & 56570.6854 & 0.0003 & 0&0005 & 19 \\
2 & 56570.7060 & 0.0002 & 0&0012 & 19 \\
3 & 56570.7253 & 0.0003 & 0&0006 & 19 \\
4 & 56570.7452 & 0.0003 & 0&0006 & 19 \\
5 & 56570.7649 & 0.0003 & 0&0004 & 19 \\
6 & 56570.7841 & 0.0003 & $-$0&0003 & 15 \\
7 & 56570.8050 & 0.0003 & 0&0007 & 38 \\
8 & 56570.8250 & 0.0003 & 0&0008 & 40 \\
9 & 56570.8439 & 0.0006 & $-$0&0002 & 24 \\
10 & 56570.8640 & 0.0005 & $-$0&0000 & 21 \\
11 & 56570.8836 & 0.0012 & $-$0&0002 & 21 \\
12 & 56570.9029 & 0.0005 & $-$0&0009 & 21 \\
13 & 56570.9217 & 0.0006 & $-$0&0019 & 21 \\
14 & 56570.9446 & 0.0006 & 0&0010 & 20 \\
15 & 56570.9634 & 0.0004 & $-$0&0001 & 21 \\
56 & 56571.7790 & 0.0006 & $-$0&0003 & 15 \\
57 & 56571.7999 & 0.0005 & 0&0008 & 15 \\
58 & 56571.8186 & 0.0010 & $-$0&0005 & 14 \\
59 & 56571.8378 & 0.0011 & $-$0&0012 & 15 \\
60 & 56571.8587 & 0.0005 & $-$0&0001 & 19 \\
61 & 56571.8780 & 0.0004 & $-$0&0008 & 32 \\
62 & 56571.8975 & 0.0005 & $-$0&0011 & 19 \\
63 & 56571.9172 & 0.0004 & $-$0&0013 & 14 \\
64 & 56571.9380 & 0.0006 & $-$0&0004 & 14 \\
65 & 56571.9584 & 0.0013 & 0&0001 & 14 \\
66 & 56571.9779 & 0.0008 & $-$0&0003 & 11 \\
103 & 56572.7142 & 0.0004 & $-$0&0002 & 19 \\
104 & 56572.7344 & 0.0004 & 0&0001 & 20 \\
105 & 56572.7540 & 0.0005 & $-$0&0002 & 20 \\
106 & 56572.7750 & 0.0004 & 0&0009 & 18 \\
107 & 56572.7948 & 0.0004 & 0&0008 & 20 \\
108 & 56572.8139 & 0.0004 & 0&0000 & 19 \\
109 & 56572.8340 & 0.0003 & 0&0002 & 20 \\
110 & 56572.8547 & 0.0003 & 0&0011 & 19 \\
111 & 56572.8737 & 0.0008 & 0&0002 & 17 \\
\hline
  \multicolumn{6}{l}{\commenta BJD$-$2400000.} \\
  \multicolumn{6}{l}{\commentb Against max $= 2456570.6650 + 0.019897 E$.} \\
  \multicolumn{6}{l}{\commentc Number of points used to determine the maximum.} \\
\end{tabular}
\end{center}
\end{table}

\begin{table}
\caption{Superhump maxima of CP Eri (2013) (after the dip)}\label{tab:cperioc2013b}
\begin{center}
\begin{tabular}{rp{55pt}p{40pt}r@{.}lr}
\hline
\multicolumn{1}{c}{$E$} & \multicolumn{1}{c}{max\commenta} & \multicolumn{1}{c}{error} & \multicolumn{2}{c}{$O-C$\commentb} & \multicolumn{1}{c}{$N$\commentc} \\
\hline
0 & 56576.7433 & 0.0018 & 0&0013 & 19 \\
1 & 56576.7615 & 0.0011 & $-$0&0003 & 18 \\
2 & 56576.7835 & 0.0030 & 0&0019 & 18 \\
3 & 56576.7996 & 0.0006 & $-$0&0017 & 20 \\
4 & 56576.8203 & 0.0007 & $-$0&0007 & 19 \\
5 & 56576.8401 & 0.0006 & $-$0&0007 & 19 \\
6 & 56576.8609 & 0.0005 & 0&0003 & 20 \\
7 & 56576.8804 & 0.0008 & 0&0001 & 19 \\
51 & 56577.7487 & 0.0026 & $-$0&0009 & 10 \\
52 & 56577.7670 & 0.0010 & $-$0&0024 & 16 \\
53 & 56577.7869 & 0.0006 & $-$0&0022 & 20 \\
54 & 56577.8047 & 0.0011 & $-$0&0042 & 20 \\
102 & 56578.7619 & 0.0013 & 0&0047 & 17 \\
103 & 56578.7822 & 0.0011 & 0&0052 & 19 \\
104 & 56578.8009 & 0.0024 & 0&0042 & 19 \\
203 & 56580.7519 & 0.0007 & $-$0&0007 & 17 \\
206 & 56580.8114 & 0.0009 & $-$0&0006 & 19 \\
207 & 56580.8308 & 0.0004 & $-$0&0009 & 19 \\
208 & 56580.8489 & 0.0007 & $-$0&0025 & 19 \\
209 & 56580.8715 & 0.0018 & 0&0003 & 19 \\
\hline
  \multicolumn{6}{l}{\commenta BJD$-$2400000.} \\
  \multicolumn{6}{l}{\commentb Against max $= 2456576.7420 + 0.019757 E$.} \\
  \multicolumn{6}{l}{\commentc Number of points used to determine the maximum.} \\
\end{tabular}
\end{center}
\end{table}

\subsection{V1239 Herculis}\label{obj:v1239her}

   This object is an eclipsing SU UMa-type
dwarf nova in the period gap (\cite{boy06j1702}; 
\cite{lit06j1702}).  The 2005 and 2011 superoutbursts were
reported in \citet{boy06j1702}, \citet{Pdot} and \citet{Pdot4},
respectively.  On 2013 September 26, another outburst
was reported (vsnet-alert 16462; also vsnet-alert 16464).
Although this outburst was likely a superoutburst, only
single-night observation covering an eclipse was reported.
By using these observations, we refined an ephemeris of
\begin{equation}
{\rm Min(BJD)} = 2453648.23651(1) + 0.1000822137(7) E
\label{equ:v1239herecl}
\end{equation}
using the MCMC modeling introduced in \citet{Pdot4}.
This ephemeris supersedes the one reported in \citet{Pdot4}
which was determined by the traditional minimum finding method.

\subsection{CT Hydrae}\label{obj:cthya}

   CT Hya was discovered as a dwarf nova (=AN 114.1936)
with a photographic range of 14.5 to fainter than 16.5
by \citet{hof36an25937}.  \citet{hof36an25937} recorded
four outbursts between 1929 and 1934.  The finding chart
was published in \citet{hof57MVS245}.  \citet{vog82atlas}
presented a photographic chart and identified the quiescent
counterpart.  The first secure outburst since the discovery
was reported on 1995 February 22 by CCD observations 
by M. Iida (VSOLJ).  Iida observed the object on the subsequent
night and detected variations compatible with superhumps.
The same outburst was observed by \citet{nog96cthya}, who
reported the detection of superhumps.
In \citet{Pdot}, superoutbursts in 1999, 2000, 2002 (two
superoutbursts) and 2009 were reported.  In \citet{Pdot2},
another superoutburst in 2010 was reported.

\begin{figure}
  \begin{center}
    \FigureFile(88mm,70mm){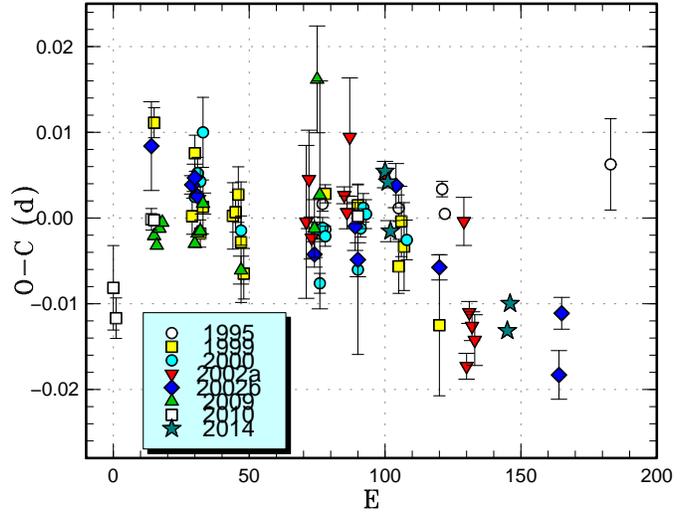}
  \end{center}
  \caption{Comparison of $O-C$ diagrams of CT Hya between different
  superoutbursts.  A period of 0.06650~d was used to draw this figure.
  Approximate cycle counts ($E$) after the maximum of the superoutburst
  were used.  This figure is updated from the corresponding one
  in \citet{Pdot2}, and includes the 1995 and 2014 observations.
  Since the start of the 2014 superoutburst
  was not well constrained, we shifted the $O-C$ diagram
  to best fit the others.  The shift value for the 2002a superoutburst
  was corrected.
  }
  \label{fig:cthyacomp3}
\end{figure}

   The 2014 superoutburst was detected by CRTS (cf. vsnet-alert
16926) and observations on two nights were obtained.
The times of superhump maxima are listed in table
\ref{tab:cthyaoc2014}.  The observation likely detected
stage C superhumps (cf. figure \ref{fig:cthyacomp3}).

\begin{table}
\caption{Superhump maxima of CT Hya (2014)}\label{tab:cthyaoc2014}
\begin{center}
\begin{tabular}{rp{55pt}p{40pt}r@{.}lr}
\hline
\multicolumn{1}{c}{$E$} & \multicolumn{1}{c}{max\commenta} & \multicolumn{1}{c}{error} & \multicolumn{2}{c}{$O-C$\commentb} & \multicolumn{1}{c}{$N$\commentc} \\
\hline
0 & 56708.1208 & 0.0004 & 0&0024 & 134 \\
1 & 56708.1860 & 0.0004 & 0&0015 & 133 \\
2 & 56708.2468 & 0.0012 & $-$0&0039 & 123 \\
45 & 56711.0947 & 0.0011 & $-$0&0017 & 28 \\
46 & 56711.1644 & 0.0007 & 0&0018 & 38 \\
\hline
  \multicolumn{6}{l}{\commenta BJD$-$2400000.} \\
  \multicolumn{6}{l}{\commentb Against max $= 2456708.1184 + 0.066178 E$.} \\
  \multicolumn{6}{l}{\commentc Number of points used to determine the maximum.} \\
\end{tabular}
\end{center}
\end{table}

\subsection{VW Hydri}\label{obj:vwhyi}

   We observed the 2012 November--December superoutburst
of this famous SU UMa-type dwarf nova.
By using the 2011--2012 data, we have determined the orbital
period to be 0.0742705(1)~d and the mean epoch of the
maximum of BJD 2456116.7250(1)~d.  Combined with the ephemeris
by \citet{vog74vwhyi}, we have obtained an updated ephemeris of
\begin{equation}
{\rm Max(BJD)} = 2456116.7250(1) + 0.074271061(4) E.
\label{equ:vwhyiorb}
\end{equation}

   The times of superhump maxima during the superoutburst plateau
are listed in table \ref{tab:vwhyioc2012}.
Although the evolution of superhumps was similar to the one
in 2011 \citep{Pdot4}, stage A superhumps were not
well observed in the 2012 superoutburst.  The precursor
was not as apparent as in the 2011 superoutburst.
During the rapid fading phase from the superoutburst,
an $\sim$0.5 phase jump was observed as in 2011.
These superhumps can be interpreted as ``traditional''
late superhumps.  The times of these post-superoutburst 
superhumps were determined after subtraction of
the mean orbital profile (table \ref{tab:vwhyioc2012b}).

   A comparison of the $O-C$ diagrams between the 2011 \citep{Pdot4}
and 2012 superoutbursts shows a slight difference in
the curvature of the $O-C$ diagram during the superoutburst
plateau.  Although the times of post-superoutburst superhumps
for the 2012 superoutburst were shown only before the next
normal outburst, the signal remained detectable until
$\sim$40~d after the termination of the superoutburst by
the PDM method.  The resultant periods in 5-d intervals
are listed in table \ref{tab:vwhyipost}.

   The detection of negative superhumps in this object
is discussed in subsection \ref{sec:vwhyinegsh}.

\begin{figure}
  \begin{center}
    \FigureFile(88mm,70mm){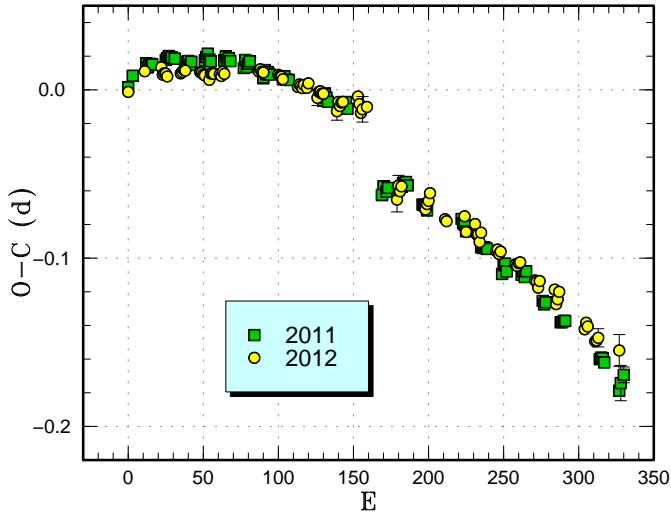}
  \end{center}
  \caption{Comparison of $O-C$ diagrams of VW Hyi between different
  superoutbursts.  A period of 0.076914~d was used to draw this figure.
  Approximate cycle counts ($E$) after the maximum of the superoutburst
  were used.
  }
  \label{fig:vwhyicomp}
\end{figure}

\begin{table}
\caption{Superhump maxima of VW Hyi (2012)}\label{tab:vwhyioc2012}
\begin{center}
\begin{tabular}{rp{55pt}p{40pt}r@{.}lr}
\hline
\multicolumn{1}{c}{$E$} & \multicolumn{1}{c}{max\commenta} & \multicolumn{1}{c}{error} & \multicolumn{2}{c}{$O-C$\commentb} & \multicolumn{1}{c}{$N$\commentc} \\
\hline
0 & 56254.8352 & 0.0012 & $-$0&0172 & 14 \\
11 & 56255.6936 & 0.0022 & $-$0&0034 & 21 \\
22 & 56256.5420 & 0.0019 & 0&0005 & 22 \\
23 & 56256.6145 & 0.0007 & $-$0&0037 & 31 \\
24 & 56256.6923 & 0.0004 & $-$0&0028 & 29 \\
25 & 56256.7690 & 0.0004 & $-$0&0028 & 31 \\
26 & 56256.8442 & 0.0017 & $-$0&0044 & 11 \\
35 & 56257.5382 & 0.0005 & $-$0&0013 & 29 \\
36 & 56257.6160 & 0.0005 & $-$0&0003 & 31 \\
37 & 56257.6942 & 0.0005 & 0&0011 & 28 \\
38 & 56257.7707 & 0.0004 & 0&0008 & 31 \\
48 & 56258.5388 & 0.0006 & 0&0012 & 29 \\
49 & 56258.6154 & 0.0006 & 0&0011 & 31 \\
50 & 56258.6913 & 0.0006 & 0&0001 & 28 \\
51 & 56258.7676 & 0.0004 & $-$0&0003 & 31 \\
54 & 56258.9958 & 0.0007 & $-$0&0025 & 97 \\
55 & 56259.0764 & 0.0003 & 0&0014 & 163 \\
56 & 56259.1536 & 0.0003 & 0&0018 & 167 \\
57 & 56259.2301 & 0.0003 & 0&0015 & 170 \\
61 & 56259.5375 & 0.0013 & 0&0019 & 20 \\
62 & 56259.6134 & 0.0009 & 0&0010 & 22 \\
63 & 56259.6928 & 0.0007 & 0&0036 & 21 \\
64 & 56259.7685 & 0.0009 & 0&0025 & 31 \\
87 & 56261.5389 & 0.0010 & 0&0071 & 23 \\
88 & 56261.6171 & 0.0009 & 0&0086 & 23 \\
89 & 56261.6927 & 0.0010 & 0&0073 & 22 \\
90 & 56261.7691 & 0.0008 & 0&0070 & 31 \\
100 & 56262.5368 & 0.0011 & 0&0069 & 23 \\
101 & 56262.6130 & 0.0010 & 0&0063 & 25 \\
102 & 56262.6896 & 0.0008 & 0&0061 & 25 \\
103 & 56262.7649 & 0.0007 & 0&0047 & 36 \\
113 & 56263.5295 & 0.0014 & 0&0015 & 19 \\
\hline
  \multicolumn{6}{l}{\commenta BJD$-$2400000.} \\
  \multicolumn{6}{l}{\commentb Against max $= 2456254.8524 + 0.076774 E$.} \\
  \multicolumn{6}{l}{\commentc Number of points used to determine the maximum.} \\
\end{tabular}
\end{center}
\end{table}

\addtocounter{table}{-1}
\begin{table}
\caption{Superhump maxima of VW Hyi (2012) (continued)}
\begin{center}
\begin{tabular}{rp{55pt}p{40pt}r@{.}lr}
\hline
\multicolumn{1}{c}{$E$} & \multicolumn{1}{c}{max\commenta} & \multicolumn{1}{c}{error} & \multicolumn{2}{c}{$O-C$\commentb} & \multicolumn{1}{c}{$N$\commentc} \\
\hline
114 & 56263.6079 & 0.0010 & 0&0032 & 25 \\
115 & 56263.6839 & 0.0009 & 0&0024 & 24 \\
116 & 56263.7595 & 0.0006 & 0&0012 & 37 \\
117 & 56263.8383 & 0.0024 & 0&0033 & 14 \\
119 & 56263.9904 & 0.0004 & 0&0018 & 127 \\
120 & 56264.0700 & 0.0005 & 0&0047 & 145 \\
126 & 56264.5229 & 0.0046 & $-$0&0031 & 16 \\
127 & 56264.6034 & 0.0010 & 0&0007 & 25 \\
128 & 56264.6806 & 0.0010 & 0&0011 & 24 \\
129 & 56264.7561 & 0.0008 & $-$0&0002 & 38 \\
130 & 56264.8330 & 0.0023 & $-$0&0001 & 17 \\
139 & 56265.5148 & 0.0053 & $-$0&0093 & 15 \\
140 & 56265.5973 & 0.0013 & $-$0&0036 & 25 \\
141 & 56265.6717 & 0.0009 & $-$0&0059 & 25 \\
142 & 56265.7507 & 0.0009 & $-$0&0037 & 37 \\
143 & 56265.8280 & 0.0025 & $-$0&0031 & 22 \\
153 & 56266.6003 & 0.0032 & 0&0013 & 25 \\
154 & 56266.6727 & 0.0018 & $-$0&0030 & 24 \\
155 & 56266.7443 & 0.0015 & $-$0&0081 & 37 \\
156 & 56266.8234 & 0.0078 & $-$0&0059 & 25 \\
158 & 56266.9779 & 0.0100 & $-$0&0049 & 68 \\
159 & 56267.0557 & 0.0011 & $-$0&0039 & 70 \\
\hline
  \multicolumn{6}{l}{\commenta BJD$-$2400000.} \\
  \multicolumn{6}{l}{\commentb Against max $= 2456254.8524 + 0.076774 E$.} \\
  \multicolumn{6}{l}{\commentc Number of points used to determine the maximum.} \\
\end{tabular}
\end{center}
\end{table}

\begin{table}
\caption{Superhump maxima of VW Hyi (2012) (post-superoutburst)}\label{tab:vwhyioc2012b}
\begin{center}
\begin{tabular}{rp{55pt}p{40pt}r@{.}lr}
\hline
\multicolumn{1}{c}{$E$} & \multicolumn{1}{c}{max\commenta} & \multicolumn{1}{c}{error} & \multicolumn{2}{c}{$O-C$\commentb} & \multicolumn{1}{c}{$N$\commentc} \\
\hline
0 & 56268.5388 & 0.0073 & $-$0&0124 & 17 \\
1 & 56268.6241 & 0.0061 & $-$0&0033 & 25 \\
2 & 56268.6976 & 0.0011 & $-$0&0061 & 30 \\
3 & 56268.7775 & 0.0014 & $-$0&0025 & 37 \\
19 & 56269.9944 & 0.0009 & $-$0&0054 & 113 \\
20 & 56270.0744 & 0.0006 & $-$0&0016 & 162 \\
21 & 56270.1533 & 0.0006 & 0&0010 & 164 \\
22 & 56270.2348 & 0.0009 & 0&0062 & 170 \\
32 & 56270.9885 & 0.0004 & $-$0&0025 & 115 \\
33 & 56271.0642 & 0.0006 & $-$0&0030 & 108 \\
45 & 56271.9901 & 0.0006 & 0&0080 & 110 \\
46 & 56272.0576 & 0.0004 & $-$0&0007 & 128 \\
52 & 56272.5238 & 0.0019 & 0&0080 & 15 \\
53 & 56272.5946 & 0.0010 & 0&0026 & 25 \\
54 & 56272.6717 & 0.0013 & 0&0034 & 26 \\
55 & 56272.7439 & 0.0012 & $-$0&0006 & 36 \\
56 & 56272.8264 & 0.0011 & 0&0056 & 21 \\
66 & 56273.5848 & 0.0007 & 0&0016 & 25 \\
67 & 56273.6624 & 0.0007 & 0&0030 & 25 \\
68 & 56273.7367 & 0.0006 & 0&0010 & 37 \\
69 & 56273.8150 & 0.0008 & 0&0031 & 28 \\
79 & 56274.5765 & 0.0016 & 0&0022 & 25 \\
80 & 56274.6526 & 0.0018 & 0&0020 & 24 \\
81 & 56274.7311 & 0.0009 & 0&0043 & 37 \\
82 & 56274.8085 & 0.0015 & 0&0055 & 33 \\
92 & 56275.5669 & 0.0018 & 0&0014 & 25 \\
93 & 56275.6431 & 0.0020 & 0&0013 & 25 \\
94 & 56275.7164 & 0.0010 & $-$0&0015 & 38 \\
95 & 56275.7973 & 0.0013 & 0&0031 & 36 \\
105 & 56276.5613 & 0.0020 & 0&0047 & 15 \\
106 & 56276.6296 & 0.0009 & $-$0&0033 & 23 \\
107 & 56276.7094 & 0.0009 & 0&0003 & 37 \\
\hline
  \multicolumn{6}{l}{\commenta BJD$-$2400000.} \\
  \multicolumn{6}{l}{\commentb Against max $= 2456268.5512 + 0.076242 E$.} \\
  \multicolumn{6}{l}{\commentc Number of points used to determine the maximum.} \\
\end{tabular}
\end{center}
\end{table}

\addtocounter{table}{-1}
\begin{table}
\caption{Superhump maxima of VW Hyi (2012) (post-superoutburst, continued)}
\begin{center}
\begin{tabular}{rp{55pt}p{40pt}r@{.}lr}
\hline
\multicolumn{1}{c}{$E$} & \multicolumn{1}{c}{max\commenta} & \multicolumn{1}{c}{error} & \multicolumn{2}{c}{$O-C$\commentb} & \multicolumn{1}{c}{$N$\commentc} \\
\hline
108 & 56276.7908 & 0.0011 & 0&0054 & 37 \\
125 & 56278.0760 & 0.0006 & $-$0&0055 & 163 \\
126 & 56278.1569 & 0.0007 & $-$0&0009 & 170 \\
127 & 56278.2315 & 0.0008 & $-$0&0025 & 164 \\
132 & 56278.6073 & 0.0027 & $-$0&0079 & 25 \\
133 & 56278.6847 & 0.0024 & $-$0&0068 & 22 \\
134 & 56278.7632 & 0.0054 & $-$0&0045 & 20 \\
148 & 56279.8325 & 0.0094 & $-$0&0026 & 22 \\
\hline
  \multicolumn{6}{l}{\commenta BJD$-$2400000.} \\
  \multicolumn{6}{l}{\commentb Against max $= 2456268.5512 + 0.076242 E$.} \\
  \multicolumn{6}{l}{\commentc Number of points used to determine the maximum.} \\
\end{tabular}
\end{center}
\end{table}

\begin{table}
\caption{Post-superoutburst superhumps in VW Hyi (2012)}\label{tab:vwhyipost}
\begin{center}
\begin{tabular}{cccc}
\hline
JD$-$2400000 & Period & Error & Amplitude \\
             & (d)    & (d)   & (mag) \\
\hline
56270--56275 & 0.07623 & 0.00001 & 0.33 \\
56275--56280 & 0.07584 & 0.00004 & 0.36 \\
56280--56285 & 0.07732 & 0.00011 & 0.08 \\
56285--56290 & 0.07615 & 0.00014 & 0.13 \\
56290--56295 & 0.07600 & 0.00005 & 0.15 \\
56295--56300 & 0.07586 & 0.00005 & 0.07 \\
56300--56305 & --      & --      & -- \\
56305--56310 & 0.07712 & 0.00013 & 0.08 \\
\hline
\\
\end{tabular}
\end{center}
\end{table}

\subsection{WX Hydri}\label{obj:wxhyi}

   WX Hyi was originally discovered as a variable star
(=AN 9.1932) with a range of 10.7--14.2 (photographic scale
at that time) by \citet{luy32wxhyi}.  \citet{hof49newvar}
classified the object as a Mira-type variable (also \cite{GCVS3}).
\citet{phi71wxhyiiauc2308} noted its blue color, 
rapid light variations and an emission-line spectrum on 
a low-dispersion objective-prism plate.
\citet{kuk71wxhyiiauc2319} suggested the variable to be
either a dwarf nova or a symbiotic object, not a Mira
as originally proposed.  \citet{fis71wxhyiiauc2348}
communicated $UBV$ and visual observations
giving a range (visual and $V$) of 11.5--14.73.
The blue color and variation was incompatible with
the Mira-type classification.  This object became recognized 
as a dwarf nova.  \citet{spl71wxhyi} also reported 
the detection of two outbursts.

   Amateur observers (particularly RASNZ members) started
visual observations since 1971 April.  \citet{bat76wxhyi}
suggested the SU UMa-type classification
based on the presence of superoutbursts detected by
visual observations.  \citet{wal76wxhyi} reported
the detection of superhumps by photoelectric photometry.
The reported period was 0.0783~d based on two-night observation.
\citet{wal76wxhyi} also reported a period of 0.0749~d,
which was suggested to be the orbital period.
\citet{san76wxhyispec} reported a spectrum showing
Balmer lines in emission, which is typical for a dwarf nova.
\citet{bai79wxhyiv436cen} reported high-speed photometry
both in superoutburst and in quiescence.  Using observations
on four consecutive nights in 1977 December, \citet{bai79wxhyiv436cen}
derived a superhump period of 0.07737~d.  This observation
corresponded to the middle part of the superoutburst.
In contrast to \citet{wal76wxhyi}, \citet{bai79wxhyiv436cen}
could not detect orbital modulations in quiescence
(also Bailey 1979, unpublished, see \cite{sch81vwhyiwxhyi}).
\citet{sch81vwhyiwxhyi} obtained high-time resolution spectroscopy
and determined the orbital period to be 0.0748134(2)~d.
\citet{pre06DNOQPO} reported the detection of quasi-periodic
oscillations (QPOs) in quiescence.

   The identification of this object as a dwarf nova
led to a suggestion that some of Mira-type variables
could be misclassified SU UMa-type dwarf novae
\citep{vog80suumastars}.  DH Aql (\cite{tse69dhaqlgmasql};
\cite{nog95dhaql}), SY Cap \citep{Pdot} and FQ Mon
(vsnet-chat 3063, 3066; \cite{Pdot}) are indeed
such objects.
Despite that WX Hyi is a well-known SU UMa-type dwarf nova,
the listed set of literature was probably the last published
observation of superhumps before this paper.

   Our 2014 January--February observation started after 
the detection of a bright outburst on January 27 by S. Hovell
and R. Stubbings (the start of the outburst was on
January 25).  Observations on January 30 detected
fully grown superhumps and the subsequent evolution
was observed (vsnet-alert 16851, 16868, 16904).
The times of superhump maxima are listed in table
\ref{tab:wxhyioc2014}, which includes post-superoutburst
observations.

\begin{figure}
  \begin{center}
    \FigureFile(88mm,70mm){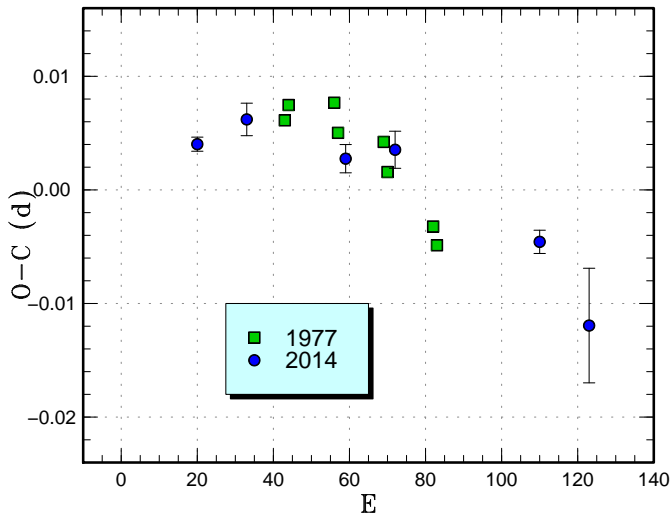}
  \end{center}
  \caption{Comparison of $O-C$ diagrams of WX Hyi between different
  superoutbursts.  A period of 0.07765~d was used to draw this figure.
  Approximate cycle counts ($E$) after the maximum of the superoutburst
  were used.
  }
  \label{fig:wxhyicomp}
\end{figure}

   A comparison of the $O-C$ diagrams (figure \ref{fig:wxhyicomp})
suggests that the first two nights of \citet{sch81vwhyiwxhyi}
recorded stage B superhumps and the last two nights stage C
superhumps (since both superoutbursts started with a precursor,
we used the maximum which was easier to define).
The resultant period of stage B is in good
agreement with the value by \citet{wal76wxhyi}, who reported
an early part of the superoutburst.
We listed the estimated periods in this interpretation
in table \ref{tab:perlist}.

\begin{table}
\caption{Superhump maxima of WX Hyi (2014)}\label{tab:wxhyioc2014}
\begin{center}
\begin{tabular}{rp{55pt}p{40pt}r@{.}lr}
\hline
\multicolumn{1}{c}{$E$} & \multicolumn{1}{c}{max\commenta} & \multicolumn{1}{c}{error} & \multicolumn{2}{c}{$O-C$\commentb} & \multicolumn{1}{c}{$N$\commentc} \\
\hline
0 & 56687.5605 & 0.0006 & $-$0&0035 & 47 \\
13 & 56688.5721 & 0.0014 & 0&0007 & 28 \\
39 & 56690.5876 & 0.0012 & 0&0012 & 19 \\
52 & 56691.5978 & 0.0016 & 0&0039 & 13 \\
90 & 56694.5404 & 0.0010 & 0&0015 & 21 \\
103 & 56695.5425 & 0.0050 & $-$0&0039 & 21 \\
\hline
  \multicolumn{6}{l}{\commenta BJD$-$2400000.} \\
  \multicolumn{6}{l}{\commentb Against max $= 2456687.5640 + 0.077499 E$.} \\
  \multicolumn{6}{l}{\commentc Number of points used to determine the maximum.} \\
\end{tabular}
\end{center}
\end{table}

\subsection{AY Lyrae}\label{obj:aylyr}

   Observations of this well-known SU UMa-type dwarf nova
were performed only on two nights in 2013 August.
The times of superhump maxima are listed
in table \ref{tab:aylyroc2013}.

\begin{table}
\caption{Superhump maxima of AY Lyr (2013)}\label{tab:aylyroc2013}
\begin{center}
\begin{tabular}{rp{55pt}p{40pt}r@{.}lr}
\hline
\multicolumn{1}{c}{$E$} & \multicolumn{1}{c}{max\commenta} & \multicolumn{1}{c}{error} & \multicolumn{2}{c}{$O-C$\commentb} & \multicolumn{1}{c}{$N$\commentc} \\
\hline
0 & 56533.0467 & 0.0007 & $-$0&0000 & 84 \\
1 & 56533.1221 & 0.0005 & $-$0&0007 & 79 \\
2 & 56533.1996 & 0.0008 & 0&0008 & 55 \\
14 & 56534.1116 & 0.0006 & $-$0&0001 & 39 \\
\hline
  \multicolumn{6}{l}{\commenta BJD$-$2400000.} \\
  \multicolumn{6}{l}{\commentb Against max $= 2456533.0467 + 0.076064 E$.} \\
  \multicolumn{6}{l}{\commentc Number of points used to determine the maximum.} \\
\end{tabular}
\end{center}
\end{table}

\subsection{AO Octantis}\label{obj:aooct}

   Due to the large outburst amplitude (7.5 mag) listed in
\citet{GCVS}, this object had long been considered as
a candidate WZ Sge-type dwarf nova (\cite{dow90wxcet};
\cite{how90highgalCV}; \cite{odo91wzsge}; \cite{kat01hvvir}).
Although \citet{how91faintCV4} observed this object in quiescence,
no orbital modulation was detected.
\citet{mas03faintCV} obtained a spectrum in quiescence,
which was typical for a dwarf nova with a low mass-transfer
rate but was not so extreme as a WZ Sge-type dwarf nova.
\citet{pat03suumas} observed the 2000 September outburst
and obtained a superhump period of 0.06557(13)~d.
\citet{wou04CV4} obtained time-resolved photometry
in quiescence and detected an orbital modulation
with a period of 0.065345(15)~d.  The presence of the orbital
modulation appears to be consistent with the relatively
broad emission lines in \citet{mas03faintCV}.

\begin{figure}
  \begin{center}
    \FigureFile(88mm,110mm){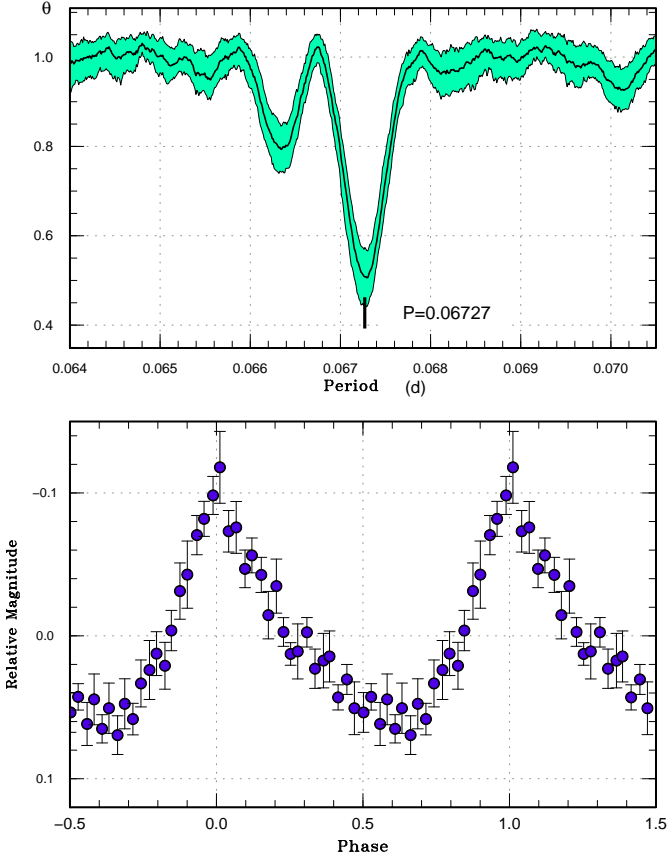}
  \end{center}
  \caption{Superhumps in AO Oct (2013).  (Upper): PDM analysis.
     (Lower): Phase-averaged profile.}
  \label{fig:aooctshpdm}
\end{figure}

   The 2013 superoutburst of AO Oct was detected by
R. Stubbings (vsnet-alert 16376).  Subsequent observations
detected superhumps (vsnet-alert 16388, 16396, 16411;
figure \ref{fig:aooctshpdm}).
The times of superhump maxima are listed in table
\ref{tab:aooctoc2013}.  The stages B and C were clearly
present.  The large $P_{\rm dot}$ of $+19(6) \times 10^{-5}$
for stage B superhumps is typical for this $P_{\rm orb}$.

   According to \citet{wou04CV4}, the maximum magnitude
of 13.5 in \citet{GCVS} was probably a typographical error
of 15.3 based on the discovery paper \citep{vonges74southvar}.
The object, however, has been detected as bright as 14.2
(visual magnitude) in outburst several times.
The true range of variability can be regarded as 14.2--20.9,
where the minimum magnitude is taken from \citet{wou04CV4}.
Considering the supercycle of $\sim$300~d, this outburst
amplitude is indeed slightly too large for this intermediate
length of supercycle.  Although TV Crv was reported
to have similar parameters (cf. table 1 in \cite{nog97sxlmi}),
the maximum magnitude of TV Crv was probably an overestimate,
since recent magnitudes of superoutbursts only reach 13.0.
AO Oct apparently deserves a further study for its rather
unusual outburst parameters.

\begin{table}
\caption{Superhump maxima of AO Oct (2013)}\label{tab:aooctoc2013}
\begin{center}
\begin{tabular}{rp{55pt}p{40pt}r@{.}lr}
\hline
\multicolumn{1}{c}{$E$} & \multicolumn{1}{c}{max\commenta} & \multicolumn{1}{c}{error} & \multicolumn{2}{c}{$O-C$\commentb} & \multicolumn{1}{c}{$N$\commentc} \\
\hline
0 & 56545.5712 & 0.0006 & 0&0003 & 29 \\
1 & 56545.6371 & 0.0007 & $-$0&0011 & 12 \\
2 & 56545.7043 & 0.0021 & $-$0&0011 & 9 \\
3 & 56545.7744 & 0.0016 & 0&0017 & 8 \\
14 & 56546.5113 & 0.0009 & $-$0&0011 & 22 \\
15 & 56546.5781 & 0.0011 & $-$0&0015 & 28 \\
16 & 56546.6466 & 0.0012 & $-$0&0002 & 15 \\
17 & 56546.7126 & 0.0009 & $-$0&0015 & 15 \\
29 & 56547.5202 & 0.0011 & $-$0&0007 & 26 \\
45 & 56548.5933 & 0.0066 & $-$0&0035 & 10 \\
46 & 56548.6664 & 0.0014 & 0&0024 & 13 \\
59 & 56549.5480 & 0.0011 & 0&0099 & 27 \\
75 & 56550.6161 & 0.0012 & 0&0022 & 25 \\
76 & 56550.6872 & 0.0032 & 0&0061 & 12 \\
89 & 56551.5518 & 0.0020 & $-$0&0035 & 27 \\
90 & 56551.6183 & 0.0052 & $-$0&0043 & 12 \\
91 & 56551.6855 & 0.0019 & $-$0&0042 & 14 \\
\hline
  \multicolumn{6}{l}{\commenta BJD$-$2400000.} \\
  \multicolumn{6}{l}{\commentb Against max $= 2456545.5710 + 0.067240 E$.} \\
  \multicolumn{6}{l}{\commentc Number of points used to determine the maximum.} \\
\end{tabular}
\end{center}
\end{table}

\subsection{DT Octantis}\label{obj:dtoct}

   This object was discovered as a variable star (=BV 966)
with a large amplitude (11.2 to fainter than 15.0 in
photographic magnitudes) \citep{kni67dtoct}.
\citet{kat02gzcncnsv10934} noticed the identification with 
a bright ROSAT source and suggested that the object is a
cataclysmic variable.  \citet{kat02gzcncnsv10934} detected 
multiple outbursts upon this suggestion.  Although
\citet{kat02gzcncnsv10934} initially suggested that these
outbursts may be outbursts in an intermediate polar,
\citet{kat04nsv10934mmscoabnorcal86} detected superhumps
during the 2003 January outburst.  DT Oct was thus recognized
as an SU UMa-type dwarf nova.  \citet{Pdot} further studied
another superoutburst in 2003 November and the superoutburst
in 2008.

\begin{figure}
  \begin{center}
    \FigureFile(88mm,70mm){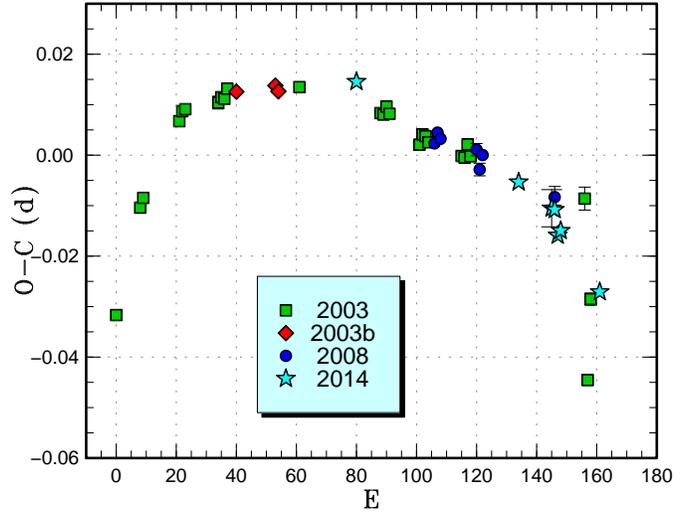}
  \end{center}
  \caption{Comparison of $O-C$ diagrams of DT Oct between different
  superoutbursts.  A period of 0.07485~d was used to draw this figure.
  Approximate cycle counts ($E$) after the start of the superoutburst
  were used.  Since the start of the 2014 superoutburst
  was not well constrained, we shifted the $O-C$ diagram
  to best fit the others.
  }
  \label{fig:dtoctcomp2}
\end{figure}

   The 2014 superoutburst was detected by R. Stubbings
(cf. vsnet-alert 16892) and the later part of the superoutburst
was observed.  The times of superhump maxima are listed
in table \ref{tab:dtoctoc2014}.  The data mostly recorded
stage C superhumps (figure \ref{fig:dtoctcomp2}).

   Using the near quiescent data in 2013 (by OkC,
BJD 2456380--2456445), we have obtained a possible orbital
signal of 0.072707(5)~d with a mean amplitude of 0.16 mag.
This period was adopted in table \ref{tab:perlist}.
The $\varepsilon^*$ value for stage A superhumps 
[0.050(2)] in 2003 \citep{Pdot} corresponds to
$q$=0.147(7).

\begin{table}
\caption{Superhump maxima of DT Oct (2014)}\label{tab:dtoctoc2014}
\begin{center}
\begin{tabular}{rp{55pt}p{40pt}r@{.}lr}
\hline
\multicolumn{1}{c}{$E$} & \multicolumn{1}{c}{max\commenta} & \multicolumn{1}{c}{error} & \multicolumn{2}{c}{$O-C$\commentb} & \multicolumn{1}{c}{$N$\commentc} \\
\hline
0 & 56696.7567 & 0.0008 & $-$0&0020 & 83 \\
54 & 56700.7787 & 0.0006 & 0&0031 & 119 \\
65 & 56701.5969 & 0.0037 & 0&0031 & 35 \\
66 & 56701.6714 & 0.0007 & 0&0032 & 66 \\
67 & 56701.7412 & 0.0006 & $-$0&0014 & 80 \\
68 & 56701.8170 & 0.0014 & 0&0000 & 63 \\
81 & 56702.7779 & 0.0011 & $-$0&0061 & 70 \\
\hline
  \multicolumn{6}{l}{\commenta BJD$-$2400000.} \\
  \multicolumn{6}{l}{\commentb Against max $= 2456696.7587 + 0.074386 E$.} \\
  \multicolumn{6}{l}{\commentc Number of points used to determine the maximum.} \\
\end{tabular}
\end{center}
\end{table}

\subsection{V521 Pegasi}\label{obj:v521peg}

   This object (=HS 2219$+$1824) is a dwarf nova which was
reported in \citet{rod05hs2219}.  Although \citet{rod05hs2219}
reported the detection of superhumps and likely orbital
modulation, subsequent superoutbursts occurred in poor
seasonal condition, and it was only in 2012 when we
succeeded in obtaining the superhump period \citep{Pdot5}.

   The 2013 superoutburst was detected by the ASAS-SN team
(vsnet-alert 16093).  K. Wenzel also reported the outburst
detection before the ASAS-SN detection.
Subsequent observations detected
superhumps (vsnet-alert 16121, 16129, 16134).
After rapid fading from the superoutburst, the object
continued to show superhumps (vsnet-alert 16146, 16149, 16166,
16186).  The times of superhumps are listed in table
\ref{tab:v521pegoc2013}, which includes post-superoutburst
observations.  Stages B and C were observed.
During the phase of the rapid fading, the superhump
profile became doubly humped.  The maxima which are
on the smooth extension of the rest of the data
were selected in the table.

   A comparison of the $O-C$ diagram clarified that
the 2012 observation recorded stage C superhumps,
rather stage B superhump identified in \citet{Pdot5}.

\begin{figure}
  \begin{center}
    \FigureFile(88mm,70mm){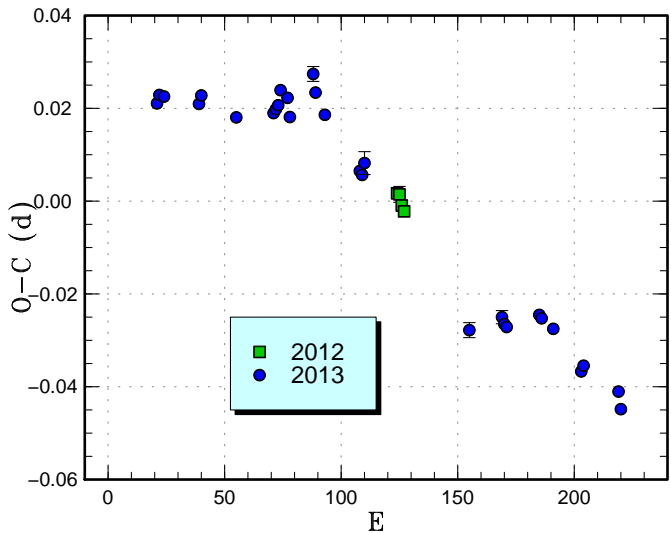}
  \end{center}
  \caption{Comparison of $O-C$ diagrams of V521 Peg between different
  superoutbursts.  A period of 0.06150~d was used to draw this figure.
  Approximate cycle counts ($E$) after the start of the superoutburst
  were used.
  }
  \label{fig:v521pegcomp}
\end{figure}

\begin{table}
\caption{Superhump maxima of V521 Peg (2013)}\label{tab:v521pegoc2013}
\begin{center}
\begin{tabular}{rp{55pt}p{40pt}r@{.}lr}
\hline
\multicolumn{1}{c}{$E$} & \multicolumn{1}{c}{max\commenta} & \multicolumn{1}{c}{error} & \multicolumn{2}{c}{$O-C$\commentb} & \multicolumn{1}{c}{$N$\commentc} \\
\hline
0 & 56507.0874 & 0.0009 & $-$0&0139 & 44 \\
1 & 56507.1507 & 0.0006 & $-$0&0118 & 66 \\
3 & 56507.2734 & 0.0005 & $-$0&0114 & 63 \\
18 & 56508.1942 & 0.0007 & $-$0&0074 & 70 \\
19 & 56508.2575 & 0.0007 & $-$0&0052 & 67 \\
34 & 56509.1753 & 0.0006 & $-$0&0043 & 40 \\
50 & 56510.1601 & 0.0004 & 0&0026 & 91 \\
51 & 56510.2223 & 0.0008 & 0&0037 & 86 \\
52 & 56510.2848 & 0.0005 & 0&0050 & 96 \\
53 & 56510.3497 & 0.0011 & 0&0088 & 41 \\
56 & 56510.5325 & 0.0007 & 0&0082 & 66 \\
57 & 56510.5899 & 0.0012 & 0&0045 & 28 \\
67 & 56511.2146 & 0.0016 & 0&0179 & 63 \\
68 & 56511.2718 & 0.0009 & 0&0140 & 62 \\
72 & 56511.5129 & 0.0003 & 0&0106 & 47 \\
87 & 56512.4234 & 0.0008 & 0&0042 & 52 \\
88 & 56512.4840 & 0.0015 & 0&0037 & 159 \\
89 & 56512.5469 & 0.0030 & 0&0055 & 211 \\
134 & 56515.2792 & 0.0016 & $-$0&0128 & 123 \\
148 & 56516.1433 & 0.0014 & $-$0&0044 & 113 \\
149 & 56516.2033 & 0.0009 & $-$0&0055 & 230 \\
150 & 56516.2645 & 0.0012 & $-$0&0054 & 226 \\
164 & 56517.1278 & 0.0006 & 0&0021 & 80 \\
165 & 56517.1885 & 0.0008 & 0&0017 & 128 \\
170 & 56517.4937 & 0.0004 & 0&0013 & 54 \\
182 & 56518.2226 & 0.0006 & $-$0&0034 & 82 \\
183 & 56518.2853 & 0.0006 & $-$0&0018 & 125 \\
198 & 56519.2022 & 0.0011 & $-$0&0017 & 126 \\
199 & 56519.2599 & 0.0012 & $-$0&0051 & 120 \\
\hline
  \multicolumn{6}{l}{\commenta BJD$-$2400000.} \\
  \multicolumn{6}{l}{\commentb Against max $= 2456507.1013 + 0.061124 E$.} \\
  \multicolumn{6}{l}{\commentc Number of points used to determine the maximum.} \\
\end{tabular}
\end{center}
\end{table}

\subsection{TY Piscium}\label{obj:typsc}

   Although the SU UMa-type nature of TY Psa had been long
known, the information in the literature was very limited
(\cite{szk88DNnovaIR}; \cite{kun01typsc}).  \citet{Pdot}
was the first to determine the superhump period precisely
during the 2005 and 2008 superoutbursts.  \citet{tho96Porb}
determined the orbital period by a radial-velocity study.

\begin{figure}
  \begin{center}
    \FigureFile(88mm,70mm){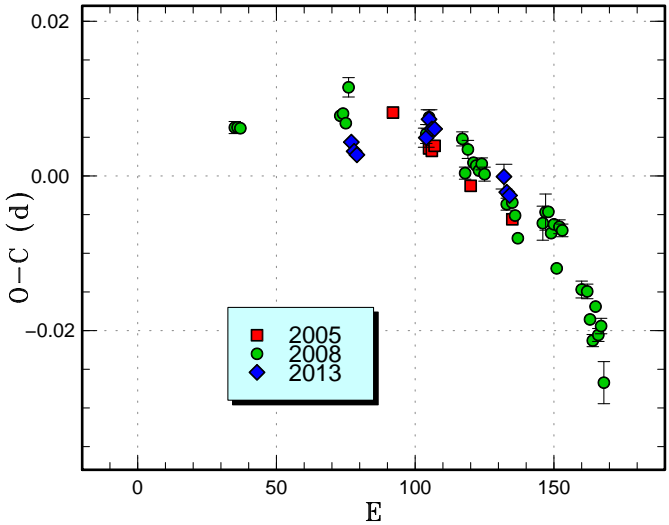}
  \end{center}
  \caption{Comparison of $O-C$ diagrams of TY Psc between different
  superoutbursts.  A period of 0.07066~d was used to draw this figure.
  Approximate cycle counts ($E$) after the start of the superoutburst
  were used.  Since the start of the 2013 superoutburst
  was not well constrained, we shifted the $O-C$ diagram
  to best fit the others.
  }
  \label{fig:typsccomp2}
\end{figure}

   The 2013 December superoutburst was detected by
Kyoto and Kiso Wide-field Survey (KWS) (vsnet-alert 16682)
and was observed for three nights.
The times of superhump maxima are listed in
table \ref{tab:typscoc2013}.  The observation apparently
detected stage B-C transition.  This identification is
confirmed by a comparison of the $O-C$ diagrams
(figure \ref{fig:typsccomp2}.  Also note that the figure
in \citet{Pdot} used the period of stage C superhumps,
rather than that of stage B superhumps, making 
the impression of the figure different).

\begin{table}
\caption{Superhump maxima of TY Psc (2013)}\label{tab:typscoc2013}
\begin{center}
\begin{tabular}{rp{55pt}p{40pt}r@{.}lr}
\hline
\multicolumn{1}{c}{$E$} & \multicolumn{1}{c}{max\commenta} & \multicolumn{1}{c}{error} & \multicolumn{2}{c}{$O-C$\commentb} & \multicolumn{1}{c}{$N$\commentc} \\
\hline
0 & 56635.0415 & 0.0006 & $-$0&0012 & 91 \\
1 & 56635.1110 & 0.0004 & $-$0&0023 & 130 \\
2 & 56635.1812 & 0.0005 & $-$0&0027 & 109 \\
27 & 56636.9499 & 0.0012 & 0&0018 & 25 \\
28 & 56637.0230 & 0.0012 & 0&0043 & 46 \\
29 & 56637.0924 & 0.0006 & 0&0031 & 63 \\
30 & 56637.1630 & 0.0009 & 0&0032 & 47 \\
55 & 56638.9234 & 0.0016 & $-$0&0007 & 52 \\
56 & 56638.9920 & 0.0008 & $-$0&0026 & 52 \\
57 & 56639.0622 & 0.0008 & $-$0&0029 & 101 \\
\hline
  \multicolumn{6}{l}{\commenta BJD$-$2400000.} \\
  \multicolumn{6}{l}{\commentb Against max $= 2456635.0428 + 0.070569 E$.} \\
  \multicolumn{6}{l}{\commentc Number of points used to determine the maximum.} \\
\end{tabular}
\end{center}
\end{table}

\subsection{V893 Scorpii}\label{obj:v893sco}

   V893 Sco was discovered as a variable star (=SVS 1772) by
\citet{sat72v893sco}.  Since this object is located in the region of
the Scorpius T1 association, \citet{sat82v893sco} classified this
object as a rapid irregular variable of InSF type (object normally
in faint states with occasional brightenings up to 3 mag)
according to the classification scheme by \citet{fil75scot1}.
This classification corresponded to ``RWF''-type RW Aur-type
in \citet{tse73rwaurbook}.  Around this time, the RW Aur-type
or ``In''-type (irregular, nebular, cf. \cite{GCVS}) referred
to pre-main sequence variables.  It is apparent that 
\citet{sat82v893sco} considered this variable as a 
pre-main sequence variable.  In \citet{GCVS}, however,
the object was reclassified as a dwarf nova probably based
on the published light curve.

   The finding chart in this discovery article was interchanged
with that of a different star, and this bright dwarf nova remained
virtually ``lost'' for a long time (cf. \cite{DownesCVatlas2}).
The presented light curves in \citet{fil75scot1} and
\citet{sat82v893sco}, however, were so characteristic of
a dwarf nova and at least two outbursts
may be attributed to a superoutburst due to its long duration,
this object attracted amateur astronomers (particularly VSOLJ
members) since the late 1980s.
Despite the high potentiality of being a bright SU UMa-type 
dwarf nova, all attempts (visually watching for the nominal
position and photographic searches) to recover this variable 
had been unsuccessful.

   In 1998, K. Haseda reported a detection of a transient
object near the catalog position of V893 Sco, and this object
was readily identified with an ROSAT X-ray source.
A search for the plate collections at the time of
observations of \citet{sat82v893sco} clarified that this
outbursting object is indeed V893 Sco.  This rediscovery
was reported in \citet{kat98v893sco}.

   \citet{tho99v893sco} confirmed spectroscopically that 
this object is indeed a dwarf nova and obtained an orbital
period of 0.0760~d.  This short orbital period strengthened
the suggestion that this object belongs to the SU UMa-type
dwarf novae.  \citet{tho99v893sco} also measured a large
proper motion, implying a nearby object.

   Since 1998, this object has been regularly monitored
by amateur observers, and outbursts reaching $\sim$12 mag
were recorded.  During the observation in 1999, the group
by K. Matsumoto clarified that this object is a grazing
eclipsing dwarf nova below the period gap (vsnet-alert 3432,
announced on 1999 September 2).  \citet{bru00v893sco} independently
reached the same conclusion and submitted a paper
on 1999 September 11.  The result of the former research
was published as \citet{mat00v893sco}.

   Although \citet{bru00v893sco} mentioned that V893 Sco
cannot be an ER UMa-type dwarf nova, \citet{mas01v893sco}
suggested that it is an ER UMa-type dwarf nova by demonstrating
their new Doppler tomograms.  \citet{kat02v893sco}
explained that this object cannot be an ER UMa-type
dwarf nova.

   Such a confusion apparently comes from the lack of
a definite superoutburst, despite that its existence has
been expected for the short orbital period.  Although
there have been a number of possible detections of
``slightly brighter'' outbursts, none of them had been
confirmed to be a genuine superoutburst until 2013.

\begin{figure}
  \begin{center}
    \FigureFile(88mm,70mm){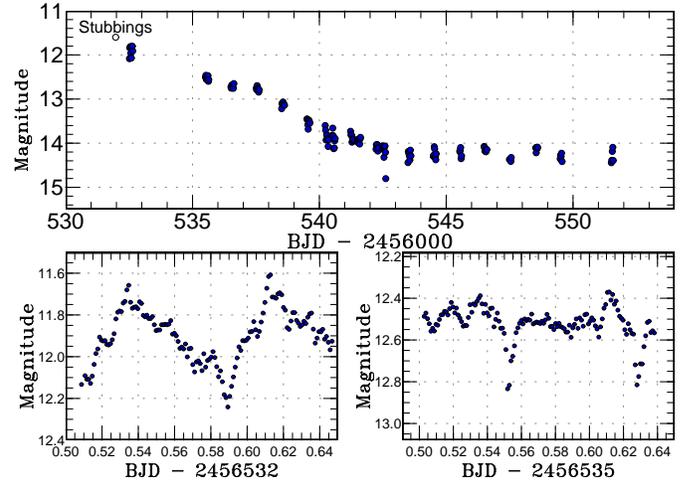}
  \end{center}
  \caption{Light curve of V893 Sco during the superoutburst (2013).
     (Upper): Overall light curve.
     (Lower panels): Examples of nightly observations.  Both
     eclipses and superhumps were detected.}
  \label{fig:v893scolc}
\end{figure}

\begin{figure}
  \begin{center}
    \FigureFile(88mm,110mm){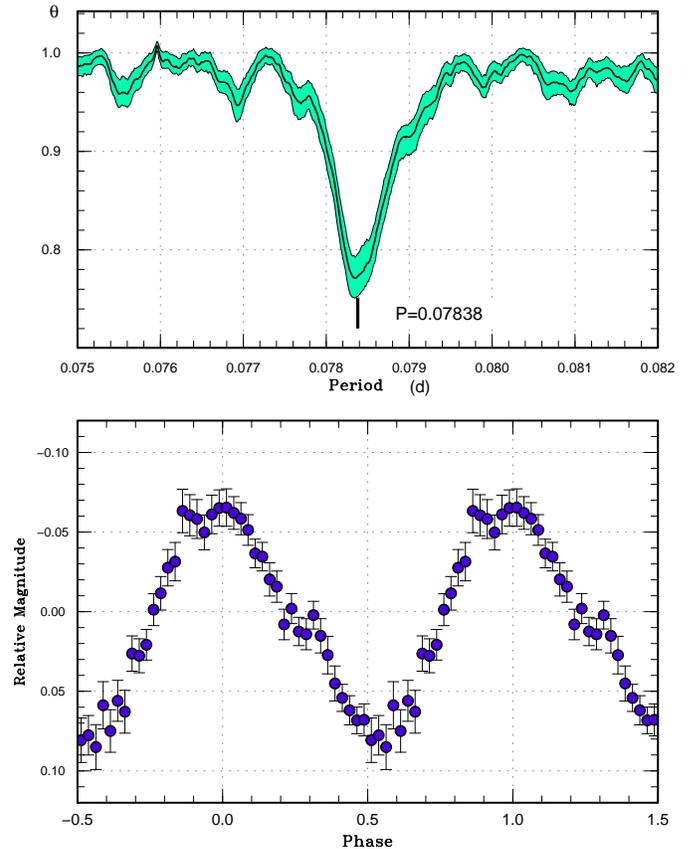}
  \end{center}
  \caption{Superhumps in V893 Sco after subtracting the orbital
     modulation (2013). (Upper): PDM analysis.
     (Lower): Phase-averaged profile.}
  \label{fig:v893scoshpdm}
\end{figure}

   On 2013 August 27, R. Stubbings reported a bright
(11.6 mag) outburst (vsnet-alert 16276;
figure \ref{fig:v893scolc}).
Subsequent observation of this outburst finally confirmed 
the presence of superhumps (vsnet-alert 16315; 
figure \ref{fig:v893scoshpdm}).
We observed this superoutburst and report the result here.

   Since the eclipse ephemeris by \citet{bru00v893sco} does
not fit modern observations \citep{muk09v893scoSuzaku},
we first refined the eclipse ephemeris.
Since the white dwarf is partially eclipsed
\citep{muk09v893scoSuzaku}, we used outburst observations
in which the central part of the accretion disk is expected
to be eclipsed, and the times of minima are expected to be
close to the center of eclipse of the white dwarf.
We used the combined set of the 2007 data (MLF and OKU data),
2008 data (GBo data), 2010 data (GBo and OKU data)
and the present data.  All observations other than the 2013
data were obtained during normal outbursts.
We obtained an ephemeris of
\begin{equation}
{\rm Min(BJD)} = 2454173.3030(4) + 0.0759614614(18) E
\label{equ:v893scoecl}
\end{equation}
using the MCMC modeling \citet{Pdot4}.
This orbital period is in good agreement with
\citet{muk09v893scoSuzaku}, suggesting that the original
orbital period by \citet{bru00v893sco} was systematically
too long.

   The times of superhump maxima after subtracting
the mean orbital modulations were determined outside
the eclipses (orbital phase 0.07--0.93)
(table \ref{tab:v893scooc2013}).  Although some of
superhumps were visible in the light curve, some of
the times of maxima could not be determined because
the superhump maxima coincided the eclipses.
These superhumps were not included in table \ref{tab:v893scooc2013}.
We identified stages B and C and gave the measured
periods in table \ref{tab:perlist}.

   As mentioned in \citet{bru00v893sco}, such a bright
dwarf nova escaped detection of nova searches.
This may have been the chance coincidence of superoutbursts
occurring in unfavorable seasonal condition before
modern CCD-based search became popular.  In table
\ref{tab:v893scosuper}, we list the possible superoutburst
in modern observations.  Except for the 2013 one, all
suspected superoutburst occurred near the solar conjunction.
It may be that the supercycle is close to one year,
and all superoutbursts in the late 1990s and 2000s
could not be observed due to the solar conjunction.
The relatively small outburst amplitude is likely
a result of the high system inclination.  It was also
likely that the magnitude scale in \citet{sat82v893sco}
was $\sim$1 mag brighter than the modern one.

\begin{table}
\caption{Superhump maxima of V893 Sco (2013)}\label{tab:v893scooc2013}
\begin{center}
\begin{tabular}{rp{50pt}p{30pt}r@{.}lcr}
\hline
$E$ & max\commenta & error & \multicolumn{2}{c}{$O-C$\commentb} & phase\commentc & $N$\commentd \\
\hline
0 & 56532.5397 & 0.0008 & $-$0&0073 & 0.34 & 43 \\
1 & 56532.6179 & 0.0010 & $-$0&0075 & 0.37 & 45 \\
38 & 56535.5283 & 0.0013 & 0&0032 & 0.68 & 42 \\
39 & 56535.6098 & 0.0028 & 0&0063 & 0.76 & 36 \\
51 & 56536.5526 & 0.0007 & 0&0087 & 0.17 & 36 \\
52 & 56536.6294 & 0.0012 & 0&0071 & 0.18 & 15 \\
89 & 56539.5270 & 0.0016 & 0&0050 & 0.32 & 40 \\
99 & 56540.3015 & 0.0009 & $-$0&0041 & 0.52 & 136 \\
102 & 56540.5296 & 0.0005 & $-$0&0111 & 0.52 & 41 \\
128 & 56542.5793 & 0.0013 & 0&0009 & 0.51 & 47 \\
140 & 56543.5208 & 0.0010 & 0&0020 & 0.90 & 36 \\
141 & 56543.5975 & 0.0013 & 0&0003 & 0.91 & 35 \\
153 & 56544.5343 & 0.0066 & $-$0&0034 & 0.24 & 24 \\
\hline
  \multicolumn{7}{l}{\commenta BJD$-$2400000.} \\
  \multicolumn{7}{l}{\commentb Against max $= 2456532.5471 + 0.078370 E$.} \\
  \multicolumn{7}{l}{\commentc Orbital phase.} \\
  \multicolumn{7}{l}{\commentd Number of points used to determine the maximum.} \\
\end{tabular}
\end{center}
\end{table}

\begin{table}
\caption{Possible superoutbursts of V893 Sco}\label{tab:v893scosuper}
\begin{center}
\begin{tabular}{cccc}
\hline
JD & Date & maximum & duration (d) \\
\hline
2454883 & 2009 February 20 & 11.7 & $>$8 \\
2455975 & 2012 February 17 & 12.0 & $>$4 \\
2456532 & 2013 August 27 & 11.6 & $>$8\commenta \\
\hline
  \multicolumn{4}{l}{\commenta Confirmed superoutburst.} \\
\end{tabular}
\end{center}
\end{table}

\subsection{RZ Sagittae}\label{obj:rzsge}

   RZ Sge has long been known as a dwarf nova with a long
cycle length (e.g. \cite{pet56uvper}).  \citet{bon82rzsge}
reported the detection of superhumps during the 1981 October
outburst.  Retrospective examination of the past visual
observation also clarified a number of superoutbursts
in the 1970s \citep{bon82rzsge}.  \citet{kat96rzsge} and
\citet{sem97rzsge} reported observations of superhumps
during the 1994 and 1996 superoutburst, respectively.
Although both \citet{kat96rzsge} and \citet{sem97rzsge}
resulted in global negative $P_{\rm dot}$ of about
$-10 \times 10^{-5}$, \citet{Pdot} considered that
this is a result of stage B-C transition and that
$P_{\rm dot}$ for stage B can be positive.
\citet{pat03suumas} also reported the 1996 superoutburst
and detection of the photometric orbital period in 1999.

\begin{figure}
  \begin{center}
    \FigureFile(88mm,70mm){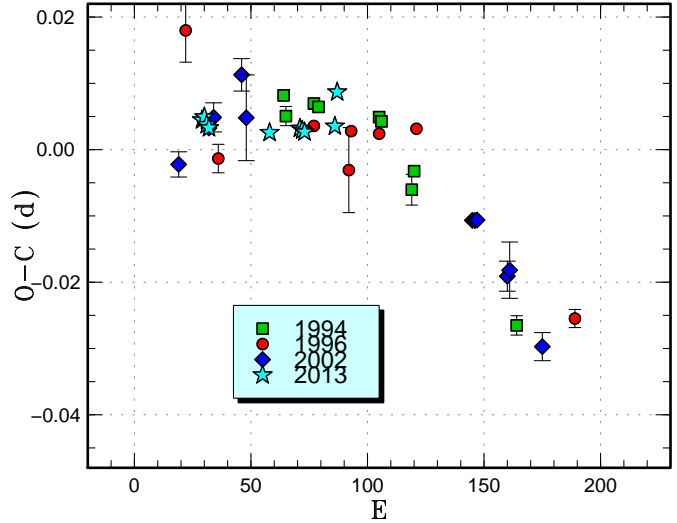}
  \end{center}
  \caption{Comparison of $O-C$ diagrams of RZ Sge between different
  superoutbursts.  A period of 0.07063~d was used to draw this figure.
  Approximate cycle counts ($E$) after the start of the superoutburst
  were used.}
  \label{fig:rzsgecomp2}
\end{figure}

   We observed the 2013 superoutburst, which was detected
in relatively early phase (vsnet-alert 16326).
The times of superhump maxima are listed in table
\ref{tab:rzsgeoc2013}.  The present data suggest a
positive $P_{\rm dot}$ (the data for $E=58$ was better
than $E=57$, and the positive $O-C$ for $E=58$ appears
to be real).  The $O-C$ values of present and past superoutbursts
can be well expressed by the stage B and C
(figure \ref{fig:rzsgecomp2}).

\begin{table}
\caption{Superhump maxima of RZ Sge (2013)}\label{tab:rzsgeoc2013}
\begin{center}
\begin{tabular}{rp{55pt}p{40pt}r@{.}lr}
\hline
\multicolumn{1}{c}{$E$} & \multicolumn{1}{c}{max\commenta} & \multicolumn{1}{c}{error} & \multicolumn{2}{c}{$O-C$\commentb} & \multicolumn{1}{c}{$N$\commentc} \\
\hline
0 & 56539.3871 & 0.0002 & 0&0008 & 90 \\
1 & 56539.4582 & 0.0002 & 0&0013 & 140 \\
2 & 56539.5277 & 0.0002 & 0&0001 & 136 \\
3 & 56539.5978 & 0.0013 & $-$0&0004 & 51 \\
29 & 56541.4335 & 0.0003 & $-$0&0014 & 89 \\
42 & 56542.3523 & 0.0003 & $-$0&0010 & 90 \\
43 & 56542.4227 & 0.0004 & $-$0&0012 & 90 \\
44 & 56542.4930 & 0.0002 & $-$0&0015 & 77 \\
57 & 56543.4121 & 0.0005 & $-$0&0008 & 14 \\
58 & 56543.4879 & 0.0006 & 0&0043 & 22 \\
\hline
  \multicolumn{6}{l}{\commenta BJD$-$2400000.} \\
  \multicolumn{6}{l}{\commentb Against max $= 2456539.3863 + 0.070642 E$.} \\
  \multicolumn{6}{l}{\commentc Number of points used to determine the maximum.} \\
\end{tabular}
\end{center}
\end{table}

\subsection{AW Sagittae}\label{obj:awsge}

\begin{figure}
  \begin{center}
    \FigureFile(88mm,70mm){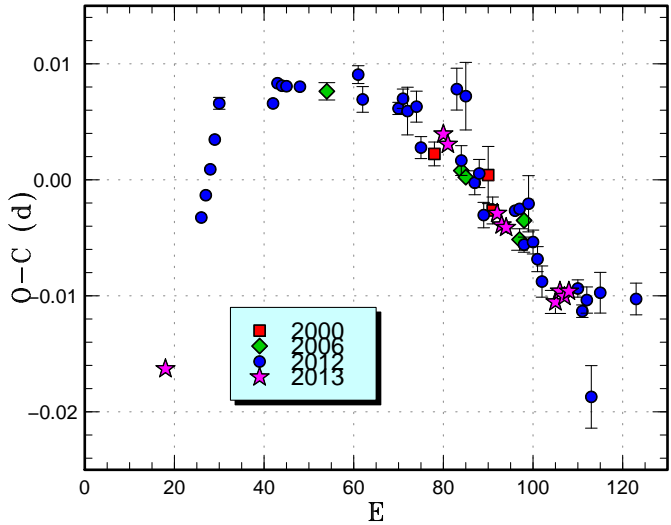}
  \end{center}
  \caption{Comparison of $O-C$ diagrams of AW Sge between different
  superoutbursts.  A period of 0.07480~d was used to draw this figure.
  Approximate cycle counts ($E$) after the start of the superoutburst
  were used.
  Since the start of the superoutburst was best constrained
  in 2013, we shifted the $O-C$ diagrams to fit the 2013 one.
  This made a 18 cycle correction to the figure in \citet{Pdot5}.
  }
  \label{fig:awsgecomp2}
\end{figure}

   AW Sge was discovered as a dwarf nova by \citet{wol06awsge}.
The 2000 and 2006 superoutbursts were reported in \citet{Pdot}.
\citet{Pdot5} further reported the best observed 2012 superoutburst.
On 2013 October 6, R. Stubbings detected another likely
superoutburst (vsnet-alert 16512).  This outburst was independently
detected several hours earlier by AAVSO observers.
On the very night of this
detection, no superhumps were detected.  On the next night,
a developing superhump was detected.  There was a 4-d gap
after these observations.  The times of superhump maxima
are listed in table \ref{tab:awsgeoc2013}.  The epoch $E=0$
was given a cycle count assuming that this was a stage A
superhump with a longer period.  A comparison of the period
for $E \le 62$ indicated that they are stage C superhumps.
This identification was confirmed by comparison of
$O-C$ diagrams (figure \ref{fig:awsgecomp2}).
The negative detection of the outburst
on October 3 could constrain the growth time of superhumps
no larger than 3.5~d.

\begin{table}
\caption{Superhump maxima of AW Sge (2013)}\label{tab:awsgeoc2013}
\begin{center}
\begin{tabular}{rp{55pt}p{40pt}r@{.}lr}
\hline
\multicolumn{1}{c}{$E$} & \multicolumn{1}{c}{max\commenta} & \multicolumn{1}{c}{error} & \multicolumn{2}{c}{$O-C$\commentb} & \multicolumn{1}{c}{$N$\commentc} \\
\hline
0 & 56572.6575 & 0.0003 & $-$0&0068 & 52 \\
62 & 56577.3153 & 0.0004 & 0&0104 & 64 \\
63 & 56577.3892 & 0.0005 & 0&0094 & 64 \\
74 & 56578.2061 & 0.0003 & 0&0029 & 69 \\
75 & 56578.2798 & 0.0004 & 0&0019 & 72 \\
76 & 56578.3544 & 0.0005 & 0&0016 & 72 \\
87 & 56579.1708 & 0.0010 & $-$0&0054 & 45 \\
88 & 56579.2465 & 0.0005 & $-$0&0045 & 71 \\
89 & 56579.3209 & 0.0006 & $-$0&0050 & 69 \\
90 & 56579.3961 & 0.0009 & $-$0&0046 & 48 \\
\hline
  \multicolumn{6}{l}{\commenta BJD$-$2400000.} \\
  \multicolumn{6}{l}{\commentb Against max $= 2456572.6642 + 0.074850 E$.} \\
  \multicolumn{6}{l}{\commentc Number of points used to determine the maximum.} \\
\end{tabular}
\end{center}
\end{table}

\subsection{V1265 Tauri}\label{obj:v1265tau}

   V1265 Tau was originally detected as an optical transient
\citep{skv06j0329cbet701}.  \citet{sha07j0329} studied this
object and detected short-period [0.053394(7)~d] superhumps,
which has been the shortest known superhump period in
classical SU UMa-type dwarf novae.

   On 2013 August 4, ASAS-SN team detect this object again
in outburst (vsnet-alert 16120).  The outburst was detected
sufficiently early to follow the early evolution of
the superhumps.  On the first night of the observation,
superhumps were already detected (vsnet-alert 16137).
This observations has confirmed that V1265 Tau bears
no characteristics of a WZ Sge-type dwarf nova, such as
the existence of early superhumps, despite its very short
superhump period.

   The times of superhump maxima are listed in table
\ref{tab:v1265tauoc2013}.  Due to the faintness ($\sim$16 mag)
of the object, the errors in the times of superhump maxima
were relatively large.  We could determine the
$P_{\rm dot}$ of $+1.9(1.9) \times 10^{-5}$ for stage B
superhumps.  This value is consistent with
$P_{\rm dot} = +2.1(0.8) \times 10^{-5}$ reported by
\citet{skv06j0329cbet701}.  This value of period derivative
is small for an ordinary SU UMa-type dwarf nova with
a short superhump period.  After BJD 2456520, the object
apparently showed a stage B-C transition.  Since the
measured times of superhump maxima during the late stage
of the superoutburst were noisy, we determined the period
with the PDM method to be 0.05309(4)~d, which is adopted
in table \ref{tab:perlist}.

   This unusual short-period system is unlike other systems
with similar short superhump periods in that it neither
exhibited WZ Sge-type characteristics nor a large positive
$P_{\rm dot}$ as in a non-WZ Sge-type short-period system,
V844 Her (\cite{oiz07v844her}; \cite{Pdot}; \cite{Pdot4}).
The evolution of superhumps was more similar to ER UMa-type
dwarf novae such as RZ LMi (\cite{ole08rzlmi}; \cite{Pdot4})
although this object does not show frequent outbursts
as in ER UMa-type dwarf novae.  Further study is needed
to clarify the unusual nature of this object.

\begin{table}
\caption{Superhump maxima of V1265 Tau (2013)}\label{tab:v1265tauoc2013}
\begin{center}
\begin{tabular}{rp{55pt}p{40pt}r@{.}lr}
\hline
\multicolumn{1}{c}{$E$} & \multicolumn{1}{c}{max\commenta} & \multicolumn{1}{c}{error} & \multicolumn{2}{c}{$O-C$\commentb} & \multicolumn{1}{c}{$N$\commentc} \\
\hline
0 & 56509.9538 & 0.0016 & $-$0&0052 & 25 \\
1 & 56510.0132 & 0.0024 & 0&0008 & 11 \\
37 & 56511.9444 & 0.0045 & 0&0100 & 16 \\
38 & 56511.9844 & 0.0004 & $-$0&0034 & 37 \\
56 & 56512.9464 & 0.0025 & $-$0&0024 & 30 \\
57 & 56512.9894 & 0.0025 & $-$0&0128 & 24 \\
94 & 56514.9731 & 0.0008 & $-$0&0046 & 48 \\
113 & 56515.9901 & 0.0012 & $-$0&0019 & 32 \\
131 & 56516.9515 & 0.0013 & $-$0&0016 & 35 \\
148 & 56517.8658 & 0.0031 & 0&0050 & 28 \\
149 & 56517.9275 & 0.0015 & 0&0134 & 8 \\
150 & 56517.9660 & 0.0065 & $-$0&0015 & 49 \\
167 & 56518.8748 & 0.0021 & $-$0&0003 & 28 \\
168 & 56518.9290 & 0.0049 & 0&0005 & 12 \\
186 & 56519.8959 & 0.0028 & 0&0063 & 19 \\
187 & 56519.9481 & 0.0019 & 0&0052 & 31 \\
204 & 56520.8559 & 0.0014 & 0&0053 & 28 \\
205 & 56520.9085 & 0.0056 & 0&0046 & 11 \\
206 & 56520.9582 & 0.0007 & 0&0009 & 45 \\
222 & 56521.8086 & 0.0038 & $-$0&0030 & 14 \\
223 & 56521.8692 & 0.0016 & 0&0042 & 25 \\
260 & 56523.8356 & 0.0033 & $-$0&0048 & 25 \\
261 & 56523.8818 & 0.0013 & $-$0&0120 & 23 \\
278 & 56524.7991 & 0.0096 & $-$0&0024 & 10 \\
279 & 56524.8669 & 0.0023 & 0&0120 & 18 \\
297 & 56525.8170 & 0.0104 & 0&0011 & 13 \\
305 & 56526.2474 & 0.0089 & 0&0044 & 63 \\
319 & 56526.9728 & 0.0041 & $-$0&0177 & 35 \\
\hline
  \multicolumn{6}{l}{\commenta BJD$-$2400000.} \\
  \multicolumn{6}{l}{\commentb Against max $= 2456509.9590 + 0.053390 E$.} \\
  \multicolumn{6}{l}{\commentc Number of points used to determine the maximum.} \\
\end{tabular}
\end{center}
\end{table}

\subsection{SU Ursae Majoris}\label{obj:suuma}

   We observed this ``prototype'' of SU UMa-type dwarf novae
\citep{uda90suuma} during the late stage of the 2013
November superoutburst.  The times of superhump maxima
(table \ref{tab:suumaoc2013}) also includes first two
nights of post-superoutburst state.  The resultant period
of stage C superhump is in agreement with previous
observations (1989, 1999: \cite{Pdot}).

\begin{table}
\caption{Superhump maxima of SU UMa (2013)}\label{tab:suumaoc2013}
\begin{center}
\begin{tabular}{rp{55pt}p{40pt}r@{.}lr}
\hline
\multicolumn{1}{c}{$E$} & \multicolumn{1}{c}{max\commenta} & \multicolumn{1}{c}{error} & \multicolumn{2}{c}{$O-C$\commentb} & \multicolumn{1}{c}{$N$\commentc} \\
\hline
0 & 56627.2982 & 0.0011 & 0&0051 & 127 \\
1 & 56627.3712 & 0.0029 & $-$0&0007 & 83 \\
10 & 56628.0796 & 0.0013 & $-$0&0013 & 45 \\
11 & 56628.1626 & 0.0007 & 0&0030 & 198 \\
12 & 56628.2401 & 0.0009 & 0&0017 & 85 \\
13 & 56628.3141 & 0.0009 & $-$0&0031 & 85 \\
23 & 56629.1012 & 0.0010 & $-$0&0037 & 13 \\
24 & 56629.1703 & 0.0032 & $-$0&0135 & 8 \\
25 & 56629.2622 & 0.0014 & $-$0&0003 & 86 \\
26 & 56629.3573 & 0.0008 & 0&0160 & 111 \\
27 & 56629.4194 & 0.0009 & $-$0&0007 & 79 \\
28 & 56629.4999 & 0.0007 & 0&0011 & 69 \\
29 & 56629.5701 & 0.0016 & $-$0&0075 & 50 \\
38 & 56630.2856 & 0.0036 & $-$0&0010 & 54 \\
51 & 56631.3153 & 0.0009 & 0&0047 & 86 \\
\hline
  \multicolumn{6}{l}{\commenta BJD$-$2400000.} \\
  \multicolumn{6}{l}{\commentb Against max $= 2456627.2931 + 0.078775 E$.} \\
  \multicolumn{6}{l}{\commentc Number of points used to determine the maximum.} \\
\end{tabular}
\end{center}
\end{table}

\subsection{SS Ursae Minoris}\label{obj:ssumi}

   SS UMi was originally discovered as the optical counterpart
(dwarf nova) of the X-ray source E 1551$+$718 \citep{mas82ssumi}.
This object was also selected as a CV by Palomer Green survey
\citep{gre82PGsurveyCV}.  \citet{and86ssumi} reported variations
with a period of 127~min, which was considered to be the orbital
period.  \citet{ric89ssumi} studied the behavior of this object and 
found that the mean cycle length is 30--48~d.
Amateur astronomers also started observations of this object
in 1987 and recorded frequent outbursts with cycle lengths
as short as $\sim$10~d.

   \citet{uda90ssumi} observed this object and reported that
the orbital period is much longer (likely 6.8~hr) in contrast to
\citet{and86ssumi}.  This contradiction between observations
was resolved by the detection of superhumps with a period
of 101~min \citep{che91ssumi}: neither \citet{and86ssumi}
nor \citet{uda90ssumi} turned out to be correct.
\citet{kat98ssumi} also reported observations of superhumps.

   This object started receiving attention because of its
high frequency of outbursts.  \citet{kat00ssumi} clarified
that the supercycle of this object is 84.7~d, the shortest
known supercycle in ordinary SU UMa-type dwarf novae
[note that this object was not classified as an ER UMa-type
dwarf nova in this reference; see also \citet{kat01bfara}
for the similar case of BF Ara].  \citet{ole06ssumi} reported
that the supercycle lengthened to 197~d in 2004.
\citet{ole06ssumi} also reported development of superhumps.
\citet{Pdot4} also studied the 2012 superoutburst.
The supercycle as short as 84.7~d has never been convincingly
recorded in the recent decade.

   We observed the final part of the superoutburst in
2003 August--September (the early part of this superoutburst
was likely missed).  The times of superhump maxima are
listed in table \ref{tab:ssumioc2013}.  The times for
$E \ge 29$ were obtained after the rapid fading from
the superoutburst.  Although these epochs of maxima
could be expressed without a phase jump from those for
$E \le 3$, the identification of the phase was not
complete due to the gap in the observation.

\begin{table}
\caption{Superhump maxima of SS UMi (2013)}\label{tab:ssumioc2013}
\begin{center}
\begin{tabular}{rp{55pt}p{40pt}r@{.}lr}
\hline
\multicolumn{1}{c}{$E$} & \multicolumn{1}{c}{max\commenta} & \multicolumn{1}{c}{error} & \multicolumn{2}{c}{$O-C$\commentb} & \multicolumn{1}{c}{$N$\commentc} \\
\hline
0 & 56540.3119 & 0.0010 & 0&0008 & 53 \\
1 & 56540.3717 & 0.0012 & $-$0&0092 & 77 \\
2 & 56540.4451 & 0.0012 & $-$0&0058 & 78 \\
3 & 56540.5170 & 0.0008 & $-$0&0038 & 77 \\
29 & 56542.3524 & 0.0010 & 0&0132 & 73 \\
30 & 56542.4149 & 0.0008 & 0&0058 & 33 \\
31 & 56542.4847 & 0.0005 & 0&0056 & 77 \\
32 & 56542.5575 & 0.0006 & 0&0085 & 70 \\
33 & 56542.6244 & 0.0012 & 0&0054 & 41 \\
43 & 56543.3123 & 0.0008 & $-$0&0060 & 33 \\
44 & 56543.3922 & 0.0015 & 0&0039 & 37 \\
57 & 56544.2931 & 0.0008 & $-$0&0043 & 38 \\
58 & 56544.3588 & 0.0021 & $-$0&0085 & 35 \\
59 & 56544.4318 & 0.0030 & $-$0&0055 & 22 \\
\hline
  \multicolumn{6}{l}{\commenta BJD$-$2400000.} \\
  \multicolumn{6}{l}{\commentb Against max $= 2456540.3111 + 0.069936 E$.} \\
  \multicolumn{6}{l}{\commentc Number of points used to determine the maximum.} \\
\end{tabular}
\end{center}
\end{table}

\subsection{CU Velorum}\label{obj:cuvel}

   Although CU Vel is a well-known SU UMa-type dwarf nova,
only limited amount of information has been published
(\cite{vog80suumastars}; \cite{men96cuvel}).
We reported observations of the 2002 superoutburst
in \citet{Pdot}.

   The 2013 superoutburst was detected
on November 25 (vsnet-alert 16648), but the actual start
of the superoutburst must have been several days earlier
as later shown in the comparison of the $O-C$ diagrams.
The 2013 observation, however, well recorded the
post-superoutburst state.

   The times of superhump maxima during the plateau phase
are listed in table \ref{tab:cuveloc2013}.  These
superhumps are likely stage C superhumps since they
were observed during the late stage of the superoutburst.
A comparison of the $O-C$ diagrams supports this
identification (figure \ref{fig:cuvelcomp}).

\begin{figure}
  \begin{center}
    \FigureFile(88mm,70mm){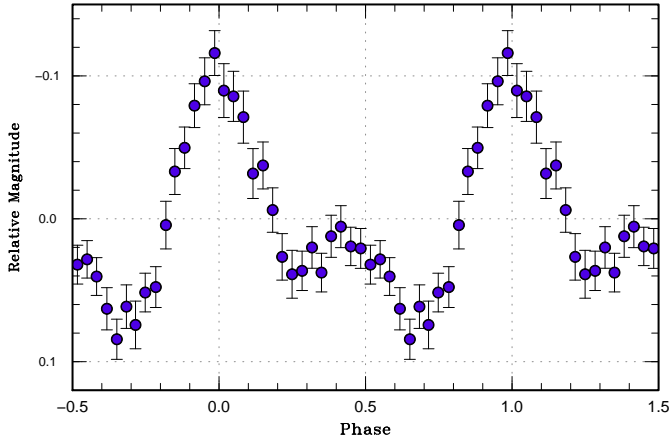}
  \end{center}
  \caption{Orbital variation of CU Vel in quiescence.}
  \label{fig:cuvelorb}
\end{figure}

\begin{figure}
  \begin{center}
    \FigureFile(88mm,70mm){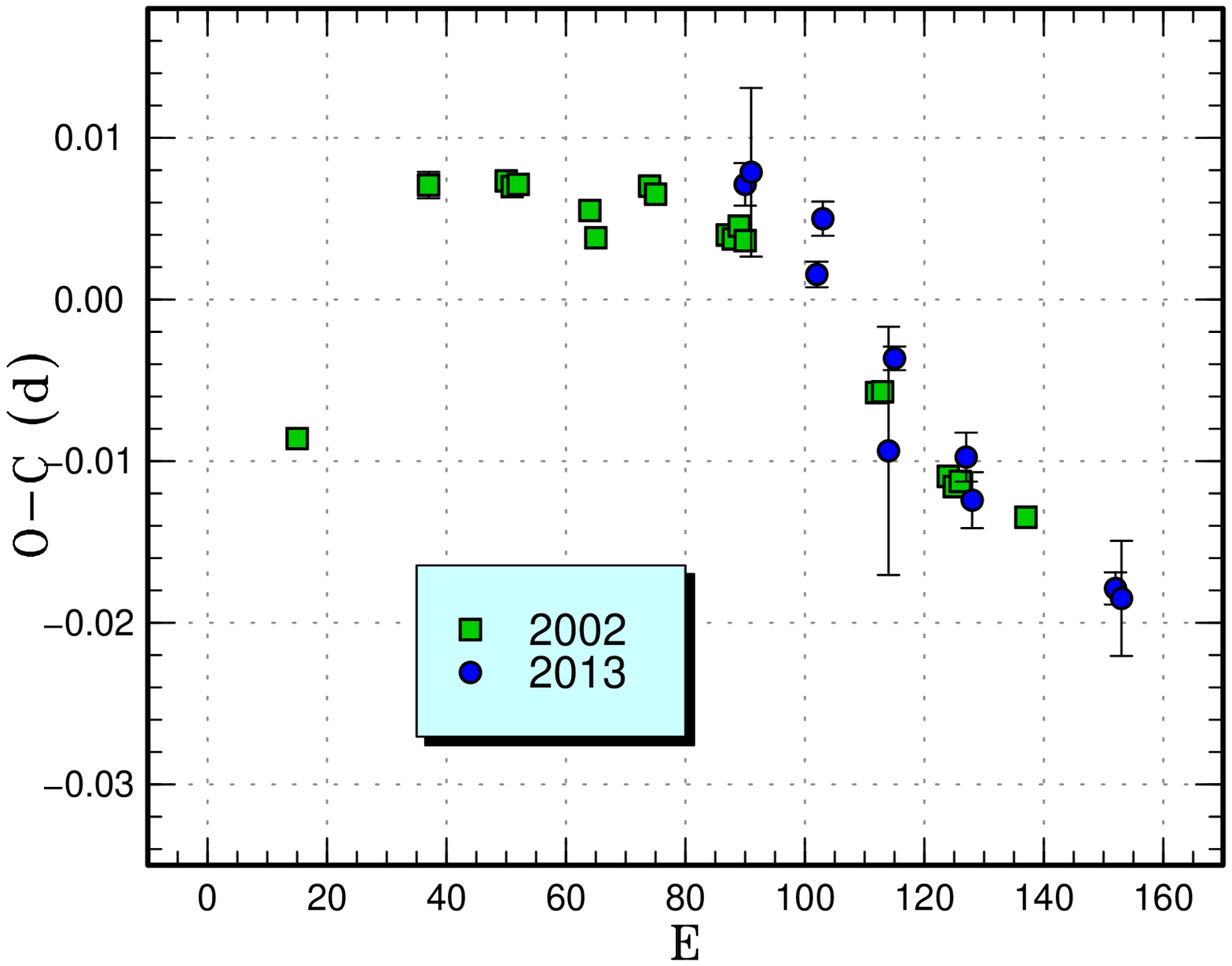}
  \end{center}
  \caption{Comparison of $O-C$ diagrams of CU Vel between different
  superoutbursts.  A period of 0.08100~d was used to draw this figure.
  Approximate cycle counts ($E$) after the start of the superoutburst
  were used.  Since the start of the 2013 superoutburst
  was not well constrained, we shifted the $O-C$ diagram
  to best fit the 2002 one.}
  \label{fig:cuvelcomp}
\end{figure}

   Since orbital modulations are clearly seen in the light curve, 
we refined the orbital period by using observations
around quiescence.  The 2013--2014 data after the 2013 November
superoutburst yielded a period of 0.078043(5)~d.
The 2013 January data (Nel and HaC) yielded a period of 0.07805(4)~d.
Using the combined data set, we selected a period of 0.0780541(3)~d.
The resultant profile (figure \ref{fig:cuvelorb}) showed two
maxima of different amplitudes.  This feature of double maxima
is to some extent similar to WZ Sge-type dwarf nova in quiescence
(e.g. \cite{pat98wzsge}; \cite{ara05v455and}).
CU Vel may bear intermediate characteristics between
ordinary SU UMa-type dwarf novae and WZ Sge-type dwarf novae.

\begin{figure}
  \begin{center}
    \FigureFile(88mm,70mm){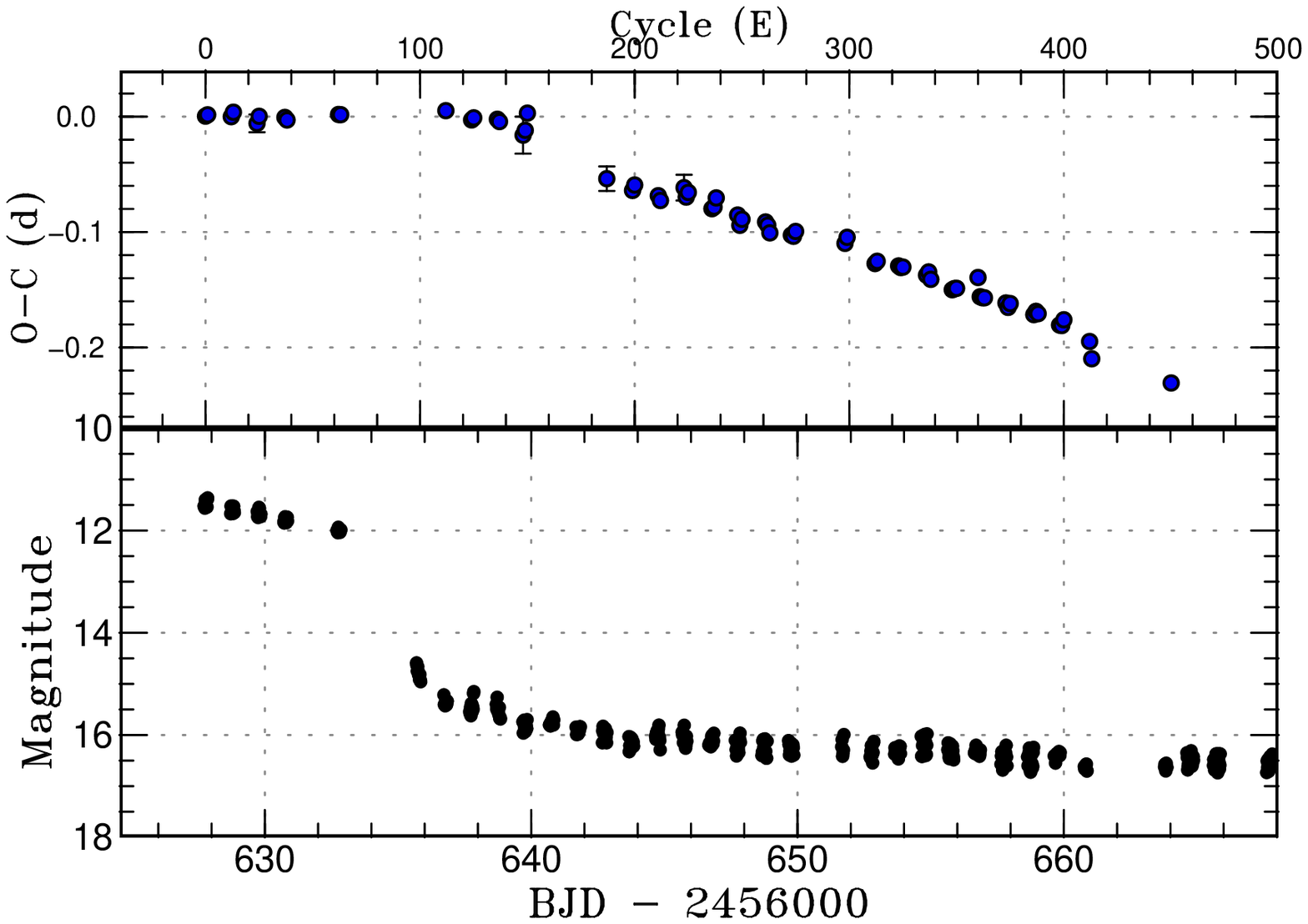}
  \end{center}
  \caption{$O-C$ diagram of superhumps in CU Vel (2013).
     (Upper): $O-C$ diagram.  A period of 0.080573~d
     was used to draw this figure.
     (Lower): Light curve.  The observations were binned to 0.016~d.}
  \label{fig:cuvel2013humpall}
\end{figure}

\begin{figure}
  \begin{center}
    \FigureFile(88mm,95mm){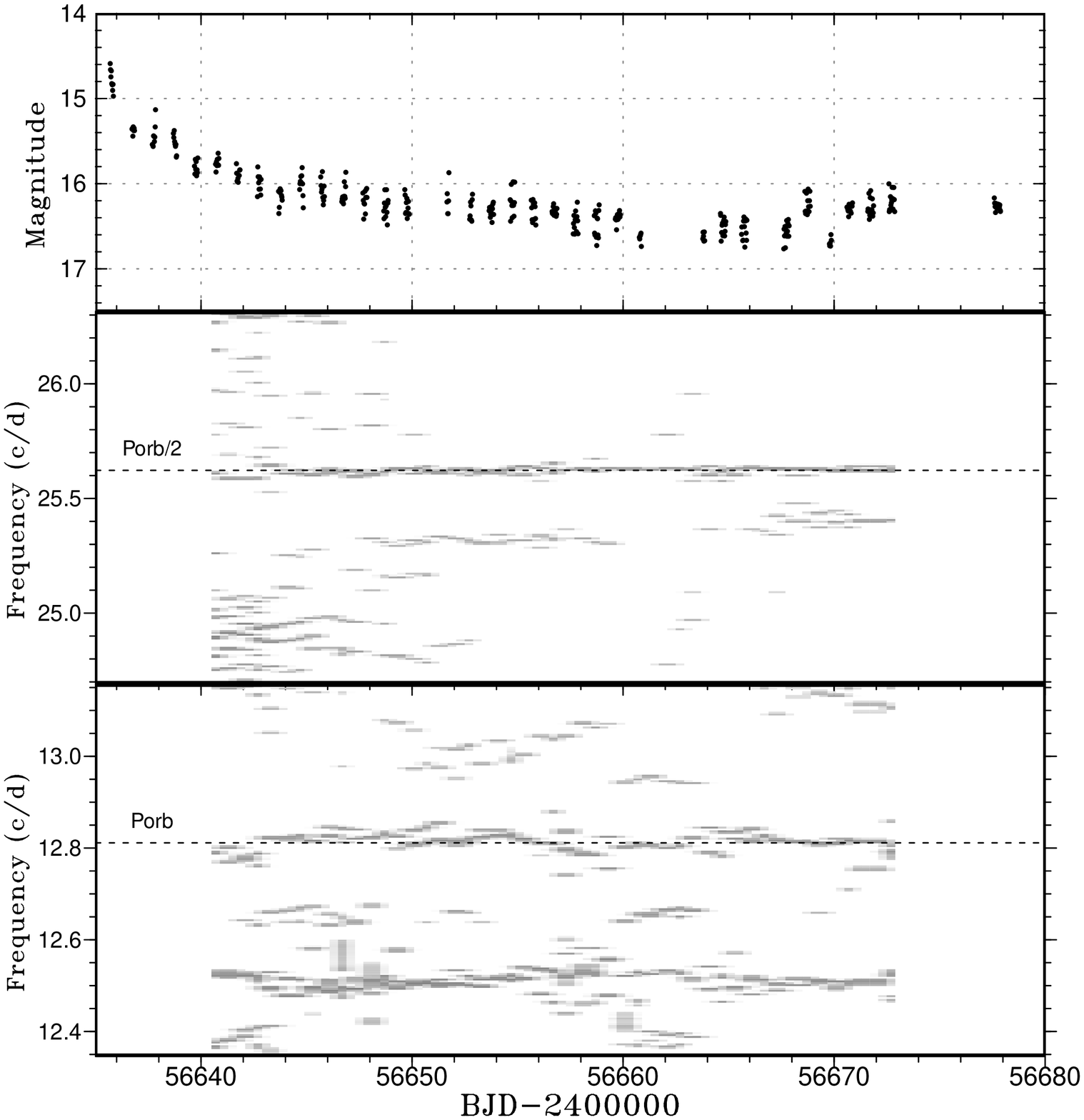}
  \end{center}
  \caption{Two-dimensional Lasso period analysis of CU Vel
  in the post-superoutburst stage (2013).
  The superoutburst plateau was excluded from the analysis
  because it did not yield a meaningful spectrum due to
  the shortness of the observation.
  (Upper:) Light curve.  The data were binned to 0.01~d.
  (Middle:) First harmonics of the superhump and orbital signals.
  (Lower:) Fundamental of the superhump and the orbital signal.
  The orbital signal was present both in the fundamental and
  the first harmonic.  The signal of (positive) superhumps with
  increasing frequency was recorded during the post-superoutburst
  stage.
  No indication of negative superhump was present.
  $\log \lambda=-8.5$ was used.
  The width of the sliding window and the time step used are
  10~d and 0.4~d, respectively.
  }
  \label{fig:cuvellasso}
\end{figure}

   We subtracted the mean orbital light curve from the
post-superoutburst light curve.  A PDM analysis yielded
a strong signal at 0.079906(4)~d.  The times of these
post-superoutburst superhumps are listed in table
\ref{tab:cuveloc2013b}.  The $O-C$ diagram
(figure \ref{fig:cuvel2013humpall}) indicates that the
superhump phase was continuous before and after the
rapid fading from the superoutburst.  There was
a decrease in period after $E=110$ (after the rapid fading)
and the period was almost constant at least until $E=400$.
Both the signals of superhumps and orbital period
are clearly seen in the Lasso power spectrum
(figure \ref{fig:cuvellasso}).  The post-superoutburst
superhumps survived at least 30~d after the fading.

\begin{table}
\caption{Superhump maxima of CU Vel (2013)}\label{tab:cuveloc2013}
\begin{center}
\begin{tabular}{rp{55pt}p{40pt}r@{.}lr}
\hline
\multicolumn{1}{c}{$E$} & \multicolumn{1}{c}{max\commenta} & \multicolumn{1}{c}{error} & \multicolumn{2}{c}{$O-C$\commentb} & \multicolumn{1}{c}{$N$\commentc} \\
\hline
0 & 56627.7714 & 0.0013 & 0&0004 & 17 \\
1 & 56627.8531 & 0.0052 & 0&0016 & 8 \\
12 & 56628.7378 & 0.0008 & $-$0&0001 & 16 \\
13 & 56628.8223 & 0.0011 & 0&0038 & 17 \\
24 & 56629.6989 & 0.0077 & $-$0&0059 & 6 \\
25 & 56629.7856 & 0.0007 & 0&0003 & 17 \\
37 & 56630.7515 & 0.0015 & $-$0&0007 & 17 \\
38 & 56630.8298 & 0.0017 & $-$0&0029 & 15 \\
62 & 56632.7684 & 0.0010 & 0&0019 & 17 \\
63 & 56632.8488 & 0.0036 & 0&0017 & 10 \\
\hline
  \multicolumn{6}{l}{\commenta BJD$-$2400000.} \\
  \multicolumn{6}{l}{\commentb Against max $= 2456627.7710 + 0.080573 E$.} \\
  \multicolumn{6}{l}{\commentc Number of points used to determine the maximum.} \\
\end{tabular}
\end{center}
\end{table}

\begin{table}
\caption{Superhump maxima of CU Vel in (2013) (post-superoutburst)}\label{tab:cuveloc2013b}
\begin{center}
\begin{tabular}{rp{55pt}p{40pt}r@{.}lr}
\hline
\multicolumn{1}{c}{$E$} & \multicolumn{1}{c}{max\commenta} & \multicolumn{1}{c}{error} & \multicolumn{2}{c}{$O-C$\commentb} & \multicolumn{1}{c}{$N$\commentc} \\
\hline
0 & 56636.8003 & 0.0026 & $-$0&0013 & 17 \\
12 & 56637.7592 & 0.0021 & $-$0&0014 & 15 \\
13 & 56637.8416 & 0.0023 & 0&0010 & 10 \\
24 & 56638.7266 & 0.0037 & 0&0070 & 16 \\
25 & 56638.8052 & 0.0027 & 0&0057 & 16 \\
36 & 56639.6797 & 0.0160 & 0&0011 & 8 \\
37 & 56639.7645 & 0.0063 & 0&0060 & 15 \\
38 & 56639.8600 & 0.0028 & 0&0215 & 7 \\
75 & 56642.7844 & 0.0106 & $-$0&0109 & 15 \\
87 & 56643.7411 & 0.0017 & $-$0&0133 & 8 \\
88 & 56643.8264 & 0.0048 & $-$0&0079 & 11 \\
99 & 56644.7034 & 0.0057 & $-$0&0099 & 13 \\
100 & 56644.7798 & 0.0033 & $-$0&0134 & 15 \\
111 & 56645.6772 & 0.0112 & 0&0049 & 10 \\
112 & 56645.7496 & 0.0016 & $-$0&0026 & 14 \\
113 & 56645.8343 & 0.0017 & 0&0021 & 15 \\
124 & 56646.7064 & 0.0039 & $-$0&0048 & 14 \\
125 & 56646.7885 & 0.0016 & $-$0&0027 & 17 \\
126 & 56646.8770 & 0.0049 & 0&0059 & 6 \\
136 & 56647.6679 & 0.0028 & $-$0&0023 & 9 \\
137 & 56647.7395 & 0.0043 & $-$0&0106 & 13 \\
138 & 56647.8254 & 0.0021 & $-$0&0047 & 20 \\
149 & 56648.7093 & 0.0032 & 0&0002 & 14 \\
150 & 56648.7870 & 0.0016 & $-$0&0020 & 17 \\
151 & 56648.8610 & 0.0024 & $-$0&0080 & 12 \\
161 & 56649.6647 & 0.0022 & $-$0&0034 & 9 \\
162 & 56649.7443 & 0.0012 & $-$0&0037 & 13 \\
163 & 56649.8291 & 0.0018 & 0&0012 & 20 \\
186 & 56651.6719 & 0.0027 & 0&0059 & 9 \\
187 & 56651.7578 & 0.0048 & 0&0118 & 6 \\
200 & 56652.7825 & 0.0016 & $-$0&0024 & 19 \\
201 & 56652.8650 & 0.0036 & 0&0003 & 11 \\
\hline
  \multicolumn{6}{l}{\commenta BJD$-$2400000.} \\
  \multicolumn{6}{l}{\commentb Against max $= 2456636.8016 + 0.079916 E$.} \\
  \multicolumn{6}{l}{\commentc Number of points used to determine the maximum.} \\
\end{tabular}
\end{center}
\end{table}

\addtocounter{table}{-1}
\begin{table}
\caption{Superhump maxima of CU Vel in (2013) (post-superoutburst, continued)}
\begin{center}
\begin{tabular}{rp{55pt}p{40pt}r@{.}lr}
\hline
\multicolumn{1}{c}{$E$} & \multicolumn{1}{c}{max\commenta} & \multicolumn{1}{c}{error} & \multicolumn{2}{c}{$O-C$\commentb} & \multicolumn{1}{c}{$N$\commentc} \\
\hline
211 & 56653.6668 & 0.0017 & 0&0028 & 12 \\
212 & 56653.7459 & 0.0023 & 0&0020 & 14 \\
213 & 56653.8269 & 0.0013 & 0&0031 & 19 \\
224 & 56654.7062 & 0.0020 & 0&0034 & 9 \\
225 & 56654.7896 & 0.0018 & 0&0068 & 19 \\
226 & 56654.8637 & 0.0027 & 0&0010 & 11 \\
236 & 56655.6604 & 0.0012 & $-$0&0014 & 12 \\
237 & 56655.7420 & 0.0020 & 0&0003 & 14 \\
238 & 56655.8228 & 0.0014 & 0&0011 & 20 \\
248 & 56656.6379 & 0.0061 & 0&0171 & 11 \\
249 & 56656.7018 & 0.0013 & 0&0011 & 19 \\
250 & 56656.7815 & 0.0019 & 0&0009 & 19 \\
251 & 56656.8621 & 0.0031 & 0&0015 & 16 \\
261 & 56657.6633 & 0.0038 & 0&0036 & 13 \\
262 & 56657.7400 & 0.0034 & 0&0004 & 16 \\
263 & 56657.8239 & 0.0028 & 0&0044 & 22 \\
274 & 56658.7006 & 0.0016 & 0&0020 & 15 \\
275 & 56658.7841 & 0.0025 & 0&0055 & 20 \\
276 & 56658.8626 & 0.0027 & 0&0042 & 18 \\
286 & 56659.6585 & 0.0026 & 0&0008 & 18 \\
287 & 56659.7386 & 0.0030 & 0&0011 & 18 \\
288 & 56659.8242 & 0.0031 & 0&0067 & 22 \\
300 & 56660.7723 & 0.0020 & $-$0&0042 & 32 \\
301 & 56660.8379 & 0.0023 & $-$0&0185 & 34 \\
338 & 56663.7981 & 0.0028 & $-$0&0152 & 25 \\
\hline
  \multicolumn{6}{l}{\commenta BJD$-$2400000.} \\
  \multicolumn{6}{l}{\commentb Against max $= 2456636.8016 + 0.079916 E$.} \\
  \multicolumn{6}{l}{\commentc Number of points used to determine the maximum.} \\
\end{tabular}
\end{center}
\end{table}

\subsection{1RXS J231935.0$+$364705}\label{obj:j231935}

   This object (hereafter 1RXS J231935) was selected as a variable
star (=DDE 8, likely a dwarf nova) during the course of identification
of the ROSAT sources \citep{den11ROSATCVs}.  There
was a well-observed superoutburst in 2011 \citep{Pdot4}.
D. Denisenko detected a new outburst on 2013 September 27
(vsnet-alert 16460).  This outburst turned out to be
a superoutburst.  Only single superhump maximum of
BJD 2456565.5851(18) ($N=19$) was obtained during
the observation on two nights.

\subsection{ASAS J224349$+$0809.5}\label{obj:asas2243}

   This dwarf nova (hereafter ASAS J224349) was selected by
P. Wils (cf. \cite{she11asas2243}).  There was one well-recorded
outburst (superoutburst) in the ASAS data in 2005 October--November.
The outburst in 2009 October was well observed, and
superhumps were detected (\cite{Pdot2}; \cite{she11asas2243}).
Although there was another superoutburst in 2011 June,
the outburst was observed only for two nights \citep{Pdot4}.

\begin{figure}
  \begin{center}
    \FigureFile(88mm,70mm){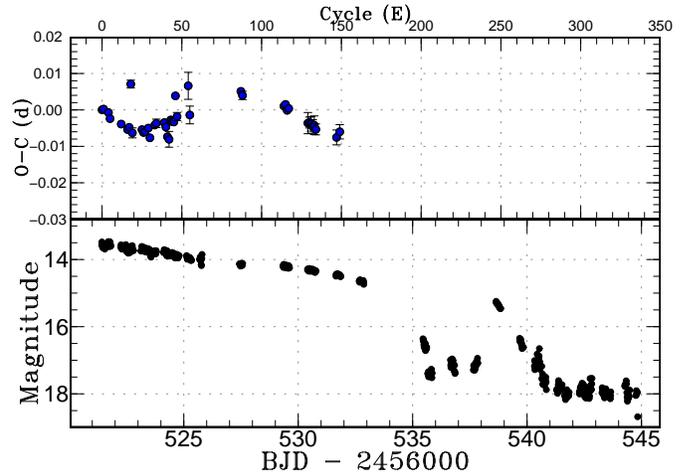}
  \end{center}
  \caption{$O-C$ diagram of superhumps in ASAS J2243 (2013).
     (Upper): $O-C$ diagram.  A period of 0.069715~d
     was used to draw this figure.
     (Lower): Light curve.  The observations were binned to 0.014~d.}
  \label{fig:asas2243humpall}
\end{figure}

\begin{figure}
  \begin{center}
    \FigureFile(88mm,70mm){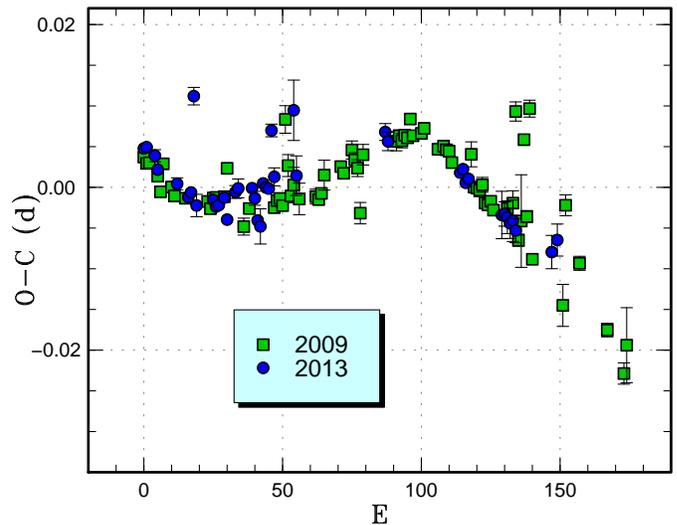}
  \end{center}
  \caption{Comparison of $O-C$ diagrams of ASAS J2243 between different
  superoutbursts.  A period of 0.06975~d was used to draw this figure.
  Approximate cycle counts ($E$) after the start of the superoutburst
  were used.
  The coincidence in cycle counts between two superoutbursts
  was by chance.
  }
  \label{fig:asas2243comp}
\end{figure}

   On 2013 August 14, ASAS-SN team detected another outburst
(vsnet-alert 16197).  This outburst turned out to be
a superoutburst, and stages B and C were well recorded
(table \ref{tab:asas2243oc2013}).  This superoutburst
was followed by one post-superoutburst rebrightening
4~d later than the rapid fading from the plateau phase
(figure \ref{fig:asas2243humpall}).
This interval was rather short for an ordinary
SU UMa-type dwarf nova.
As in the 2009 superoutburst, a definitely positive
$P_{\rm dot}$ was recorded during stage B.
The $O-C$ diagrams in the two superoutbursts was
very similar (figure \ref{fig:asas2243comp}).
Although the coincidence in cycle counts between two superoutbursts
was by chance, the 2013 superoutburst was confirmed to be
detected sufficiently early (vsnet-alert 16207, within
3~d of the start of the outburst).

\begin{table}
\caption{Superhump maxima of ASAS J224349 (2013).}\label{tab:asas2243oc2013}
\begin{center}
\begin{tabular}{rp{55pt}p{40pt}r@{.}lr}
\hline
\multicolumn{1}{c}{$E$} & \multicolumn{1}{c}{max\commenta} & \multicolumn{1}{c}{error} & \multicolumn{2}{c}{$O-C$\commentb} & \multicolumn{1}{c}{$N$\commentc} \\
\hline
0 & 56521.4291 & 0.0002 & 0&0027 & 116 \\
1 & 56521.4990 & 0.0003 & 0&0028 & 146 \\
4 & 56521.7072 & 0.0007 & 0&0019 & 19 \\
5 & 56521.7753 & 0.0006 & 0&0002 & 24 \\
12 & 56522.2618 & 0.0007 & $-$0&0013 & 45 \\
16 & 56522.5392 & 0.0004 & $-$0&0028 & 123 \\
17 & 56522.6095 & 0.0003 & $-$0&0022 & 142 \\
18 & 56522.6911 & 0.0011 & 0&0097 & 40 \\
19 & 56522.7474 & 0.0014 & $-$0&0037 & 20 \\
25 & 56523.1666 & 0.0006 & $-$0&0028 & 57 \\
26 & 56523.2355 & 0.0004 & $-$0&0036 & 65 \\
27 & 56523.3054 & 0.0003 & $-$0&0034 & 134 \\
29 & 56523.4459 & 0.0003 & $-$0&0024 & 130 \\
30 & 56523.5129 & 0.0006 & $-$0&0050 & 143 \\
33 & 56523.7255 & 0.0008 & $-$0&0016 & 22 \\
34 & 56523.7957 & 0.0012 & $-$0&0011 & 17 \\
39 & 56524.1445 & 0.0003 & $-$0&0009 & 125 \\
40 & 56524.2130 & 0.0009 & $-$0&0021 & 66 \\
41 & 56524.2801 & 0.0006 & $-$0&0048 & 68 \\
42 & 56524.3491 & 0.0022 & $-$0&0055 & 86 \\
43 & 56524.4241 & 0.0005 & $-$0&0002 & 143 \\
44 & 56524.4934 & 0.0004 & $-$0&0006 & 139 \\
45 & 56524.5630 & 0.0004 & $-$0&0007 & 142 \\
46 & 56524.6398 & 0.0008 & 0&0065 & 29 \\
47 & 56524.7039 & 0.0011 & 0&0008 & 33 \\
54 & 56525.2003 & 0.0037 & 0&0092 & 61 \\
55 & 56525.2620 & 0.0024 & 0&0012 & 22 \\
87 & 56527.4994 & 0.0010 & 0&0077 & 49 \\
88 & 56527.5680 & 0.0011 & 0&0066 & 39 \\
114 & 56529.3777 & 0.0004 & 0&0036 & 136 \\
115 & 56529.4478 & 0.0004 & 0&0041 & 158 \\
116 & 56529.5159 & 0.0005 & 0&0025 & 154 \\
117 & 56529.5861 & 0.0006 & 0&0030 & 138 \\
\hline
  \multicolumn{6}{l}{\commenta BJD$-$2400000.} \\
  \multicolumn{6}{l}{\commentb Against max $= 2456521.4265 + 0.069715 E$.} \\
  \multicolumn{6}{l}{\commentc Number of points used to determine the maximum.} \\
\end{tabular}
\end{center}
\end{table}

\addtocounter{table}{-1}
\begin{table}
\caption{Superhump maxima of ASAS J224349 (2013) (continued).}
\begin{center}
\begin{tabular}{rp{55pt}p{40pt}r@{.}lr}
\hline
\multicolumn{1}{c}{$E$} & \multicolumn{1}{c}{max\commenta} & \multicolumn{1}{c}{error} & \multicolumn{2}{c}{$O-C$\commentb} & \multicolumn{1}{c}{$N$\commentc} \\
\hline
129 & 56530.4187 & 0.0029 & $-$0&0010 & 53 \\
130 & 56530.4886 & 0.0015 & $-$0&0009 & 58 \\
131 & 56530.5579 & 0.0020 & $-$0&0013 & 66 \\
132 & 56530.6269 & 0.0019 & $-$0&0020 & 43 \\
133 & 56530.6970 & 0.0025 & $-$0&0016 & 17 \\
134 & 56530.7656 & 0.0015 & $-$0&0028 & 17 \\
147 & 56531.6697 & 0.0020 & $-$0&0050 & 23 \\
149 & 56531.8106 & 0.0020 & $-$0&0034 & 25 \\
\hline
  \multicolumn{6}{l}{\commenta BJD$-$2400000.} \\
  \multicolumn{6}{l}{\commentb Against max $= 2456521.4265 + 0.069715 E$.} \\
  \multicolumn{6}{l}{\commentc Number of points used to determine the maximum.} \\
\end{tabular}
\end{center}
\end{table}

\subsection{ASASSN-13cf}\label{obj:asassn13cf}

\begin{figure}
  \begin{center}
    \FigureFile(88mm,110mm){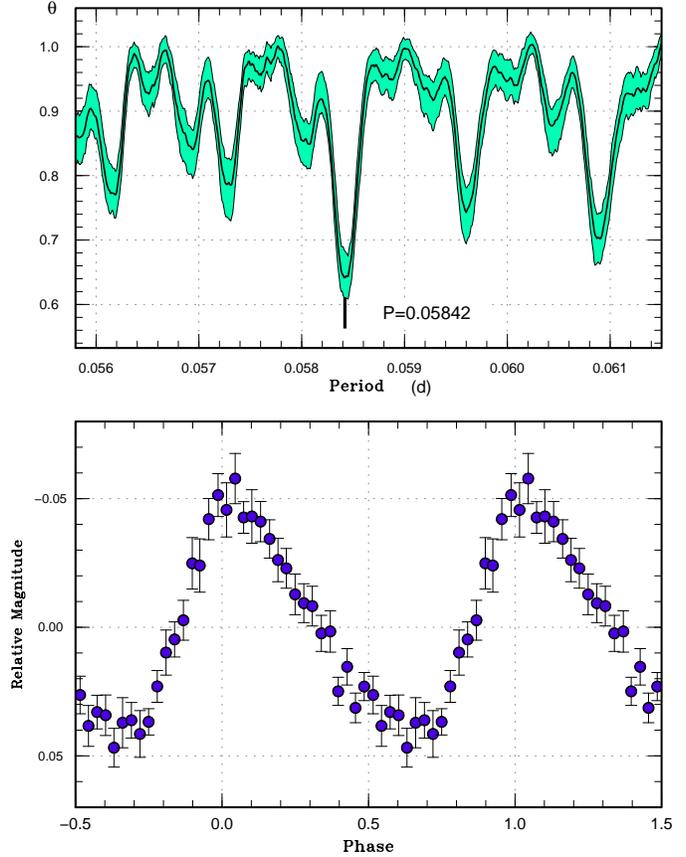}
  \end{center}
  \caption{Superhumps in ASASSN-13cf (2013). (Upper): PDM analysis.
     (Lower): Phase-averaged profile.}
  \label{fig:asassn13cfshpdm}
\end{figure}

   This object was discovered by ASAS-SN survey on 2013
August 24 (vsnet-alert 16261).  The coordinates are
\timeform{21h 55m 12.76s}, \timeform{+27D 41' 18.9''}.
One previous outburst was detected in the CRTS data
and another outburst was recorded Palomar quick-V plate
(D. Denisenko, vsnet-alert 16263).
Subsequent observations detected superhumps
(vsnet-alert 16284, 16308; figure \ref{fig:asassn13cfshpdm}).
The times of superhump maxima are listed in table
\ref{tab:asassn13cfoc2013}.  A positive $P_{\rm dot}$
of $+7.1(1.9) \times 10^{-5}$, characteristic to this
short superhump period, was obtained.

\begin{table}
\caption{Superhump maxima of ASASSN-13cf (2013)}\label{tab:asassn13cfoc2013}
\begin{center}
\begin{tabular}{rp{55pt}p{40pt}r@{.}lr}
\hline
\multicolumn{1}{c}{$E$} & \multicolumn{1}{c}{max\commenta} & \multicolumn{1}{c}{error} & \multicolumn{2}{c}{$O-C$\commentb} & \multicolumn{1}{c}{$N$\commentc} \\
\hline
0 & 56529.7208 & 0.0008 & 0&0007 & 101 \\
1 & 56529.7839 & 0.0013 & 0&0054 & 61 \\
2 & 56529.8388 & 0.0007 & 0&0018 & 96 \\
29 & 56531.4116 & 0.0028 & $-$0&0026 & 27 \\
48 & 56532.5201 & 0.0004 & $-$0&0040 & 100 \\
49 & 56532.5772 & 0.0007 & $-$0&0053 & 65 \\
98 & 56535.4447 & 0.0012 & $-$0&0002 & 52 \\
114 & 56536.3780 & 0.0020 & $-$0&0016 & 40 \\
115 & 56536.4419 & 0.0015 & 0&0039 & 43 \\
149 & 56538.4257 & 0.0015 & 0&0015 & 63 \\
150 & 56538.4828 & 0.0011 & 0&0002 & 64 \\
\hline
  \multicolumn{6}{l}{\commenta BJD$-$2400000.} \\
  \multicolumn{6}{l}{\commentb Against max $= 2456529.7201 + 0.058416 E$.} \\
  \multicolumn{6}{l}{\commentc Number of points used to determine the maximum.} \\
\end{tabular}
\end{center}
\end{table}

\subsection{ASASSN-13cg}\label{obj:asassn13cg}

\begin{figure}
  \begin{center}
    \FigureFile(88mm,70mm){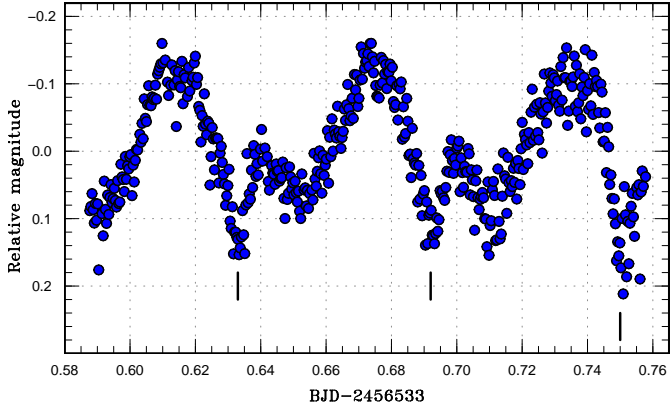}
  \end{center}
  \caption{Superhumps and possible eclipses in ASASSN-13cg
  on 2013 August 29.  The vertical ticks represent
  possible shallow eclipses.}
  \label{fig:asassn13cglc}
\end{figure}

\begin{figure}
  \begin{center}
    \FigureFile(88mm,110mm){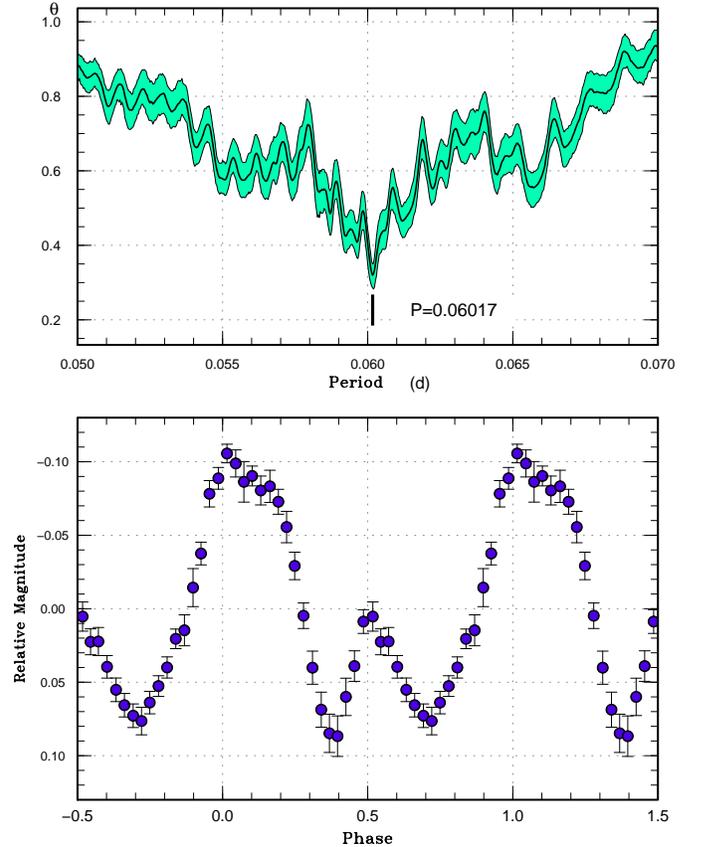}
  \end{center}
  \caption{Superhumps in ASASSN-13cg (2013). (Upper): PDM analysis.
     (Lower): Phase-averaged profile.  The dip around phase 0.4
     is a result of eclipse-like feature recorded on the first night.}
  \label{fig:asassn13cgshpdm}
\end{figure}

   This object was discovered by ASAS-SN survey on 2013
August 27.  The coordinates are
\timeform{20h 52m 52.74s}, \timeform{-02D 39' 53.0''}.
This object attracted attention because it has a very
blue ($u-g=-0.2$) SDSS color (vsnet-alert 16280).
It also has an X-ray counterpart of 1RXS J205252.1$-$023952.
Time-resolved photometry detected superhumps and possible
shallow eclipses (vsnet-alert 16302; figures \ref{fig:asassn13cglc},
\ref{fig:asassn13cgshpdm}).
These eclipse-like fadings were not recorded on later nights 
and we could not determine its period.
The reality of the eclipses requires future observations.
The times of superhump maxima
are listed in table \ref{tab:asassn13cgoc2013},
which clearly shows a positive $P_{\rm dot}$.

\begin{table}
\caption{Superhump maxima of ASASSN-13cg (2013)}\label{tab:asassn13cgoc2013}
\begin{center}
\begin{tabular}{rp{55pt}p{40pt}r@{.}lr}
\hline
\multicolumn{1}{c}{$E$} & \multicolumn{1}{c}{max\commenta} & \multicolumn{1}{c}{error} & \multicolumn{2}{c}{$O-C$\commentb} & \multicolumn{1}{c}{$N$\commentc} \\
\hline
0 & 56533.6131 & 0.0003 & 0&0013 & 132 \\
1 & 56533.6735 & 0.0004 & 0&0015 & 138 \\
2 & 56533.7326 & 0.0005 & 0&0003 & 140 \\
12 & 56534.3338 & 0.0008 & $-$0&0008 & 23 \\
13 & 56534.3939 & 0.0013 & $-$0&0009 & 23 \\
46 & 56536.3759 & 0.0020 & $-$0&0064 & 21 \\
61 & 56537.2850 & 0.0021 & $-$0&0007 & 22 \\
62 & 56537.3490 & 0.0012 & 0&0031 & 23 \\
63 & 56537.4088 & 0.0030 & 0&0026 & 23 \\
\hline
  \multicolumn{6}{l}{\commenta BJD$-$2400000.} \\
  \multicolumn{6}{l}{\commentb Against max $= 2456533.6118 + 0.060228 E$.} \\
  \multicolumn{6}{l}{\commentc Number of points used to determine the maximum.} \\
\end{tabular}
\end{center}
\end{table}

\subsection{ASASSN-13ck}\label{obj:asassn13ck}

   This object was discovered by ASAS-SN survey on 2013
August 29 (vsnet-alert 16303).  The coordinates are
\timeform{00h 11m 33.71s}, \timeform{+04D 51' 23.0''}.
The object had a blue SDSS counterpart ($g=20.8$) and its
outburst amplitude immediately suggested a WZ Sge-type
dwarf nova.

\begin{figure}
  \begin{center}
    \FigureFile(88mm,110mm){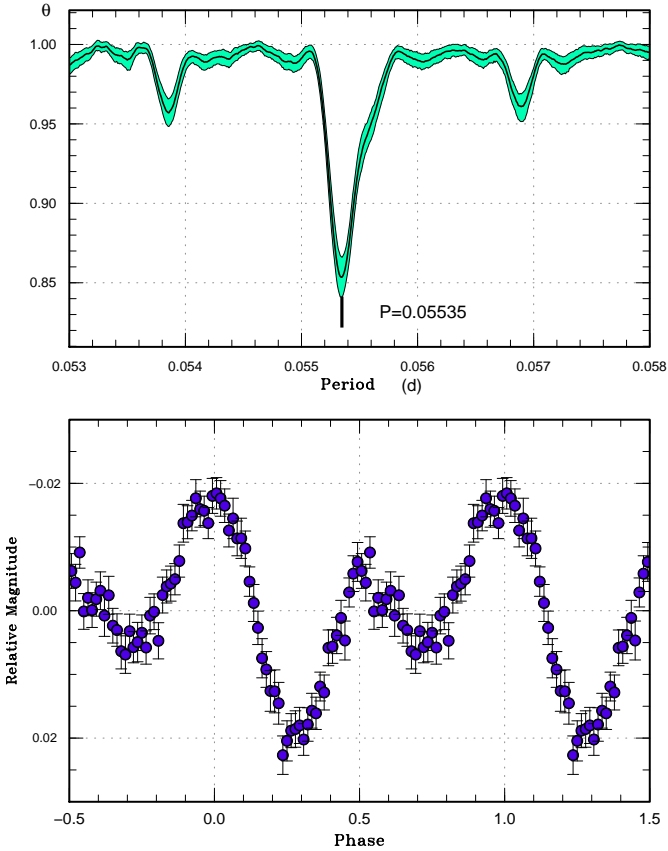}
  \end{center}
  \caption{Early superhumps in ASASSN-13ck (2013). (Upper): PDM analysis.
     (Lower): Phase-averaged profile.}
  \label{fig:asassn13ckeshpdm}
\end{figure}

\begin{figure}
  \begin{center}
    \FigureFile(88mm,110mm){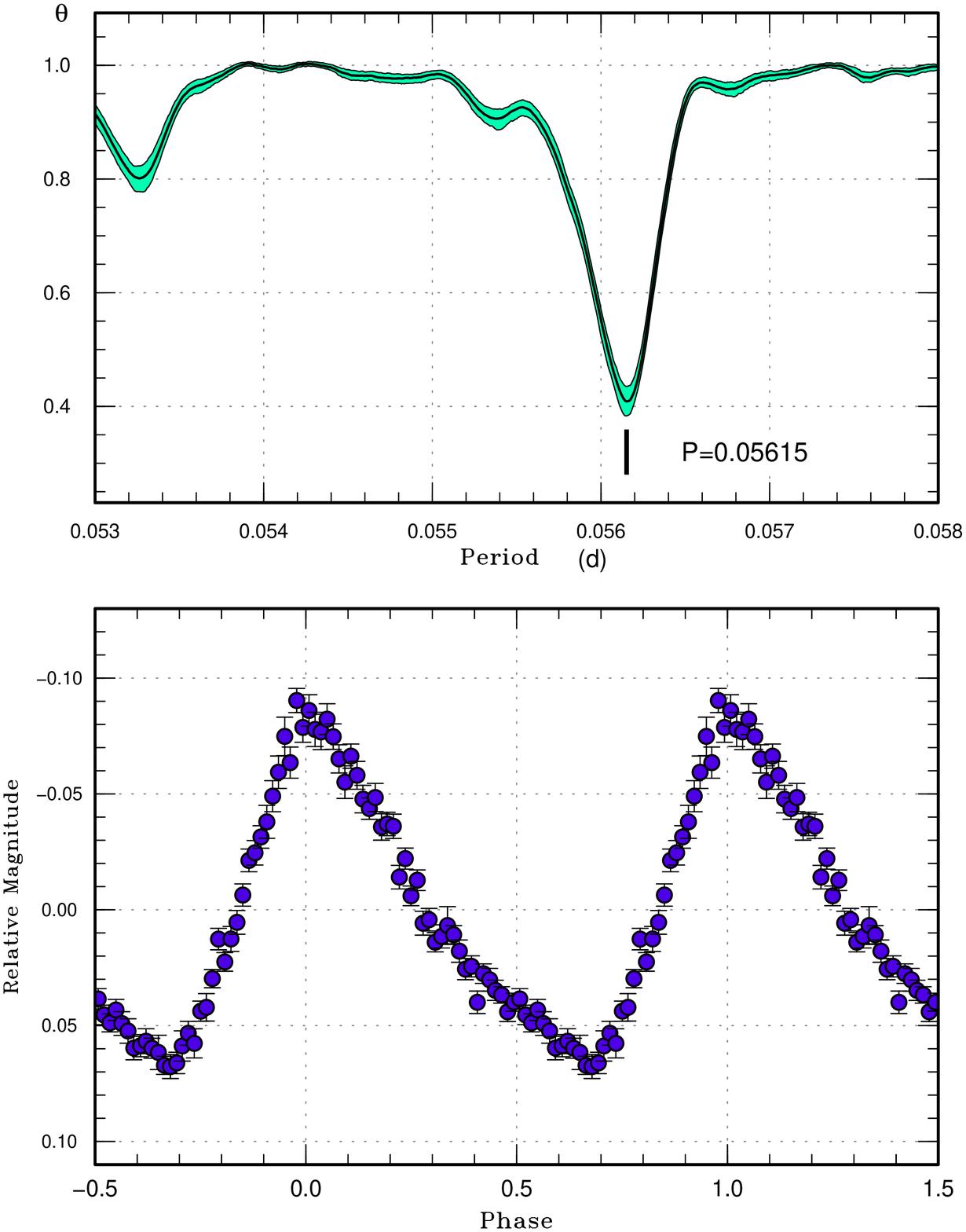}
  \end{center}
  \caption{Ordinary superhumps in ASASSN-13ck (2013). (Upper): PDM analysis.
     (Lower): Phase-averaged profile.}
  \label{fig:asassn13ckshpdm}
\end{figure}

   Subsequent observations recorded early superhumps
(vsnet-alert 16307, 16309, 16313, 16314, 16332, 16368; 
figure \ref{fig:asassn13ckeshpdm}).  Ten days after the
outburst detection, ordinary superhumps grew
(vsnet-alert 16370, 16374, 16375, 16385; figure
\ref{fig:asassn13ckshpdm}).  As judged from the evolution
of superhumps, the outburst of this object was detected
sufficiently early.

\begin{figure}
  \begin{center}
    \FigureFile(88mm,70mm){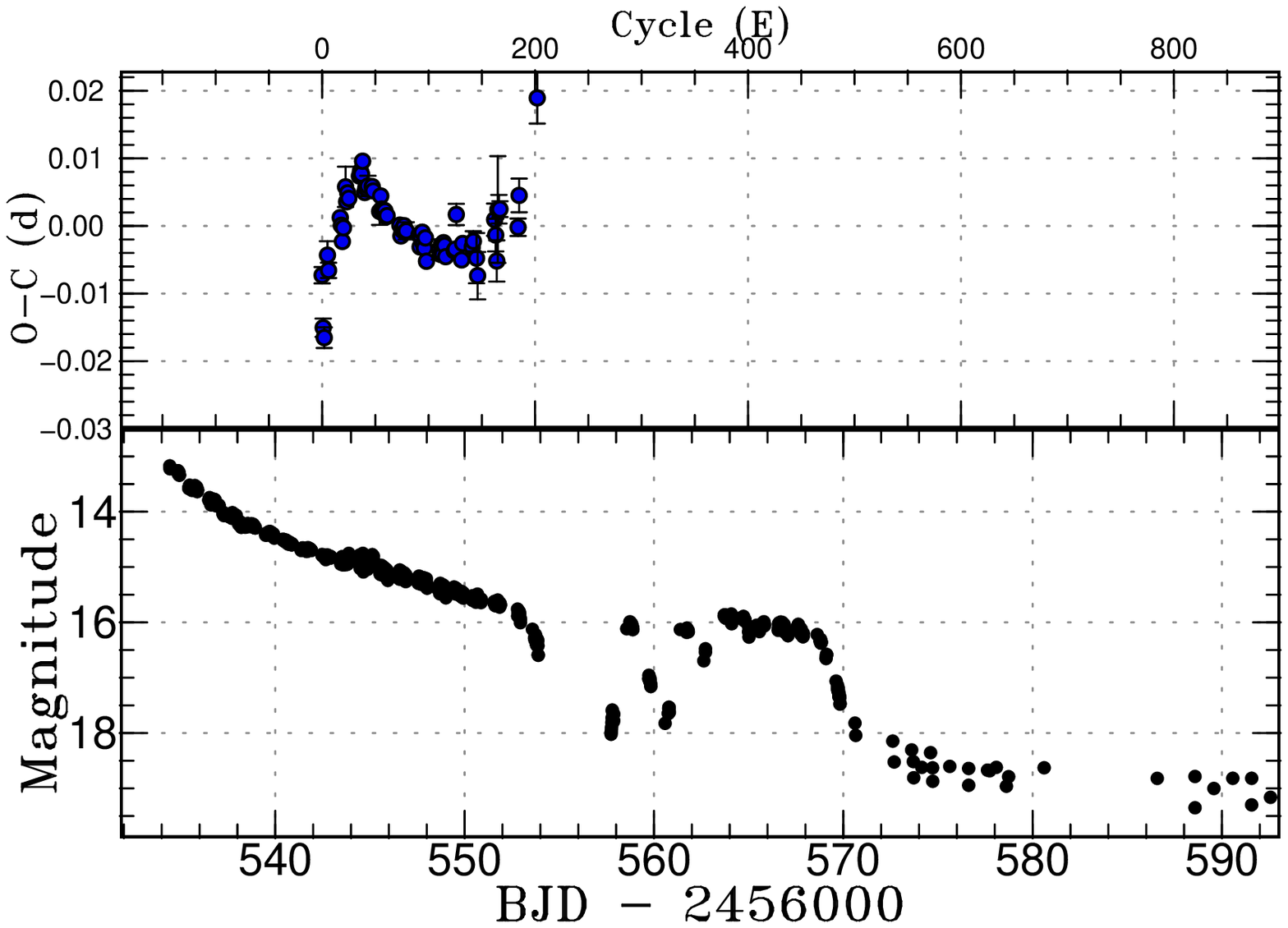}
  \end{center}
  \caption{$O-C$ diagram of superhumps in ASASSN-13ck (2013).
     (Upper): $O-C$ diagram.  A period of 0.056238~d
     was used to draw this figure.
     (Lower): Light curve.  The observations were binned to 0.012~d.}
  \label{fig:asassn13ckhumpall}
\end{figure}

   The times of superhump maxima during the superoutburst
plateau are listed in table \ref{tab:asassn13ckoc2013}.
Clear stages A and B can be recognized as in many WZ Sge-type
dwarf novae (figure \ref{fig:asassn13ckhumpall}).
The last point ($E=202$) was obtained during
the fading branch of the superoutburst, and its large
positive $O-C$ probably reflects the decrease of
the pressure effect (cf. \cite{nak13j2112j2037}).
This maximum was not used to determine $P_{\rm dot}$
for stage B.  As in many WZ Sge-type dwarf novae,
this object did not show a marked transition to
stage C superhumps.

   Five days after the rapid fading, the object showed
a short rebrightening (September 23, BJD 2456558.5).
The object showed another rebrightening (September 26,
BJD 2456561.7), which served as a precursor outburst 
to the second plateau phase.  This second plateau
phase lasted for 6~d (figure \ref{fig:asassn13ckhumpall}).
During the second plateau phase, 
superhumps were also present.  The mean superhump period
during this phase was 0.056172(14)~d, indicating that
the precession rate was smaller than in the main
superoutburst.  This smaller precession rate can be
interpreted as a result of a smaller disk radius.

   The pattern of rebrightening was like a ``hybrid''
type between long-lasting plateau type (type A rebrightening
in \cite{Pdot}) and distinct repetitive rebrightenings
(type B rebrightening in \cite{Pdot}).  There appears
to be a smooth transition between these types of
rebrightenings.  The presence of a precursor outburst
in the second plateau is also intriguing.  Such a precursor
was also recorded in AL Com in 1995 (cf. \cite{nog97alcom};
alsp subsection \ref{obj:alcom}).
It is likely that a normal outburst triggered the
second plateau phase (second superoutburst) as in
ordinary SU UMa-type dwarf novae \citep{osa89suuma}.

\begin{figure}
  \begin{center}
    \FigureFile(88mm,100mm){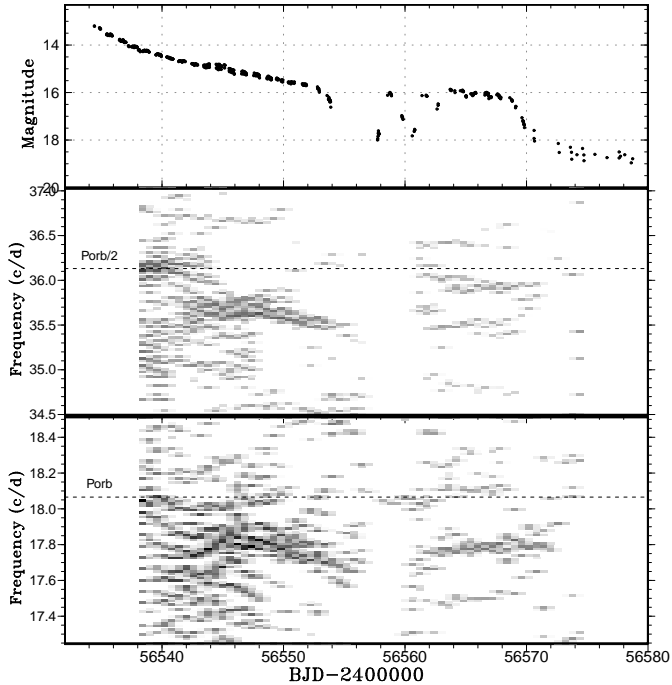}
  \end{center}
  \caption{Lasso analysis of ASASSN-13ck (2013).
  (Upper:) Light curve.  The data were binned to 0.02~d.
  (Middle:) First harmonics of the superhump and orbital signals.
  (Lower:) Fundamental of the superhump and the orbital signal.
  The orbital signal was present both in the fundamental and
  the first harmonic during the earliest phase (early superhumps).
  The signal of (positive) superhumps with variable frequency was 
  recorded during the superoutburst plateau.
  No indication of negative superhump was present.
  $\log \lambda=-8.8$ was used.
  The width of the sliding window and the time step used are
  8~d and 0.6~d, respectively.
  }
  \label{fig:asassn13cklasso}
\end{figure}

   A two-dimensional Lasso analysis is presented in
figure \ref{fig:asassn13cklasso}.  The orbital signal
was only present during the stage of early superhumps.
The signal of (positive) superhumps showed a decrease
in frequency (increase in period) during the superoutburst
plateau.  The superhumps appeared with slightly higher
frequencies during the rebrightening plateau.

\begin{table}
\caption{Superhump maxima of ASASSN-13ck (2013)}\label{tab:asassn13ckoc2013}
\begin{center}
\begin{tabular}{rp{55pt}p{40pt}r@{.}lr}
\hline
\multicolumn{1}{c}{$E$} & \multicolumn{1}{c}{max\commenta} & \multicolumn{1}{c}{error} & \multicolumn{2}{c}{$O-C$\commentb} & \multicolumn{1}{c}{$N$\commentc} \\
\hline
0 & 56542.4669 & 0.0012 & $-$0&0073 & 50 \\
1 & 56542.5154 & 0.0014 & $-$0&0150 & 57 \\
2 & 56542.5701 & 0.0015 & $-$0&0165 & 58 \\
5 & 56542.7511 & 0.0021 & $-$0&0043 & 15 \\
6 & 56542.8051 & 0.0011 & $-$0&0066 & 70 \\
17 & 56543.4315 & 0.0006 & 0&0013 & 52 \\
18 & 56543.4866 & 0.0003 & 0&0001 & 111 \\
19 & 56543.5404 & 0.0002 & $-$0&0023 & 104 \\
20 & 56543.5987 & 0.0006 & $-$0&0003 & 55 \\
22 & 56543.7172 & 0.0030 & 0&0058 & 21 \\
23 & 56543.7713 & 0.0002 & 0&0036 & 54 \\
24 & 56543.8288 & 0.0003 & 0&0049 & 64 \\
25 & 56543.8842 & 0.0005 & 0&0041 & 30 \\
35 & 56544.4499 & 0.0006 & 0&0074 & 33 \\
36 & 56544.5069 & 0.0002 & 0&0081 & 82 \\
37 & 56544.5627 & 0.0002 & 0&0077 & 73 \\
38 & 56544.6208 & 0.0010 & 0&0096 & 37 \\
40 & 56544.7287 & 0.0006 & 0&0050 & 26 \\
41 & 56544.7856 & 0.0003 & 0&0056 & 70 \\
42 & 56544.8417 & 0.0004 & 0&0055 & 65 \\
43 & 56544.8983 & 0.0015 & 0&0059 & 20 \\
47 & 56545.1232 & 0.0002 & 0&0058 & 73 \\
48 & 56545.1788 & 0.0004 & 0&0052 & 76 \\
54 & 56545.5132 & 0.0020 & 0&0022 & 26 \\
55 & 56545.5717 & 0.0003 & 0&0045 & 106 \\
56 & 56545.6260 & 0.0003 & 0&0025 & 115 \\
57 & 56545.6820 & 0.0003 & 0&0022 & 99 \\
58 & 56545.7383 & 0.0004 & 0&0024 & 63 \\
59 & 56545.7945 & 0.0003 & 0&0022 & 73 \\
60 & 56545.8498 & 0.0004 & 0&0014 & 76 \\
61 & 56545.9062 & 0.0007 & 0&0016 & 36 \\
73 & 56546.5797 & 0.0004 & 0&0002 & 58 \\
74 & 56546.6343 & 0.0005 & $-$0&0015 & 69 \\
\hline
  \multicolumn{6}{l}{\commenta BJD$-$2400000.} \\
  \multicolumn{6}{l}{\commentb Against max $= 2456542.4742 + 0.056238 E$.} \\
  \multicolumn{6}{l}{\commentc Number of points used to determine the maximum.} \\
\end{tabular}
\end{center}
\end{table}

\addtocounter{table}{-1}
\begin{table}
\caption{Superhump maxima of ASASSN-13ck (2013) (continued)}
\begin{center}
\begin{tabular}{rp{55pt}p{40pt}r@{.}lr}
\hline
\multicolumn{1}{c}{$E$} & \multicolumn{1}{c}{max\commenta} & \multicolumn{1}{c}{error} & \multicolumn{2}{c}{$O-C$\commentb} & \multicolumn{1}{c}{$N$\commentc} \\
\hline
75 & 56546.6915 & 0.0005 & $-$0&0005 & 74 \\
76 & 56546.7476 & 0.0005 & $-$0&0007 & 62 \\
77 & 56546.8046 & 0.0003 & 0&0001 & 77 \\
78 & 56546.8598 & 0.0006 & $-$0&0009 & 53 \\
79 & 56546.9163 & 0.0013 & $-$0&0007 & 31 \\
91 & 56547.5905 & 0.0005 & $-$0&0013 & 45 \\
92 & 56547.6450 & 0.0005 & $-$0&0031 & 66 \\
93 & 56547.7026 & 0.0005 & $-$0&0017 & 69 \\
94 & 56547.7597 & 0.0005 & $-$0&0009 & 55 \\
95 & 56547.8142 & 0.0008 & $-$0&0026 & 18 \\
96 & 56547.8698 & 0.0011 & $-$0&0032 & 22 \\
97 & 56547.9275 & 0.0006 & $-$0&0017 & 58 \\
98 & 56547.9803 & 0.0007 & $-$0&0052 & 57 \\
110 & 56548.6569 & 0.0015 & $-$0&0034 & 13 \\
111 & 56548.7124 & 0.0005 & $-$0&0041 & 48 \\
112 & 56548.7701 & 0.0005 & $-$0&0028 & 48 \\
113 & 56548.8248 & 0.0009 & $-$0&0042 & 35 \\
114 & 56548.8829 & 0.0005 & $-$0&0024 & 51 \\
115 & 56548.9387 & 0.0006 & $-$0&0028 & 76 \\
116 & 56548.9933 & 0.0009 & $-$0&0045 & 52 \\
124 & 56549.4440 & 0.0004 & $-$0&0036 & 59 \\
125 & 56549.5005 & 0.0004 & $-$0&0034 & 53 \\
126 & 56549.5619 & 0.0016 & 0&0017 & 18 \\
131 & 56549.8364 & 0.0009 & $-$0&0050 & 50 \\
132 & 56549.8950 & 0.0009 & $-$0&0025 & 40 \\
141 & 56550.4006 & 0.0007 & $-$0&0031 & 33 \\
142 & 56550.4577 & 0.0015 & $-$0&0022 & 18 \\
145 & 56550.6239 & 0.0037 & $-$0&0047 & 15 \\
146 & 56550.6776 & 0.0035 & $-$0&0073 & 17 \\
162 & 56551.5857 & 0.0024 & 0&0010 & 53 \\
163 & 56551.6396 & 0.0024 & $-$0&0013 & 59 \\
164 & 56551.6921 & 0.0031 & $-$0&0051 & 12 \\
\hline
  \multicolumn{6}{l}{\commenta BJD$-$2400000.} \\
  \multicolumn{6}{l}{\commentb Against max $= 2456542.4742 + 0.056238 E$.} \\
  \multicolumn{6}{l}{\commentc Number of points used to determine the maximum.} \\
\end{tabular}
\end{center}
\end{table}

\addtocounter{table}{-1}
\begin{table}
\caption{Superhump maxima of ASASSN-13ck (2013) (continued)}
\begin{center}
\begin{tabular}{rp{55pt}p{40pt}r@{.}lr}
\hline
\multicolumn{1}{c}{$E$} & \multicolumn{1}{c}{max\commenta} & \multicolumn{1}{c}{error} & \multicolumn{2}{c}{$O-C$\commentb} & \multicolumn{1}{c}{$N$\commentc} \\
\hline
165 & 56551.7559 & 0.0079 & 0&0025 & 42 \\
166 & 56551.8122 & 0.0021 & 0&0025 & 71 \\
167 & 56551.8684 & 0.0012 & 0&0026 & 64 \\
184 & 56552.8218 & 0.0013 & $-$0&0001 & 28 \\
185 & 56552.8827 & 0.0025 & 0&0046 & 26 \\
202 & 56553.8532 & 0.0038 & 0&0190 & 61 \\
\hline
  \multicolumn{6}{l}{\commenta BJD$-$2400000.} \\
  \multicolumn{6}{l}{\commentb Against max $= 2456542.4742 + 0.056238 E$.} \\
  \multicolumn{6}{l}{\commentc Number of points used to determine the maximum.} \\
\end{tabular}
\end{center}
\end{table}

\subsection{ASASSN-13cv}\label{obj:asassn13cv}

   This object was discovered by ASAS-SN survey on 2013
September 5 (vsnet-alert 16303).  The coordinates are
\timeform{22h 10m 25.24s}, \timeform{+30D 46' 06.9''}.
Although the quiescent counterpart was not listed in
the photometric catalog of SDSS (vsnet-alert 16354), 
the Guide Star Catalog (GSC), Version 2.3.2 has an object 
with a 21.4 mag.
GALEX \citep{GALEX} also has a ultraviolet counterpart
within \timeform{1''} [NUV and FUV magnitudes 22.07(3) and
21.6(4) mag, respectively].

\begin{figure}
  \begin{center}
    \FigureFile(88mm,70mm){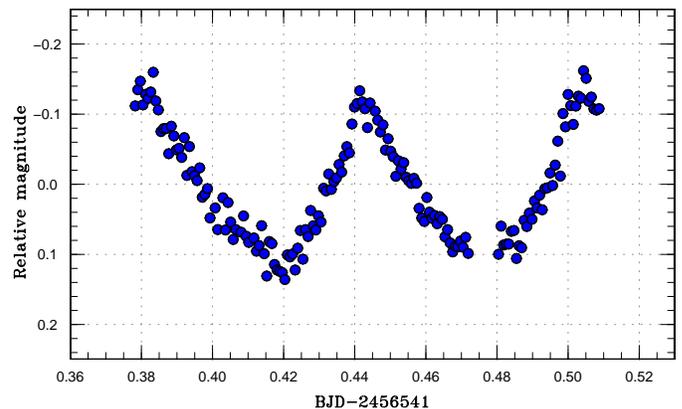}
  \end{center}
  \caption{Superhumps in ASASSN-13cv on 2013 September 5.}
  \label{fig:asassn13cvlc}
\end{figure}

   We obtained a single-night observation of this object
(vsnet-alert 16364; figure \ref{fig:asassn13cvlc}).
Three superhump maxima were obtained:
BJD 2456541.3818(7) ($N=37$), 2456541.4441(2) ($N=66$)
and 2456541.5072(6) ($N=36$).  The best period by the PDM
method is 0.0607(2)~d.

\subsection{ASASSN-13cz}\label{obj:asassn13cz}

\begin{figure}
  \begin{center}
    \FigureFile(88mm,110mm){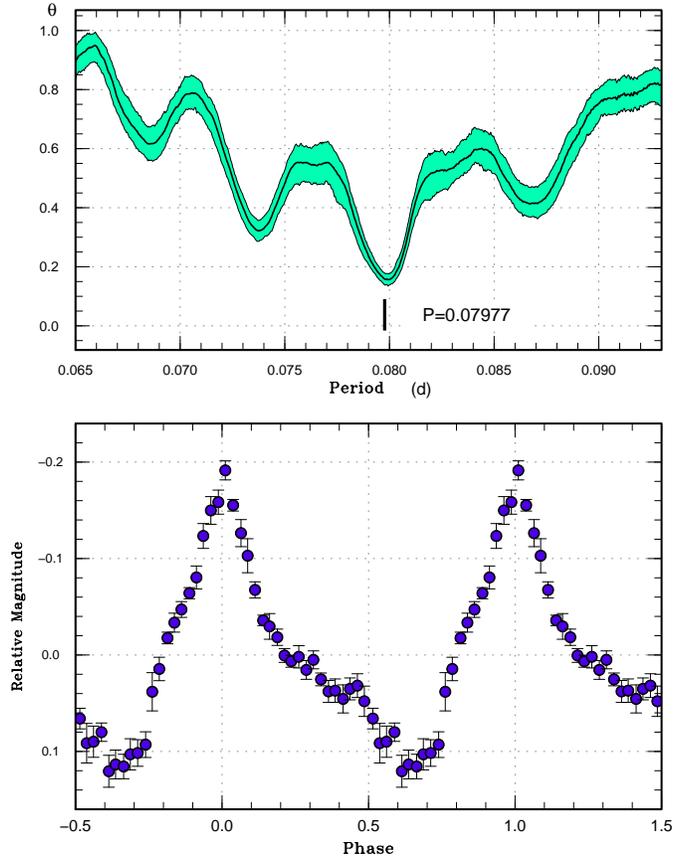}
  \end{center}
  \caption{Superhumps in ASASSN-13cz (2013). (Upper): PDM analysis.
     (Lower): Phase-averaged profile.}
  \label{fig:asassn13czshpdm}
\end{figure}

   This object was discovered by ASAS-SN survey on 2013
September 14 (vsnet-alert 16401).  The coordinates are
\timeform{15h 27m 55.3s}, \timeform{+63D 27' 53.4''}.
Subsequent observations detected superhumps (vsnet-alert 16405;
figure \ref{fig:asassn13czshpdm}).
The times of superhump maxima are listed in table
\ref{tab:asassn13czoc2013}.  The period shown in table
\ref{tab:perlist} is a result of the PDM analysis.

\begin{table}
\caption{Superhump maxima of ASASSN-13cz (2013)}\label{tab:asassn13czoc2013}
\begin{center}
\begin{tabular}{rp{55pt}p{40pt}r@{.}lr}
\hline
\multicolumn{1}{c}{$E$} & \multicolumn{1}{c}{max\commenta} & \multicolumn{1}{c}{error} & \multicolumn{2}{c}{$O-C$\commentb} & \multicolumn{1}{c}{$N$\commentc} \\
\hline
0 & 56550.3610 & 0.0004 & 0&0003 & 86 \\
1 & 56550.4407 & 0.0004 & 0&0002 & 81 \\
2 & 56550.5198 & 0.0005 & $-$0&0006 & 82 \\
13 & 56551.3986 & 0.0006 & 0&0001 & 50 \\
\hline
  \multicolumn{6}{l}{\commenta BJD$-$2400000.} \\
  \multicolumn{6}{l}{\commentb Against max $= 2456550.3607 + 0.079834 E$.} \\
  \multicolumn{6}{l}{\commentc Number of points used to determine the maximum.} \\
\end{tabular}
\end{center}
\end{table}

\subsection{ASASSN-13da}\label{obj:asassn13da}

\begin{figure}
  \begin{center}
    \FigureFile(88mm,110mm){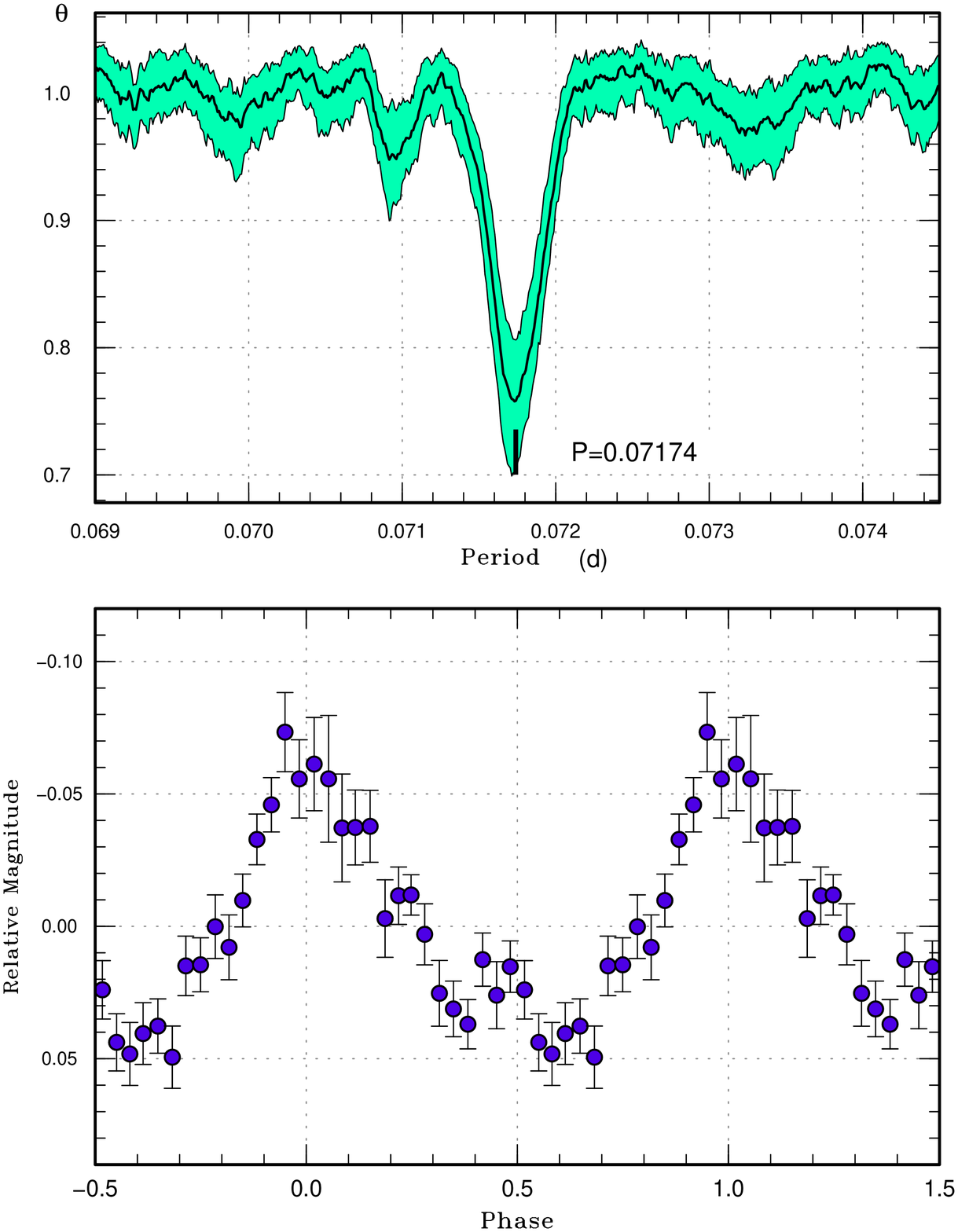}
  \end{center}
  \caption{Superhumps in ASASSN-13da (2013). (Upper): PDM analysis.
     (Lower): Phase-averaged profile.}
  \label{fig:asassn13dashpdm}
\end{figure}

   This object was discovered by ASAS-SN survey on 2013
September 20 (vsnet-alert 16426).  The coordinates are
\timeform{19h 59m 18.03s}, \timeform{-18D 33' 31.4''}.
The quiescent counterpart is a 21 mag object in CRTS.
On the first three nights, only low-amplitude modulations
were detected.  Four days after the detection, fully grown
superhumps appeared (vsnet-alert 16494; figure
\ref{fig:asassn13dashpdm}).  This growth
of superhumps was associated with the brightening of
the system brightness.
The times of superhump maxima are listed in table
\ref{tab:asassn13daoc2013}.  The times for $E \le 29$
were those for superhumps in the growing stage and
they are likely stage A superhumps.  Cycle counts
for these maxima are uncertain.  Although a PDM analysis 
of this segment yielded a possible period of 0.07300(5)~d,
this value was not included in table \ref{tab:perlist}
due to its uncertainty.  The decrease in $O-C$
for $E \ge 153$ suggests stage B-C transition.

\begin{table}
\caption{Superhump maxima of ASASSN-13da (2013)}\label{tab:asassn13daoc2013}
\begin{center}
\begin{tabular}{rp{55pt}p{40pt}r@{.}lr}
\hline
\multicolumn{1}{c}{$E$} & \multicolumn{1}{c}{max\commenta} & \multicolumn{1}{c}{error} & \multicolumn{2}{c}{$O-C$\commentb} & \multicolumn{1}{c}{$N$\commentc} \\
\hline
0 & 56556.5090 & 0.0025 & 0&0017 & 19 \\
1 & 56556.5711 & 0.0047 & $-$0&0079 & 20 \\
29 & 56558.5871 & 0.0048 & $-$0&0002 & 17 \\
56 & 56560.5208 & 0.0010 & $-$0&0032 & 19 \\
57 & 56560.5976 & 0.0012 & 0&0019 & 20 \\
70 & 56561.5317 & 0.0009 & 0&0036 & 19 \\
71 & 56561.6015 & 0.0007 & 0&0016 & 19 \\
98 & 56563.5360 & 0.0015 & $-$0&0006 & 19 \\
99 & 56563.6082 & 0.0014 & $-$0&0001 & 15 \\
112 & 56564.5350 & 0.0011 & $-$0&0057 & 24 \\
113 & 56564.6136 & 0.0021 & 0&0011 & 17 \\
126 & 56565.5511 & 0.0016 & 0&0062 & 21 \\
139 & 56566.4829 & 0.0111 & 0&0055 & 9 \\
140 & 56566.5555 & 0.0018 & 0&0065 & 25 \\
153 & 56567.4878 & 0.0212 & 0&0062 & 10 \\
154 & 56567.5563 & 0.0010 & 0&0031 & 25 \\
168 & 56568.5571 & 0.0014 & $-$0&0004 & 26 \\
181 & 56569.4737 & 0.0046 & $-$0&0162 & 6 \\
182 & 56569.5585 & 0.0024 & $-$0&0031 & 26 \\
\hline
  \multicolumn{6}{l}{\commenta BJD$-$2400000.} \\
  \multicolumn{6}{l}{\commentb Against max $= 2456556.5072 + 0.071728 E$.} \\
  \multicolumn{6}{l}{\commentc Number of points used to determine the maximum.} \\
\end{tabular}
\end{center}
\end{table}

\subsection{ASASSN-14ac}\label{obj:asassn14ac}

\begin{figure}
  \begin{center}
    \FigureFile(88mm,110mm){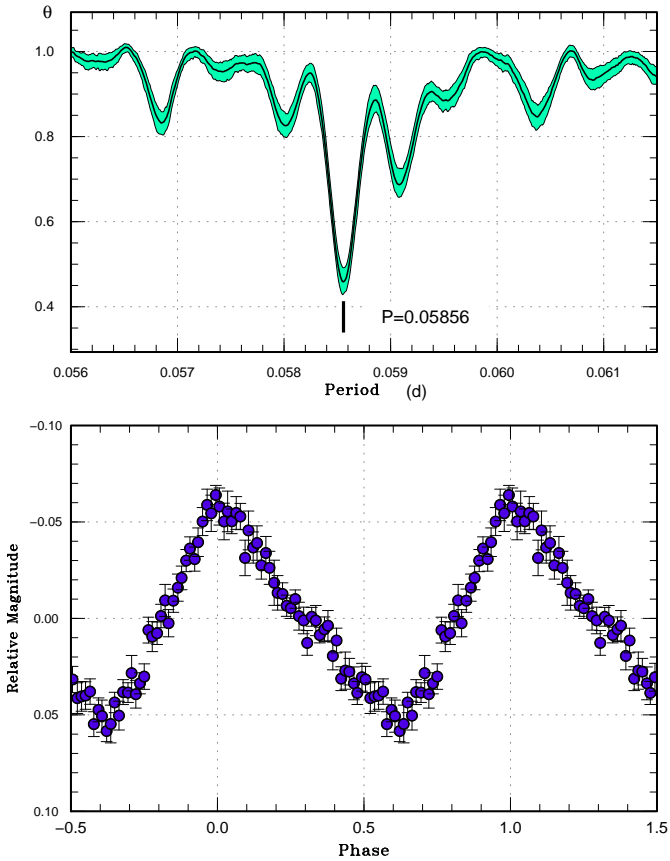}
  \end{center}
  \caption{Superhumps in ASASSN-14ac (2014). (Upper): PDM analysis
     for the interval BJD 2456692--2456701.
     (Lower): Phase-averaged profile.}
  \label{fig:asassn14acshpdm}
\end{figure}

   This object was discovered by ASAS-SN survey on 2014
January 18 \citep{sha14asassn14acatel5775}.  The coordinates are
\timeform{07h 52m 54.9s}, \timeform{+53D 05' 31.2''}.
The large outburst amplitude ($\sim$7 mag) and the blue
SDSS counterpart ($u-g=-0.3$) was suggestive of
a WZ Sge-type dwarf nova (vsnet-alert 16794).
Although subsequent observations recorded some low
amplitude modulations (vsnet-alert 16823, 16830),
no distinct period was obtained.  The object started to show
ordinary superhumps 14~d after the discovery
(vsnet-alert 16880, 16895; figure \ref{fig:asassn14acshpdm}).

\begin{figure}
  \begin{center}
    \FigureFile(88mm,70mm){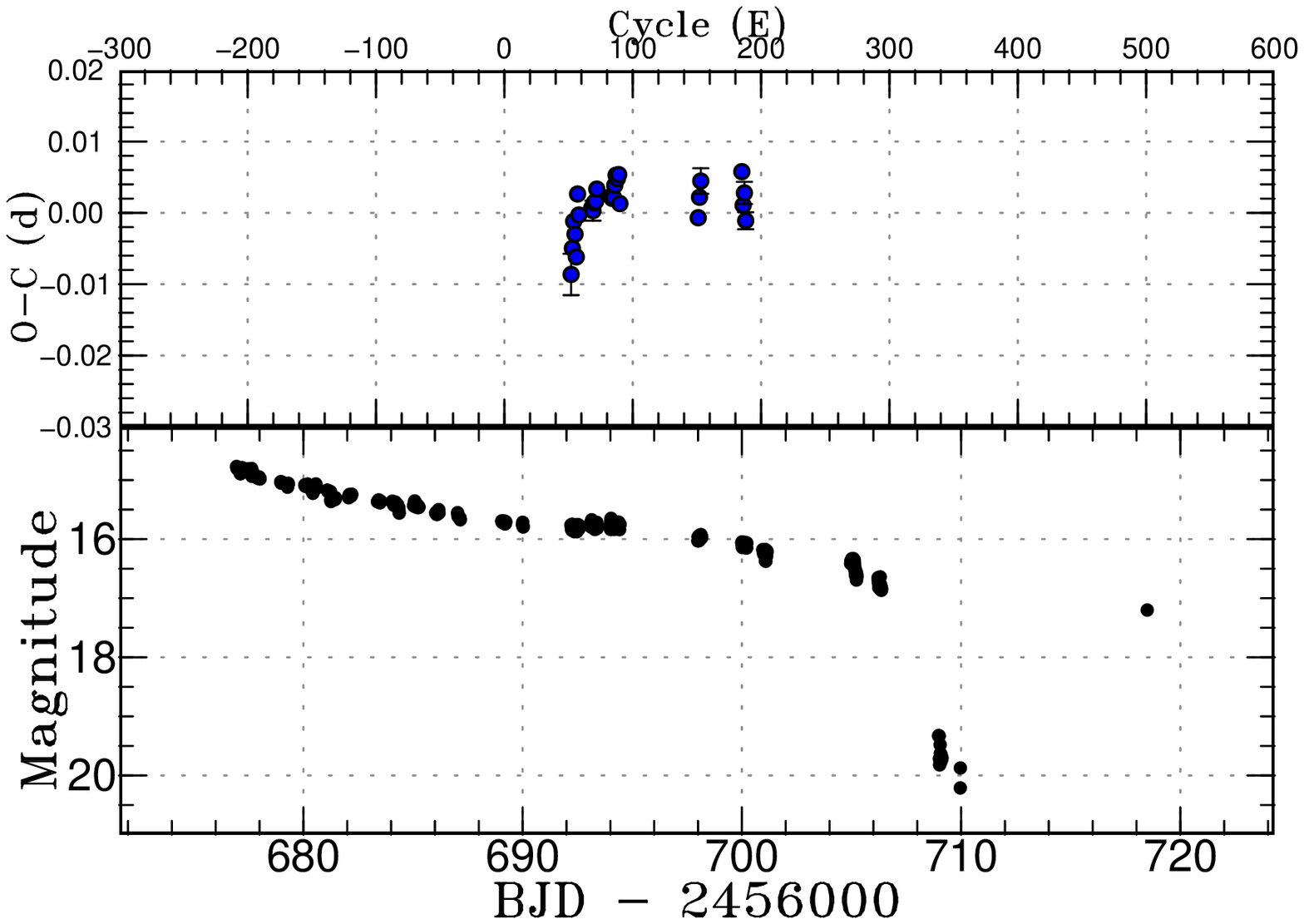}
  \end{center}
  \caption{$O-C$ diagram of superhumps in ASASSN-14ac (2014).
     (Upper): $O-C$ diagram.  A period of 0.05855~d
     was used to draw this figure.
     (Lower): Light curve.  The observations were binned to 0.012~d.}
  \label{fig:asassn14achumpall}
\end{figure}

   The times of superhump maxima are listed in table
\ref{tab:asassn14acoc2014}.  The early part of the observation
clearly recorded stage A superhumps.  The global
light curve showed systematic brightening associated
with the appearance of superhumps (see also figure
\ref{fig:asassn14achumpall}).  Although the later 
part of the superoutburst was not very well recorded, 
the $P_{\rm dot}$ of stage B superhumps was not
strongly positive.  The object started rapid fading
from the plateau stage on February 17 (30~d after
the discovery).

   On March 1, E. Muyllaert recorded the object
at 17.2 mag (unfiltered CCD magnitude), which appeared
to be a post-superoutburst rebrightening.
The type of the rebrightening, however, could not be
determined.

   Since the stage A superhumps of this object are 
well established, determination of the orbital period 
in quiescence is desired to estimate the mass ratio.

\begin{table}
\caption{Superhump maxima of ASASSN-14ac (2014)}\label{tab:asassn14acoc2014}
\begin{center}
\begin{tabular}{rp{55pt}p{40pt}r@{.}lr}
\hline
\multicolumn{1}{c}{$E$} & \multicolumn{1}{c}{max\commenta} & \multicolumn{1}{c}{error} & \multicolumn{2}{c}{$O-C$\commentb} & \multicolumn{1}{c}{$N$\commentc} \\
\hline
0 & 56689.1105 & 0.0023 & $-$0&0364 & 43 \\
1 & 56689.1686 & 0.0016 & $-$0&0370 & 42 \\
15 & 56690.0034 & 0.0006 & $-$0&0245 & 32 \\
52 & 56692.2034 & 0.0029 & 0&0023 & 17 \\
53 & 56692.2656 & 0.0006 & 0&0058 & 31 \\
54 & 56692.3279 & 0.0010 & 0&0094 & 31 \\
55 & 56692.3847 & 0.0003 & 0&0074 & 24 \\
56 & 56692.4400 & 0.0010 & 0&0040 & 23 \\
57 & 56692.5074 & 0.0006 & 0&0127 & 31 \\
58 & 56692.5630 & 0.0006 & 0&0096 & 30 \\
68 & 56693.1495 & 0.0007 & 0&0087 & 25 \\
69 & 56693.2077 & 0.0014 & 0&0081 & 39 \\
70 & 56693.2673 & 0.0008 & 0&0090 & 47 \\
71 & 56693.3261 & 0.0005 & 0&0091 & 31 \\
72 & 56693.3863 & 0.0004 & 0&0106 & 32 \\
83 & 56694.0294 & 0.0004 & 0&0076 & 42 \\
84 & 56694.0876 & 0.0006 & 0&0071 & 42 \\
85 & 56694.1463 & 0.0004 & 0&0070 & 42 \\
86 & 56694.2066 & 0.0007 & 0&0086 & 50 \\
87 & 56694.2665 & 0.0010 & 0&0098 & 42 \\
88 & 56694.3246 & 0.0006 & 0&0091 & 31 \\
89 & 56694.3837 & 0.0007 & 0&0095 & 31 \\
90 & 56694.4382 & 0.0007 & 0&0053 & 18 \\
151 & 56698.0078 & 0.0008 & $-$0&0079 & 56 \\
152 & 56698.0692 & 0.0010 & $-$0&0052 & 50 \\
153 & 56698.1300 & 0.0018 & $-$0&0032 & 56 \\
185 & 56700.0049 & 0.0009 & $-$0&0077 & 38 \\
186 & 56700.0588 & 0.0008 & $-$0&0126 & 61 \\
187 & 56700.1190 & 0.0016 & $-$0&0111 & 61 \\
188 & 56700.1737 & 0.0012 & $-$0&0151 & 60 \\
\hline
  \multicolumn{6}{l}{\commenta BJD$-$2400000.} \\
  \multicolumn{6}{l}{\commentb Against max $= 2456689.1469 + 0.058734 E$.} \\
  \multicolumn{6}{l}{\commentc Number of points used to determine the maximum.} \\
\end{tabular}
\end{center}
\end{table}

\subsection{CSS J024354.0$-$160314}\label{obj:j024354}

\begin{figure}
  \begin{center}
    \FigureFile(88mm,110mm){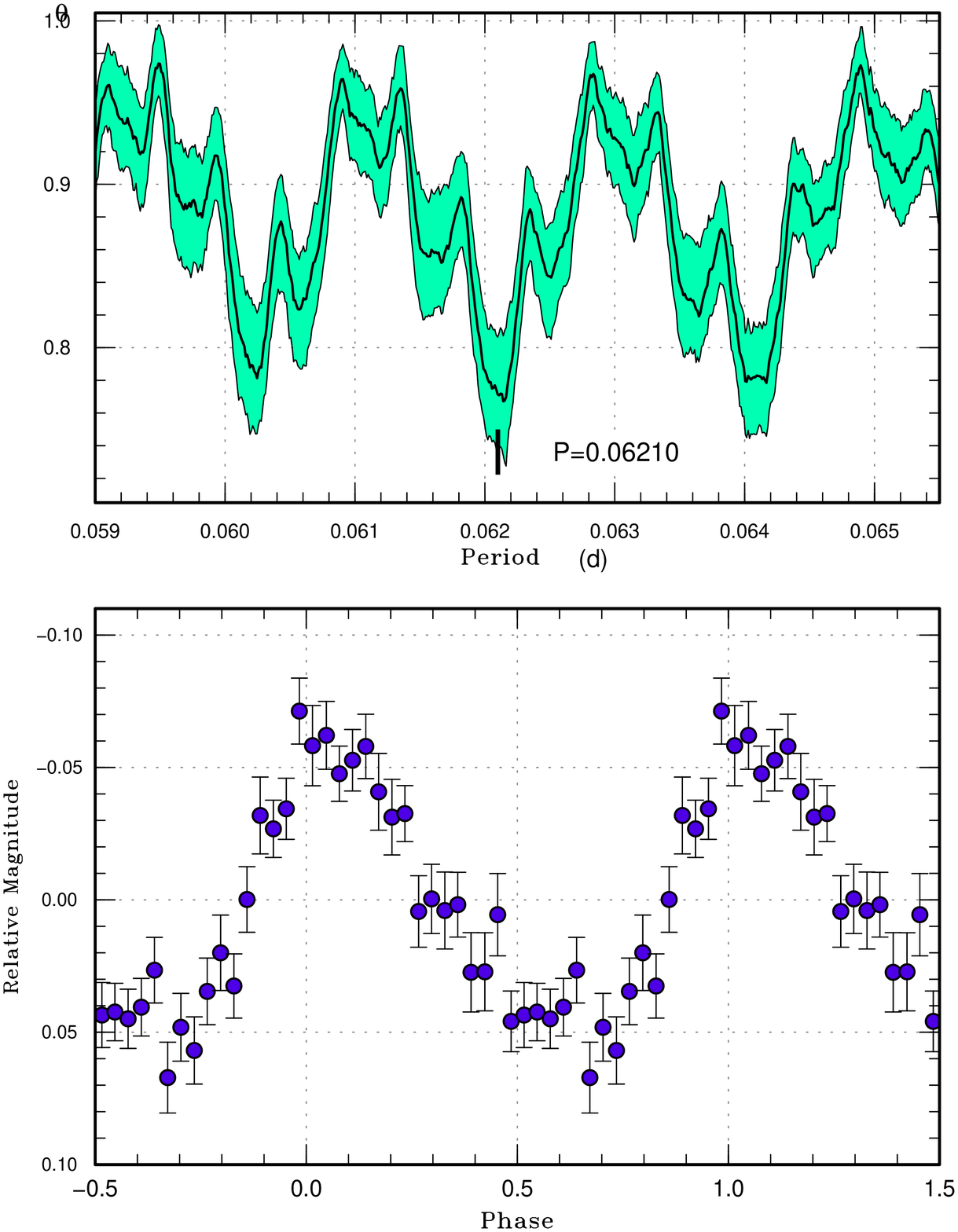}
  \end{center}
  \caption{Superhumps in CSS J024354 (2013). (Upper): PDM analysis.
     (Lower): Phase-averaged profile.}
  \label{fig:j0243shpdm}
\end{figure}

   This object (=CSS131026:024354$-$160314, hereafter CSS J024354)
was detected as a large-amplitude dwarf nova by CRTS on 
2013 October 26 (vsnet-alert 16564).  Subsequent observations
detected superhumps (vsnet-alert 16600; figure \ref{fig:j0243shpdm}).
The times of superhump maxima are listed in table
\ref{tab:j024354oc2013}.

\begin{table}
\caption{Superhump maxima of CSS J024354 (2013)}\label{tab:j024354oc2013}
\begin{center}
\begin{tabular}{rp{55pt}p{40pt}r@{.}lr}
\hline
\multicolumn{1}{c}{$E$} & \multicolumn{1}{c}{max\commenta} & \multicolumn{1}{c}{error} & \multicolumn{2}{c}{$O-C$\commentb} & \multicolumn{1}{c}{$N$\commentc} \\
\hline
0 & 56594.5208 & 0.0033 & 0&0094 & 143 \\
1 & 56594.5645 & 0.0024 & $-$0&0089 & 140 \\
94 & 56600.3445 & 0.0009 & $-$0&0020 & 80 \\
126 & 56602.3336 & 0.0007 & 0&0007 & 121 \\
127 & 56602.3953 & 0.0008 & 0&0003 & 143 \\
128 & 56602.4548 & 0.0011 & $-$0&0022 & 142 \\
129 & 56602.5220 & 0.0093 & 0&0028 & 50 \\
\hline
  \multicolumn{6}{l}{\commenta BJD$-$2400000.} \\
  \multicolumn{6}{l}{\commentb Against max $= 2456594.5114 + 0.062076 E$.} \\
  \multicolumn{6}{l}{\commentc Number of points used to determine the maximum.} \\
\end{tabular}
\end{center}
\end{table}

\subsection{DDE 31}\label{obj:dde31}

   This dwarf nova was discovered by D. Denisenko in outburst
(16.3 mag; all magnitudes for this object are unfiltered CCD
magnitudes) on 2012 October 15 (vsnet-alert 15007).
The coordinates are
\timeform{02h 13m 17.18s}, \timeform{+46D 06' 03.4''}.
On the next night, the object faded to 17.7 mag (B. Staels).
On 2012 December 15, D. Denisenko again detected this
object in outburst (16.2 mag) and reported a previous
outburst on 2012 September 14 (17.0 mag, vsnet-alert 15170).

\begin{figure}
  \begin{center}
    \FigureFile(88mm,70mm){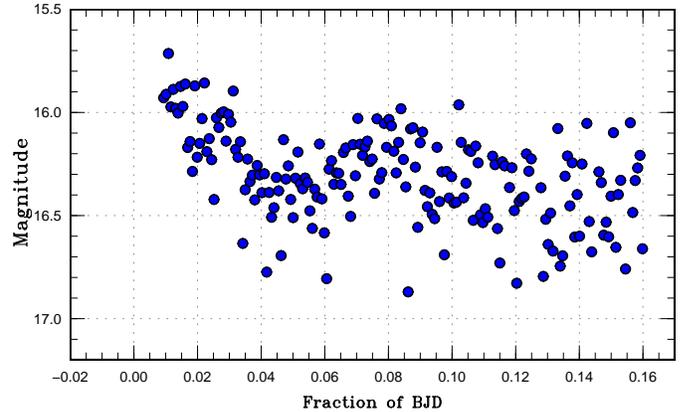}
  \end{center}
  \caption{Possible superhumps in DDE 31.}
  \label{fig:dde31shlc}
\end{figure}

   On 2014 January 4, D. Denisenko reported a brighter
outburst (15.3 mag, vsnet-alert 16757).  Time-resolved
observations detected modulations resembling superoutburst
(figure \ref{fig:dde31shlc}).  Although a period of 0.073(3)~d
was inferred, the outburst faded rapidly ($\sim$1 mag d$^{-1}$)
and the object was not detected brighter than 18.2 mag
five nights later.  This behavior was unusual for a superoutburst.
Furthermore, an analysis of the SDSS colors \citep{kat12DNSDSS}
yielded an expected orbital period longer than 0.1~d
(it was also likely that these SDSS observations were not
obtained in true quiescence, cf. vsnet-alert 15170).
We therefore should wait another outburst to clarify
the classification of this object.

\subsection{MASTER OT J004527.52$+$503213.8}\label{obj:j004527}

   This object (hereafter MASTER J004527) was detected
as a bright (12.5 mag) transient by the MASTER network \citep{MASTER}
on 2013 September 17 \citep{den13j0045atel5399}.
The object has an 18--19 mag quiescent counterpart and
the large outburst amplitude suggested a WZ Sge-type
dwarf nova.  The last non-detection observation prior
to the outburst was reported on September 13
(fainter than 18.5, vsnet-alert 16422).

\begin{figure}
  \begin{center}
    \FigureFile(88mm,70mm){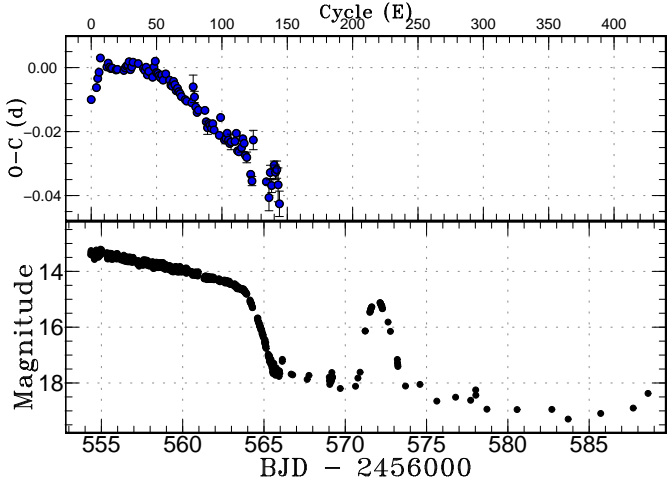}
  \end{center}
  \caption{$O-C$ diagram of superhumps in MASTER J004527 (2013).
     (Upper): $O-C$ diagram.  A period of 0.08039~d
     was used to draw this figure.
     (Lower): Light curve.  The observations were binned to 0.016~d.}
  \label{fig:j0045humpall}
\end{figure}

   Subsequent observations immediately detected superhumps
(vsnet-alert 16423).  The long ($\sim$0.081~d) superhump period,
however, was not well compatible with the suggested WZ Sge-type
classification (vsnet-alert 16425).  Later observations
yielded a slightly shorter superhump period (vsnet-alert
16433, 16434, 16459).  A further decrease in the superhump
period was reported on September 22 (vsnet-alert 16463, 16497).
The object showed a single post-superoutburst rebrightening
on October 6 (vsnet-alert 16513; figure \ref{fig:j0045humpall}).

\begin{figure}
  \begin{center}
    \FigureFile(88mm,110mm){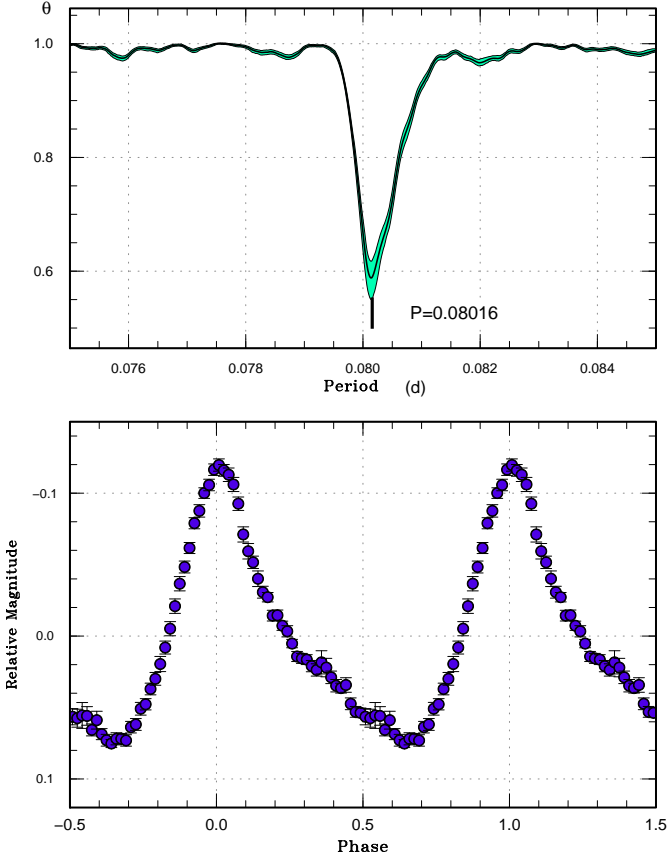}
  \end{center}
  \caption{Superhumps in MASTER J004527 (2013). (Upper): PDM analysis.
     (Lower): Phase-averaged profile.}
  \label{fig:j0045shpdm}
\end{figure}

   The times of superhump maxima are listed in table
\ref{tab:j004527oc2013}.  The decrease in the period
around September 22 ($E \sim 50$; figure \ref{fig:j0045humpall})
may be either attributed
to stage B-C transition (as in ordinary SU UMa-type
dwarf novae) or stage A-B transition (as reported in
likely period bouncers SSS J122221.7$-$311523,
\cite{kat13j1222}; OT J075418.7$+$381225 and
OT J230425.8$+$062546, C. Nakata, in preparation).
Since the large outburst amplitude of MASTER J004527
suggested the WZ Sge-type classification, the second
possibility would deserve consideration.  We consider
that the former interpretation is more likely for
several reasons: (1) The difference in the periods
before and after the transition was 0.5\%, which is
typical for stage B-C transition (e.g. \cite{Pdot}),
but is smaller than stage A-B transition (1.0--1.5\%,
cf. \cite{nak13j2112j2037}).  (2) There was a phase
with a longer superhump period ($E \le 7$), which can
be considered as stage A.  (3) The amplitudes of
superhumps were much larger 
(0.2--0.3 mag; figure \ref{fig:j0045shpdm}) than in 
superhumps in likely period bouncers (e.g. \cite{kat13j1222}).
We list the periods based on the former interpretation
in table \ref{tab:perlist}.

   Following this interpretation of the superhump stages,
the object can be recognized as an SU UMa-type dwarf nova
(not a period bouncer) with a long orbital period and
infrequent, large-amplitude outbursts.  The object
may resemble V1251 Cyg (\cite{kat95v1251cyg}; \cite{Pdot})
or QY Per \citep{Pdot}.  Future monitoring of outbursts
would be helpful in identifying the supercycle.

\begin{table}
\caption{Superhump maxima of MASTER J004527 (2013)}\label{tab:j004527oc2013}
\begin{center}
\begin{tabular}{rp{55pt}p{40pt}r@{.}lr}
\hline
\multicolumn{1}{c}{$E$} & \multicolumn{1}{c}{max\commenta} & \multicolumn{1}{c}{error} & \multicolumn{2}{c}{$O-C$\commentb} & \multicolumn{1}{c}{$N$\commentc} \\
\hline
0 & 56554.3649 & 0.0003 & $-$0&0181 & 95 \\
4 & 56554.6902 & 0.0002 & $-$0&0132 & 226 \\
5 & 56554.7735 & 0.0002 & $-$0&0101 & 147 \\
6 & 56554.8558 & 0.0002 & $-$0&0079 & 124 \\
7 & 56554.9406 & 0.0008 & $-$0&0031 & 73 \\
12 & 56555.3398 & 0.0002 & $-$0&0045 & 197 \\
13 & 56555.4214 & 0.0001 & $-$0&0030 & 292 \\
14 & 56555.5005 & 0.0001 & $-$0&0040 & 386 \\
15 & 56555.5804 & 0.0001 & $-$0&0042 & 305 \\
16 & 56555.6610 & 0.0002 & $-$0&0037 & 310 \\
19 & 56555.9016 & 0.0004 & $-$0&0035 & 63 \\
20 & 56555.9821 & 0.0003 & $-$0&0030 & 59 \\
25 & 56556.3838 & 0.0002 & $-$0&0019 & 279 \\
26 & 56556.4649 & 0.0002 & $-$0&0009 & 279 \\
27 & 56556.5459 & 0.0002 & 0&0000 & 171 \\
28 & 56556.6266 & 0.0002 & 0&0006 & 202 \\
29 & 56556.7080 & 0.0003 & 0&0019 & 190 \\
30 & 56556.7859 & 0.0004 & $-$0&0003 & 186 \\
31 & 56556.8672 & 0.0003 & 0&0009 & 197 \\
32 & 56556.9491 & 0.0006 & 0&0027 & 52 \\
36 & 56557.2702 & 0.0003 & 0&0033 & 312 \\
40 & 56557.5902 & 0.0006 & 0&0030 & 62 \\
41 & 56557.6700 & 0.0004 & 0&0026 & 69 \\
42 & 56557.7514 & 0.0005 & 0&0039 & 61 \\
43 & 56557.8294 & 0.0005 & 0&0018 & 71 \\
44 & 56557.9109 & 0.0005 & 0&0032 & 71 \\
47 & 56558.1502 & 0.0004 & 0&0022 & 125 \\
48 & 56558.2338 & 0.0003 & 0&0057 & 212 \\
49 & 56558.3160 & 0.0003 & 0&0078 & 227 \\
50 & 56558.3929 & 0.0003 & 0&0046 & 187 \\
51 & 56558.4730 & 0.0003 & 0&0046 & 196 \\
52 & 56558.5527 & 0.0003 & 0&0041 & 225 \\
\hline
  \multicolumn{6}{l}{\commenta BJD$-$2400000.} \\
  \multicolumn{6}{l}{\commentb Against max $= 2456554.3830 + 0.080106 E$.} \\
  \multicolumn{6}{l}{\commentc Number of points used to determine the maximum.} \\
\end{tabular}
\end{center}
\end{table}

\addtocounter{table}{-1}
\begin{table}
\caption{Superhump maxima of MASTER J004527 (2013) (continued)}
\begin{center}
\begin{tabular}{rp{55pt}p{40pt}r@{.}lr}
\hline
\multicolumn{1}{c}{$E$} & \multicolumn{1}{c}{max\commenta} & \multicolumn{1}{c}{error} & \multicolumn{2}{c}{$O-C$\commentb} & \multicolumn{1}{c}{$N$\commentc} \\
\hline
53 & 56558.6324 & 0.0003 & 0&0037 & 256 \\
54 & 56558.7136 & 0.0004 & 0&0049 & 130 \\
55 & 56558.7924 & 0.0005 & 0&0036 & 61 \\
57 & 56558.9552 & 0.0011 & 0&0061 & 24 \\
60 & 56559.1944 & 0.0002 & 0&0050 & 209 \\
61 & 56559.2731 & 0.0002 & 0&0037 & 263 \\
62 & 56559.3533 & 0.0002 & 0&0037 & 144 \\
63 & 56559.4352 & 0.0003 & 0&0055 & 128 \\
64 & 56559.5145 & 0.0003 & 0&0047 & 169 \\
65 & 56559.5930 & 0.0004 & 0&0031 & 66 \\
66 & 56559.6741 & 0.0003 & 0&0041 & 206 \\
67 & 56559.7535 & 0.0002 & 0&0034 & 188 \\
68 & 56559.8333 & 0.0003 & 0&0030 & 68 \\
69 & 56559.9127 & 0.0005 & 0&0024 & 70 \\
72 & 56560.1529 & 0.0003 & 0&0023 & 120 \\
73 & 56560.2329 & 0.0002 & 0&0021 & 137 \\
77 & 56560.5539 & 0.0006 & 0&0027 & 56 \\
78 & 56560.6393 & 0.0037 & 0&0080 & 42 \\
79 & 56560.7165 & 0.0018 & 0&0051 & 28 \\
80 & 56560.7938 & 0.0007 & 0&0023 & 61 \\
81 & 56560.8725 & 0.0006 & 0&0009 & 69 \\
82 & 56560.9536 & 0.0009 & 0&0019 & 37 \\
87 & 56561.3555 & 0.0009 & 0&0032 & 104 \\
88 & 56561.4323 & 0.0003 & $-$0&0001 & 108 \\
89 & 56561.5108 & 0.0021 & $-$0&0017 & 30 \\
90 & 56561.5925 & 0.0008 & $-$0&0000 & 70 \\
91 & 56561.6724 & 0.0005 & $-$0&0002 & 69 \\
92 & 56561.7518 & 0.0008 & $-$0&0010 & 57 \\
93 & 56561.8336 & 0.0007 & 0&0007 & 72 \\
94 & 56561.9121 & 0.0008 & $-$0&0009 & 68 \\
98 & 56562.2320 & 0.0004 & $-$0&0015 & 154 \\
99 & 56562.3179 & 0.0006 & 0&0044 & 113 \\
\hline
  \multicolumn{6}{l}{\commenta BJD$-$2400000.} \\
  \multicolumn{6}{l}{\commentb Against max $= 2456554.3830 + 0.080106 E$.} \\
  \multicolumn{6}{l}{\commentc Number of points used to determine the maximum.} \\
\end{tabular}
\end{center}
\end{table}

\addtocounter{table}{-1}
\begin{table}
\caption{Superhump maxima of MASTER J004527 (2013) (continued)}
\begin{center}
\begin{tabular}{rp{55pt}p{40pt}r@{.}lr}
\hline
\multicolumn{1}{c}{$E$} & \multicolumn{1}{c}{max\commenta} & \multicolumn{1}{c}{error} & \multicolumn{2}{c}{$O-C$\commentb} & \multicolumn{1}{c}{$N$\commentc} \\
\hline
102 & 56562.5520 & 0.0008 & $-$0&0018 & 61 \\
103 & 56562.6329 & 0.0008 & $-$0&0010 & 65 \\
104 & 56562.7149 & 0.0008 & 0&0009 & 52 \\
105 & 56562.7932 & 0.0009 & $-$0&0010 & 63 \\
106 & 56562.8725 & 0.0006 & $-$0&0018 & 70 \\
107 & 56562.9534 & 0.0024 & $-$0&0010 & 21 \\
110 & 56563.1947 & 0.0005 & 0&0001 & 75 \\
111 & 56563.2776 & 0.0007 & 0&0028 & 81 \\
112 & 56563.3525 & 0.0006 & $-$0&0024 & 149 \\
113 & 56563.4326 & 0.0005 & $-$0&0024 & 93 \\
115 & 56563.5947 & 0.0014 & $-$0&0005 & 64 \\
116 & 56563.6779 & 0.0010 & 0&0026 & 124 \\
117 & 56563.7569 & 0.0009 & 0&0014 & 87 \\
118 & 56563.8335 & 0.0011 & $-$0&0020 & 71 \\
119 & 56563.9132 & 0.0017 & $-$0&0024 & 66 \\
122 & 56564.1491 & 0.0004 & $-$0&0068 & 63 \\
123 & 56564.2274 & 0.0014 & $-$0&0087 & 84 \\
124 & 56564.3206 & 0.0030 & 0&0045 & 17 \\
134 & 56565.1115 & 0.0011 & $-$0&0058 & 58 \\
136 & 56565.2673 & 0.0041 & $-$0&0101 & 27 \\
137 & 56565.3555 & 0.0023 & $-$0&0020 & 36 \\
138 & 56565.4319 & 0.0022 & $-$0&0058 & 35 \\
140 & 56565.5990 & 0.0013 & 0&0012 & 71 \\
141 & 56565.6772 & 0.0023 & $-$0&0008 & 72 \\
142 & 56565.7584 & 0.0027 & 0&0003 & 69 \\
143 & 56565.8340 & 0.0053 & $-$0&0042 & 73 \\
144 & 56565.9085 & 0.0040 & $-$0&0098 & 67 \\
\hline
  \multicolumn{6}{l}{\commenta BJD$-$2400000.} \\
  \multicolumn{6}{l}{\commentb Against max $= 2456554.3830 + 0.080106 E$.} \\
  \multicolumn{6}{l}{\commentc Number of points used to determine the maximum.} \\
\end{tabular}
\end{center}
\end{table}

\subsection{MASTER OT J005740.99$+$443101.5}\label{obj:j005740}

\begin{figure}
  \begin{center}
    \FigureFile(88mm,70mm){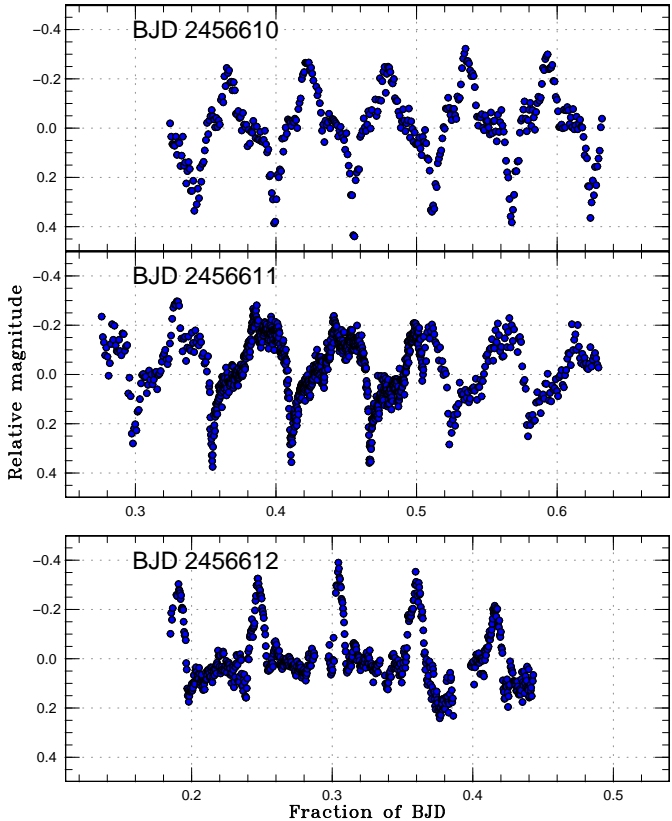}
  \end{center}
  \caption{Superhumps and eclipses in MASTER J005740 (2013).
  On the first night, eclipses were sharply detected.
  On the second night, eclipses became less apparent
  as the superhump maximum approaches the eclipses.
  On the third night, eclipses became inapparent.}
  \label{fig:j0057shlc}
\end{figure}

   This object (hereafter MASTER J005740) was detected
as a transient by the MASTER network
on 2013 November 6 \citep{bal13j0057atel5555}.
Subsequent observations Immediately detected early
superhumps (vsnet-alert 16603, 16606, 16609).
The large amplitude (0.4 mag) of early superhumps
suggested a high inclination.  There was also eclipse-like
feature in the light curve (vsnet-alert 16614; see also
vsnet-alert 16603).  On November 12--13, ordinary superhumps
started to appear, which were accompanied by eclipses
(vsnet-alert 16624; figure \ref{fig:j0057shlc}).
After fading from the plateau,
the eclipses became deeper ($\sim$1 mag, vsnet-alert 16640),
implying that the white dwarf is eclipsed.  This object
became the first candidate WZ Sge-type dwarf nova showing
the eclipse of the white dwarf.

   We first obtained the eclipse ephemeris using observations
other than the phase of early superhumps, since the profile
of early superhumps is similar to, but known to be different
from that of the eclipse (cf. \cite{uem12ESHrecon}).
We obtained the following ephemeris
\begin{equation}
{\rm Min(BJD)} = 2456617.36772(4) + 0.0561904(3) E
\label{equ:j0057ecl}
\end{equation}
by using the MCMC modelling as in V893 Sco.
Note that this ephemeris is not intended for long-term
prediction, in contrast to other eclipsing dwarf novae
treated in this paper, of eclipses since the eclipse profile 
is strongly affected by the varying superhumps and
systematic variations in relation to the system
brightness.  The errors given in equation (\ref{equ:j0057ecl})
is formal statistic ones, and the actual errors are
expected to be larger due to the systematic errors.

   The times of superhump maxima, determined after removing
the within 0.07 orbital phases of eclipses, are listed in
table \ref{tab:j005740oc2013}.  The $O-C$ diagram
is presented in figure \ref{fig:j0057humpall}.
The maxima for $E < 14$ are stage A superhumps with
growing amplitudes.  The maxima for $14 \le E \le 144$
are stage B superhumps.  The times for $144 < E < 245$
were not well determined because the amplitudes of superhumps
became smaller and the profile was difficult to detect
due to the orbital modulation.  After $E=245$, the amplitudes
of superhumps grew again.  These superhumps can be identified
as stage C superhumps, which are not usually seen in
WZ Sge-type dwarf novae (cf. \cite{Pdot}).

\begin{figure}
  \begin{center}
    \FigureFile(88mm,110mm){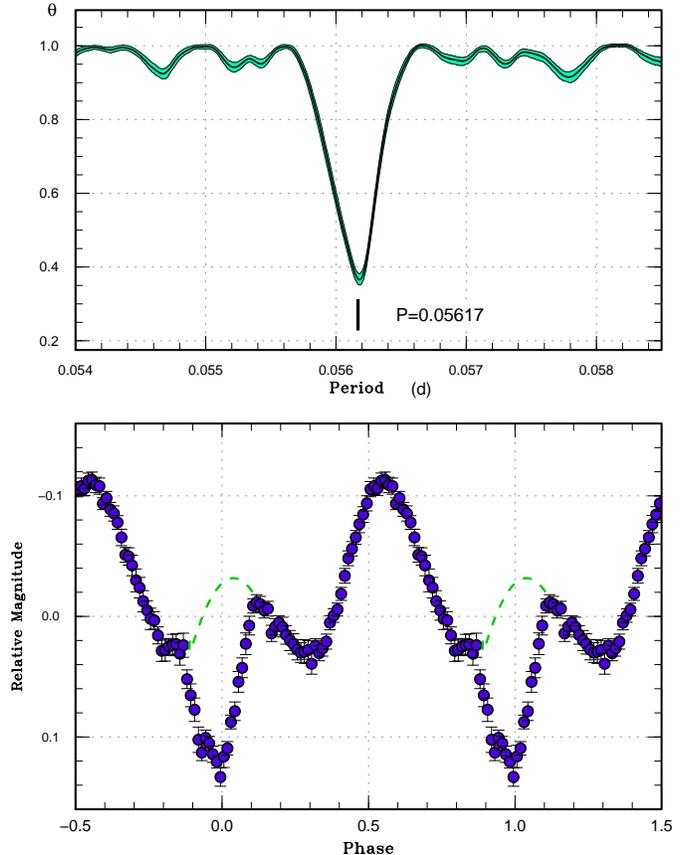}
  \end{center}
  \caption{Early superhumps in MASTER J005740 (2013).
     (Upper): PDM analysis.
     (Lower): Phase-averaged profile.  The phase is relative
     to the eclipse ephemeris (see text for details).
     The dashed line represents a hypothetical hump maximum
     without an eclipse.}
  \label{fig:j0057eshpdm}
\end{figure}

   The period of early superhump by the PDM method was
0.056169(3)~d (figure \ref{fig:j0057eshpdm}), which is
0.04\% shorter than the orbital period.  A summary of
comparison of periods of early superhumps and orbital periods
in various WZ Sge-type dwarf novae is given in subsection
\ref{sec:earlysh}.  The zero epoch in this figure is based 
on the ephemeris equation (\ref{equ:j0057ecl}).  Since 
the period of early superhumps and orbital period are
very slightly different, we used the eclipse center
nearest to the center of observation of early superhumps.
The period used for phase-averaging is the period of
early superhumps.  The profile of these early superhumps
and its implication is discussed in subsection \ref{sec:earlyshecl}.

\begin{figure}
  \begin{center}
    \FigureFile(88mm,110mm){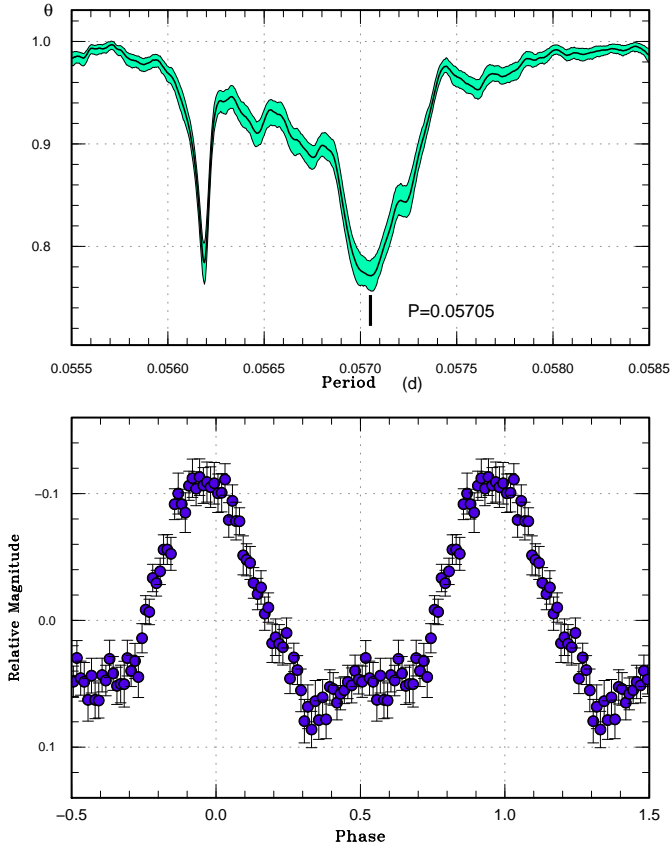}
  \end{center}
  \caption{Ordinary superhumps in MASTER J005740 (2013).
     The segment of stage B superhumps was used.
     (Upper): PDM analysis.
     (Lower): Phase-averaged profile.  The sharp signal
     at 0.05619~d is the orbital period.}
  \label{fig:j0057shpdm}
\end{figure}

\begin{figure}
  \begin{center}
    \FigureFile(88mm,70mm){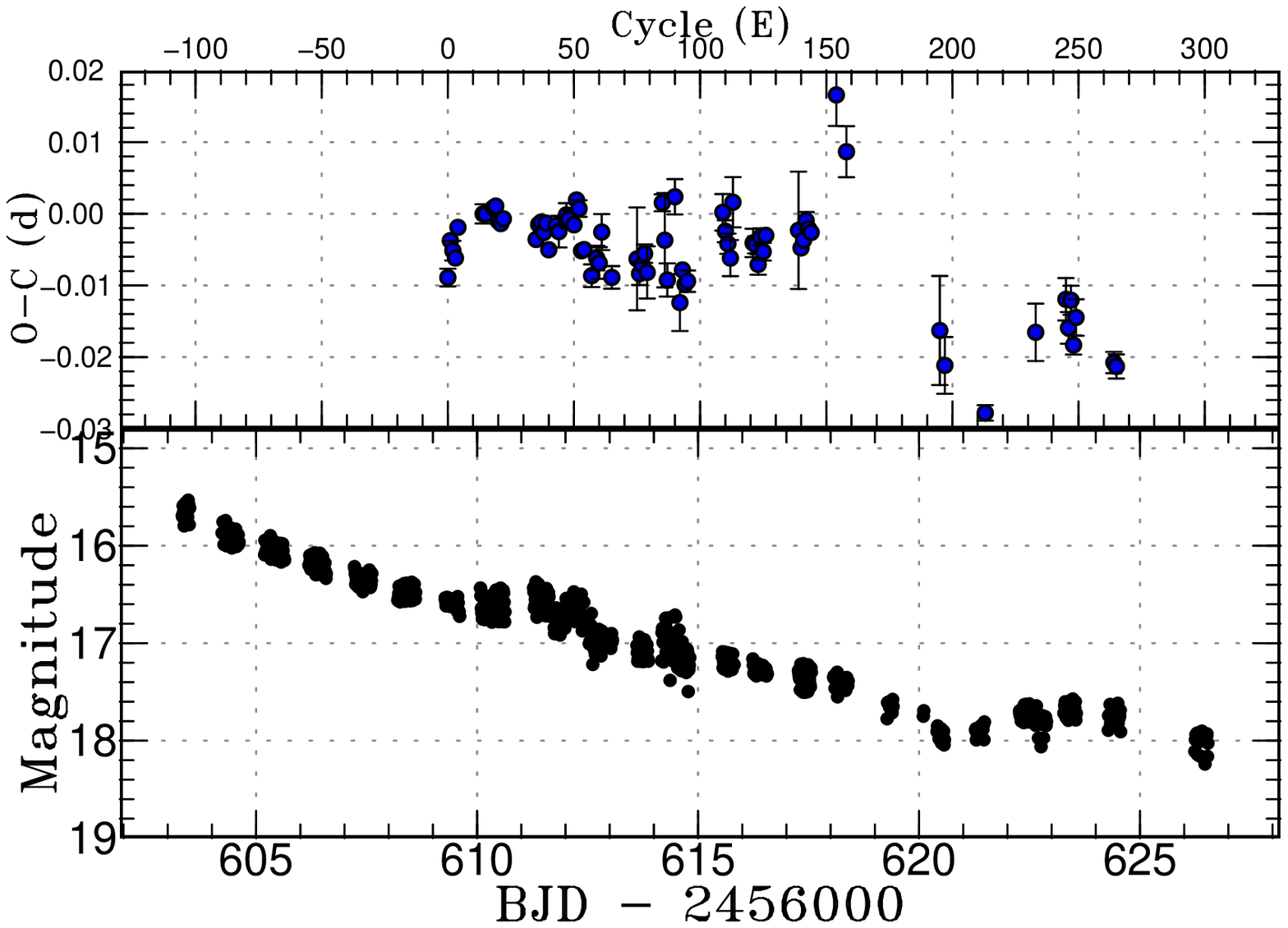}
  \end{center}
  \caption{$O-C$ diagram of superhumps in MASTER J005740 (2013).
     (Upper): $O-C$ diagram.  A period of 0.05709~d
     was used to draw this figure.
     (Lower): Light curve.  The observations were binned to 0.011~d.}
  \label{fig:j0057humpall}
\end{figure}

   The mean profile of stage B superhumps is shown in
figure \ref{fig:j0057shpdm}.  Since the times of superhumps
during the growing stage (stage A) were difficult to determine
due to the orbital modulation, we also measured the period
by the PDM for the interval BJD 2456608--2456610.
The resultant period was 0.05783(3)~d.  This value is
slightly different from that of the $O-C$ analysis
[0.05758(19)~d, $E \le 14$], which was likely more affected by
the shorter period close to stage B.  We therefore adopted
the former period as the representative period of stage A.
The resultant $\varepsilon^*$ of 0.028(5) corresponds to
$q$=0.076(16).

\begin{figure}
  \begin{center}
    \FigureFile(88mm,100mm){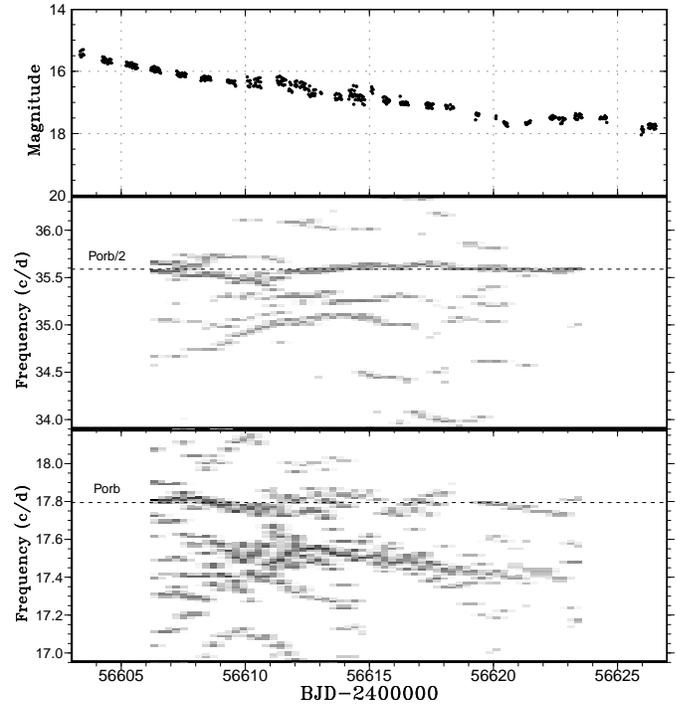}
  \end{center}
  \caption{Lasso analysis of MASTER J005740 (2013).
  (Upper:) Light curve.  The data were binned to 0.02~d.
  (Middle:) First harmonics of the superhump and orbital signals.
  (Lower:) Fundamental of the superhump and the orbital signal.
  The orbital signal was present both in the fundamental and
  the first harmonic.  The signal of (positive) superhumps with
  variable frequency was recorded during the superoutburst
  plateau.
  No indication of negative superhump was present.
  $\log \lambda=-8.7$ was used.
  The width of the sliding window and the time step used are
  6~d and 0.3~d, respectively.
  }
  \label{fig:j0057lasso}
\end{figure}

   A two-dimensional Lasso analysis is presented in
figure \ref{fig:j0057lasso}.  As in the eclipsing WZ Sge-type
dwarf nova WZ Sge \citep{Pdot5}, the orbital signal was
continuously seen.  The superhump signal with decreasing
frequencies (increasing periods) during the plateau phase
of the superoutburst is also clearly visible.

\begin{table}
\caption{Superhump maxima of MASTER J005740 (2013)}\label{tab:j005740oc2013}
\begin{center}
\begin{tabular}{rp{50pt}p{30pt}r@{.}lcr}
\hline
$E$ & max\commenta & error & \multicolumn{2}{c}{$O-C$\commentb} & phase\commentc & $N$\commentd \\
\hline
0 & 56609.3288 & 0.0012 & $-$0&0089 & 70 \\
1 & 56609.3910 & 0.0008 & $-$0&0036 & 71 \\
2 & 56609.4467 & 0.0014 & $-$0&0050 & 64 \\
3 & 56609.5028 & 0.0008 & $-$0&0060 & 54 \\
4 & 56609.5642 & 0.0008 & $-$0&0016 & 46 \\
14 & 56610.1370 & 0.0013 & 0&0008 & 35 \\
15 & 56610.1940 & 0.0007 & 0&0008 & 35 \\
18 & 56610.3661 & 0.0005 & 0&0019 & 50 \\
19 & 56610.4235 & 0.0005 & 0&0022 & 50 \\
20 & 56610.4785 & 0.0005 & 0&0002 & 48 \\
21 & 56610.5352 & 0.0005 & $-$0&0001 & 51 \\
22 & 56610.5930 & 0.0007 & 0&0006 & 53 \\
35 & 56611.3323 & 0.0007 & $-$0&0016 & 51 \\
36 & 56611.3915 & 0.0003 & 0&0006 & 193 \\
37 & 56611.4489 & 0.0005 & 0&0009 & 198 \\
38 & 56611.5045 & 0.0005 & $-$0&0005 & 139 \\
39 & 56611.5629 & 0.0008 & 0&0009 & 45 \\
40 & 56611.6163 & 0.0008 & $-$0&0028 & 44 \\
43 & 56611.7909 & 0.0010 & 0&0008 & 37 \\
44 & 56611.8472 & 0.0022 & 0&0000 & 35 \\
47 & 56612.0208 & 0.0016 & 0&0025 & 31 \\
48 & 56612.0773 & 0.0013 & 0&0019 & 25 \\
50 & 56612.1907 & 0.0006 & 0&0013 & 46 \\
51 & 56612.2512 & 0.0005 & 0&0048 & 103 \\
52 & 56612.3071 & 0.0012 & 0&0037 & 75 \\
53 & 56612.3583 & 0.0006 & $-$0&0022 & 95 \\
54 & 56612.4155 & 0.0005 & $-$0&0020 & 86 \\
57 & 56612.5832 & 0.0016 & $-$0&0055 & 30 \\
59 & 56612.6998 & 0.0017 & $-$0&0029 & 28 \\
60 & 56612.7562 & 0.0022 & $-$0&0035 & 30 \\
61 & 56612.8176 & 0.0025 & 0&0009 & 17 \\
\hline
  \multicolumn{7}{l}{\commenta BJD$-$2400000.} \\
  \multicolumn{7}{l}{\commentb Against max $= 2456609.3376 + 0.057035 E$.} \\
  \multicolumn{7}{l}{\commentc Orbital phase.} \\
  \multicolumn{7}{l}{\commentd Number of points used to determine the maximum.} \\
\end{tabular}
\end{center}
\end{table}

\addtocounter{table}{-1}
\begin{table}
\caption{Superhump maxima of MASTER J005740 (2013) (continued)}
\begin{center}
\begin{tabular}{rp{50pt}p{30pt}r@{.}lcr}
\hline
$E$ & max\commenta & error & \multicolumn{2}{c}{$O-C$\commentb} & phase\commentc & $N$\commentd \\
\hline
65 & 56613.0396 & 0.0016 & $-$0&0053 & 28 \\
75 & 56613.6132 & 0.0072 & $-$0&0021 & 29 \\
76 & 56613.6682 & 0.0016 & $-$0&0041 & 50 \\
77 & 56613.7265 & 0.0009 & $-$0&0028 & 50 \\
78 & 56613.7852 & 0.0013 & $-$0&0012 & 50 \\
79 & 56613.8396 & 0.0036 & $-$0&0038 & 50 \\
85 & 56614.1919 & 0.0012 & 0&0062 & 31 \\
86 & 56614.2438 & 0.0066 & 0&0011 & 25 \\
87 & 56614.2953 & 0.0023 & $-$0&0044 & 24 \\
90 & 56614.4781 & 0.0025 & 0&0073 & 27 \\
92 & 56614.5781 & 0.0034 & $-$0&0067 & 8 \\
93 & 56614.6390 & 0.0010 & $-$0&0030 & 40 \\
94 & 56614.6940 & 0.0010 & $-$0&0049 & 73 \\
95 & 56614.7516 & 0.0015 & $-$0&0044 & 63 \\
101 & 56615.0939 & 0.0027 & $-$0&0043 & 83 \\
109 & 56615.5609 & 0.0026 & 0&0064 & 27 \\
110 & 56615.6153 & 0.0015 & 0&0038 & 63 \\
111 & 56615.6708 & 0.0014 & 0&0022 & 65 \\
112 & 56615.7257 & 0.0025 & 0&0001 & 31 \\
113 & 56615.7905 & 0.0035 & 0&0079 & 26 \\
121 & 56616.2415 & 0.0020 & 0&0026 & 17 \\
122 & 56616.2984 & 0.0012 & 0&0024 & 69 \\
123 & 56616.3527 & 0.0014 & $-$0&0002 & 85 \\
124 & 56616.4136 & 0.0012 & 0&0036 & 69 \\
125 & 56616.4687 & 0.0012 & 0&0016 & 37 \\
126 & 56616.5280 & 0.0010 & 0&0039 & 37 \\
139 & 56617.2709 & 0.0082 & 0&0054 & 19 \\
140 & 56617.3255 & 0.0009 & 0&0030 & 38 \\
141 & 56617.3837 & 0.0009 & 0&0041 & 47 \\
142 & 56617.4435 & 0.0012 & 0&0069 & 45 \\
143 & 56617.4995 & 0.0006 & 0&0058 & 47 \\
\hline
  \multicolumn{7}{l}{\commenta BJD$-$2400000.} \\
  \multicolumn{7}{l}{\commentb Against max $= 2456609.3376 + 0.057035 E$.} \\
  \multicolumn{7}{l}{\commentc Orbital phase.} \\
  \multicolumn{7}{l}{\commentd Number of points used to determine the maximum.} \\
\end{tabular}
\end{center}
\end{table}

\addtocounter{table}{-1}
\begin{table}
\caption{Superhump maxima of MASTER J005740 (2013) (continued)}
\begin{center}
\begin{tabular}{rp{50pt}p{30pt}r@{.}lcr}
\hline
$E$ & max\commenta & error & \multicolumn{2}{c}{$O-C$\commentb} & phase\commentc & $N$\commentd \\
\hline
144 & 56617.5561 & 0.0010 & 0&0053 & 33 \\
154 & 56618.1461 & 0.0043 & 0&0251 & 23 \\
158 & 56618.3666 & 0.0036 & 0&0174 & 24 \\
195 & 56620.4539 & 0.0076 & $-$0&0056 & 45 \\
197 & 56620.5633 & 0.0040 & $-$0&0103 & 34 \\
213 & 56621.4701 & 0.0011 & $-$0&0161 & 23 \\
233 & 56622.6231 & 0.0040 & $-$0&0038 & 26 \\
245 & 56623.3129 & 0.0029 & 0&0016 & 45 \\
246 & 56623.3660 & 0.0022 & $-$0&0023 & 53 \\
247 & 56623.4269 & 0.0020 & 0&0016 & 84 \\
248 & 56623.4778 & 0.0013 & $-$0&0046 & 65 \\
249 & 56623.5386 & 0.0025 & $-$0&0008 & 46 \\
264 & 56624.3887 & 0.0015 & $-$0&0063 & 46 \\
265 & 56624.4452 & 0.0017 & $-$0&0068 & 45 \\
\hline
  \multicolumn{7}{l}{\commenta BJD$-$2400000.} \\
  \multicolumn{7}{l}{\commentb Against max $= 2456609.3377 + 0.057035 E$.} \\
  \multicolumn{7}{l}{\commentc Orbital phase.} \\
  \multicolumn{7}{l}{\commentd Number of points used to determine the maximum.} \\
\end{tabular}
\end{center}
\end{table}

\subsection{MASTER OT J024847.86$+$501239.7}\label{obj:j024847}

\begin{figure}
  \begin{center}
    \FigureFile(88mm,70mm){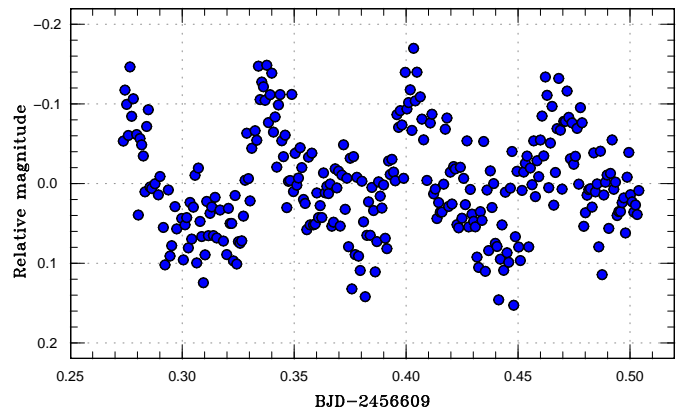}
  \end{center}
  \caption{Superhumps in MASTER J024847.}
  \label{fig:j0248shlc}
\end{figure}

   This object (hereafter MASTER J024847) was detected
as a transient by the MASTER network
on 2013 November 11 \citep{den13j0248atel5572}.
Subsequent observations immediately detected superhumps
(vsnet-alert 16619; figure \ref{fig:j0248shlc}).
A single-night observation yielded the following times of
superhump maxima: BJD 2456609.3402(7) ($E=62$),
2456609.4043(8) ($E=62$), 2456609.4683(12) ($E=65$).
A PDM analysis yielded a period of 0.0644(3)~d.

\subsection{MASTER OT J061335.30$+$395714.7}\label{obj:j061335}

\begin{figure}
  \begin{center}
    \FigureFile(88mm,110mm){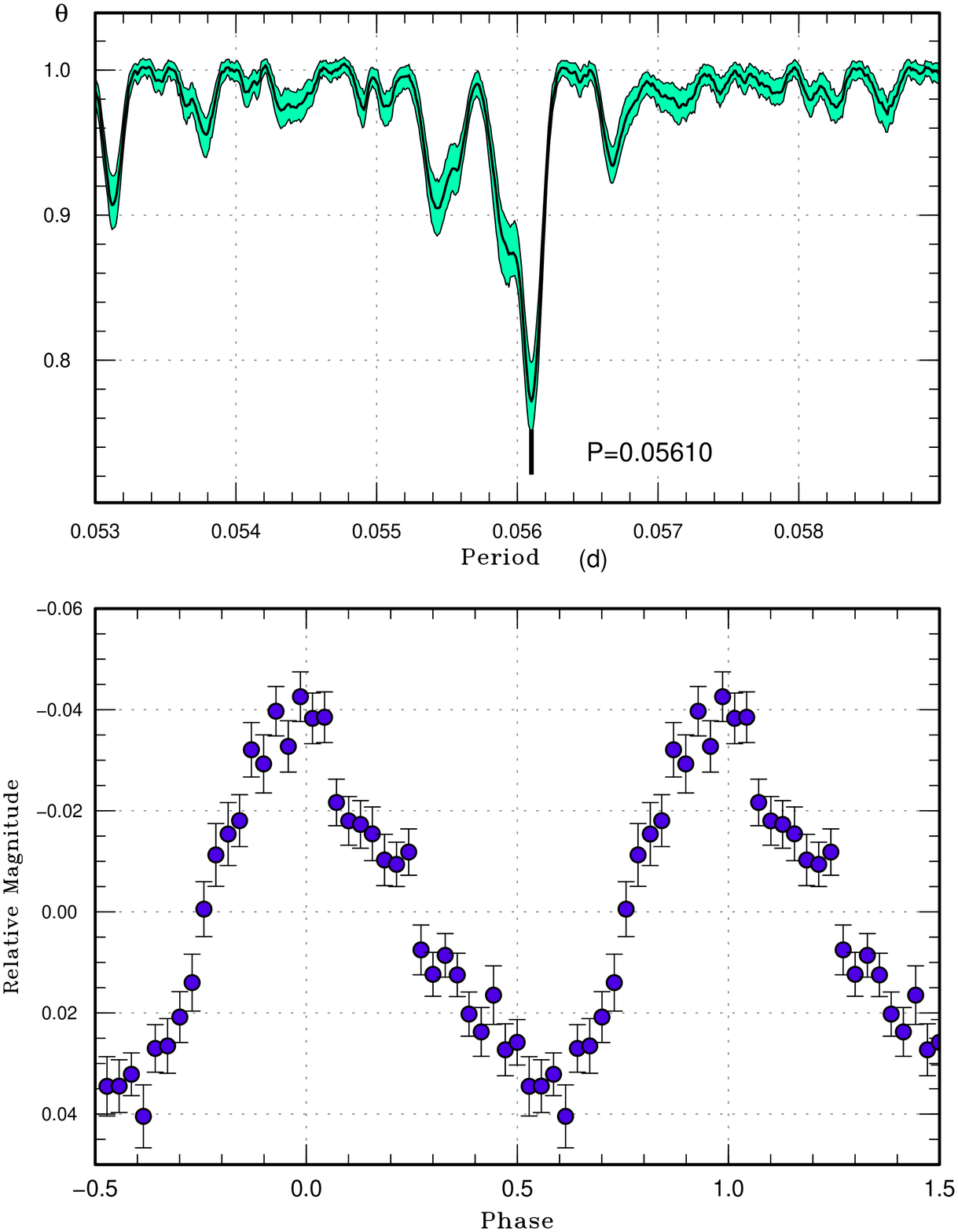}
  \end{center}
  \caption{Possible early superhumps in MASTER J061335 (2013).
     (Upper): PDM analysis.
     (Lower): Phase-averaged profile.}
  \label{fig:j0613shpdm}
\end{figure}

\begin{figure}
  \begin{center}
    \FigureFile(88mm,70mm){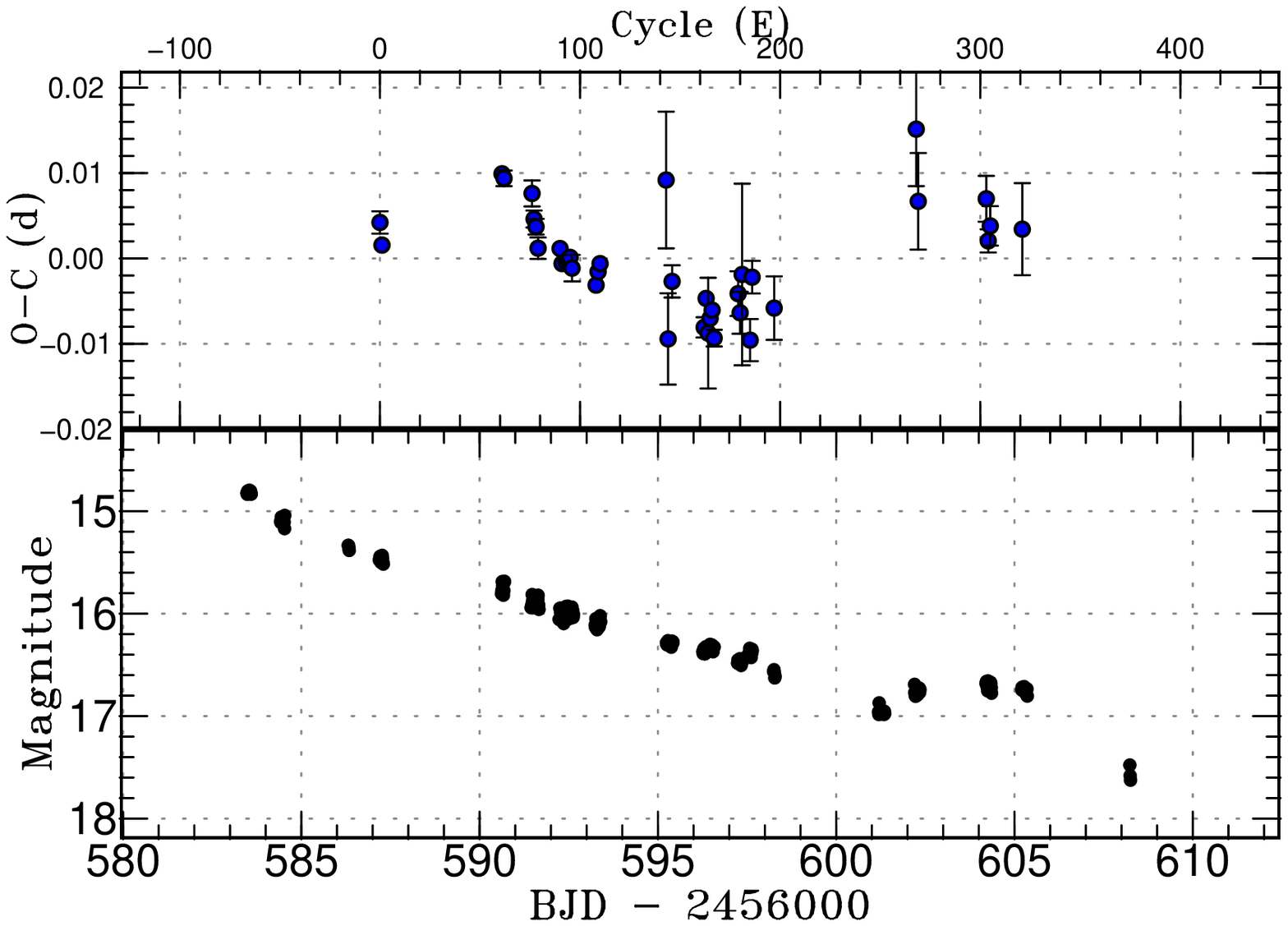}
  \end{center}
  \caption{$O-C$ diagram of superhumps in MASTER J061335 (2013).
     (Upper): $O-C$ diagram.  A period of 0.056114~d
     was used to draw this figure.
     (Lower): Light curve.  The observations were binned to 0.011~d.}
  \label{fig:j0613humpall}
\end{figure}

   This object (hereafter MASTER J061335) was detected
as a bright (14.2 mag) transient by the MASTER network
on 2013 October 15 \citep{vla13j0613atel5481}.
After a period without strong modulations,
growing superhumps were detected (vsnet-alert 16554, 16555,
16556, 16563, 16567; figure \ref{fig:j0613shpdm}).
The times of superhump maxima are
listed in table \ref{tab:j061335oc2013}.
There are well-defined stages A-B-C
(figure \ref{fig:j0613humpall}), although the period
of stage A superhumps was not determined.
Although early superhumps were potentially present during the first
two nights, we could not detect the period due to the
shortness of the observation.
There was significant brightening around the stage B-C
transition (figure \ref{fig:j0613humpall}).

\begin{table}
\caption{Superhump maxima of MASTER J061335 (2013)}\label{tab:j061335oc2013}
\begin{center}
\begin{tabular}{rp{55pt}p{40pt}r@{.}lr}
\hline
\multicolumn{1}{c}{$E$} & \multicolumn{1}{c}{max\commenta} & \multicolumn{1}{c}{error} & \multicolumn{2}{c}{$O-C$\commentb} & \multicolumn{1}{c}{$N$\commentc} \\
\hline
0 & 56587.2100 & 0.0013 & 0&0042 & 119 \\
1 & 56587.2635 & 0.0007 & 0&0016 & 167 \\
61 & 56590.6387 & 0.0007 & 0&0099 & 56 \\
62 & 56590.6942 & 0.0009 & 0&0094 & 37 \\
76 & 56591.4781 & 0.0015 & 0&0076 & 32 \\
77 & 56591.5312 & 0.0010 & 0&0047 & 31 \\
78 & 56591.5864 & 0.0009 & 0&0038 & 30 \\
79 & 56591.6400 & 0.0013 & 0&0012 & 21 \\
90 & 56592.2572 & 0.0008 & 0&0012 & 41 \\
91 & 56592.3116 & 0.0008 & $-$0&0006 & 41 \\
93 & 56592.4240 & 0.0007 & $-$0&0004 & 35 \\
94 & 56592.4804 & 0.0007 & $-$0&0001 & 90 \\
95 & 56592.5368 & 0.0007 & 0&0002 & 74 \\
96 & 56592.5916 & 0.0015 & $-$0&0011 & 30 \\
108 & 56593.2630 & 0.0009 & $-$0&0031 & 46 \\
109 & 56593.3207 & 0.0006 & $-$0&0015 & 101 \\
110 & 56593.3777 & 0.0004 & $-$0&0006 & 51 \\
143 & 56595.2393 & 0.0080 & 0&0093 & 36 \\
144 & 56595.2768 & 0.0053 & $-$0&0094 & 33 \\
146 & 56595.3958 & 0.0019 & $-$0&0026 & 31 \\
162 & 56596.2882 & 0.0012 & $-$0&0080 & 58 \\
163 & 56596.3477 & 0.0008 & $-$0&0046 & 95 \\
164 & 56596.3997 & 0.0065 & $-$0&0087 & 27 \\
165 & 56596.4576 & 0.0006 & $-$0&0069 & 67 \\
166 & 56596.5147 & 0.0005 & $-$0&0060 & 77 \\
167 & 56596.5675 & 0.0010 & $-$0&0092 & 39 \\
179 & 56597.2461 & 0.0026 & $-$0&0040 & 35 \\
180 & 56597.3000 & 0.0025 & $-$0&0063 & 41 \\
181 & 56597.3606 & 0.0106 & $-$0&0018 & 20 \\
185 & 56597.5773 & 0.0025 & $-$0&0095 & 23 \\
186 & 56597.6408 & 0.0019 & $-$0&0021 & 19 \\
197 & 56598.2544 & 0.0037 & $-$0&0057 & 50 \\
268 & 56602.2595 & 0.0067 & 0&0153 & 41 \\
269 & 56602.3072 & 0.0057 & 0&0068 & 40 \\
303 & 56604.2153 & 0.0027 & 0&0071 & 36 \\
304 & 56604.2665 & 0.0014 & 0&0022 & 41 \\
305 & 56604.3244 & 0.0023 & 0&0040 & 41 \\
321 & 56605.2218 & 0.0054 & 0&0036 & 44 \\
\hline
  \multicolumn{6}{l}{\commenta BJD$-$2400000.} \\
  \multicolumn{6}{l}{\commentb Against max $= 2456587.2058 + 0.056114 E$.} \\
  \multicolumn{6}{l}{\commentc Number of points used to determine the maximum.} \\
\end{tabular}
\end{center}
\end{table}

\subsection{MASTER OT J073208.11$+$064149.5}\label{obj:j073208}

   This object (hereafter MASTER J073208) was detected
as a large-amplitude ($\sim$7 mag) transient by the MASTER network
on 2013 December 29 \citep{bal13j0732atel5708}.
The object indeed showed short-period superhumps
(vsnet-alert 16747, 16756).
The times of superhump maxima are listed in table
\ref{tab:j073208oc2013}.
Although a PDM analysis favors a period of
0.05878(2)~d, a shorter alias of 0.05722(2)~d
is not excluded.

\begin{figure}
  \begin{center}
    \FigureFile(88mm,110mm){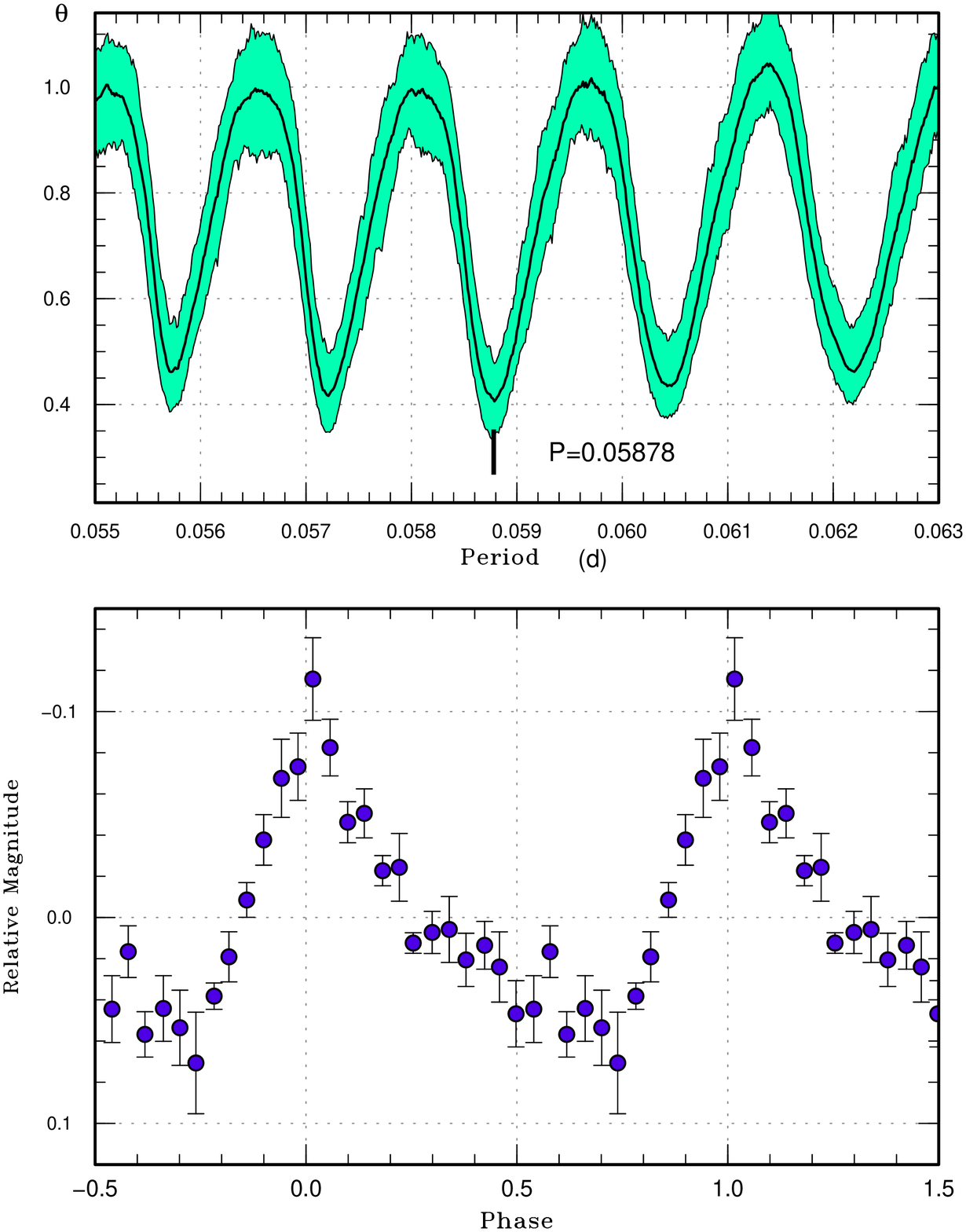}
  \end{center}
  \caption{(Early?) superhumps in MASTER J073208 (2013).
     (Upper): PDM analysis.
     (Lower): Phase-averaged profile.}
  \label{fig:j073208shpdm}
\end{figure}

\begin{table}
\caption{Superhump maxima of MASTER J073208 (2013)}\label{tab:j073208oc2013}
\begin{center}
\begin{tabular}{rp{55pt}p{40pt}r@{.}lr}
\hline
\multicolumn{1}{c}{$E$} & \multicolumn{1}{c}{max\commenta} & \multicolumn{1}{c}{error} & \multicolumn{2}{c}{$O-C$\commentb} & \multicolumn{1}{c}{$N$\commentc} \\
\hline
0 & 56658.3918 & 0.0018 & $-$0&0028 & 15 \\
1 & 56658.4553 & 0.0006 & 0&0018 & 40 \\
2 & 56658.5134 & 0.0007 & 0&0010 & 23 \\
37 & 56660.5747 & 0.0007 & 0&0031 & 20 \\
38 & 56660.6273 & 0.0011 & $-$0&0031 & 42 \\
\hline
  \multicolumn{6}{l}{\commenta BJD$-$2400000.} \\
  \multicolumn{6}{l}{\commentb Against max $= 2456658.3947 + 0.058836 E$.} \\
  \multicolumn{6}{l}{\commentc Number of points used to determine the maximum.} \\
\end{tabular}
\end{center}
\end{table}

\subsection{MASTER OT J095018.04$-$063921.9}\label{obj:j095018}

\begin{figure}
  \begin{center}
    \FigureFile(88mm,70mm){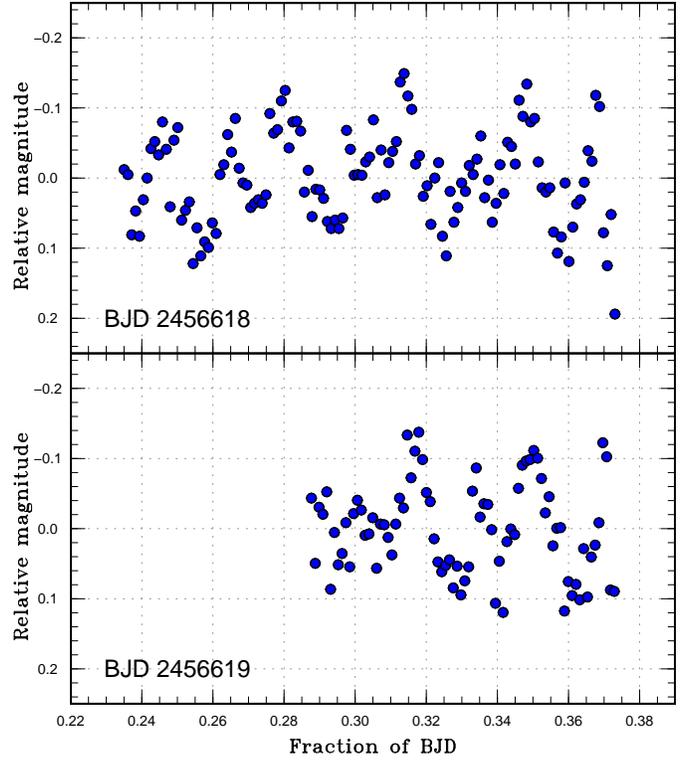}
  \end{center}
  \caption{Example of (early?) superhumps in MASTER J095018
  on two nights.}
  \label{fig:j0950shlc}
\end{figure}

\begin{figure}
  \begin{center}
    \FigureFile(88mm,110mm){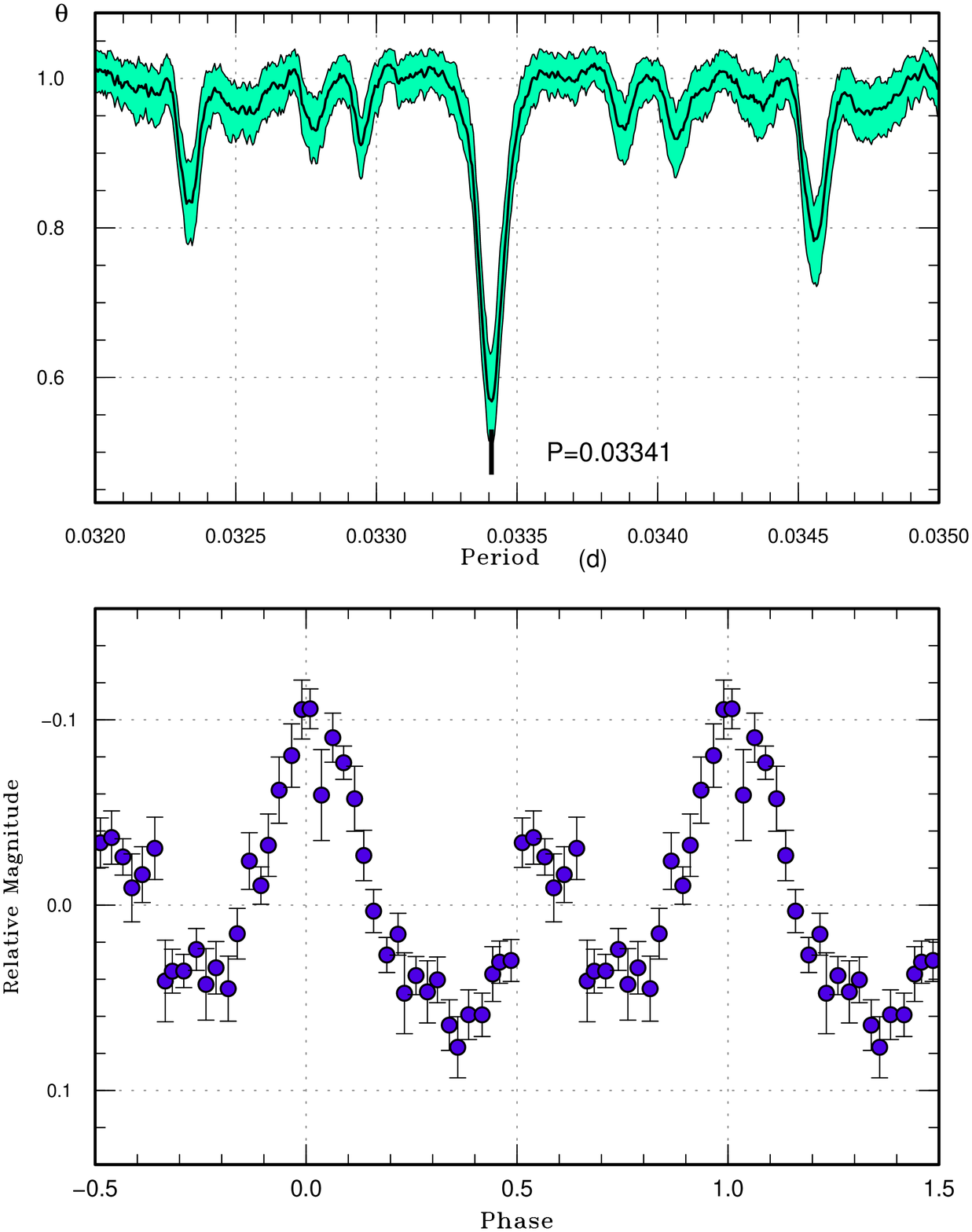}
  \end{center}
  \caption{(Early?) superhumps in MASTER J095018 (2013).
     (Upper): PDM analysis.
     (Lower): Phase-averaged profile.}
  \label{fig:j0950shpdm}
\end{figure}

   This object (hereafter MASTER J095018) was detected
as a 14.1-mag transient by the MASTER network
on 2013 November 11 \citep{ruf13j0950atel5587}.
Subsequent observations immediately detected superhumps
(vsnet-alert 16631, 16659; figure \ref{fig:j0950shlc}).
Although a period of 0.06681(3)~d was initially reported,
a re-analysis of the data clarified that this is 
the double value of the true variation (see also
figure \ref{fig:j0950shlc}).  A PDM analysis of the entire data
yielded a period of 0.033409(4)~d (figure \ref{fig:j0950shpdm}).
The profile, however, is unlike that of ordinary superhumps,
but resembles that of early superhumps with double maxima
(cf. \cite{kat02wzsgeESH}).  An $O-C$ analysis did not
show significant period variation.

   The period suggests a system with a compact, evolved 
secondary like SBS 1108$+$574 (\cite{Pdot4}; \cite{lit13sbs1108};
\cite{car13sbs1108}).  If the present variation is indeed
early superhumps, the present observation becomes the first
to detect early superhumps in such systems.
We leave this possibility for future observations
since we could not follow the later stage, when ordinary
superhumps were expected.  On the last two nights
(November 24 and 25), the double-wave modulation became
less apparent and waves with a period of $\sim$0.08~d
seemed to appear.  We were not, however, confident in
the presence of this periodicity due to the limited
signal-to-ratio for this faint object.

\subsection{MASTER OT J141143.46$+$262051.5}\label{obj:j141143}

   This object (hereafter MASTER J141143) was detected
as a transient (15.4 mag) by the MASTER network
on 2013 February 13 \citep{shu13j1141atel4814}.
The object was again detected in outburst on 2014
January 30.  The outburst was caught in the early
stage (vsnet-alert 16852).
The only available observation on February 5
showed superhumps with amplitudes of 0.15 mag
(vsnet-alert 16881).
The period was determined by the PDM method to be 0.064(1)~d.
The times of superhump maxima were BJD 2456693.5806(9) ($N=35$)
and 2456693.6466(9) ($N=33$).

\begin{figure}
  \begin{center}
    \FigureFile(88mm,70mm){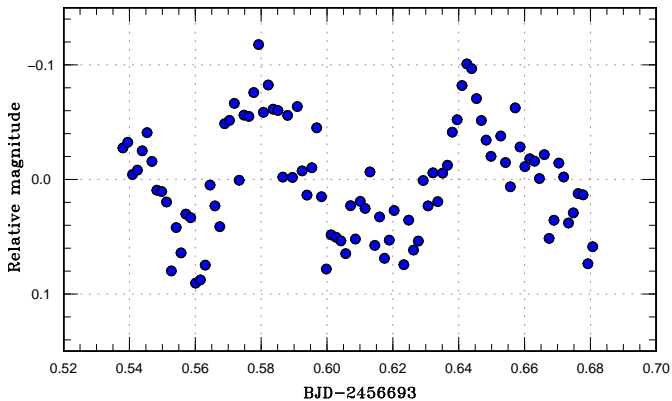}
  \end{center}
  \caption{Superhumps in MASTER J141143 on 2014 February 5.}
  \label{fig:j1411lc}
\end{figure}

\subsection{MASTER OT J162323.48$+$782603.3}\label{obj:j162323}

\begin{figure}
  \begin{center}
    \FigureFile(88mm,110mm){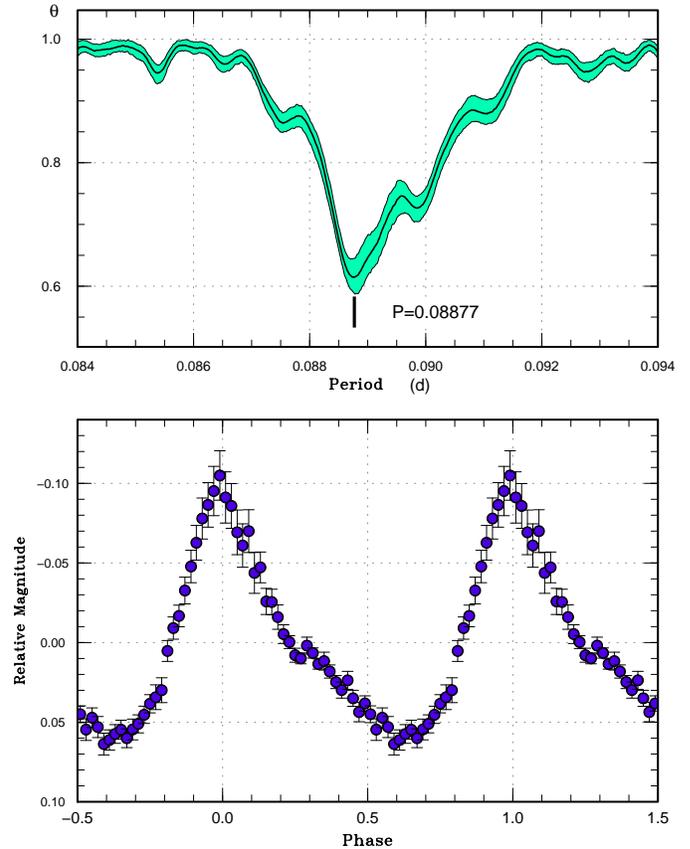}
  \end{center}
  \caption{Superhumps in MASTER J162323 during the superoutburst
     plateau (2013).
     (Upper): PDM analysis.
     (Lower): Phase-averaged profile.}
  \label{fig:j1623shpdm}
\end{figure}

   This object (hereafter MASTER J162323) was detected
as a bright (13.0-mag) transient by the MASTER network
on 2013 December 9 \citep{den13j1623atel5643}.
There were a number of past ROSAT chance observations,
recording a relatively soft source with variable
intensity (vsnet-alert 16698).
Subsequent observations immediately detected modulation
(vsnet-alert 16703) which were followed by growing superhumps
(vsnet-alert 16706, 16716, 16717, 16723; figure
\ref{fig:j1623shpdm}).
The times of superhump maxima are listed in table
\ref{tab:j162323oc2013}.  The were clear stages A and B.
There was no variation of the superhump period
around the rapid fading.  There was an apparent decrease
in the period 5~d after the fading (around $E=200$).
We interpret the superhumps up to $E=192$ to be
stage B superhumps and listed the period in table
\ref{tab:perlist}.  The resultant $P_{\rm dot}$ of
$+3.9(9) \times 10^{-5}$ for stage B again makes
another example of a positive $P_{\rm dot}$ in a
long-$P_{\rm orb}$ system [such examples include GX Cas
\citep{Pdot3}, V1239 Her, OT J145921.8$+$354806,
OT J214738.4$+$244553 \citep{Pdot4}, V444 Peg,
CSS J203937.7$-$042907, MASTER OT J212624.16$+$253827.2
\citep{Pdot5}].

   The quiescent SDSS colors suggest an orbital period
of 0.072~d based on \citet{kat12DNSDSS}.  This relatively
long orbital period for an SU UMa-type dwarf nova
is in agreement with the present observation.
The object was twice detected in outburst in 2012
by MASTER data \citep{den13j1623atel5643}.  The supercycle
appears to be less than 1~yr.  Being bright and
frequently outbursting, future observations including
the determination of the orbital period, will be
promising.

\begin{table}
\caption{Superhump maxima of MASTER J162323 (2013)}\label{tab:j162323oc2013}
\begin{center}
\begin{tabular}{rp{55pt}p{40pt}r@{.}lr}
\hline
\multicolumn{1}{c}{$E$} & \multicolumn{1}{c}{max\commenta} & \multicolumn{1}{c}{error} & \multicolumn{2}{c}{$O-C$\commentb} & \multicolumn{1}{c}{$N$\commentc} \\
\hline
0 & 56637.5690 & 0.0011 & $-$0&0542 & 94 \\
1 & 56637.6655 & 0.0020 & $-$0&0464 & 86 \\
2 & 56637.7684 & 0.0012 & $-$0&0323 & 46 \\
19 & 56639.3104 & 0.0006 & 0&0021 & 40 \\
20 & 56639.3987 & 0.0002 & 0&0017 & 177 \\
21 & 56639.4874 & 0.0003 & 0&0017 & 157 \\
22 & 56639.5749 & 0.0008 & 0&0005 & 12 \\
23 & 56639.6642 & 0.0007 & 0&0011 & 12 \\
30 & 56640.2879 & 0.0010 & 0&0040 & 17 \\
31 & 56640.3795 & 0.0002 & 0&0069 & 159 \\
32 & 56640.4681 & 0.0003 & 0&0068 & 165 \\
33 & 56640.5578 & 0.0003 & 0&0078 & 112 \\
35 & 56640.7456 & 0.0014 & 0&0182 & 9 \\
42 & 56641.3661 & 0.0028 & 0&0179 & 24 \\
44 & 56641.5352 & 0.0003 & 0&0096 & 66 \\
45 & 56641.6238 & 0.0008 & 0&0095 & 25 \\
46 & 56641.7088 & 0.0009 & 0&0059 & 12 \\
56 & 56642.6015 & 0.0007 & 0&0116 & 56 \\
57 & 56642.6897 & 0.0015 & 0&0112 & 16 \\
67 & 56643.5716 & 0.0012 & 0&0062 & 10 \\
68 & 56643.6643 & 0.0019 & 0&0102 & 15 \\
79 & 56644.6320 & 0.0031 & 0&0023 & 15 \\
98 & 56646.3168 & 0.0005 & 0&0020 & 73 \\
99 & 56646.4037 & 0.0005 & 0&0003 & 112 \\
101 & 56646.5840 & 0.0018 & 0&0032 & 12 \\
102 & 56646.6708 & 0.0010 & 0&0012 & 15 \\
135 & 56649.5986 & 0.0057 & 0&0024 & 16 \\
136 & 56649.6894 & 0.0021 & 0&0044 & 16 \\
143 & 56650.3106 & 0.0012 & 0&0048 & 44 \\
153 & 56651.2012 & 0.0055 & 0&0086 & 28 \\
154 & 56651.2825 & 0.0012 & 0&0012 & 50 \\
155 & 56651.3732 & 0.0011 & 0&0031 & 50 \\
156 & 56651.4641 & 0.0063 & 0&0053 & 13 \\
180 & 56653.5948 & 0.0027 & 0&0075 & 13 \\
181 & 56653.6807 & 0.0100 & 0&0047 & 13 \\
187 & 56654.2236 & 0.0044 & 0&0155 & 18 \\
188 & 56654.3132 & 0.0028 & 0&0164 & 13 \\
192 & 56654.6519 & 0.0030 & 0&0004 & 21 \\
223 & 56657.3771 & 0.0015 & $-$0&0238 & 30 \\
224 & 56657.4603 & 0.0018 & $-$0&0293 & 28 \\
244 & 56659.2330 & 0.0013 & $-$0&0304 & 14 \\
\hline
  \multicolumn{6}{l}{\commenta BJD$-$2400000.} \\
  \multicolumn{6}{l}{\commentb Against max $= 2456637.6232 + 0.088689 E$.} \\
  \multicolumn{6}{l}{\commentc Number of points used to determine the maximum.} \\
\end{tabular}
\end{center}
\end{table}

\subsection{MASTER OT J234843.23$+$250250.4}\label{obj:j234843}

   This object (hereafter MASTER J234843) was detected
as a 14.4-mag transient by the MASTER network
on 2013 October 29 \citep{shu13j2348atel5526}.
The object has a faint, blue ($g=20.2$, $g-r=-0.1$)
SDSS counterpart.  Five previous outbursts were recorded
by CRTS.

\begin{figure}
  \begin{center}
    \FigureFile(88mm,70mm){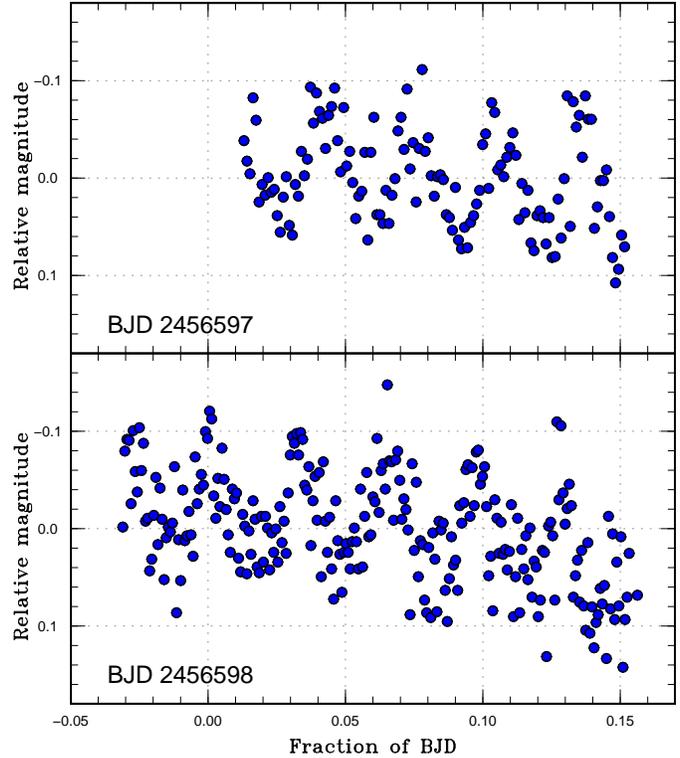}
  \end{center}
  \caption{Example of superhumps in MASTER J234843
  on two nights.}
  \label{fig:j2348shlc}
\end{figure}

\begin{figure}
  \begin{center}
    \FigureFile(88mm,110mm){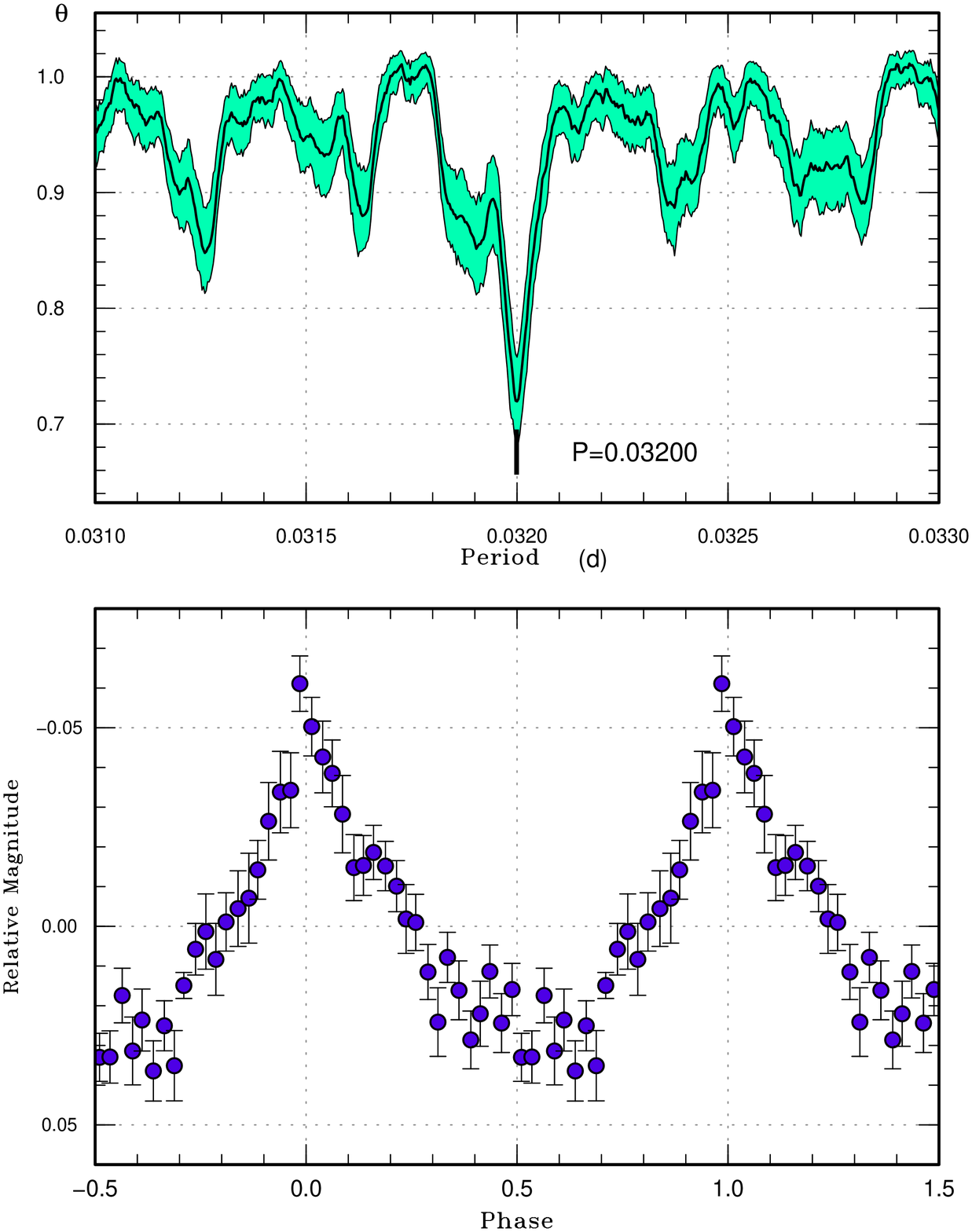}
  \end{center}
  \caption{Superhumps in MASTER J234843 during the superoutburst
     plateau (2013).
     (Upper): PDM analysis.
     (Lower): Phase-averaged profile.}
  \label{fig:j2348shpdm}
\end{figure}

\begin{figure}
  \begin{center}
    \FigureFile(88mm,70mm){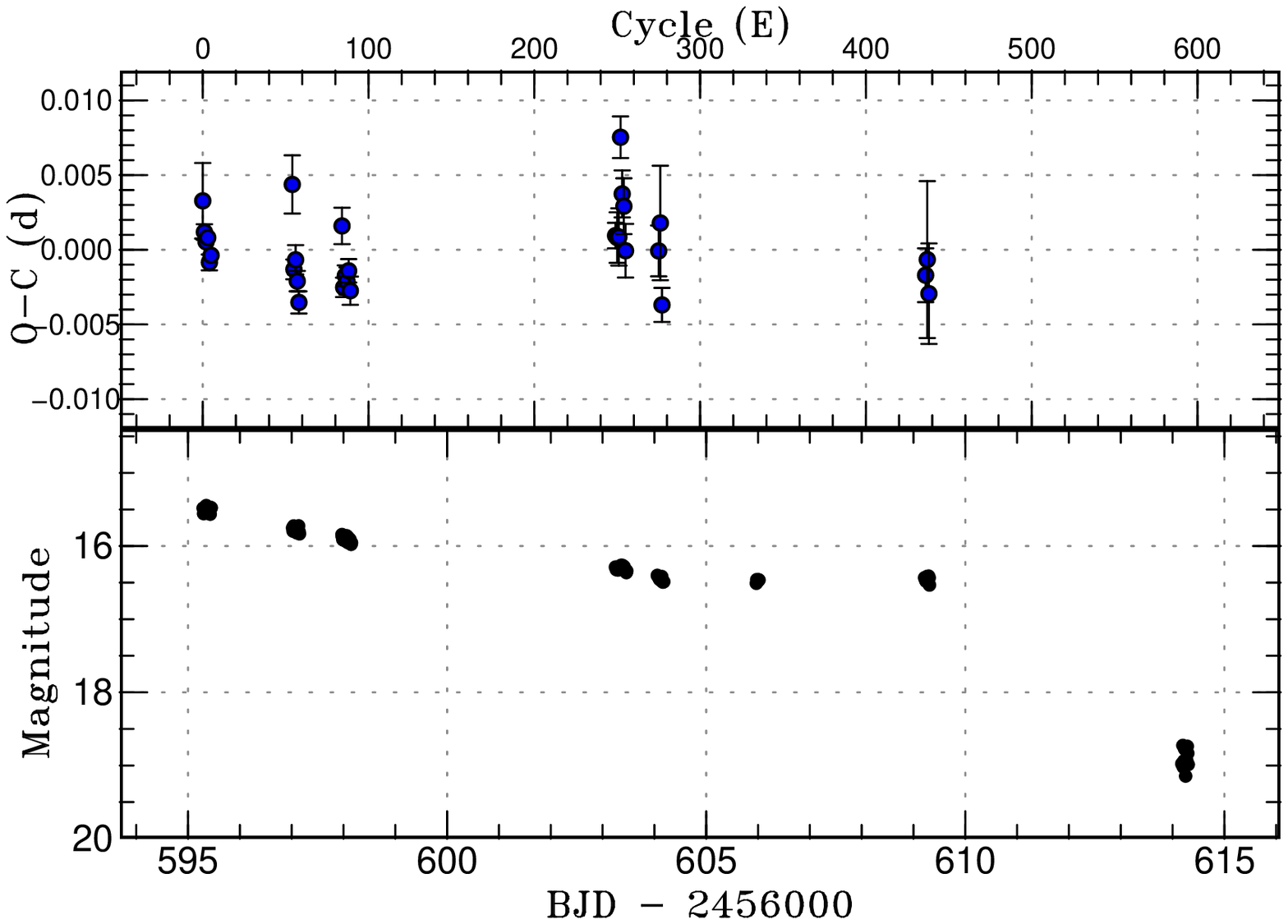}
  \end{center}
  \caption{$O-C$ diagram of superhumps in MASTER J234843 (2013).
     (Upper): $O-C$ diagram.  A period of 0.031997~d
     was used to draw this figure.
     (Lower): Light curve.  The observations were binned to 0.010~d.}
  \label{fig:j2348humpall}
\end{figure}

   Subsequent observations detected modulations
resembling early superhumps (vsnet-alert 16577, 16578).
Later observation, however, indicated that the period
was the half of what was initially suggested
(figure \ref{fig:j2348shlc}), and
the object was recognized as a CV below the period
minimum (vsnet-alert 16608; the superhump profile
is shown in figure \ref{fig:j2348shpdm}).
The times of superhump maxima based on this interpretation
are listed in table \ref{tab:j234843oc2013}.
Although the data were not sufficient, the $O-C$ values
suggest that there was a stage B-C transition
around $E=252$ as in SBS 1108$+$574, a similar 
ultrashort-$P_{\rm orb}$ dwarf nova \citep{Pdot4}.
The outburst light curve and $O-C$ diagram are presented
in figure \ref{fig:j2348humpall}.

   Although spectroscopic observation is needed to
see whether this object is hydrogen-rich or not,
we suspect that this object is a binary containing
hydrogen in the secondary with an evolved core
rather than a hydrogen-depleted AM CVn-type binary,
since the known AM CVn-type objects with this
$P_{\rm orb}$ do not show outbursts 
(cf. \cite{sol10amcvnreview}; \cite{ram12amcvnLC}).
Further detailed observations will clarify
the nature of this object.

\begin{table}
\caption{Superhump maxima of MASTER J234843 (2013)}\label{tab:j234843oc2013}
\begin{center}
\begin{tabular}{rp{55pt}p{40pt}r@{.}lr}
\hline
\multicolumn{1}{c}{$E$} & \multicolumn{1}{c}{max\commenta} & \multicolumn{1}{c}{error} & \multicolumn{2}{c}{$O-C$\commentb} & \multicolumn{1}{c}{$N$\commentc} \\
\hline
0 & 56595.2868 & 0.0025 & 0&0032 & 13 \\
1 & 56595.3167 & 0.0005 & 0&0011 & 33 \\
2 & 56595.3480 & 0.0004 & 0&0005 & 26 \\
3 & 56595.3803 & 0.0004 & 0&0008 & 34 \\
4 & 56595.4106 & 0.0005 & $-$0&0009 & 29 \\
5 & 56595.4431 & 0.0005 & $-$0&0004 & 26 \\
54 & 56597.0157 & 0.0019 & 0&0043 & 15 \\
55 & 56597.0420 & 0.0007 & $-$0&0014 & 23 \\
56 & 56597.0747 & 0.0010 & $-$0&0007 & 23 \\
57 & 56597.1052 & 0.0007 & $-$0&0021 & 23 \\
58 & 56597.1358 & 0.0007 & $-$0&0036 & 23 \\
84 & 56597.9728 & 0.0012 & 0&0015 & 22 \\
85 & 56598.0007 & 0.0006 & $-$0&0026 & 34 \\
86 & 56598.0335 & 0.0007 & $-$0&0018 & 34 \\
87 & 56598.0651 & 0.0009 & $-$0&0021 & 34 \\
88 & 56598.0978 & 0.0008 & $-$0&0015 & 35 \\
89 & 56598.1285 & 0.0009 & $-$0&0028 & 34 \\
249 & 56603.2517 & 0.0009 & 0&0009 & 22 \\
250 & 56603.2836 & 0.0017 & 0&0007 & 28 \\
251 & 56603.3156 & 0.0019 & 0&0008 & 28 \\
252 & 56603.3543 & 0.0014 & 0&0075 & 19 \\
253 & 56603.3825 & 0.0016 & 0&0037 & 26 \\
254 & 56603.4137 & 0.0019 & 0&0028 & 28 \\
255 & 56603.4427 & 0.0018 & $-$0&0001 & 28 \\
275 & 56604.0826 & 0.0017 & $-$0&0002 & 23 \\
276 & 56604.1165 & 0.0038 & 0&0017 & 22 \\
277 & 56604.1430 & 0.0011 & $-$0&0038 & 24 \\
436 & 56609.2325 & 0.0018 & $-$0&0018 & 15 \\
437 & 56609.2655 & 0.0053 & $-$0&0008 & 17 \\
438 & 56609.2952 & 0.0034 & $-$0&0030 & 16 \\
\hline
  \multicolumn{6}{l}{\commenta BJD$-$2400000.} \\
  \multicolumn{6}{l}{\commentb Against max $= 2456595.2835 + 0.031997 E$.} \\
  \multicolumn{6}{l}{\commentc Number of points used to determine the maximum.} \\
\end{tabular}
\end{center}
\end{table}

\subsection{OT J013741.1$+$220312}\label{obj:j013741}

\begin{figure}
  \begin{center}
    \FigureFile(88mm,110mm){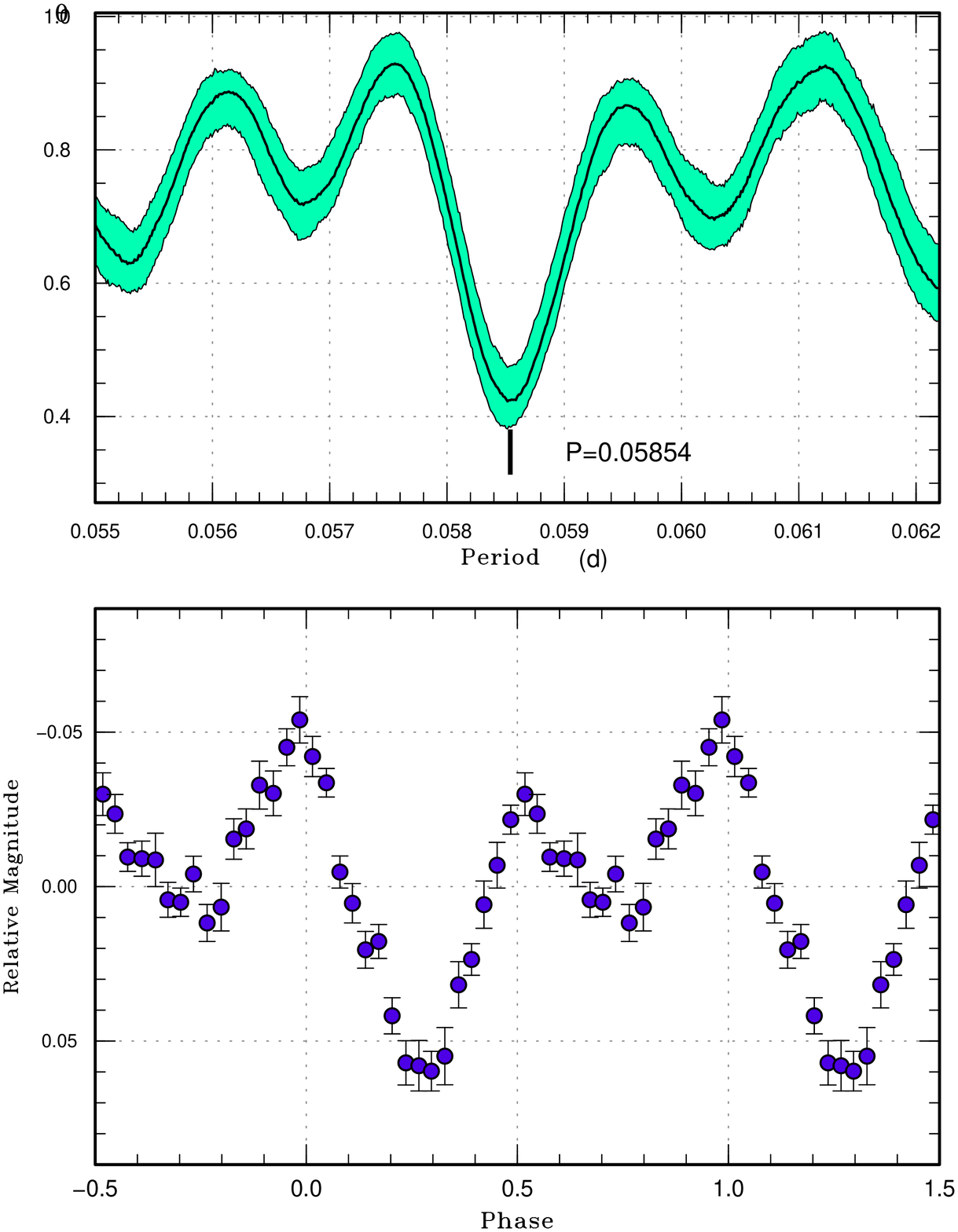}
  \end{center}
  \caption{Early superhumps in OT J013741 (2013).
     (Upper): PDM analysis.
     (Lower): Phase-averaged profile.}
  \label{fig:j013741eshpdm}
\end{figure}

   This object was detected as a transient of unknown type by CRTS
(=CSS140104:013741$+$220312, hereafter OT J013741)
on 2014 January 4.  There was no previous outburst detection
in CRTS.  Subsequent observations detected double-wave
early superhumps with a mean period of 0.05854(2)~d
(vsnet-alert 16765, 16769; figure \ref{fig:j013741eshpdm}).
Only the stage of early superhumps was observed.

\subsection{OT J210016.0$-$024258}\label{obj:j210016}

   This object was detected as a transient by CRTS
(=CSS130905:210016$-$024258, hereafter OT J210016)
on 2013 September 5.  The quiescent counterpart has an SDSS
magnitude of $g=19.9$, implying a dwarf nova with a large
outburst amplitude (vsnet-alert 16347).

\begin{figure}
  \begin{center}
    \FigureFile(88mm,110mm){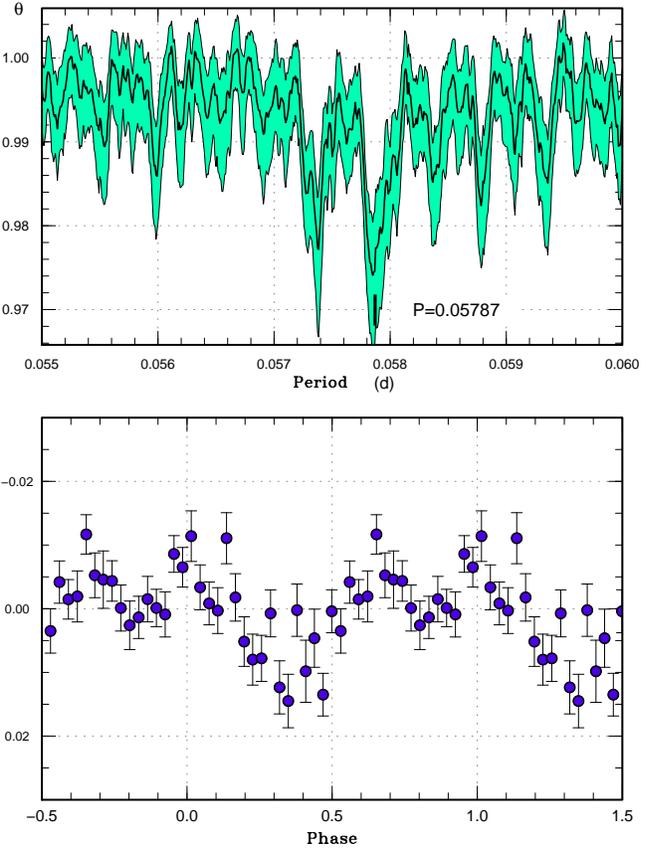}
  \end{center}
  \caption{Possible early superhumps in OT J210016 (2013).
     (Upper): PDM analysis.  The rejection rate for bootstraping
     was reduced to 0.3.
     (Lower): Phase-averaged profile.}
  \label{fig:j210016eshpdm}
\end{figure}

\begin{figure}
  \begin{center}
    \FigureFile(88mm,110mm){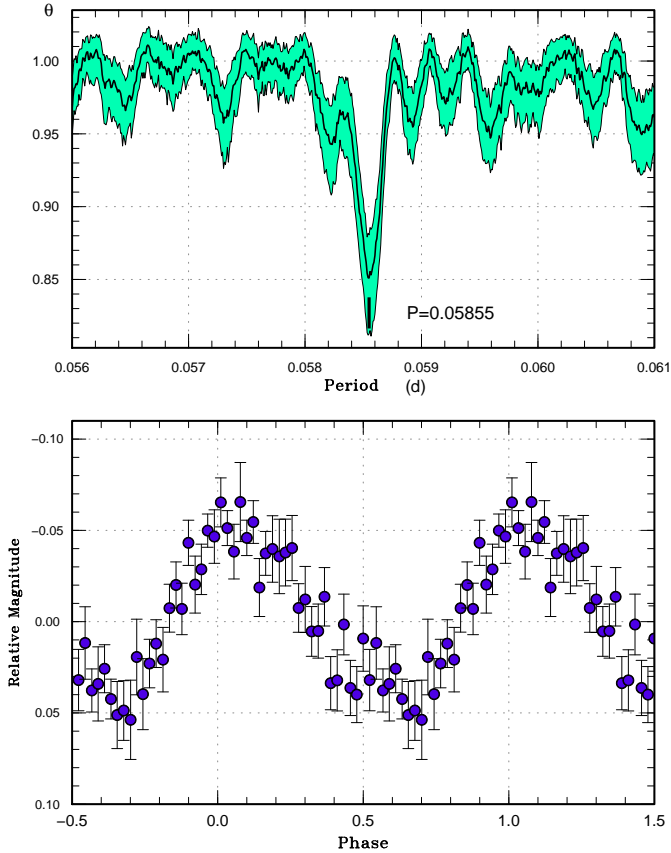}
  \end{center}
  \caption{Ordinary superhumps in OT J210016 (2013).
     (Upper): PDM analysis.
     (Lower): Phase-averaged profile.}
  \label{fig:j210016shpdm}
\end{figure}

   Low amplitude modulations resembling early superhumps
were recorded until September 12 (vsnet-alert 16362;
figure \ref{fig:j210016eshpdm}).
Starting from September 16, ordinary superhumps appeared
(vsnet-alert 16420, 16430; figure \ref{fig:j210016shpdm}).
The times of ordinary superhumps are listed in table
\ref{tab:j210016oc2013}.  Although the earliest epochs
may contain stage A superhumps, we could not convincingly
detect stage A.  Since $E=0$ and $E=1$ were likely obtained
during the growing stage of superhumps, we excluded these
epochs when determining $P_{\rm dot}$ of stage B superhumps.

   The object rapidly faded from the superoutburst plateau
on September 29.  It showed a post-superoutburst rebrightening
on October 6 (around 17.4 mag, all magnitudes during this
rebrightening were unfiltered CCD magnitudes; vsnet-alert 16525).
The snapshot CCD observations indicate that this object was 
already bright bright on October 1 (16.86 mag) and October 4
(17.55 mag).  These observations indicate that the faint
state following the superoutburst plateau lasted less than 2~d.
This fading was probably a ``dip''-type between the main
superoutburst and long rebrightening as in the WZ Sge-type
dwarf nova AL Com (1995, cf. \cite{nog97alcom}).
The WZ Sge-type nature is supported by the likely presence of 
early superhumps, the large outburst amplitude and the lack of 
past outburst detections in the CRTS data.

\begin{table}
\caption{Superhump maxima of OT J210016 (2013)}\label{tab:j210016oc2013}
\begin{center}
\begin{tabular}{rp{55pt}p{40pt}r@{.}lr}
\hline
\multicolumn{1}{c}{$E$} & \multicolumn{1}{c}{max\commenta} & \multicolumn{1}{c}{error} & \multicolumn{2}{c}{$O-C$\commentb} & \multicolumn{1}{c}{$N$\commentc} \\
\hline
0 & 56552.0076 & 0.0014 & $-$0&0070 & 40 \\
1 & 56552.0764 & 0.0018 & 0&0033 & 61 \\
17 & 56553.0093 & 0.0007 & $-$0&0001 & 60 \\
18 & 56553.0705 & 0.0006 & 0&0027 & 61 \\
34 & 56554.0039 & 0.0014 & $-$0&0002 & 47 \\
35 & 56554.0678 & 0.0014 & 0&0052 & 47 \\
57 & 56555.3469 & 0.0013 & $-$0&0030 & 47 \\
58 & 56555.4057 & 0.0010 & $-$0&0027 & 46 \\
59 & 56555.4675 & 0.0015 & 0&0006 & 47 \\
60 & 56555.5296 & 0.0030 & 0&0042 & 17 \\
108 & 56558.3312 & 0.0109 & $-$0&0030 & 17 \\
109 & 56558.3908 & 0.0008 & $-$0&0019 & 44 \\
110 & 56558.4503 & 0.0010 & $-$0&0008 & 45 \\
160 & 56561.3796 & 0.0012 & 0&0027 & 31 \\
\hline
  \multicolumn{6}{l}{\commenta BJD$-$2400000.} \\
  \multicolumn{6}{l}{\commentb Against max $= 2456552.0146 + 0.058514 E$.} \\
  \multicolumn{6}{l}{\commentc Number of points used to determine the maximum.} \\
\end{tabular}
\end{center}
\end{table}

\subsection{PNV J19150199$+$0719471}\label{obj:j191501}

   This object (hereafter PNV J191501) was detected as
a possible nova of 10.8 mag on 2013 May 31.5974 UT
\citep{ita13j1915cbet3554}.\footnote{
See also
$<$http://www.cbat.eps.harvard.edu/unconf/\\
followups/J19150199+0719471.html$>$.
}
The object was detected at 9.8 mag on May 30.721 UT by T. Kojima.
The object was identified as an H$\alpha$ emission
object IPHAS J191502.09$+$071947.6 ($r$=18.503, \cite{IPHAS}),
and the color and large proper motion suggested
a dwarf nova rather than a classical nova (vsnet-alert 15768).
The object was then regarded as a good candidate for
a WZ Sge-type dwarf nova (vsnet-alert 15776).
Although subsequent observations detected small variations
suggestive of early superhumps (vsnet-alert 15778, 15785,
15788), the period was difficult to determine due to the
low amplitude.  We will deal with this issue later.

   In the meantime, low-resolution spectra confirmed
the dwarf nova-type classification (vsnet-alert 15779).
The spectrum showed double-peak H$\beta$ emission line
and a C\textsc{iii}/N\textsc{iii} emission lune,
suggesting that this object is a WZ Sge-type dwarf nova
with a moderate inclination (vsnet-alert 15782).
Further spectroscopic observation were also reported
(vsnet-alert 15787, 15800).  The latter spectrum
showed Balmer series in absorption with emission cores
in H$\alpha$ and H$\beta$.  \citet{nak13j1915atel5253}
also reported a UV spectrum taken by Swift satellite.

\begin{figure}
  \begin{center}
    \FigureFile(88mm,110mm){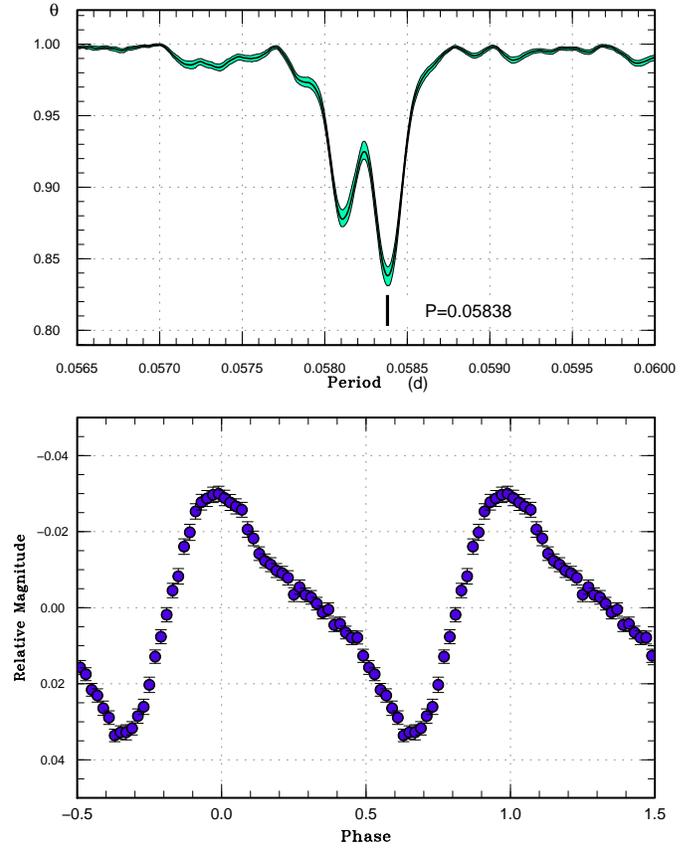}
  \end{center}
  \caption{Ordinary superhumps in PNV J191501 (2013).
     (Upper): PDM analysis.
     (Lower): Phase-averaged profile.}
  \label{fig:j1915shpdm}
\end{figure}

   Six days after the discovery, modulations suggesting
growing superhumps appeared (vsnet-alert 15815; the actual
variations could be detected two days earlier).
The times of superhump maxima during the superoutburst
plateau are listed in table \ref{tab:j191501oc2013}.
The profile is shown in figure \ref{fig:j1915shpdm}.
There were remarkably well-sampled stage B and C.
The existence of stage C in a WZ Sge-type dwarf nova
is rather exceptional.  Although stage A was detected,
the lack of observation for 1~d due to the bad weather
prevented us from measuring the period of stage A superhumps
from the $O-C$ analysis.  A PDM analysis for an interval
of BJD 2456449--2456452 yielded a period of 0.05883(6)~d
with an amplitude of 0.015 mag.

\begin{figure}
  \begin{center}
    \FigureFile(88mm,70mm){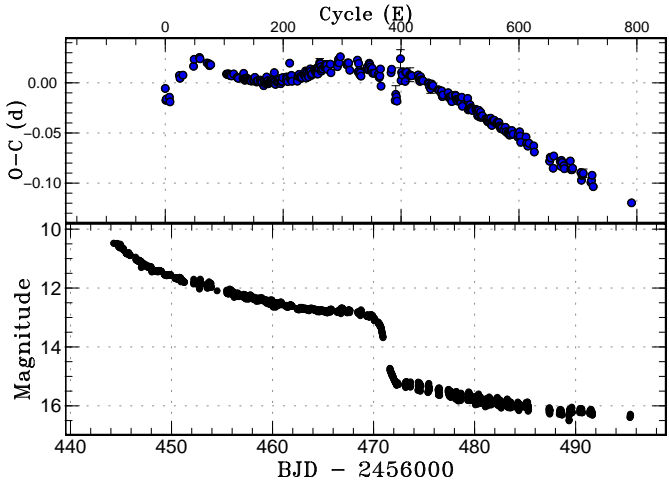}
  \end{center}
  \caption{$O-C$ diagram of superhumps in PNV J191501 (2013).
     (Upper): $O-C$ diagram.  A period of 0.05835~d
     was used to draw this figure.
     (Lower): Light curve.  The observations were binned to 0.010~d.}
  \label{fig:j1915humpall}
\end{figure}

   After rapid fading from the superoutburst, the superhump
signal became strong again.  The times of these post-superoutburst
superhumps are listed in table \ref{tab:j191501oc2013b}.
The data indicate that there was a phase $\sim$0.5 jump
between BJD 2456472.295 and BJD 2456472.688.
The combined $O-C$ diagram (figure \ref{fig:j1915humpall})
indicate, however, the superhumps in the post-superoutburst
stage are on a smooth extension of stage C superhumps.
The signal of reversed phases (likely corresponding to
``traditional'' late superhumps) appeared only briefly
near the rapid fading.

\begin{figure}
  \begin{center}
    \FigureFile(88mm,110mm){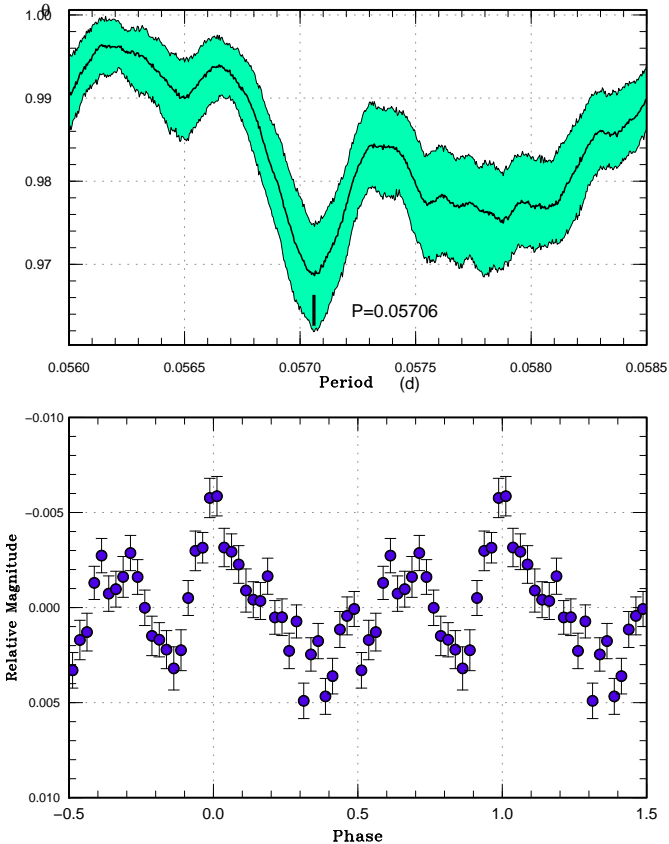}
  \end{center}
  \caption{Possible early superhumps in PNV J191501 (2013).
     (Upper): PDM analysis.
     (Lower): Phase-averaged profile.}
  \label{fig:j1915eshpdm}
\end{figure}

   The signal of early superhumps was very weak and
the period of 0.05706(2)~d is the only candidate period
in the region of the period of early superhumps
expected from the superhump period
(figure \ref{fig:j1915eshpdm}).
Although the fractional excess of stage A superhumps
based on the $O-C$ analysis corresponds to $q$=0.095(4),
this measurement suffers from uncertainties arising from 
the low amplitudes of both early superhumps and 
stage A superhumps.  Future determination of the orbital
period will be necessary to confirm this value.

\begin{figure}
  \begin{center}
    \FigureFile(88mm,100mm){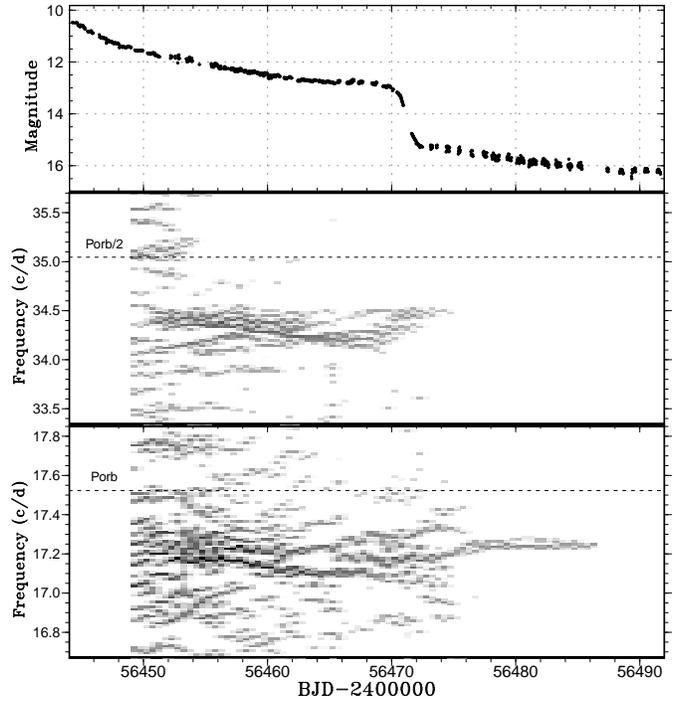}
  \end{center}
  \caption{Lasso analysis of PNV J191501 (2013).
  (Upper:) Light curve.  The data were binned to 0.02~d.
  (Middle:) First harmonics of the superhump and possible
  orbital signals.
  (Lower:) Fundamental of the superhump and the possible
  orbital signal.  The orbital signal was present only
  in the initial part of the outburst.
  The signal of (positive) superhumps with
  variable frequency was recorded during the superoutburst
  plateau and post-superoutburst stage.
  No indication of negative superhump was present.
  $\log \lambda=-8.0$ was used.
  The width of the sliding window and the time step used are
  10~d and 0.5~d, respectively.
  }
  \label{fig:j1915lasso}
\end{figure}

   A two dimensional Lasso analysis is presented in
figure \ref{fig:j1915lasso}.  The superhumps with
increasing frequencies (decreasing periods) after the
superoutburst are clearly seen.  The overall behavior
resembles that of an SU UMa-type dwarf nova rather than
an extreme WZ Sge-type dwarf nova (cf. WZ Sge in
\cite{Pdot5}) except for the possible presence of
early superhumps.

\begin{table}
\caption{Superhump maxima of PNV J191501 (2013)}\label{tab:j191501oc2013}
\begin{center}
\begin{tabular}{rp{55pt}p{40pt}r@{.}lr}
\hline
\multicolumn{1}{c}{$E$} & \multicolumn{1}{c}{max\commenta} & \multicolumn{1}{c}{error} & \multicolumn{2}{c}{$O-C$\commentb} & \multicolumn{1}{c}{$N$\commentc} \\
\hline
0 & 56449.3764 & 0.0010 & $-$0&0056 & 221 \\
1 & 56449.4229 & 0.0025 & $-$0&0175 & 222 \\
2 & 56449.4826 & 0.0010 & $-$0&0162 & 211 \\
7 & 56449.7761 & 0.0018 & $-$0&0146 & 64 \\
8 & 56449.8298 & 0.0012 & $-$0&0193 & 313 \\
24 & 56450.7898 & 0.0009 & 0&0064 & 124 \\
25 & 56450.8453 & 0.0019 & 0&0035 & 154 \\
30 & 56451.1402 & 0.0006 & 0&0065 & 784 \\
48 & 56452.1993 & 0.0005 & 0&0146 & 98 \\
49 & 56452.2646 & 0.0009 & 0&0215 & 126 \\
58 & 56452.7916 & 0.0005 & 0&0230 & 32 \\
59 & 56452.8486 & 0.0002 & 0&0217 & 51 \\
71 & 56453.5446 & 0.0002 & 0&0170 & 63 \\
74 & 56453.7190 & 0.0003 & 0&0162 & 38 \\
75 & 56453.7754 & 0.0003 & 0&0142 & 53 \\
76 & 56453.8347 & 0.0003 & 0&0151 & 89 \\
77 & 56453.8929 & 0.0002 & 0&0150 & 99 \\
104 & 56455.4592 & 0.0014 & 0&0048 & 45 \\
105 & 56455.5171 & 0.0002 & 0&0042 & 92 \\
106 & 56455.5754 & 0.0002 & 0&0042 & 92 \\
107 & 56455.6345 & 0.0003 & 0&0049 & 108 \\
109 & 56455.7499 & 0.0002 & 0&0035 & 97 \\
110 & 56455.8079 & 0.0002 & 0&0031 & 114 \\
111 & 56455.8665 & 0.0001 & 0&0033 & 113 \\
112 & 56455.9241 & 0.0002 & 0&0025 & 113 \\
116 & 56456.1595 & 0.0003 & 0&0044 & 651 \\
122 & 56456.5059 & 0.0003 & 0&0005 & 181 \\
123 & 56456.5645 & 0.0003 & 0&0006 & 101 \\
124 & 56456.6218 & 0.0003 & $-$0&0004 & 89 \\
132 & 56457.0887 & 0.0002 & $-$0&0007 & 238 \\
133 & 56457.1520 & 0.0030 & 0&0043 & 183 \\
134 & 56457.2050 & 0.0001 & $-$0&0011 & 586 \\
\hline
  \multicolumn{6}{l}{\commenta BJD$-$2400000.} \\
  \multicolumn{6}{l}{\commentb Against max $= 2456449.3820 + 0.058389 E$.} \\
  \multicolumn{6}{l}{\commentc Number of points used to determine the maximum.} \\
\end{tabular}
\end{center}
\end{table}

\addtocounter{table}{-1}
\begin{table}
\caption{Superhump maxima of PNV J191501 (2013) (continued)}
\begin{center}
\begin{tabular}{rp{55pt}p{40pt}r@{.}lr}
\hline
\multicolumn{1}{c}{$E$} & \multicolumn{1}{c}{max\commenta} & \multicolumn{1}{c}{error} & \multicolumn{2}{c}{$O-C$\commentb} & \multicolumn{1}{c}{$N$\commentc} \\
\hline
135 & 56457.2618 & 0.0001 & $-$0&0027 & 566 \\
136 & 56457.3217 & 0.0002 & $-$0&0012 & 97 \\
137 & 56457.3803 & 0.0003 & $-$0&0010 & 192 \\
138 & 56457.4391 & 0.0003 & $-$0&0006 & 221 \\
139 & 56457.4958 & 0.0002 & $-$0&0022 & 303 \\
140 & 56457.5550 & 0.0003 & $-$0&0015 & 121 \\
141 & 56457.6122 & 0.0002 & $-$0&0026 & 105 \\
143 & 56457.7287 & 0.0003 & $-$0&0029 & 132 \\
144 & 56457.7872 & 0.0002 & $-$0&0028 & 156 \\
145 & 56457.8450 & 0.0002 & $-$0&0034 & 163 \\
146 & 56457.9028 & 0.0002 & $-$0&0040 & 124 \\
150 & 56458.1365 & 0.0004 & $-$0&0039 & 234 \\
151 & 56458.1950 & 0.0004 & $-$0&0037 & 394 \\
152 & 56458.2567 & 0.0009 & $-$0&0004 & 77 \\
155 & 56458.4271 & 0.0003 & $-$0&0052 & 44 \\
156 & 56458.4866 & 0.0005 & $-$0&0041 & 33 \\
157 & 56458.5451 & 0.0003 & $-$0&0039 & 66 \\
161 & 56458.7791 & 0.0003 & $-$0&0035 & 137 \\
162 & 56458.8365 & 0.0003 & $-$0&0045 & 183 \\
163 & 56458.8933 & 0.0003 & $-$0&0061 & 180 \\
164 & 56458.9516 & 0.0003 & $-$0&0062 & 98 \\
167 & 56459.1236 & 0.0025 & $-$0&0094 & 83 \\
168 & 56459.1862 & 0.0002 & $-$0&0051 & 199 \\
169 & 56459.2426 & 0.0002 & $-$0&0071 & 204 \\
170 & 56459.3015 & 0.0002 & $-$0&0066 & 202 \\
171 & 56459.3604 & 0.0003 & $-$0&0061 & 201 \\
172 & 56459.4240 & 0.0007 & $-$0&0009 & 72 \\
173 & 56459.4773 & 0.0003 & $-$0&0059 & 172 \\
174 & 56459.5371 & 0.0003 & $-$0&0046 & 181 \\
175 & 56459.5924 & 0.0002 & $-$0&0077 & 183 \\
176 & 56459.6519 & 0.0009 & $-$0&0066 & 71 \\
177 & 56459.7101 & 0.0010 & $-$0&0067 & 15 \\
178 & 56459.7702 & 0.0006 & $-$0&0051 & 142 \\
\hline
  \multicolumn{6}{l}{\commenta BJD$-$2400000.} \\
  \multicolumn{6}{l}{\commentb Against max $= 2456449.3820 + 0.058389 E$.} \\
  \multicolumn{6}{l}{\commentc Number of points used to determine the maximum.} \\
\end{tabular}
\end{center}
\end{table}

\addtocounter{table}{-1}
\begin{table}
\caption{Superhump maxima of PNV J191501 (2013) (continued)}
\begin{center}
\begin{tabular}{rp{55pt}p{40pt}r@{.}lr}
\hline
\multicolumn{1}{c}{$E$} & \multicolumn{1}{c}{max\commenta} & \multicolumn{1}{c}{error} & \multicolumn{2}{c}{$O-C$\commentb} & \multicolumn{1}{c}{$N$\commentc} \\
\hline
179 & 56459.8276 & 0.0004 & $-$0&0060 & 147 \\
180 & 56459.8857 & 0.0004 & $-$0&0063 & 147 \\
181 & 56459.9445 & 0.0005 & $-$0&0059 & 129 \\
183 & 56460.0628 & 0.0004 & $-$0&0043 & 81 \\
184 & 56460.1196 & 0.0003 & $-$0&0059 & 94 \\
185 & 56460.1869 & 0.0008 & 0&0030 & 421 \\
186 & 56460.2359 & 0.0005 & $-$0&0065 & 614 \\
187 & 56460.2919 & 0.0010 & $-$0&0088 & 370 \\
190 & 56460.4692 & 0.0002 & $-$0&0067 & 228 \\
191 & 56460.5278 & 0.0003 & $-$0&0064 & 208 \\
192 & 56460.5861 & 0.0003 & $-$0&0065 & 176 \\
193 & 56460.6474 & 0.0008 & $-$0&0036 & 108 \\
195 & 56460.7645 & 0.0005 & $-$0&0033 & 91 \\
196 & 56460.8177 & 0.0007 & $-$0&0085 & 232 \\
197 & 56460.8783 & 0.0005 & $-$0&0063 & 241 \\
198 & 56460.9369 & 0.0006 & $-$0&0061 & 79 \\
199 & 56460.9989 & 0.0006 & $-$0&0025 & 22 \\
205 & 56461.3464 & 0.0009 & $-$0&0053 & 70 \\
206 & 56461.4080 & 0.0023 & $-$0&0021 & 176 \\
207 & 56461.4647 & 0.0004 & $-$0&0038 & 207 \\
208 & 56461.5200 & 0.0005 & $-$0&0069 & 196 \\
211 & 56461.7135 & 0.0024 & 0&0114 & 72 \\
212 & 56461.7574 & 0.0006 & $-$0&0031 & 161 \\
213 & 56461.8138 & 0.0009 & $-$0&0050 & 179 \\
214 & 56461.8763 & 0.0013 & $-$0&0009 & 247 \\
215 & 56461.9288 & 0.0019 & $-$0&0068 & 146 \\
224 & 56462.4541 & 0.0014 & $-$0&0070 & 76 \\
225 & 56462.5174 & 0.0010 & $-$0&0021 & 92 \\
226 & 56462.5758 & 0.0011 & $-$0&0021 & 87 \\
227 & 56462.6359 & 0.0012 & $-$0&0004 & 120 \\
229 & 56462.7502 & 0.0007 & $-$0&0029 & 153 \\
230 & 56462.8088 & 0.0008 & $-$0&0026 & 195 \\
231 & 56462.8654 & 0.0004 & $-$0&0045 & 251 \\
\hline
  \multicolumn{6}{l}{\commenta BJD$-$2400000.} \\
  \multicolumn{6}{l}{\commentb Against max $= 2456449.3820 + 0.058389 E$.} \\
  \multicolumn{6}{l}{\commentc Number of points used to determine the maximum.} \\
\end{tabular}
\end{center}
\end{table}

\addtocounter{table}{-1}
\begin{table}
\caption{Superhump maxima of PNV J191501 (2013) (continued)}
\begin{center}
\begin{tabular}{rp{55pt}p{40pt}r@{.}lr}
\hline
\multicolumn{1}{c}{$E$} & \multicolumn{1}{c}{max\commenta} & \multicolumn{1}{c}{error} & \multicolumn{2}{c}{$O-C$\commentb} & \multicolumn{1}{c}{$N$\commentc} \\
\hline
232 & 56462.9244 & 0.0010 & $-$0&0038 & 89 \\
236 & 56463.1602 & 0.0005 & $-$0&0016 & 208 \\
237 & 56463.2184 & 0.0006 & $-$0&0018 & 210 \\
238 & 56463.2811 & 0.0012 & 0&0026 & 208 \\
239 & 56463.3324 & 0.0009 & $-$0&0045 & 257 \\
240 & 56463.3952 & 0.0008 & $-$0&0001 & 251 \\
241 & 56463.4557 & 0.0011 & 0&0020 & 84 \\
242 & 56463.5102 & 0.0007 & $-$0&0019 & 63 \\
243 & 56463.5670 & 0.0007 & $-$0&0035 & 61 \\
244 & 56463.6315 & 0.0020 & 0&0026 & 102 \\
246 & 56463.7500 & 0.0014 & 0&0043 & 64 \\
247 & 56463.8061 & 0.0018 & 0&0020 & 59 \\
248 & 56463.8603 & 0.0010 & $-$0&0021 & 112 \\
249 & 56463.9201 & 0.0018 & $-$0&0007 & 43 \\
253 & 56464.1579 & 0.0007 & 0&0035 & 206 \\
254 & 56464.2148 & 0.0008 & 0&0020 & 206 \\
255 & 56464.2735 & 0.0007 & 0&0024 & 204 \\
256 & 56464.3298 & 0.0006 & 0&0002 & 209 \\
257 & 56464.3869 & 0.0008 & $-$0&0011 & 248 \\
258 & 56464.4514 & 0.0010 & 0&0051 & 151 \\
259 & 56464.5034 & 0.0023 & $-$0&0013 & 131 \\
260 & 56464.5692 & 0.0012 & 0&0061 & 61 \\
261 & 56464.6256 & 0.0011 & 0&0041 & 98 \\
262 & 56464.6899 & 0.0040 & 0&0100 & 65 \\
263 & 56464.7471 & 0.0014 & 0&0089 & 104 \\
264 & 56464.8025 & 0.0028 & 0&0058 & 162 \\
265 & 56464.8564 & 0.0008 & 0&0014 & 191 \\
266 & 56464.9168 & 0.0015 & 0&0033 & 84 \\
267 & 56464.9753 & 0.0009 & 0&0035 & 76 \\
270 & 56465.1522 & 0.0011 & 0&0053 & 117 \\
271 & 56465.2116 & 0.0012 & 0&0062 & 115 \\
272 & 56465.2715 & 0.0008 & 0&0078 & 121 \\
273 & 56465.3290 & 0.0015 & 0&0069 & 226 \\
\hline
  \multicolumn{6}{l}{\commenta BJD$-$2400000.} \\
  \multicolumn{6}{l}{\commentb Against max $= 2456449.3820 + 0.058389 E$.} \\
  \multicolumn{6}{l}{\commentc Number of points used to determine the maximum.} \\
\end{tabular}
\end{center}
\end{table}

\addtocounter{table}{-1}
\begin{table}
\caption{Superhump maxima of PNV J191501 (2013) (continued)}
\begin{center}
\begin{tabular}{rp{55pt}p{40pt}r@{.}lr}
\hline
\multicolumn{1}{c}{$E$} & \multicolumn{1}{c}{max\commenta} & \multicolumn{1}{c}{error} & \multicolumn{2}{c}{$O-C$\commentb} & \multicolumn{1}{c}{$N$\commentc} \\
\hline
274 & 56465.3880 & 0.0016 & 0&0075 & 112 \\
275 & 56465.4417 & 0.0014 & 0&0028 & 112 \\
280 & 56465.7386 & 0.0033 & 0&0077 & 15 \\
281 & 56465.7884 & 0.0019 & $-$0&0008 & 15 \\
292 & 56466.4363 & 0.0010 & 0&0048 & 135 \\
293 & 56466.4991 & 0.0005 & 0&0092 & 156 \\
294 & 56466.5604 & 0.0006 & 0&0121 & 136 \\
295 & 56466.6151 & 0.0004 & 0&0085 & 118 \\
297 & 56466.7380 & 0.0028 & 0&0145 & 16 \\
308 & 56467.3717 & 0.0008 & 0&0060 & 111 \\
309 & 56467.4320 & 0.0012 & 0&0079 & 136 \\
310 & 56467.4883 & 0.0007 & 0&0058 & 88 \\
311 & 56467.5487 & 0.0009 & 0&0078 & 63 \\
313 & 56467.6581 & 0.0006 & 0&0004 & 34 \\
326 & 56468.4267 & 0.0007 & 0&0100 & 62 \\
327 & 56468.4734 & 0.0006 & $-$0&0017 & 124 \\
328 & 56468.5401 & 0.0010 & 0&0066 & 87 \\
329 & 56468.5915 & 0.0009 & $-$0&0004 & 60 \\
330 & 56468.6457 & 0.0008 & $-$0&0046 & 66 \\
331 & 56468.7059 & 0.0041 & $-$0&0028 & 15 \\
332 & 56468.7609 & 0.0020 & $-$0&0062 & 17 \\
342 & 56469.3528 & 0.0013 & 0&0019 & 79 \\
343 & 56469.4120 & 0.0008 & 0&0027 & 194 \\
344 & 56469.4674 & 0.0008 & $-$0&0003 & 224 \\
345 & 56469.5263 & 0.0005 & 0&0002 & 151 \\
346 & 56469.5906 & 0.0009 & 0&0061 & 151 \\
347 & 56469.6476 & 0.0015 & 0&0047 & 64 \\
348 & 56469.7037 & 0.0013 & 0&0024 & 96 \\
349 & 56469.7592 & 0.0007 & $-$0&0005 & 127 \\
350 & 56469.8181 & 0.0008 & 0&0000 & 172 \\
351 & 56469.8773 & 0.0010 & 0&0009 & 161 \\
352 & 56469.9307 & 0.0016 & $-$0&0041 & 64 \\
355 & 56470.1060 & 0.0009 & $-$0&0041 & 77 \\
\hline
  \multicolumn{6}{l}{\commenta BJD$-$2400000.} \\
  \multicolumn{6}{l}{\commentb Against max $= 2456449.3820 + 0.058389 E$.} \\
  \multicolumn{6}{l}{\commentc Number of points used to determine the maximum.} \\
\end{tabular}
\end{center}
\end{table}

\addtocounter{table}{-1}
\begin{table}
\caption{Superhump maxima of PNV J191501 (2013) (continued)}
\begin{center}
\begin{tabular}{rp{55pt}p{40pt}r@{.}lr}
\hline
\multicolumn{1}{c}{$E$} & \multicolumn{1}{c}{max\commenta} & \multicolumn{1}{c}{error} & \multicolumn{2}{c}{$O-C$\commentb} & \multicolumn{1}{c}{$N$\commentc} \\
\hline
361 & 56470.4542 & 0.0013 & $-$0&0062 & 63 \\
362 & 56470.5139 & 0.0015 & $-$0&0049 & 61 \\
363 & 56470.5683 & 0.0009 & $-$0&0088 & 59 \\
364 & 56470.6282 & 0.0035 & $-$0&0073 & 64 \\
365 & 56470.6935 & 0.0014 & $-$0&0004 & 93 \\
366 & 56470.7348 & 0.0012 & $-$0&0175 & 180 \\
\hline
  \multicolumn{6}{l}{\commenta BJD$-$2400000.} \\
  \multicolumn{6}{l}{\commentb Against max $= 2456449.3820 + 0.058389 E$.} \\
  \multicolumn{6}{l}{\commentc Number of points used to determine the maximum.} \\
\end{tabular}
\end{center}
\end{table}

\begin{table}
\caption{Superhump maxima of PNV J191501 (2013) (post-superoutburst)}\label{tab:j191501oc2013b}
\begin{center}
\begin{tabular}{rp{55pt}p{40pt}r@{.}lr}
\hline
\multicolumn{1}{c}{$E$} & \multicolumn{1}{c}{max\commenta} & \multicolumn{1}{c}{error} & \multicolumn{2}{c}{$O-C$\commentb} & \multicolumn{1}{c}{$N$\commentc} \\
\hline
0 & 56471.7401 & 0.0018 & $-$0&0068 & 185 \\
1 & 56471.8019 & 0.0026 & $-$0&0030 & 210 \\
7 & 56472.1209 & 0.0024 & $-$0&0322 & 61 \\
8 & 56472.1853 & 0.0086 & $-$0&0258 & 62 \\
9 & 56472.2412 & 0.0021 & $-$0&0280 & 62 \\
10 & 56472.2953 & 0.0011 & $-$0&0319 & 77 \\
16 & 56472.6877 & 0.0090 & 0&0124 & 23 \\
17 & 56472.7242 & 0.0029 & $-$0&0091 & 84 \\
18 & 56472.7908 & 0.0019 & $-$0&0005 & 224 \\
19 & 56472.8465 & 0.0006 & $-$0&0029 & 202 \\
24 & 56473.1319 & 0.0020 & $-$0&0076 & 59 \\
25 & 56473.1993 & 0.0004 & 0&0018 & 123 \\
26 & 56473.2578 & 0.0008 & 0&0023 & 119 \\
32 & 56473.6053 & 0.0069 & 0&0017 & 27 \\
33 & 56473.6628 & 0.0014 & 0&0012 & 50 \\
34 & 56473.7212 & 0.0013 & 0&0015 & 56 \\
35 & 56473.7793 & 0.0004 & 0&0016 & 125 \\
46 & 56474.4220 & 0.0005 & 0&0061 & 84 \\
47 & 56474.4768 & 0.0006 & 0&0028 & 104 \\
48 & 56474.5360 & 0.0004 & 0&0041 & 108 \\
49 & 56474.5906 & 0.0009 & 0&0006 & 116 \\
50 & 56474.6537 & 0.0022 & 0&0057 & 86 \\
51 & 56474.7087 & 0.0006 & 0&0027 & 104 \\
52 & 56474.7671 & 0.0004 & 0&0031 & 161 \\
54 & 56474.8831 & 0.0004 & 0&0030 & 184 \\
63 & 56475.4060 & 0.0006 & 0&0038 & 28 \\
64 & 56475.4634 & 0.0003 & 0&0031 & 77 \\
65 & 56475.5192 & 0.0004 & 0&0009 & 65 \\
67 & 56475.6335 & 0.0044 & $-$0&0009 & 20 \\
68 & 56475.6953 & 0.0015 & 0&0030 & 51 \\
69 & 56475.7551 & 0.0006 & 0&0047 & 103 \\
70 & 56475.8115 & 0.0004 & 0&0030 & 133 \\
71 & 56475.8702 & 0.0003 & 0&0037 & 138 \\
\hline
  \multicolumn{6}{l}{\commenta BJD$-$2400000.} \\
  \multicolumn{6}{l}{\commentb Against max $= 2456471.7470 + 0.058021 E$.} \\
  \multicolumn{6}{l}{\commentc Number of points used to determine the maximum.} \\
\end{tabular}
\end{center}
\end{table}

\addtocounter{table}{-1}
\begin{table}
\caption{Superhump maxima of PNV J191501 (2013) (post-superoutburst) (continued)}
\begin{center}
\begin{tabular}{rp{55pt}p{40pt}r@{.}lr}
\hline
\multicolumn{1}{c}{$E$} & \multicolumn{1}{c}{max\commenta} & \multicolumn{1}{c}{error} & \multicolumn{2}{c}{$O-C$\commentb} & \multicolumn{1}{c}{$N$\commentc} \\
\hline
80 & 56476.3910 & 0.0013 & 0&0024 & 20 \\
81 & 56476.4493 & 0.0004 & 0&0027 & 63 \\
82 & 56476.5068 & 0.0005 & 0&0022 & 54 \\
85 & 56476.6795 & 0.0015 & 0&0008 & 18 \\
86 & 56476.7401 & 0.0005 & 0&0034 & 73 \\
87 & 56476.7928 & 0.0008 & $-$0&0020 & 135 \\
88 & 56476.8527 & 0.0009 & $-$0&0000 & 128 \\
97 & 56477.3801 & 0.0007 & 0&0052 & 42 \\
98 & 56477.4338 & 0.0005 & 0&0008 & 63 \\
99 & 56477.4919 & 0.0014 & 0&0009 & 69 \\
100 & 56477.5498 & 0.0004 & 0&0008 & 45 \\
101 & 56477.6096 & 0.0006 & 0&0026 & 62 \\
102 & 56477.6669 & 0.0003 & 0&0018 & 95 \\
103 & 56477.7214 & 0.0009 & $-$0&0017 & 125 \\
104 & 56477.7859 & 0.0005 & 0&0048 & 97 \\
109 & 56478.0779 & 0.0013 & 0&0066 & 44 \\
110 & 56478.1357 & 0.0009 & 0&0064 & 60 \\
111 & 56478.1905 & 0.0006 & 0&0032 & 62 \\
112 & 56478.2476 & 0.0008 & 0&0023 & 63 \\
113 & 56478.3054 & 0.0007 & 0&0021 & 62 \\
115 & 56478.4241 & 0.0006 & 0&0048 & 102 \\
116 & 56478.4828 & 0.0008 & 0&0054 & 85 \\
117 & 56478.5384 & 0.0009 & 0&0030 & 83 \\
118 & 56478.5980 & 0.0007 & 0&0046 & 85 \\
119 & 56478.6540 & 0.0006 & 0&0025 & 44 \\
120 & 56478.7177 & 0.0006 & 0&0082 & 54 \\
121 & 56478.7671 & 0.0008 & $-$0&0004 & 91 \\
130 & 56479.3000 & 0.0016 & 0&0104 & 26 \\
131 & 56479.3469 & 0.0006 & $-$0&0007 & 39 \\
132 & 56479.4057 & 0.0005 & 0&0000 & 99 \\
133 & 56479.4658 & 0.0006 & 0&0021 & 97 \\
134 & 56479.5271 & 0.0018 & 0&0054 & 83 \\
135 & 56479.5821 & 0.0005 & 0&0023 & 85 \\
\hline
  \multicolumn{6}{l}{\commenta BJD$-$2400000.} \\
  \multicolumn{6}{l}{\commentb Against max $= 2456471.7470 + 0.058021 E$.} \\
  \multicolumn{6}{l}{\commentc Number of points used to determine the maximum.} \\
\end{tabular}
\end{center}
\end{table}

\addtocounter{table}{-1}
\begin{table}
\caption{Superhump maxima of PNV J191501 (2013) (post-superoutburst) (continued)}
\begin{center}
\begin{tabular}{rp{55pt}p{40pt}r@{.}lr}
\hline
\multicolumn{1}{c}{$E$} & \multicolumn{1}{c}{max\commenta} & \multicolumn{1}{c}{error} & \multicolumn{2}{c}{$O-C$\commentb} & \multicolumn{1}{c}{$N$\commentc} \\
\hline
136 & 56479.6407 & 0.0013 & 0&0029 & 74 \\
137 & 56479.6981 & 0.0021 & 0&0023 & 28 \\
138 & 56479.7548 & 0.0008 & 0&0010 & 68 \\
144 & 56480.1060 & 0.0010 & 0&0041 & 60 \\
145 & 56480.1632 & 0.0006 & 0&0032 & 61 \\
146 & 56480.2200 & 0.0005 & 0&0020 & 62 \\
147 & 56480.2746 & 0.0005 & $-$0&0014 & 60 \\
148 & 56480.3384 & 0.0024 & 0&0044 & 31 \\
149 & 56480.3968 & 0.0006 & 0&0047 & 50 \\
150 & 56480.4486 & 0.0012 & $-$0&0015 & 67 \\
151 & 56480.5103 & 0.0005 & 0&0022 & 71 \\
152 & 56480.5700 & 0.0006 & 0&0039 & 64 \\
153 & 56480.6232 & 0.0007 & $-$0&0009 & 63 \\
154 & 56480.6828 & 0.0008 & 0&0007 & 20 \\
155 & 56480.7413 & 0.0006 & 0&0011 & 79 \\
161 & 56481.0886 & 0.0007 & 0&0003 & 61 \\
162 & 56481.1461 & 0.0008 & $-$0&0002 & 63 \\
163 & 56481.2036 & 0.0009 & $-$0&0007 & 62 \\
164 & 56481.2606 & 0.0010 & $-$0&0017 & 60 \\
165 & 56481.3203 & 0.0007 & $-$0&0001 & 52 \\
166 & 56481.3779 & 0.0011 & $-$0&0005 & 23 \\
167 & 56481.4387 & 0.0006 & 0&0022 & 51 \\
168 & 56481.4971 & 0.0005 & 0&0027 & 69 \\
169 & 56481.5559 & 0.0008 & 0&0035 & 47 \\
170 & 56481.6079 & 0.0007 & $-$0&0026 & 43 \\
171 & 56481.6654 & 0.0020 & $-$0&0031 & 46 \\
172 & 56481.7281 & 0.0010 & 0&0016 & 58 \\
173 & 56481.7872 & 0.0015 & 0&0027 & 63 \\
174 & 56481.8456 & 0.0008 & 0&0030 & 65 \\
175 & 56481.9040 & 0.0011 & 0&0034 & 67 \\
179 & 56482.1338 & 0.0006 & 0&0011 & 77 \\
180 & 56482.1938 & 0.0006 & 0&0031 & 109 \\
181 & 56482.2501 & 0.0008 & 0&0014 & 125 \\
\hline
  \multicolumn{6}{l}{\commenta BJD$-$2400000.} \\
  \multicolumn{6}{l}{\commentb Against max $= 2456471.7470 + 0.058021 E$.} \\
  \multicolumn{6}{l}{\commentc Number of points used to determine the maximum.} \\
\end{tabular}
\end{center}
\end{table}

\addtocounter{table}{-1}
\begin{table}
\caption{Superhump maxima of PNV J191501 (2013) (post-superoutburst) (continued)}
\begin{center}
\begin{tabular}{rp{55pt}p{40pt}r@{.}lr}
\hline
\multicolumn{1}{c}{$E$} & \multicolumn{1}{c}{max\commenta} & \multicolumn{1}{c}{error} & \multicolumn{2}{c}{$O-C$\commentb} & \multicolumn{1}{c}{$N$\commentc} \\
\hline
183 & 56482.3686 & 0.0010 & 0&0038 & 69 \\
184 & 56482.4235 & 0.0007 & 0&0007 & 111 \\
185 & 56482.4814 & 0.0007 & 0&0006 & 108 \\
186 & 56482.5375 & 0.0006 & $-$0&0013 & 57 \\
187 & 56482.5969 & 0.0009 & 0&0001 & 38 \\
188 & 56482.6550 & 0.0011 & 0&0001 & 28 \\
191 & 56482.8280 & 0.0003 & $-$0&0009 & 100 \\
194 & 56483.0056 & 0.0016 & 0&0026 & 35 \\
195 & 56483.0595 & 0.0014 & $-$0&0015 & 27 \\
198 & 56483.2330 & 0.0007 & $-$0&0021 & 117 \\
199 & 56483.2923 & 0.0026 & $-$0&0008 & 66 \\
200 & 56483.3513 & 0.0009 & 0&0002 & 39 \\
201 & 56483.4062 & 0.0007 & $-$0&0029 & 108 \\
202 & 56483.4676 & 0.0004 & 0&0005 & 108 \\
203 & 56483.5236 & 0.0017 & $-$0&0016 & 61 \\
204 & 56483.5855 & 0.0006 & 0&0023 & 39 \\
205 & 56483.6442 & 0.0025 & 0&0030 & 25 \\
207 & 56483.7573 & 0.0006 & 0&0000 & 94 \\
215 & 56484.2213 & 0.0024 & $-$0&0001 & 111 \\
216 & 56484.2824 & 0.0046 & 0&0030 & 55 \\
218 & 56484.4016 & 0.0014 & 0&0061 & 21 \\
219 & 56484.4556 & 0.0015 & 0&0021 & 24 \\
220 & 56484.5078 & 0.0013 & $-$0&0037 & 24 \\
229 & 56485.0381 & 0.0019 & 0&0044 & 55 \\
230 & 56485.0901 & 0.0011 & $-$0&0016 & 56 \\
232 & 56485.2041 & 0.0027 & $-$0&0036 & 99 \\
234 & 56485.3237 & 0.0010 & $-$0&0001 & 38 \\
242 & 56485.7879 & 0.0005 & $-$0&0001 & 110 \\
243 & 56485.8401 & 0.0010 & $-$0&0059 & 108 \\
269 & 56487.3480 & 0.0019 & $-$0&0065 & 24 \\
270 & 56487.4103 & 0.0006 & $-$0&0022 & 40 \\
271 & 56487.4680 & 0.0017 & $-$0&0026 & 25 \\
\hline
  \multicolumn{6}{l}{\commenta BJD$-$2400000.} \\
  \multicolumn{6}{l}{\commentb Against max $= 2456471.7470 + 0.058021 E$.} \\
  \multicolumn{6}{l}{\commentc Number of points used to determine the maximum.} \\
\end{tabular}
\end{center}
\end{table}

\addtocounter{table}{-1}
\begin{table}
\caption{Superhump maxima of PNV J191501 (2013) (post-superoutburst) (continued)}
\begin{center}
\begin{tabular}{rp{55pt}p{40pt}r@{.}lr}
\hline
\multicolumn{1}{c}{$E$} & \multicolumn{1}{c}{max\commenta} & \multicolumn{1}{c}{error} & \multicolumn{2}{c}{$O-C$\commentb} & \multicolumn{1}{c}{$N$\commentc} \\
\hline
275 & 56487.6911 & 0.0012 & $-$0&0115 & 60 \\
276 & 56487.7617 & 0.0005 & 0&0010 & 110 \\
287 & 56488.3981 & 0.0010 & $-$0&0008 & 42 \\
288 & 56488.4516 & 0.0005 & $-$0&0053 & 41 \\
289 & 56488.5159 & 0.0006 & 0&0009 & 45 \\
290 & 56488.5721 & 0.0006 & $-$0&0008 & 45 \\
293 & 56488.7439 & 0.0006 & $-$0&0031 & 93 \\
294 & 56488.7993 & 0.0007 & $-$0&0057 & 88 \\
304 & 56489.3905 & 0.0017 & 0&0053 & 70 \\
305 & 56489.4405 & 0.0012 & $-$0&0027 & 70 \\
306 & 56489.4985 & 0.0006 & $-$0&0028 & 45 \\
307 & 56489.5569 & 0.0004 & $-$0&0024 & 46 \\
308 & 56489.6166 & 0.0007 & $-$0&0007 & 46 \\
322 & 56490.4293 & 0.0012 & $-$0&0003 & 44 \\
323 & 56490.4799 & 0.0018 & $-$0&0077 & 48 \\
324 & 56490.5430 & 0.0008 & $-$0&0026 & 46 \\
325 & 56490.6023 & 0.0007 & $-$0&0014 & 47 \\
326 & 56490.6620 & 0.0042 & 0&0003 & 30 \\
339 & 56491.4113 & 0.0008 & $-$0&0046 & 46 \\
340 & 56491.4717 & 0.0005 & $-$0&0023 & 45 \\
341 & 56491.5352 & 0.0009 & 0&0032 & 45 \\
343 & 56491.6406 & 0.0010 & $-$0&0075 & 43 \\
408 & 56495.4172 & 0.0025 & $-$0&0022 & 40 \\
\hline
  \multicolumn{6}{l}{\commenta BJD$-$2400000.} \\
  \multicolumn{6}{l}{\commentb Against max $= 2456471.7470 + 0.058021 E$.} \\
  \multicolumn{6}{l}{\commentc Number of points used to determine the maximum.} \\
\end{tabular}
\end{center}
\end{table}

\subsection{SSS J094327.3$-$272038}\label{obj:j094327}

   This object (=SSS111226:094327$-$272039,
hereafter SSS J094327) was discovered as
a transient object by CRTS Siding Spring Survey (SSS)
on 2011 December 26 at an unfiltered CCD magnitude of 16.6.
The object, however, showed several outbursts reaching
$V=12.8$ in ASAS-3 data (vsnet-alert 14013).
There is also an X-ray counterpart 1RXS J094326.1$-$272035.
The past outburst behavior suggested an SU UMa-type
dwarf nova.  The 2011 December outburst brightened to
14.7 mag (unfiltered CCD) and was observed for one night
which possibly showed superhumps (vsnet-alert 14034).

\begin{figure}
  \begin{center}
    \FigureFile(88mm,110mm){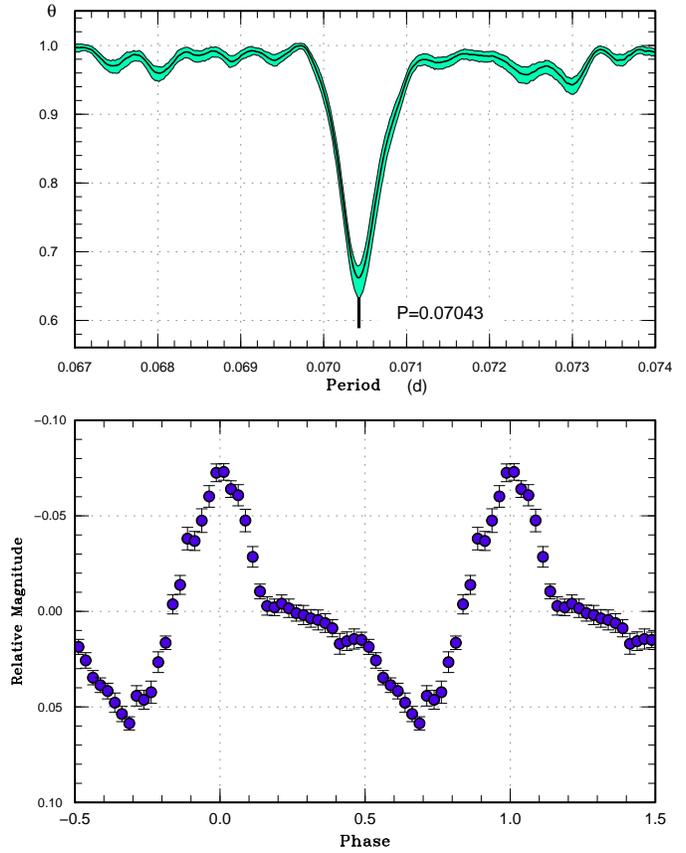}
  \end{center}
  \caption{Superhumps in SSS J094327 (2014).
     (Upper): PDM analysis.
     (Lower): Phase-averaged profile.}
  \label{fig:j0943shpdm}
\end{figure}

   The 2014 outburst was visually detected by A. Pearce on 
January 29.  The bright magnitude (13.1 mag) immediately
suggested a superoutburst (vsnet-alert 16842).  Subsequent 
observations indeed detected superhumps (vsnet-alert 16853,
16859, 16867; figure \ref{fig:j0943shpdm}).
The times of superhump maxima are liste in table
\ref{tab:j094327oc2014}.  The epochs for $E \le 2$ were
likely stage A superhumps.  There were likely stages B and C.
Despite that the object experienced a rapid fading from
the superoutburst, no phase jump (corresponding to
``traditional'' late superhumps) was observed.

   Although the period could not be determined,
the 2011 observations were compatible with a period
$\sim$0.07~d.

   In table \ref{tab:j094327out}, we summarized the past
outbursts from the ASAS-3 data.  Many of the detected
outbursts were superoutbursts and the shortest intervals
between superoutburst was $\sim$390~d.

\begin{table}
\caption{Outbursts of SSS J094327}\label{tab:j094327out}
\begin{center}
\begin{tabular}{cccc}
\hline
JD$-$2400000 & Maximum ($V$) & Duration & Type \\
\hline
51926 & l3.1 & 12 & Super \\
52440 & 12.8 & $>$3 & Super? \\
53005 & 13.8 & 1\commenta & Normal? \\
53385 & 12.8 & 11 & Super \\
53776 & 13.0 & $>$7 & Super \\
54182 & 13.1 & $>$8 & Super \\
54667 & 13.6 & 2 & Normal \\
54796 & 13.3 & 1\commenta & Normal? \\
\hline
  \multicolumn{4}{l}{\commenta Single detection.} \\
\\
\end{tabular}
\end{center}
\end{table}

\begin{table}
\caption{Superhump maxima of SSS J094327 (2014)}\label{tab:j094327oc2014}
\begin{center}
\begin{tabular}{rp{55pt}p{40pt}r@{.}lr}
\hline
\multicolumn{1}{c}{$E$} & \multicolumn{1}{c}{max\commenta} & \multicolumn{1}{c}{error} & \multicolumn{2}{c}{$O-C$\commentb} & \multicolumn{1}{c}{$N$\commentc} \\
\hline
0 & 56689.0954 & 0.0013 & $-$0&0073 & 123 \\
1 & 56689.1697 & 0.0010 & $-$0&0036 & 112 \\
14 & 56690.0889 & 0.0003 & 0&0000 & 140 \\
15 & 56690.1581 & 0.0003 & $-$0&0012 & 142 \\
16 & 56690.2310 & 0.0003 & 0&0012 & 142 \\
17 & 56690.3007 & 0.0003 & 0&0005 & 140 \\
18 & 56690.3716 & 0.0005 & 0&0009 & 81 \\
28 & 56691.0753 & 0.0004 & 0&0002 & 136 \\
29 & 56691.1462 & 0.0003 & 0&0006 & 141 \\
43 & 56692.1327 & 0.0003 & 0&0010 & 142 \\
44 & 56692.2032 & 0.0005 & 0&0010 & 142 \\
45 & 56692.2741 & 0.0004 & 0&0015 & 141 \\
46 & 56692.3434 & 0.0005 & 0&0004 & 138 \\
50 & 56692.6270 & 0.0009 & 0&0023 & 22 \\
51 & 56692.6971 & 0.0007 & 0&0019 & 24 \\
52 & 56692.7674 & 0.0009 & 0&0018 & 19 \\
53 & 56692.8385 & 0.0015 & 0&0024 & 22 \\
57 & 56693.1213 & 0.0006 & 0&0035 & 142 \\
58 & 56693.1913 & 0.0006 & 0&0030 & 143 \\
59 & 56693.2610 & 0.0010 & 0&0022 & 141 \\
60 & 56693.3325 & 0.0006 & 0&0034 & 132 \\
65 & 56693.6825 & 0.0009 & 0&0012 & 23 \\
66 & 56693.7536 & 0.0009 & 0&0018 & 19 \\
72 & 56694.1758 & 0.0031 & 0&0014 & 76 \\
78 & 56694.5904 & 0.0052 & $-$0&0067 & 11 \\
79 & 56694.6640 & 0.0011 & $-$0&0036 & 23 \\
80 & 56694.7388 & 0.0013 & 0&0008 & 20 \\
81 & 56694.8097 & 0.0016 & 0&0013 & 21 \\
82 & 56694.8773 & 0.0021 & $-$0&0016 & 16 \\
85 & 56695.0884 & 0.0012 & $-$0&0018 & 137 \\
86 & 56695.1567 & 0.0012 & $-$0&0039 & 141 \\
87 & 56695.2272 & 0.0011 & $-$0&0038 & 142 \\
88 & 56695.3007 & 0.0016 & $-$0&0008 & 78 \\
\hline
  \multicolumn{6}{l}{\commenta BJD$-$2400000.} \\
  \multicolumn{6}{l}{\commentb Against max $= 2456689.1028 + 0.070440 E$.} \\
  \multicolumn{6}{l}{\commentc Number of points used to determine the maximum.} \\
\end{tabular}
\end{center}
\end{table}

\subsection{TCP J23382254$-$2049518}\label{obj:j233822}

   This object (hereafter TCP J233822) was discovered as
a transient object by K. Itagaki on 2013 September 28
at an unfiltered CCD magnitude of 13.6.\footnote{
$<$http://www.cbat.eps.harvard.edu/unconf/\\followups/J23382254-2049518.html$>$.
}  The object is identical with a blue SDSS object
($g=21.5$, $g-r=-0.2$) and a GALEX ultraviolet source
GALEX J233822.5$-$204951 (vsnet-alert 16468).
The large outburst amplitude already suggested
the WZ Sge-type classification.

\begin{figure}
  \begin{center}
    \FigureFile(88mm,110mm){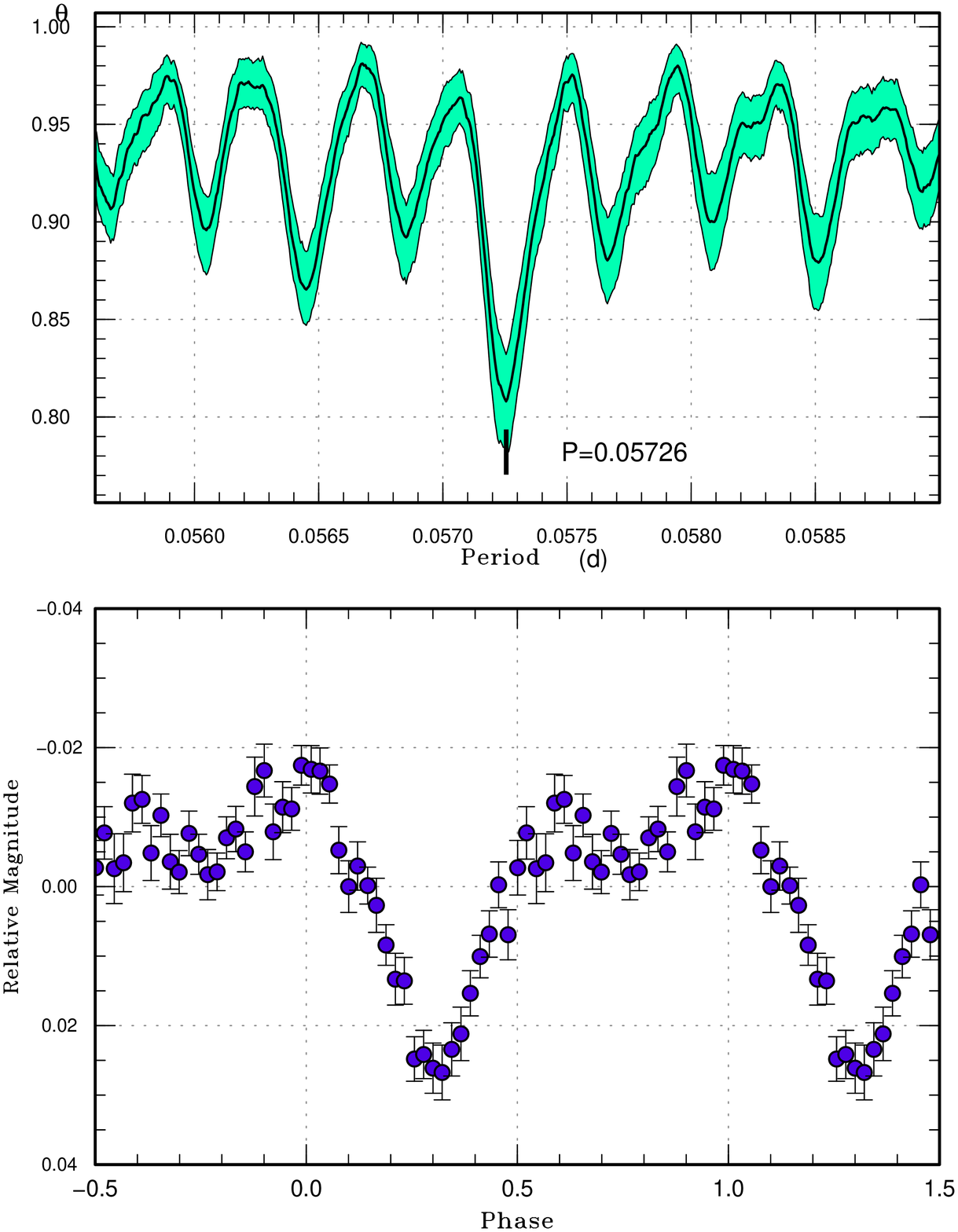}
  \end{center}
  \caption{Early superhumps in TCP J233822 (2013).
     (Upper): PDM analysis.
     (Lower): Phase-averaged profile.}
  \label{fig:j2338eshpdm}
\end{figure}

\begin{figure}
  \begin{center}
    \FigureFile(88mm,110mm){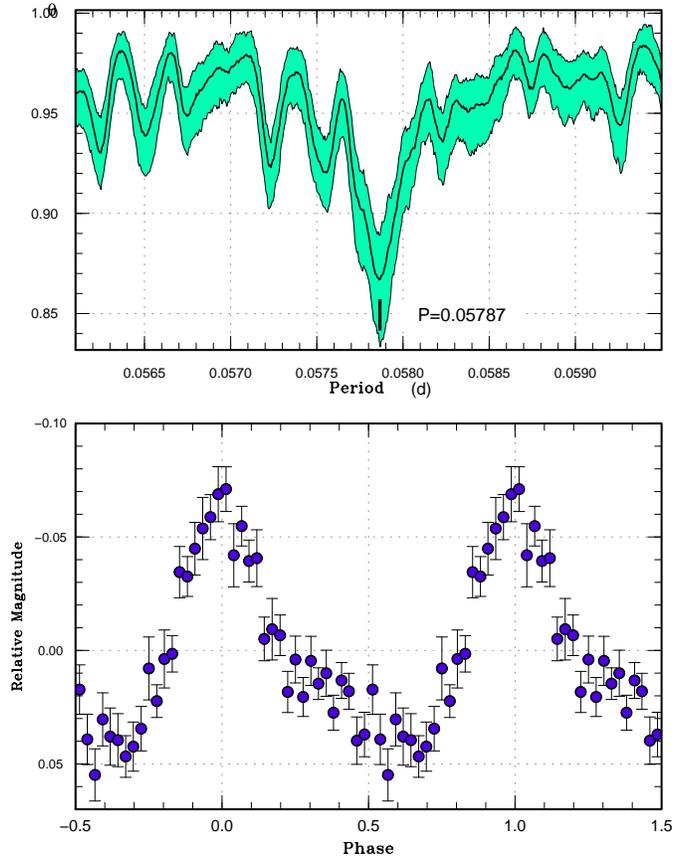}
  \end{center}
  \caption{Ordinary superhumps in TCP J233822 (2013).
     The interval of BJD 2456577.6--2456593 was used.
     (Upper): PDM analysis.
     (Lower): Phase-averaged profile.}
  \label{fig:j2338shpdm}
\end{figure}

\begin{figure}
  \begin{center}
    \FigureFile(88mm,70mm){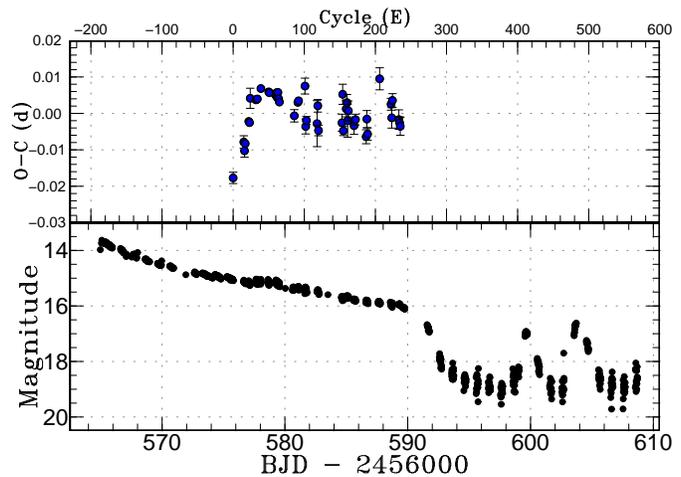}
  \end{center}
  \caption{$O-C$ diagram of superhumps in TCP J233822 (2013).
     (Upper): $O-C$ diagram.  A period of 0.057913~d
     was used to draw this figure.
     (Lower): Light curve.  The observations were binned to 0.011~d.}
  \label{fig:j2338humpall}
\end{figure}

   The object initially showed early superhumps
(vsnet-alert 16486, 16489, 16496, 16528; figure
\ref{fig:j2338eshpdm}).   Twelve days after
the discovery, the object started to show ordinary superhumps
(vsnet-alert 16520, 16529, 16532, 16536; figure
\ref{fig:j2338shpdm}).
The times of superhump maxima are listed in table
\ref{tab:j223822oc2013}, in which stage A-B transition is
clearly seen (also figure \ref{fig:j2338humpall}).
For the epochs $E \ge 222$, there was some
evidence of shortening of the period.  This part may correspond
to stage C superhumps.

   The fractional superhump excess (in frequency) for
stage A superhumps was $\varepsilon^*$=0.0231(14).
This value corresponds to $q$=0.061(4), suggesting that
the object is near the period minimum or somewhat
passed the period minimum.

   The object showed two post-superoutburst rebrightenings
(figure \ref{fig:j2338humpall}).  It is rare to see multiple
post-superoutburst rebrightenings in such a short-$P_{\rm orb}$
system (cf. \cite{nak13j2112j2037}).

\begin{table}
\caption{Superhump maxima of TCP J233822 (2013)}\label{tab:j223822oc2013}
\begin{center}
\begin{tabular}{rp{55pt}p{40pt}r@{.}lr}
\hline
\multicolumn{1}{c}{$E$} & \multicolumn{1}{c}{max\commenta} & \multicolumn{1}{c}{error} & \multicolumn{2}{c}{$O-C$\commentb} & \multicolumn{1}{c}{$N$\commentc} \\
\hline
0 & 56575.7668 & 0.0016 & $-$0&0178 & 28 \\
15 & 56576.6453 & 0.0017 & $-$0&0079 & 14 \\
16 & 56576.7008 & 0.0017 & $-$0&0103 & 15 \\
17 & 56576.7607 & 0.0010 & $-$0&0083 & 28 \\
22 & 56577.0564 & 0.0007 & $-$0&0023 & 27 \\
23 & 56577.1140 & 0.0009 & $-$0&0026 & 37 \\
24 & 56577.1786 & 0.0028 & 0&0041 & 15 \\
32 & 56577.6414 & 0.0006 & 0&0037 & 13 \\
33 & 56577.6995 & 0.0007 & 0&0038 & 14 \\
34 & 56577.7575 & 0.0005 & 0&0039 & 27 \\
39 & 56578.0499 & 0.0008 & 0&0068 & 20 \\
50 & 56578.6861 & 0.0008 & 0&0058 & 18 \\
51 & 56578.7437 & 0.0006 & 0&0056 & 25 \\
60 & 56579.2642 & 0.0004 & 0&0048 & 124 \\
61 & 56579.3229 & 0.0006 & 0&0056 & 123 \\
62 & 56579.3795 & 0.0004 & 0&0043 & 133 \\
63 & 56579.4388 & 0.0007 & 0&0057 & 134 \\
64 & 56579.4949 & 0.0008 & 0&0039 & 133 \\
65 & 56579.5520 & 0.0007 & 0&0030 & 133 \\
86 & 56580.7644 & 0.0017 & $-$0&0007 & 30 \\
91 & 56581.0576 & 0.0007 & 0&0030 & 11 \\
92 & 56581.1159 & 0.0010 & 0&0034 & 15 \\
101 & 56581.6412 & 0.0022 & 0&0074 & 13 \\
102 & 56581.6880 & 0.0021 & $-$0&0037 & 14 \\
103 & 56581.7476 & 0.0011 & $-$0&0020 & 28 \\
118 & 56582.6154 & 0.0063 & $-$0&0029 & 12 \\
119 & 56582.6782 & 0.0017 & 0&0020 & 13 \\
120 & 56582.7293 & 0.0014 & $-$0&0048 & 21 \\
153 & 56584.6426 & 0.0024 & $-$0&0027 & 19 \\
154 & 56584.7083 & 0.0028 & 0&0051 & 29 \\
155 & 56584.7562 & 0.0013 & $-$0&0049 & 72 \\
159 & 56584.9939 & 0.0022 & 0&0012 & 12 \\
\hline
  \multicolumn{6}{l}{\commenta BJD$-$2400000.} \\
  \multicolumn{6}{l}{\commentb Against max $= 2456575.7845 + 0.057913 E$.} \\
  \multicolumn{6}{l}{\commentc Number of points used to determine the maximum.} \\
\end{tabular}
\end{center}
\end{table}

\addtocounter{table}{-1}
\begin{table}
\caption{Superhump maxima of TCP J233822 (2013) (continued)}
\begin{center}
\begin{tabular}{rp{55pt}p{40pt}r@{.}lr}
\hline
\multicolumn{1}{c}{$E$} & \multicolumn{1}{c}{max\commenta} & \multicolumn{1}{c}{error} & \multicolumn{2}{c}{$O-C$\commentb} & \multicolumn{1}{c}{$N$\commentc} \\
\hline
160 & 56585.0535 & 0.0022 & 0&0029 & 15 \\
161 & 56585.1065 & 0.0045 & $-$0&0021 & 11 \\
162 & 56585.1671 & 0.0027 & 0&0006 & 15 \\
170 & 56585.6263 & 0.0024 & $-$0&0035 & 15 \\
172 & 56585.7438 & 0.0015 & $-$0&0018 & 31 \\
187 & 56586.6078 & 0.0019 & $-$0&0065 & 14 \\
188 & 56586.6706 & 0.0023 & $-$0&0017 & 13 \\
189 & 56586.7244 & 0.0017 & $-$0&0058 & 22 \\
206 & 56587.7241 & 0.0030 & 0&0094 & 23 \\
222 & 56588.6437 & 0.0016 & 0&0024 & 13 \\
223 & 56588.6979 & 0.0028 & $-$0&0013 & 17 \\
224 & 56588.7606 & 0.0019 & 0&0034 & 19 \\
233 & 56589.2765 & 0.0027 & $-$0&0018 & 133 \\
234 & 56589.3336 & 0.0016 & $-$0&0027 & 134 \\
235 & 56589.3905 & 0.0024 & $-$0&0037 & 132 \\
\hline
  \multicolumn{6}{l}{\commenta BJD$-$2400000.} \\
  \multicolumn{6}{l}{\commentb Against max $= 2456575.7845 + 0.057913 E$.} \\
  \multicolumn{6}{l}{\commentc Number of points used to determine the maximum.} \\
\end{tabular}
\end{center}
\end{table}

\section{Discussion}\label{sec:discuss}

   We first report in this section general statistical properties
of the sample together with the earlier sample as in \citet{Pdot5}.
We then deals with new topics which arose in this paper.

\subsection{Period Derivatives during Stage B}\label{sec:stagebpdot}

\begin{figure*}
  \begin{center}
    \FigureFile(160mm,110mm){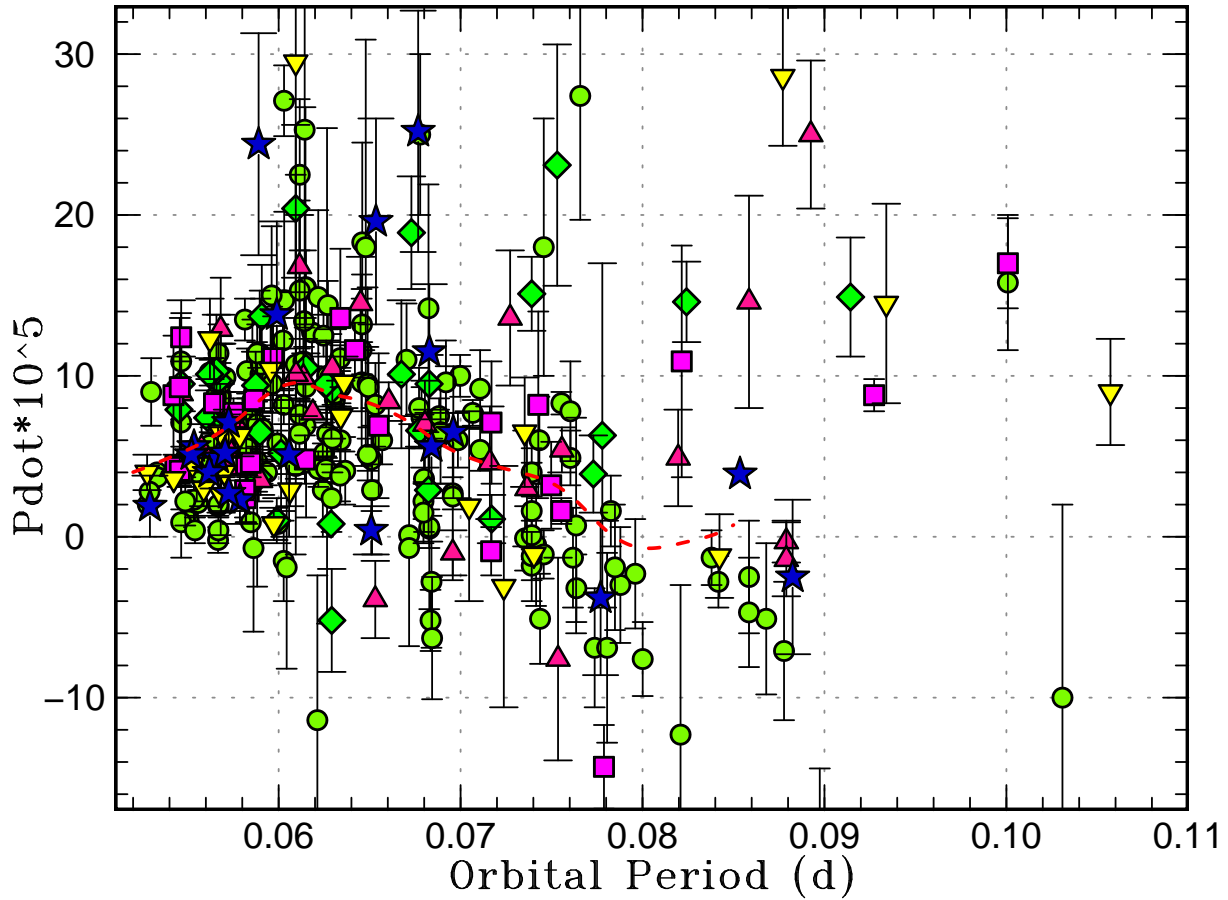}
  \end{center}
  \caption{$P_{\rm dot}$ for stage B versus $P_{\rm orb}$.
  Filled circles, filled diamonds, filled triangles, filled squares,
  filled lower-pointed triangles
  and filled stars represent samples in 
  \citet{Pdot}, \citet{Pdot2}, \citet{Pdot3},
  \citet{Pdot4}, \citet{Pdot5}  and this paper, respectively.
  The curve represents the spline-smoothed global trend.
  }
  \label{fig:pdotporb6}
\end{figure*}

   Figure \ref{fig:pdotporb6} represents updated relation
between $P_{\rm dot}$ for stage B versus $P_{\rm orb}$.
Most of the objects with $P_{\rm orb} < 0.085$~d followed
the general trend reported in \citet{Pdot4}.
Some objects with $P_{\rm orb}$ longer than 0.085~d
show negative $P_{\rm dot}$ while some of them show
large positive $P_{\rm dot}$ (as we will see in
subsection \ref{sec:stagealongp}, some of these objects
may have been contaminated by stage A superhumps).
We could add a new sample MASTER J162323 to the latter group.
We can say that most of objects $P_{\rm orb} < 0.080$~d have 
positive $P_{\rm dot}$ for stage B.

\subsection{Mass Ratios from Stage A Superhumps}\label{sec:stagea}

\begin{figure*}
  \begin{center}
    \FigureFile(160mm,110mm){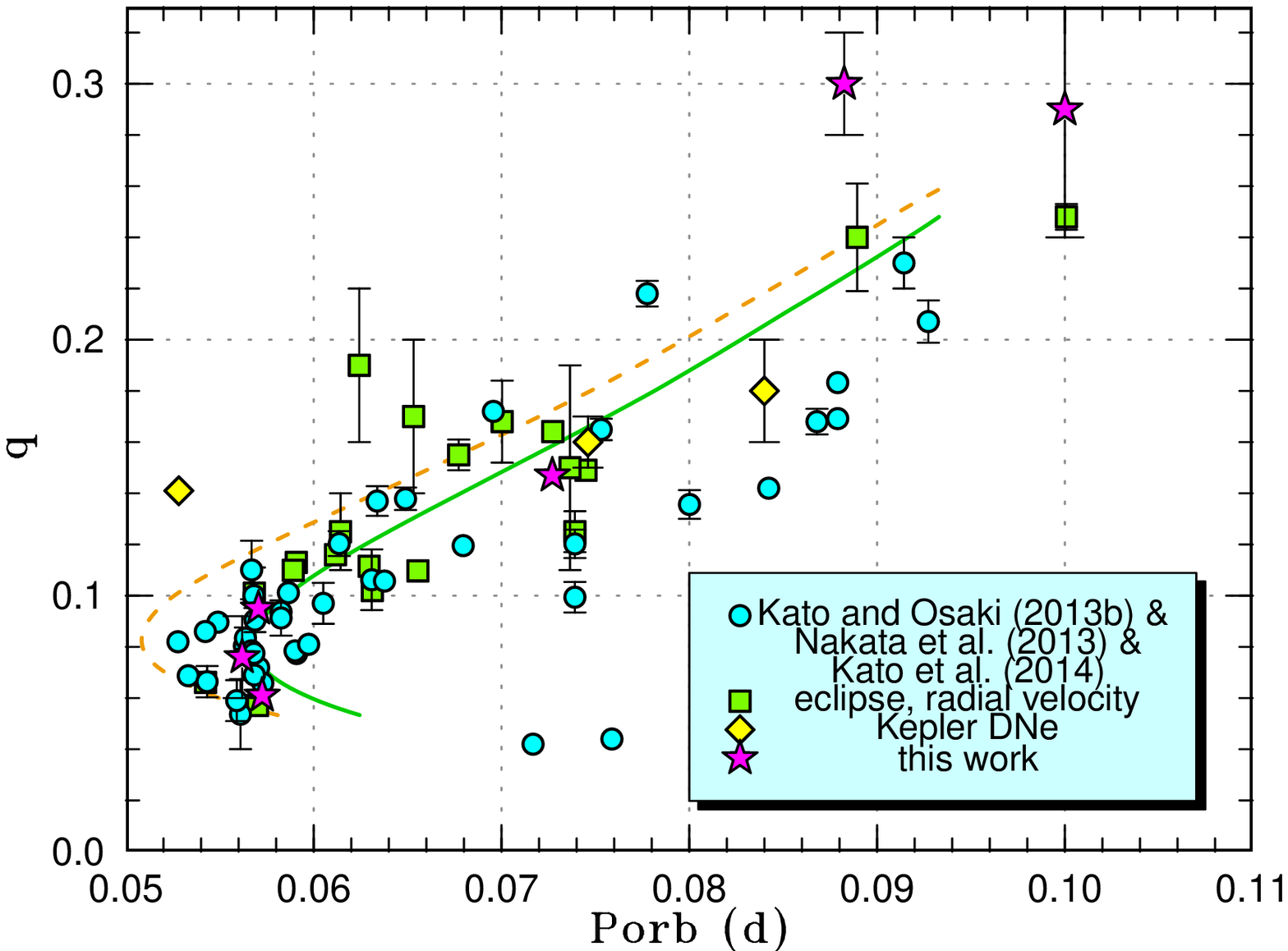}
  \end{center}
  \caption{Mass ratio versus orbital period.
  The dashed and solid curves represent the standard and optimal
  evolutionary tracks in \citet{kni11CVdonor}, respectively.
  The filled circles, filled squares, filled stars, filled diamonds
  represent $q$ values from a combination of the estimates
  from stage A superhumps published in three preceding
  sources (\cite{kat13qfromstageA}; \cite{nak13j2112j2037};
  \cite{Pdot5}), known $q$ values from quiescent eclipses or 
  radial-velocity study (see \cite{kat13qfromstageA} for
  the data source), $q$ estimated in this work and dwarf novae
  in the Kepler data (see text for the complete reference),
  respectively.}
  \label{fig:qall3}
\end{figure*}

   It has been proposed that the binary mass ratios can be
estimated from the stage A superhumps, which are considered
to reflect the dynamical precession rate at the radius
of the 3:1 resonance \citep{kat13qfromstageA}.
Stage A superhumps recorded in the present study are listed
in table \ref{tab:pera}.
In table \ref{tab:newqstageA}, we list the new estimates
of mass ratios from in this paper.
A updated summary of $q$ estimates is shown in
figure \ref{fig:qall3}, in which measurements in
\citet{nak13j2112j2037} are also included.
The Kepler DNe shown in this figures are
V516 Lyr \citep{kat13j1939v585lyrv516lyr},
KIC 7524178 \citep{kat13j1922} and the unusual
short-$P_{\rm orb}$ object GALEX J194419.33$+$491257.0
in the field of KIC 11412044 \citep{kat14j1944}
(located at $P_{\rm orb}$=0.05282~d, $q$=0.14).

\begin{table}
\caption{New estimates for the binary mass ratio from stage A superhumps}\label{tab:newqstageA}
\begin{center}
\begin{tabular}{ccc}
\hline
Object         & $\varepsilon^*$ (stage A) & $q$ from stage A \\
\hline
GZ Cnc         & 0.089(7) & 0.30(2) \\
MN Dra         & 0.078, 0.092 & 0.29(5) \\
DT Oct         & 0.050(2) & 0.147(7) \\
MASTER J005740 & 0.028(5) & 0.076(16) \\
PNV J191501    & 0.0344(12) & 0.095(4) \\
TCP J233822    & 0.0231(14) & 0.061(4) \\
\hline
\end{tabular}
\end{center}
\end{table}

\begin{table}
\caption{Superhump Periods during Stage A}\label{tab:pera}
\begin{center}
\begin{tabular}{cccc}
\hline
Object & Year & period (d) & err \\
\hline
UZ Boo & 2013 & 0.06210 & 0.00005 \\
GZ Cnc & 2014 & 0.09690 & 0.00030 \\
AL Com & 2013 & 0.05859 & 0.00009 \\
MN Dra & 2012 & 0.10993 & 0.00009 \\
MN Dra & 2013 & 0.10822 & 0.00013 \\
DT Oct & 2014 & 0.07271 & -- \\
ASASSN-13ck & 2013 & 0.05700 & 0.00010 \\
ASASSN-14ac & 2014 & 0.05952 & 0.00003 \\
MASTER J004527 & 2013 & 0.08210 & 0.00035 \\
MASTER J005740 & 2013 & 0.05783 & 0.00034 \\
MASTER J162323 & 2013 & 0.09134 & 0.00055 \\
PNV J191501 & 2013 & 0.05909 & 0.00008 \\
TCP J233822 & 2013 & 0.05861 & 0.00008 \\
\hline
\end{tabular}
\end{center}
\end{table}

\subsection{WZ Sge-Type Stars}\label{sec:wzsgestat}

\begin{figure*}
  \begin{center}
    \FigureFile(140mm,100mm){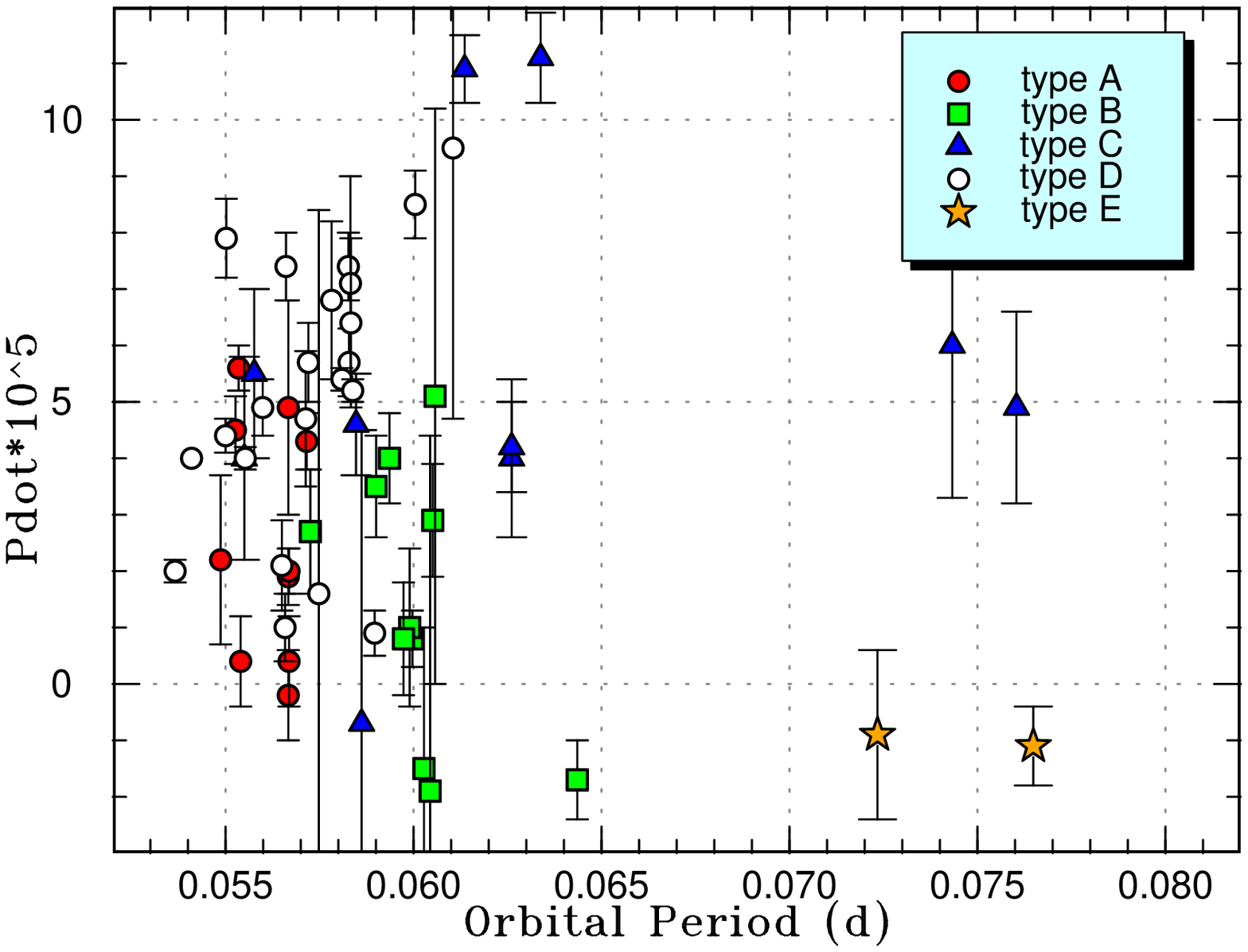}
  \end{center}
  \caption{$P_{\rm dot}$ versus $P_{\rm orb}$ for WZ Sge-type
  dwarf novae.  Symbols represent the type (cf. \cite{Pdot}) of outburst:
  type-A (filled circles), type-B (filled squares),
  type-C (filled triangles), type-D (open circles)
  and type-E (filled stars) which show double
  superoutburst as shown in \citet{kat13j1222}.
  }
  \label{fig:wzpdottype6}
\end{figure*}

   The WZ Sge-type dwarf novae in this study are listed in
table \ref{tab:wztab}.  In figure \ref{fig:wzpdottype6}, we showed
the relation between $P_{\rm dot}$ and $P_{\rm orb}$ for
the entire set of WZ Sge-type dwarf novae.
This figure is an updated version of figure 86 of \citet{Pdot5}.
We here use the types of superoutburst in terms of
rebrightenings as introduced in
\citet{ima06tss0222} and \citet{Pdot}: type-A outburst
(long-duration rebrightening), type-B outburst
(multiple discrete rebrightenings), type-C outburst
(single rebrightening) and type-D outburst (no rebrightening)
(see e.g. figure 35 in \cite{Pdot}).  The type-E outburst
(superoutburst with early superhumps and
another superoutburst with ordinary superhumps) has been
introduced since \citet{Pdot5}.
The new data have confirmed the trend in that each
subtype of the rebrightening pattern clusters in this
diagram.  Objects with type B (multiple) rebrightenings
have a high concentration around $P_{\rm orb}$=0.060~d. 
\citet{nak13j2112j2037} indicated that at least two objects
with type B rebrightening are not period bouncers but
lie on the ordinary track of CV evolution before the
period minimum.  If this interpretation applies to
the majority of the objects with type B rebrightening,
these objects may be in a stage of CV evolution between
short-period SU UMa-type dwarf novae and extreme WZ Sge-type
dwarf novae near the period minimum.

\begin{table*}
\caption{Parameters of WZ Sge-type superoutbursts.}\label{tab:wztab}
\begin{center}
\begin{tabular}{cccccccccccc}
\hline
Object & Year & $P_{\rm SH}$ & $P_{\rm orb}$ & $P_{\rm dot}$\commenta & err\commenta & $\epsilon$ & Type\commentb & $N_{\rm reb}$\commentc & delay\commentd & Max & Min \\
\hline
UZ Boo & 2013 & 0.062066 & -- & 5.1 & 5.1 & -- & B & 4 & 3 & 12.5 & 19.7 \\
AL Com & 2013 & 0.057323 & 0.056669 & 4.9 & 1.9 & 0.012 & A & 1(2) & 7 & 12.7 & 19.8 \\
ASASSN-13ck & 2013 & 0.056186 & 0.055348 & 5.6 & 0.4 & 0.015 & A & 1(3) & $\geq$8 & ]12.9 & 20.8 \\
ASASSN-14ac & 2014 & 0.058550 & -- & $-$1.7 & 1.2 & -- & -- & $\geq$1 & $\geq$14 & ]14.5 & 21.6 \\
MASTER J005740 & 2013 & 0.057067 & 0.056190 & 4.0 & 1.0 & 0.016 & -- & -- & $\geq$6 & ]15.4 & 20.9 \\
OT J210016 & 2013 & 0.058502 & 0.05787 & 2.3 & 1.5 & 0.011 & -- & -- & -- & ]14.2 & 19.9 \\
PNV J191501 & 2013 & 0.058382 & 0.05706 & 5.2 & 0.2 & 0.023 & D & 0 & $\geq$7 & ]9.8 & 18.5 \\
TCP J233822 & 2013 & 0.057868 & 0.057255 & 2.7 & 1.1 & 0.011 & B & 2 & $\geq$12 & ]13.6 & 21.5 \\
\hline
  \multicolumn{12}{l}{\commenta Unit $10^{-5}$.} \\
  \multicolumn{12}{l}{\commentb A: long-lasting rebrightening; B: multiple rebegitehnings; C: single rebrightening; D: no rebrightening.} \\
  \multicolumn{12}{l}{\commentc Number of rebrightenings.} \\
  \multicolumn{12}{l}{\commentd Days before ordinary superhumps appeared.} \\
\end{tabular}
\end{center}
\end{table*}

\subsection{Comparison of Periods of Early Superhumps and Orbital Periods}
\label{sec:earlysh}

   Although early superhumps have been documented
(e.g. \cite{kat96alcom}; \cite{pat96alcom}; \cite{kat01hvvir};
\cite{ish02wzsgeletter}; \cite{pat02wzsge}; \cite{kat02wzsgeESH})
as double-wave modulations having periods very close to
the orbital period, there has been no summary of comparison
between the period of early superhumps and the orbital period.
We therefore make a study using the WZ Sge-type dwarf novae
with well-established orbital period.

   The periods of early superhumps of AL Com (2001),
HV Vir (2008) were estimated from the data in
\citet{ish02wzsgeletter}, \citet{Pdot}, respectively.
The orbital period of BW Scl has been refined using
quiescent observations (MLF, HaC, SPE and AAVSO data)
between 2004 September and 2013 September, 14070 measurements).
The orbital period of EZ Lyn is a revised one using
post-superoutburst eclipses reported in \citet{Pdot3}.
The orbital period of AL Com is an updated one
using observations of early superhumps in three
superoutbursts (subsection \ref{obj:alcom}).

   The result is summarized in table \ref{tab:earlysh}.
Although all the objects showed statistically significant negative
$\varepsilon$ for early superhumps, the period deficit
is very small (an order of 0.05\%).  This result has confirmed
the finding in \citet{ish02wzsgeletter}.
Since the period deficit is very small, the periods of
early superhumps can be considered as the orbital periods
to an accuracy of 0.1\%.  If more accuracy is needed and
the orbital period is not known, we propose to estimate the orbital
period by assuming $\varepsilon$ of $-0.05$\%.

\begin{table*}
\caption{Periods of Early Superhumps}\label{tab:earlysh}
\begin{center}
\begin{tabular}{lr@{.}lr@{.}lr@{.}ll}
\hline
Object & \multicolumn{2}{c}{$P_{\rm orb}$ (d)} & \multicolumn{2}{c}{$P_{\rm ESH}$ (d)\commenta} & \multicolumn{2}{c}{$\varepsilon_{\rm ESH}$\commentb} & References \\
\hline
V455 And (2007) & 0&05630921(1) & 0&0562675(18) & $-$0&00074(3) & \citet{Pdot} \\
AL Com (1995)   & 0&056668589(9) & 0&05666(2)    & $-$0&0002(4)  & this work; \citet{kat96alcom} \\
AL Com (2001)   & 0&056668589(9) & 0&056660(4)   & $-$0&00015(7) & this work \\
AL Com (2013)   & 0&056668589(9) & 0&056660(8)   & $-$0&00015(14) & this work \\
EZ Lyn (2010)   & 0&05900495(3) & 0&058973(6)   & $-$0&00054(10) & \citet{Pdot3} \\
BW Scl (2011)   & 0&05432391(1) & 0&054308(2)   & $-$0&00029(4) & this work; \citet{Pdot4} \\
WZ Sge (2001)   & 0&0566878460(3) & 0&056656(2) & $-$0&00057(4) & \citet{pat02wzsge}; \citet{ish02wzsgeletter}. \\
HV Vir (1992)   & 0&057069(6)   & 0&05698(8)    & $-$0&0016(14) & \citet{pat03suumas}; \citet{kat01hvvir} \\
HV Vir (2008)   & 0&057069(6)   & 0&056991(7)   & $-$0&0014(1)  & this work \\
MASTER J005740 (2013) & 0&0561904(3) & 0&056169(3) & $-$0&0038(5) & this work \\
\hline
  \multicolumn{8}{l}{\commenta Period of early superhumps.} \\
  \multicolumn{8}{l}{\commentb Fractional excess of early superhumps.} \\
\end{tabular}
\end{center}
\end{table*}

\subsection{Eclipses during the Phase of Early Superhumps}
\label{sec:earlyshecl}

   In the light curve of early superhumps in MASTER J005740
(figure \ref{fig:j0057eshpdm}), there are sharp structures
(kinks in the light curve) around orbital phases $-$0.15 and 0.15.
They suggest that the phases
between $-$0.15 and 0.15 were affected by the eclipse.
We can propose a hypothetical uneclipsed hump structure
(assuming that larger and smaller humps are separated
equally) as the dashed line in the figure \ref{fig:j0057eshpdm}.

   It has been a mystery why eclipses are not evident during
the stage of early superhumps in such a high inclination
system such as WZ Sge \citep{pat02wzsge}, despite that eclipses
appear more strongly after the appearance of ordinary
superhumps.  While \citet{pat02wzsge} suggested the enhanced
hot spot as the origin of eclipses during the phase of
ordinary superhumps, \citet{osa03DNoutburst} suggested that
what is eclipsed is the superhump light source rather than
the enhanced hot spot.  In the interpretation
of \citet{osa03DNoutburst}, the source of early superhumps,
which \citet{osa02wzsgehump} interpret as the two-armed
dissipation pattern of the 2:1 Lindblad resonance cannot be
eclipsed because this pattern is located azimuthally far away
from the secondary star.  The present observation indicates
that the broad eclipse was located close the expected
eclipse center, which suggest that (the axisymmetric component
of) the bright disk, rather than the enhanced hot spot,
is eclipsed.  This picture smoothly fits the interpretation
by \citet{osa03DNoutburst}, and the mystery of the apparent
absence of the eclipse during the phase of early superhumps
is solved.

   In \citet{uem12ESHrecon}, the eclipse of the disk
was considered in the model.  This effect was, however,
not so large in the model parameters of V455 And, which
apparently has a lower inclination than in MASTER J005740,
and its effect was difficult to distinguish from
the uneclipsed light curve of early superhumps.
Our new observation in a system of higher inclination
now presents more convincing evidence against the greatly
increased mass-transfer in the WZ Sge-type outburst.

\subsection{Model of the Eclipse during Early Superhumps}
\label{sec:earlysheclmodel}

   We further studied whether a simple model can reproduce
the depth of the eclipse in MASTER J005740 during the
phase of early superhumps.  We adopted $q$=0.076 from our
measurement using stage A superhumps (it is well-known
that $q$ and inclination are in strong relation in modeling
the eclipse light curve, and the uncertainty in $q$ can be
reasonably neglected by allowing a free selection of
the inclination).  We assumed that the disk radius is
the radius of the 2:1 resonance (0.615$A$ in the case of
$q$=0.076).  We assumed flat and axisymmetric geometry and
a standard disk having a surface luminosity with 
a radial dependence $\propto r^{-3/4}$
(assuming that we observed the Rayleigh-Jeans
tail of the emission from the hot disk).
Although all of these assumptions are rough, they
will not seriously affect the results.  We constructed
the eclipse light curve using the Roche geometry.
We could reproduce the eclipse depth of 0.16 mag from
the assumed secondary maximum of early superhumps and 
the bottom of the eclipse (as seen from 
figure \ref{fig:j0057eshpdm})
by assuming the inclination of \timeform{81.5D}$\pm$
\timeform{0.5D}.  The total duration of the eclipse
was 0.28 binary phase, which is in agreement with
the observation.  Under these parameters, the white dwarf
is marginally eclipsed, which is also in agreement
with the observational suggestion.

\begin{figure}
  \begin{center}
    \FigureFile(88mm,70mm){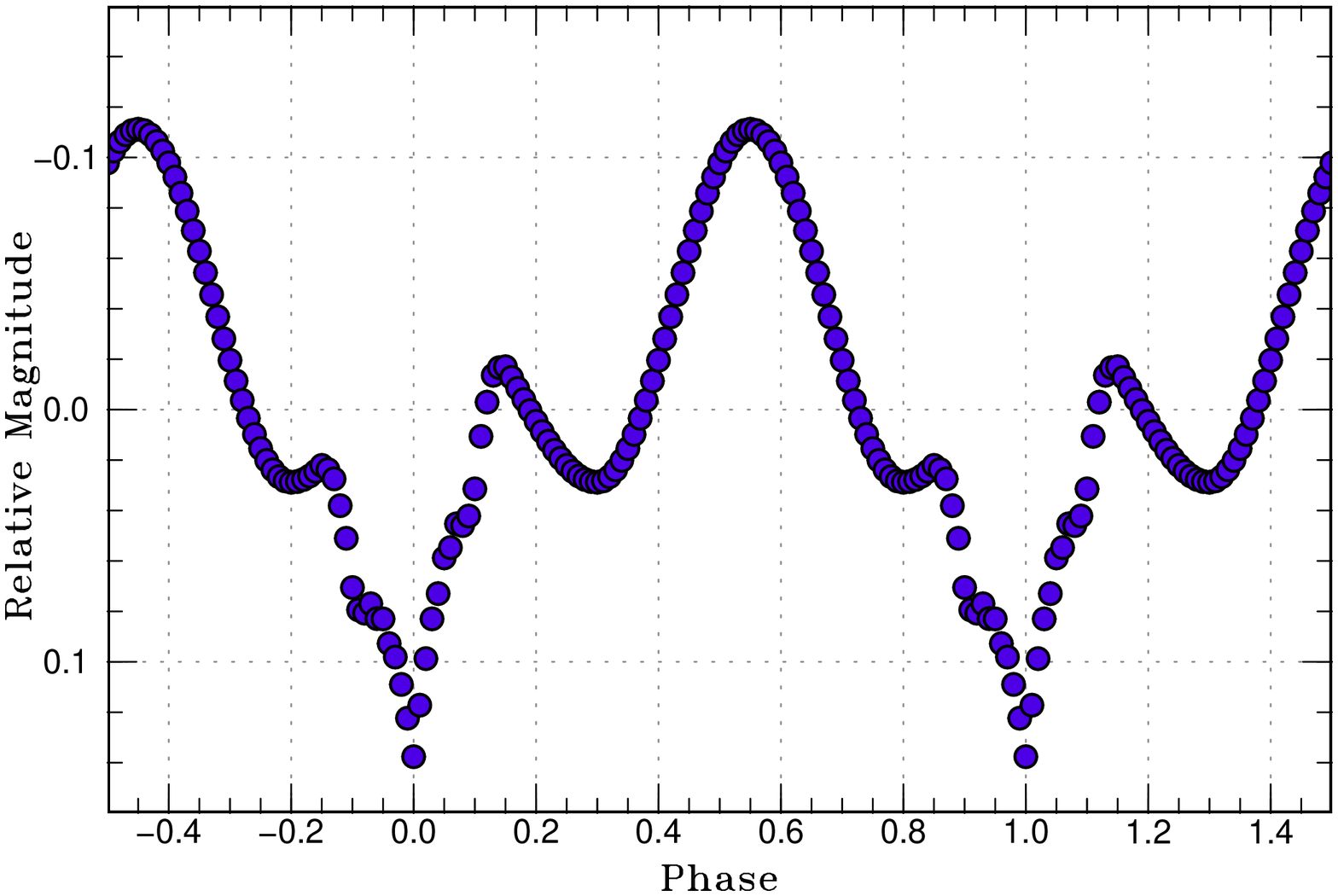}
  \end{center}
  \caption{Model light curve of the early superhump and eclipse
  of MASTER J005740.}
  \label{fig:j0057esheclmodel}
\end{figure}

   The model light curve is shown in table
\ref{fig:j0057esheclmodel}.  This model assumes that the
standard disk is eclipsed by the secondary and he early
superhumps is simply added to the light curve (i.e. we assume
that the light source of the early superhumps is not eclipsed).
The two maxima of early superhumps are approximated by sine
curves having different amplitudes (the secondary maximum
is assumed to have the half amplitude of the primary
maximum).  The resultant light curve appears to
well reproduce the basic characteristics of the observation
(figure \ref{fig:j0057eshpdm}).  The sharp minimum
around phase 0.0 reflects the eclipse of the central
part of the disk; the less clear appearance of this feature
in observation may have been the result of 
the self-obscuration of the central part of the disk.

\subsection{Stage A Superhumps in Long-Period Systems}\label{sec:stagealongp}

   In subsections \ref{obj:mndra} and \ref{obj:gzcnc},
probable stage A superhumps were detected in 
the long-$P_{\rm orb}$ systems MN Dra and GZ Cnc.
The identification of stage A superhumps in the former
system is almost certain because growing amplitudes
of the superhumps were detected.

   This finding appears to contradict the earlier
interpretation that stage A superhumps reflecting
the radius of the 3:1 resonance in high-$q$ systems are
difficult to detect because the tidal effect
is stronger in higher-$q$ systems and the eccentric region
spreads more quickly than in low-$q$ systems
\citep{kat13qfromstageA}.  In MN Dra and GZ Cnc, however,
the $q$ values are probably critically close to
the condition in which the 3:1 resonance occurs,
and the resonance may be weak enough to be confined
to the radius of the 3:1 resonance for longer time
than in ordinary SU UMa-type dwarf novae.
This possibility needs to be studied further.
Although some of objects recorded in our past study with 
long $P_{\rm orb}$ and large negative $P_{\rm dot}$ may
have been the similar ones, we could not find as convincing
case as MN Dra.  Since almost all of these systems lack
determination of the orbital period, future measurements
of the orbital periods may provide a clue to interpret
this phenomenon.

\subsection{Negative Superhumps in VW Hydri}\label{sec:vwhyinegsh}

   In recent series of papers (\cite{osa13v1504cygKepler};
\cite{osa13v344lyrv1504cyg}; \cite{osa14v1504cygv344lyrpaper3}),
it has been demonstrated using the Kepler data that the state
with negative superhumps tend to suppress normal outbursts in
V1504 Cyg and V344 Lyr.  The same phenomenon is also reported
in ER UMa (\cite{ohs12eruma}; \cite{zem13eruma}; \cite{ohs14eruma}).
This phenomenon has been interpreted as a result of
the decreased mass supply to the outermost part of
the accretion disk when the disk is tilted, thereby
reducing the occurrence of thermal instability in the outer
part of the disk (\cite{osa13v1504cygKepler}; \cite{ohs14eruma}).
\citet{osa13v1504cygKepler} called the supercycle
type S (short intervals between normal outbursts)
type L (long intervals between normal outbursts),
the nomenclature originally introduced by \citet{sma85vwhyi}
for VW Hyi.  Although the result in V1504 Cyg and V344 Lyr
suggests that the same mechanism is responsible for
type S and L supercycle in VW Hyi, this has not yet been
demonstrated by observation.

   We conducted an intensive campaign on VW Hyi in
2012--2013 to test this possibility.  Since the start
of the campaign in 2012 October 29, no outburst was
detected (for a duration of 25~d) until the next
superoutburst starting on November 23 (the superoutburst
reported in subsection \ref{obj:vwhyi}).  Although
the observations were not as dense as our CCD campaign,
visual observations by the AAVSO observers did not detect
an outburst since the last recorded outburst on September 9.
If there was no outburst between them, the interval
of normal outbursts may be as long as $\sim$70~d.
Since VW Hyi undergoes normal outbursts as frequently
as every 11--23~d in type S supercycles
(cf. \cite{sma85vwhyi}), in which intervals shorter
than 23~d were considered as type S and those
longer than $\sim$30~d were type L), the state before
the 2012 superoutburst was most likely type L.

   A PDM analysis of the observation BJD 2456229--2456254
(quiescence before the 2012 superoutburst) after subtracting 
the mean orbital variation yielded a period of 0.07265(3)~d,
2.2\% shorter than the orbital period.  Another candidate
period is 0.07829(3)~d, which is a one-day alias of
the 0.07265~d period.  Although there remains a possibility
that 0.07829~d is the true period and the 0.07265~d period
is a one-day alias signal, we consider this possibility
less likely because the similar signal was also observed
after the superoutburst (figure \ref{fig:vwhyinegshpdm2}).
In this interval, the negative superhump was the strongest signal
after subtracting the orbital modulations, and there is
no remaining ambiguity of a one-day alias.
The periods of negative superhumps determined for
different segments of the data after subtraction
of the orbital signal are listed in table \ref{tab:vwhyinegsh}.
During the superoutburst, the period of negative superhumps
could not be determined because it was located close to
the one-day alias of the (positive) superhump period.

\begin{figure}
  \begin{center}
    \FigureFile(88mm,110mm){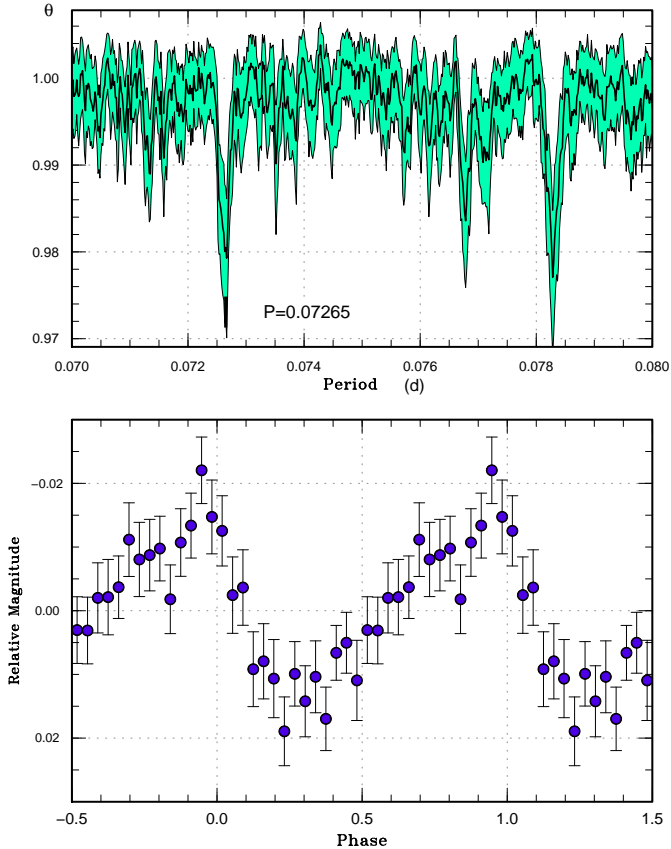}
  \end{center}
  \caption{Negative superhumps in VW Hyi (2012).
     The observation BJD 2456229--2456584 (quiescence before
     the 2012 superoutburst) was analyzed after subtracting
     the mean orbital variation.
     (Upper): PDM analysis.
     (Lower): Phase-averaged profile.}
  \label{fig:vwhyinegshpdm}
\end{figure}

\begin{figure}
  \begin{center}
    \FigureFile(88mm,110mm){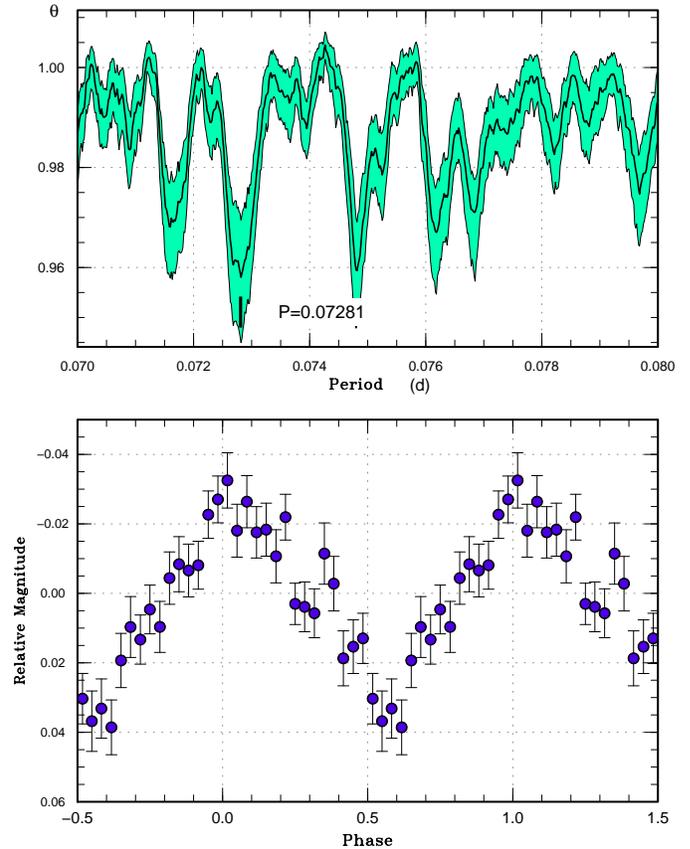}
  \end{center}
  \caption{Negative superhumps in VW Hyi (2012).
     The observation BJD 2456282--2456302 (quiescence between
     the first two normal outbursts after the superoutburst)
     was analyzed after subtracting
     the mean orbital variation.
     (Upper): PDM analysis.
     (Lower): Phase-averaged profile.}
  \label{fig:vwhyinegshpdm2}
\end{figure}

   The mean profile (figure \ref{fig:vwhyinegshpdm}) is also
characteristic of negative superhumps 
(cf. \cite{osa13v1504cygKepler}; \cite{osa13v344lyrv1504cyg})
with a slower rise to the maximum and a faster decline
to the minimum.  The co-existence of orbital humps
and negative superhumps, which is also recorded both in
V1504 Cyg and V344 Lyr, suggests that some of the stream
hits the outermost part of the disk to produce the hot spot
while some part of the stream reaches the inner disk
to produce negative superhumps (cf. \cite{woo07negSH}).

\begin{figure}
  \begin{center}
    \FigureFile(88mm,95mm){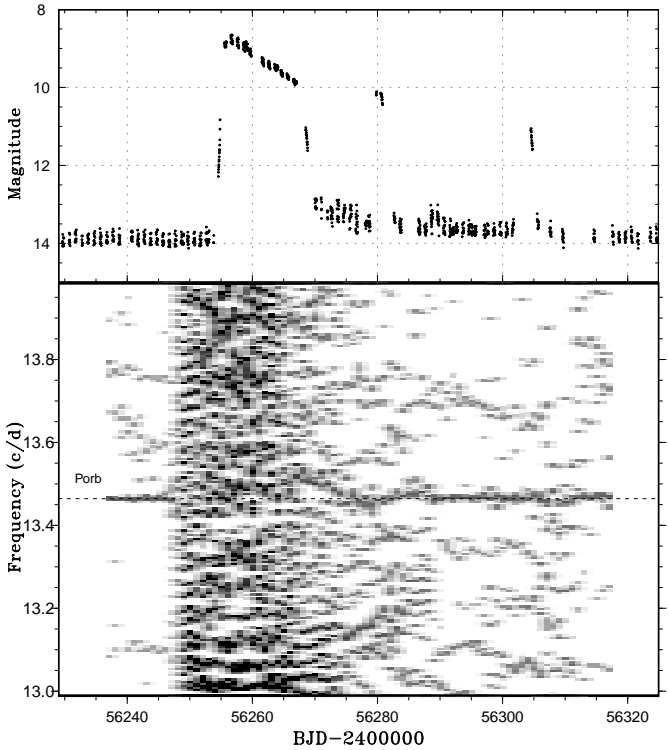}
  \end{center}
  \caption{Two-dimensional Lasso period analysis of VW Hyi
  (2012).
  (Upper:) Light curve.  The data were binned to 0.02~d.
  (Lower:) Lasso period analysis of the superhump and 
  the orbital signal.  The signal of negative superhumps with
  a frequency around 13.75 cycle d$^{-1}$ before the superoutburst
  was recorded.  A possible signal of negative superhumps
  with frequencies around 13.65--13.70 cycle d$^{-1}$ was also
  detected after the superoutburst.
  $\log \lambda=-7.8$ was used.
  The width of the sliding window and the time step used are
  15~d and 1~d, respectively.
  }
  \label{fig:vwhyinegshlasso}
\end{figure}

   A two-dimensional Lasso analysis (figure \ref{fig:vwhyinegshlasso})
shows a signal of negative superhumps with a frequency around
13.75 cycle d$^{-1}$ before the superoutburst.
A possible signal of negative superhumps with frequencies around
13.65--13.70 cycle d$^{-1}$ was also detected after
the superoutburst.  The decrease in the frequency 
(also evident as an increase in period in 
table \ref{tab:vwhyinegsh}) is compatible of the shrinkage of
the accretion disk after the superoutburst
As reported in \citet{osa13v1504cygKepler}, the precession
frequency of a tilted disk can be expressed
as follows \citep{lar98XBprecession}:
\begin{equation}
\nu_{\rm NSH}/\nu_{\rm orb} = 1+\left(\frac{3}{7} \frac{q}{\sqrt{1+q}} 
\cos \theta \right)
\left(\frac{R_{\rm d}}{A}\right)^{3/2},
\label{equ:diskradius}
\end{equation}
where $\nu_{\rm NSH}$ and $\nu_{\rm orb}$ are the frequency for 
the negative superhump and binary orbital frequency, respectively, 
$R_{\rm d}$ the disk radius, $A$ is the binary separation, 
$\theta$ is the tilt angle of the disk to the binary orbital plane. 
The smallest $|\varepsilon^*|$ (equivalent to 
$\nu_{\rm NSH}/\nu_{\rm orb}$) before the superoutburst
was 0.024 and the largest $|\varepsilon^*|$ after the superoutburst
was 0.020.  This difference can be explained by a 13 percent
shrinkage of the disk radius after the superoutburst.
Assuming $q$=0.21(2) (from radial-velocity study by
\cite{smi06vwhyi}) and a small tilt angle
($\cos \theta \sim 1$), these two values of $|\varepsilon^*|$
correspond to the disk radii of 0.42(1) $A$ and 0.39(1) $A$,
respectively.\footnote{
  \citet{smi06vwhyi} suggested that this $q$ value is highly
  insecure.  We used this value since there is no other
  direct determination of the $q$ value, and the $q$ determined
  from stage B superhumps \citep{pat05SH} has unknown
  uncertainty.
}

\begin{figure}
  \begin{center}
    \FigureFile(88mm,95mm){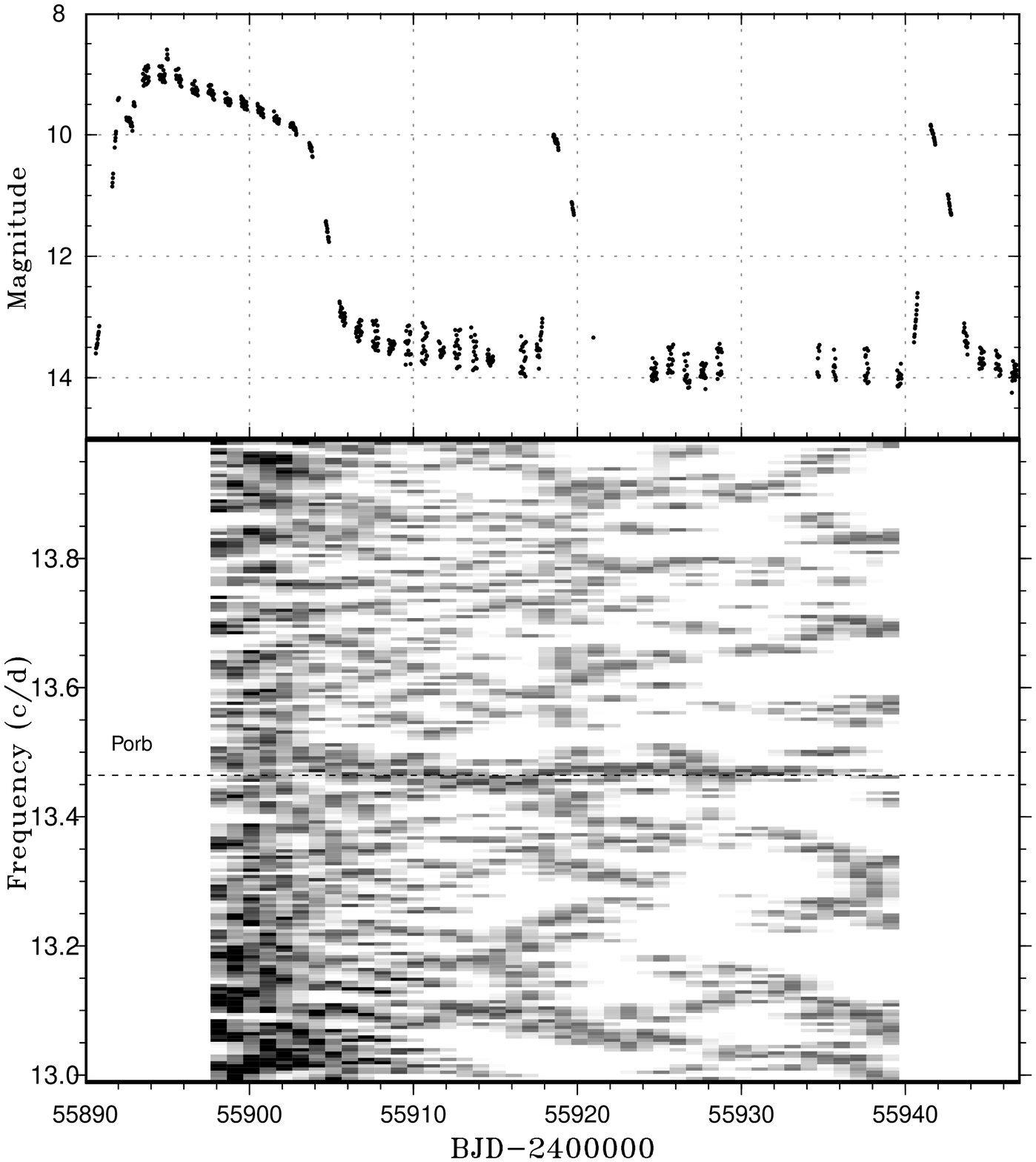}
  \end{center}
  \caption{Two-dimensional Lasso period analysis of VW Hyi
  (2011).  See figure \ref{fig:vwhyinegshlasso} for the
  explanation.  No strong signal of negative superhumps
  was detected.
  }
  \label{fig:vwhyi2011lasso}
\end{figure}

   For comparison, a two-dimensional Lasso analysis of
the 2011 data is also shown in figure \ref{fig:vwhyi2011lasso}.
No clear signal of negative superhumps was detected,
although the analysis noisy due to the poorer coverage
than in the 2012 observation and the persistence of
positive superhumps which interferes the detection of
other signals when the data are sparse.

   We conclude that type L supercycle in VW Hyi (reduced
number of normal outbursts) is associated with the presence
of negative superhumps as in V1504 Cyg and V344 Lyr, and
the tilt in the disk can be regarded as a cause of
the varying frequency of normal outbursts.

\begin{table}
\caption{Period of negative superhumps in VW Hyi (2012)}\label{tab:vwhyinegsh}
\begin{center}
\begin{tabular}{cccc}
\hline
JD$-$2400000 & Period & Error & Amplitude \\
             & (d)    & (d)   & (mag) \\
\hline
56229--56239 & 0.07252 & 0.00003 & 0.05 \\
56240--56254 & 0.07266 & 0.00002 & 0.04 \\
56269--56279 & 0.07261 & 0.00003 & 0.10 \\
56285--56302 & 0.07279 & 0.00003 & 0.07 \\
56305--56327 & 0.07275 & 0.00002 & 0.07 \\
56339--56361 & --      & --      & --   \\
56365--56374 & 0.07271 & 0.00005 & 0.05 \\
\hline
\\
\end{tabular}
\end{center}
\end{table}

\section{Summary}\label{sec:summary}

   In addition to basic data of superhumps of the objects
observed in this paper, the major findings we obtained
can be summarized as follows.

\begin{itemize}

\item We report the detection of negative superhumps
in quiescence of VW Hyi in 2012.
We conclude that type L supercycle in VW Hyi (reduced
number of normal outbursts) is associated with the presence
of negative superhumps as in V1504 Cyg and V344 Lyr.

\item MASTER J005740 is the first eclipsing WZ Sge-type
dwarf nova showing the probable eclipse of the white dwarf.
The sharp structure in the profile of the early superhumps
is interpreted as the eclipse of the accretion disk,
which has been difficult to distinguish from the profile
of the early superhump itself in other WZ Sge-type
dwarf novae.  The symmetric profile of the eclipse
indicates that the disk itself, not the enhanced hot spot,
is eclipsed.  This finding gives observational support
to \citet{osa03DNoutburst} who interpreted that the source
of the early superhumps is not the hot spot as suggested
by an enhanced mass-transfer model \citep{pat81wzsge}.
We also provided a model calculation of the eclipse light
curve of this object during the phase of early superhumps.

\item We detected stage A superhumps with growing
amplitudes in MN Dra and likely stage A (from the $O-C$ diagram)
in GZ Cnc, both of which have long orbital periods.
The stage A superhumps in these systems lasted longer
than expected.  We interpreted that the 3:1 resonance
was confined in the region of excitement because these
objects have mass-ratios critically close to the
condition in which the tidal instability occurs.
This may provide an interpretation of large negative
period derivatives recorded in the past in systems
with long orbital periods.

\item The 2013 superoutburst of UZ Boo was followed by
four post-superoutburst rebrightening as was observed
in the 2003 superoutburst.  This observation suggests
that the pattern of rebrightening tends to be reproducible
in the same object.

\item The WZ Sge-type dwarf novae AL Com and ASASSN-13ck 
showed a long-lasting (plateau-type) rebrightening.
In the early phase of the rebrightening, both objects showed
a precursor-like outburst, suggesting that the long-lasting
rebrightening is triggered by a precursor outburst.
Both objects showed small dip(s) during the rebrightening.

\item We have reviewed the observation of early superhumps
of WZ Sge-type dwarf novae and found that the fractional
superhump excesses for early superhumps have a typical
value of $-0.05$\%.

\item We have succeeded in detecting a positive period
derivative of superhumps in the helium CV CP Eri.
This object also showed oscillation-type rebrightenings.

\item We have established the long-sought superoutburst
of the eclipsing dwarf nova V893 Sco after fifteen years
since the rediscovery.

\end{itemize}

\medskip

This work was supported by the Grant-in-Aid
``Initiative for High-Dimensional Data-Driven Science through Deepening
of Sparse Modeling'' from the Ministry of Education, Culture, Sports, 
Science and Technology (MEXT) of Japan.
The authors are grateful to observers of VSNET Collaboration and
VSOLJ observers who supplied vital data.
We acknowledge with thanks the variable star
observations from the AAVSO International Database contributed by
observers worldwide and used in this research.
This work is deeply indebted to outburst detections and announcement
by a number of variable star observers worldwide, including participants of
CVNET and BAA VSS alert.
We thank Dr. Brian Skiff for making historical materials
about WX Hyi available to us.
The CCD operation of the Bronberg Observatory is partly sponsored by
the Center for Backyard Astrophysics.
The CCD operation by Peter Nelson is on loan from the AAVSO,
funded by the Curry Foundation.
We are grateful to the Catalina Real-time Transient Survey
team for making their real-time
detection of transient objects available to the public.

\end{document}